\documentclass[draft=true,universe,article,accept,moreauthors,pdftex]{Definitions/mdpi}

\firstpage{1}
\pubvolume{xx}
\issuenum{1}
\articlenumber{5}
\pubyear{2020}
\copyrightyear{2020}
\history{Received: date; Accepted: date; Published: date}

\usepackage{natbib,twoopt}
\usepackage{graphicx,units,multirow,subfigure,color,amsmath,amssymb,epsfig,makecell}
\usepackage{courier}
\usepackage{import}
\usepackage[varg]{txfonts}
\usepackage[absolute,overlay]{textpos}

\definecolor{light-gray}{gray}{0.95}
\definecolor{dark-gray}{gray}{0.4}

\newcommand{\ascii}{{\tt ASCII}}
\newcommand{\ajax}{{\tt AJAX}}
\newcommand{\crdb}{{\tt CRDB}}
\newcommand{\crdbv}[1]{{\tt CRDB~\!v{#1}}}
\newcommand{\cron}{{\tt cron}}
\newcommand{\galprop}{{\tt GALPROP}}
\newcommand{\jquery}[1]{{\tt jquery{#1}}}
\newcommand{\mysql}{{\tt MySQL}}
\newcommand{\php}{{\tt PHP}}
\newcommand{\rest}{{\tt REST}}
\newcommand{\rootcern}{{\tt ROOT}}
\newcommand{\tsorter}{{\tt table-sorter}}
\newcommand{\usine}{{\tt USINE}}
\newcommand{\xmax}{X_\text{max}}
\newcommand{\nmu}{N_\mu}
\newcommand{\code}[1]{\texttt{#1}}

\graphicspath{figures/} 

\hypersetup{breaklinks=true,%
            colorlinks=true,%
            pdfauthor={Maurin et al.},%
            pdftitle={CRDB v4.0}%
           }

\Title{Cosmic-ray database update: ultra-high energy, ultra-heavy, and antinuclei cosmic-ray data (\crdbv{4.0})}

\Author{D. Maurin$^{1}$\orcidA{}, H. Dembinski$^{2}$\orcidB{}, J. Gonzalez$^{3}$, Ioana C. Mari\c{s}$^{4}$\orcidC{}, F. Melot$^{1}$}

\AuthorNames{D. Maurin, H. Dembinski, J. Gonzalez, I. C. Maris, F. Melot}

\address{%
$^{1}$ \quad LPSC, Universit\'e Grenoble Alpes, CNRS/IN2P3, 53 avenue des Martyrs, 38026 Grenoble, France; david.maurin@lpsc.in2p3.fr\\
$^{2}$ \quad Dortmund University, Experimental Physics 5,
Otto-Hahn-Strasse 4a, 44227 Dortmund, Germany; hans.dembinski@tu-dortmund.de\\
$^{3}$ \quad Harvard-Smithsonian Center for Astrophysics, 60 Garden Street, Cambridge, MA 02138, USA\\
$^{4}$ \quad Université Libre de Bruxelles, Département de Physique, Boulevard du Triomphe, 2 CP 230, 1050 Bruxelles, Belgium}
\corres{Correspondence: david.maurin@lpsc.in2p3.fr (D.M.), hans.dembinski@tu-dortmund.de (H.D.)}

\abstract{We present an update on \crdb{} (\url{https://lpsc.in2p3.fr/crdb}), the cosmic-ray database for charged species. \crdb{} is based on \mysql{}, queried and sorted by \jquery{} and \tsorter{} libraries, and displayed via \php{} web pages through the \ajax{} protocol. We review the modifications made on the structure and outputs of the database since the first release (Maurin et al., 2014). For this update, the most important feature is the inclusion of ultra-heavy nuclei ($Z>30$), ultra-high energy nuclei (from $10^{15}$ to $10^{20}$~eV), and limits on antinuclei fluxes ($Z\leq -1$ for $A>1$); more than 100 experiments, 350 publications, and 40~000 data points are now available in \crdb{}. We also revisited and simplified how users can retrieve data and submit new ones. For questions and requests, please contact \href{mailto:crdb@lpsc.in2p3.fr}{crdb@lpsc.in2p3.fr}.}


\begin{document}

\begin{textblock*}{5cm}(15.5cm,3.2cm) 
\def\svgwidth{2.25cm}
\begingroup%
  \makeatletter%
  \providecommand\color[2][]{%
    \errmessage{(Inkscape) Color is used for the text in Inkscape, but the package 'color.sty' is not loaded}%
    \renewcommand\color[2][]{}%
  }%
  \providecommand\transparent[1]{%
    \errmessage{(Inkscape) Transparency is used (non-zero) for the text in Inkscape, but the package 'transparent.sty' is not loaded}%
    \renewcommand\transparent[1]{}%
  }%
  \providecommand\rotatebox[2]{#2}%
  \newcommand*\fsize{\dimexpr\f@size pt\relax}%
  \newcommand*\lineheight[1]{\fontsize{\fsize}{#1\fsize}\selectfont}%
  \ifx\svgwidth\undefined%
    \setlength{\unitlength}{275.06789361bp}%
    \ifx\svgscale\undefined%
      \relax%
    \else%
      \setlength{\unitlength}{\unitlength * \real{\svgscale}}%
    \fi%
  \else%
    \setlength{\unitlength}{\svgwidth}%
  \fi%
  \global\let\svgwidth\undefined%
  \global\let\svgscale\undefined%
  \makeatother%
  \begin{picture}(1,0.28075081)%
    \lineheight{1}%
    \setlength\tabcolsep{0pt}%
    \put(0,0){\includegraphics[width=\unitlength,page=1]{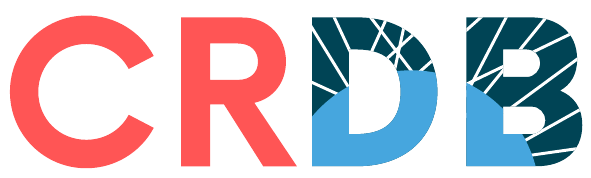}}%
  \end{picture}%
\endgroup%

\end{textblock*}

\input epsf

\maketitle

\setcounter{tocdepth}{2}
\section{Introduction}

Cosmic-Ray (CR) physics has been established more than a century ago and progress in the measurements has been performed through this entire period. Besides preserving the measurements for their historical value, the motivations for a CR database are numerous.

Firstly, GeV CRs vary with the Solar cycle, and past data are still of interest---for an illustration of the usage of CR data going back to the 50's, see \citet{2017AdSpR..60..833G}. Secondly, it is infeasible to design and build CR experiments that measure all species at all energies, so that multi-species studies over different energy ranges have to rely on many datasets from as many experiments---for an illustration of the usage of data samples of H to Ni elements over 45 years, see \citet{2019ApJ...887..132S}. Thirdly, new synergies with other fields of astrophysics make the collection of even the rarest CR data desired. For instance, ultra-heavy elements (UHCRs, $Z>30$) have very small fluxes and are difficult to measure. They require dedicated experiments and so far the data are very sparse (e.g. \citet{2012ApJ...747...40D} and \citet{2014ApJ...788...18B}). UHCRs have not been a very active topic in the last decade, but the situation is likely to change: the first detection of gravitational waves from a binary neutron star inspiral \citep{2017PhRvL.119p1101A}---and in particular optical follow-ups---indicated that neutron star mergers could be a major contributor to r-process nucleosynthesis for UHCRs \citep{2017ApJ...848L..18N,2017ApJ...848L..19C}.
Fourthly, in the last two decades, a growing number of high-precision experiments were developed to measure leptons, antimatter, and $Z<30$ nuclei in tens of MeV to hundreds of TeV range (AMS-02, PAMELA, TRACER, etc.). These balloon-borne or space-based experiments allow us to address many astrophysics questions \citep[e.g.][]{2015ARA&A..53..199G} including searches for dark matter \citep[e.g.][]{2012CRPhy..13..740L}. In the context of these searches, upper limits on antinuclei were improved \citep{2012PhRvL.108m1301A} and efforts are growing towards the first detection of CR antideuterons \citep{2016PhR...618....1A}. A few puzzling 40 GeV events compatible with $^3\overline{\rm He}$ were reported by the AMS-02 collaboration \citep{Ting}. Adding previously reported upper limits on antinuclei (on various energy ranges) in a database is thus also timely.

A central motivation for a comprehensive CR database is the energy coverage. CR experiments based on direct detection have recently shown a spectral feature in several species at a few hundreds of GV \citep{2015PhRvL.114q1103A,2015PhRvL.115u1101A,2018PhRvL.120b1101A}. Other spectral features have been observed at higher energies: the \emph{knee} (at a few PeV), the \emph{ankle} (around 5 EeV), and the \emph{toe} (flux suppression above 100\,EeV); a fainter second knee around 200\,PeV was also observed more recently \cite{2019PhRvD.100h2002A}. These features were observed with ground-based indirect CR experiments (KASCADE-Grande, Pierre Auger Observatory, Telescope Array, IceCube, TUNKA, and many more), which do not resolve CR species individually. Weak spectral features can be confirmed and enhanced by combining data from multiple experiments. Gathering data that cover the full CR spectrum can help to shed light on the Galactic to extragalactic transition (e.g. \citet{2015PhRvD..92b1302G} and \citet{2016A&A...595A..33T}) and on the possible sources for the most energetic CR events ever observed \citep[e.g.][]{2011ARA&A..49..119K}.

A central shared database offers multiple advantages for research. It enables studies, which would otherwise be too time-consuming or error-prone, by assuring data quality, completeness, and traceability. In principle, it can also allow for a pooling of human resources (from the CR community) to fill and maintain it. Electronic resources to gather and disseminate data became available in the 90's, but CR databases appeared much later. For Galactic CR (GCR) data, the momentum was provided by teams involved in the development of GCR propagation codes. The \galprop{} \footnote{\url{http://galprop.stanford.edu}} team started to distribute CR data as a simple \ascii{} file \citep{2009arXiv0907.0565S}. As part of the development of the \usine{}\footnote{\url{https://lpsc.in2p3.fr/usine}} propagation code \citep{2020CoPhC.24706942M}, our team independently gathered these data and set up the first CR database, called \crdb{} (\citet{2014A&A...569A..32M}); our database comes along with a contextualised presentation of the data and experiments, \ascii{} file exports (compliant to \galprop{} and \usine{} formats), and online plot generation. Shortly after, the Italian Space Agency (ASI) developed a CR database to host originally ASI-supported experimental data \citep{2017ICRC...35.1073D}. The design and functionality of the \crdb{}-ASI\footnote{\url{https://tools.ssdc.asi.it/CosmicRays}} was strongly influenced by the \crdb{}, following discussions between our teams. Since then, the \crdb{} and \crdb{}-ASI have been independently developed further. The \crdb{}-ASI is not as comprehensive as the current version of the \crdb{}, but it offers specialised datasets (trapped events and Solar Energetic Events) absent in our database.
For ultra high-energy CRs (UHECRs, above a PeV), the first initiative to build a common database came from the KCDC project\footnote{\url{https://kcdc.ikp.kit.edu/}} \citep{2018EPJC...78..741H}. The primary goal of KCDC is to offer researchers open access to raw data of the KASCADE and KASCADE-Grande experiments. It also contains a collection of high-energy CR flux measurements from more than 20 experiments published between 1984 and 2020 (contrarily to \crdb{}, an account is required to access KCDC data).

This paper discusses \crdbv{4.0}, a major upgrade of the database in terms of its data content, user interfaces, and data submission forms. The paper is organised as follows: Sect.~\ref{sec:db_content} recalls important definitions and the database structure; Sect.~\ref{sec:v4.0-newdata} highlights the specifics of the data added in this version; Sect.~\ref{sec:website} briefly goes over the web interfaces and outputs, focusing on the new and simpler data submission form and the updated \rest{} and Solar modulation time series interfaces. We conclude and discuss possible \crdb{} extensions in Sect.~\ref{sec:concl}. We gather in App.~\ref{app:tips} useful tips and caveats (on extracted data) for \crdb{} users, and we provide in App.~\ref{app:bibtex} sorted lists of all experiments and publications in \crdbv{4.0}.

\section{Database structure and updates}
\label{sec:db_content}

Throughout the paper, we separate the data in two broad categories, {\em data} and {\em meta-data} (data about data). Clearly separating these two categories is important to define the format to submit new data (see Sect.~\ref{sec:v4.0-newdata}). The categories comprise the following items:
\begin{itemize}
  \item {\em Data}: CR data points, including energy range and data uncertainties;
  \item {\em Meta-data:} CR data-related informations that include data taking periods, experiments that provide the data, publication from which the data were taken, etc.
\end{itemize}

The first \crdb{} publication \citep{2014A&A...569A..32M} focused on the rationale behind the design choices made for the database structure and organisation (in \mysql{}). In this update, we follow a complementary approach. We start at the data level, moving up to the experiments and then publications, underlining some salient aspects of the data themselves and how they are tied to the other meta-data tables. Along the way, we highlight the changes made between \crdbv{2.1} \citep{2014A&A...569A..32M} and \crdbv{4.0}.

\begin{figure}
\begin{center}
\includegraphics[width=0.95\textwidth]{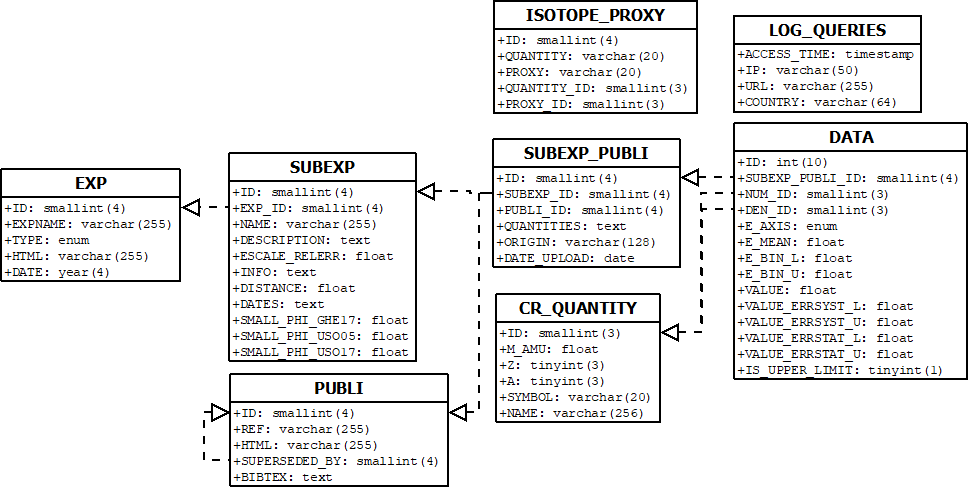}
\caption{Tables and keys of the database \mysql{}  structure. The data are stored in the \code{DATA} table (\S\ref{sec:table-data}) with CR names defined in the \code{CR\_QUANTITY} table. The meta-data related to experiments are stored in \code{EXP} (\S\ref{sec:table-exp}) and \code{SUBEXP} tables (\S\ref{sec:table-subexp}), while those related to publication are stored in the \code{PUBLI} table (\S\ref{sec:table-publi}); a bridge table, \code{SUBEXP\_PUBLI} (\S\ref{sec:table-subexppubli}), links all meta-data. The \code{LOG\_QUERIES} table is used to track the number of visits in \crdb{} (\S\ref{sec:admin-tab}). The \code{ISOTOPE\_PROXY} table is required for energy-axis conversions of CR fluxes (\S\ref{app:Econversion}).}
\label{fig:mysql}
\end{center}
\end{figure}

The database structure is shown in Fig.~\ref{fig:mysql}. By design, each entry has a unique ID in all tables, so that all tables can be accessed by other tables; this is particularly useful to build bridge tables (i.e. building associations between table entries).
The different tables, their contents, and the associations between data and meta-data tables are further detailed in the following subsections.
%

\subsection{\code{DATA} table}
\label{sec:table-data}
A data point corresponds to a quantity measurement (with uncertainties) at a given energy. We list below the keys required for a full data entry description in the \code{DATA} table (see Fig.~\ref{fig:mysql}). The only change made since \crdbv{2.1} is enabling upper limits data (new key \code{IS\_UPPER\_LIMIT}).
 \begin{itemize}

   \item \code{ID}: unique identifier for a data.

   \item \code{NUM} and \code{DEN}: a data quantity is a single item (\code{NUM}) or a ratio of two items (\code{NUM} and \code{DEN}). The former usually corresponds to a measured flux (in unit of $E_{\rm unit}^{-1}\,{\rm m}^{-2}\,{\rm s}^{-1}\,{\rm sr}^{-1}$, with $E_{\rm unit}$ in GeV, GeV/n, or GV in \crdb{}) and the latter to a ratio of fluxes (no unit). Depending on the detector charge and isotopic resolution, the measured quantities are isotopes (e.g. $^{12}$\code{C}), elements (e.g. \code{C}), or even groups of elements; the latter case mostly applies for UHCR and UHECR data (see next sections). Indirect detection CR experiments also provide shower-related quantities (with their specific units) that we plan to add in the future. To allow for these various possibilities, we define a \code{CR\_QUANTITY} table which contains a list of names including CR isotope names, element names, and names for relevant groups of elements or other significant measurable quantities (e.g. \code{AllParticles}, see Table~\ref{tab:cr_quantity} for new names in \crdbv{4.0}).

   \item \code{E-AXIS}: the database energy types are \code{ETOT}, \code{EK}, \code{R}, \code{EKN}, and \code{ETOTN}. In \crdb{}, they correspond to the total $(E_{\rm tot})$ and kinetic energy $(E_{\rm k}\!=\!E_{\rm tot}-m)$ set in GeV, rigidity $({\cal R}\!=\!p/(Ze))$ in GV, kinetic energy per nucleon $(E_{\rm k/n}\!=\!E_k/A)$ in GeV/n, and total energy per nucleon $(E_{\rm tot/n}\!=\!E_{\rm tot}/A)$ in GeV/A. Depending on the detector principle (calorimeter, spectrometer, etc.), different energy types are measured in experiments and provided in publications, and in \crdb{}, the data are always stored as published (see App.~\ref{app:Econversion} for conversion).

   \item \code{E\_MEAN}, \code{E\_BIN\_L}, and \code {E\_BIN\_U}: fluxes are measured by counting events in energy bins. In \crdb{}, we include the energy interval range $[E_{\rm lo},\, E_{\rm up}]$ when it is published. The measured flux obtained from a finite bin size neither exactly corresponds to the differential flux at the bin centre or at the mean energy $\langle E\rangle$, unless a correction is applied \citep{1995NIMPA.355..541L}; when comparing a model to the data, it is thus recommended to integrate the flux model over the bin width instead of using the value at the mean energy \citep{2019A&A...627A.158D}. If only a single value $\langle E\rangle$ is provided in the publication, we set $E_{\rm lo}\!=\!E_{\rm hi}\!=\!\langle E\rangle$ in \crdb{}; conversely, if only the bin range is provided, we set $\langle E\rangle\!=\!\sqrt{E_{\rm lo}E_{\rm hi}}$.

   \item \code{VALUE}, \code{VALUE\_ERRSYST\_L}, \dots, \code{IS\_UPPER\_LIMIT}: a measurement has a value and statistical and systematic uncertainties. In \crdb{}, we store the value and the possibly asymmetric statistical and systematic uncertainties---see App.~\ref{app:data_unc} for subtleties and caveats concerning the gathering of uncertainties from publications. In \crdbv{4.0}, upper limits with the flag \code{IS\_UPPER\_LIMIT} set to 1.

   \item \code{SUBEXP\_PUBLI\_ID}: a datum is always attached to a unique publication and a sub-experiment (see next sections). This key enables to bridge the data \code{ID} (this table) to the subexp-publi \code{ID} (\code{SUBEXP\_PUBLI} table, see Sect.~\ref{sec:table-subexp}).

\end{itemize}

\subsection{\code{EXP} and \code{SUBEXP} tables and meta-data}
\label{sec:tables-exp+subexp}

CR data are collections of data points at different energies, from an experiment measuring one or several quantities. In \crdb{}, for flexibility, we make the distinction between an experiment and sub-experiment (hereafter sub-exp for short). The experiment is defined by its name and starting date. The sub-exp is linked to the experiment, but is more closely related to the data in order to uniquely tag the dataset released by this experiment. The tag is here to inform us on (i) analyses from different data taking periods, (ii) data from the same period but from different analysis techniques or different sub-detectors, (iii) data relying on third-party model assumptions (e.g. using different Monte Carlo generators for air showers in UHECRs).

Experiment and sub-experiment data are meta-data (attached to CR data) stored in the \code{EXP} and \code{SUBEXP} tables respectively (see Fig.~\ref{fig:mysql}); they are briefly described below.

\subsubsection{\code{EXP} table}
\label{sec:table-exp}
A single change has been made for \crdbv{4.0}, related to the experience type (key \code{TYPE}, see below).

 \begin{itemize}

   \item \code{ID}: unique identifier for an experiment.

   \item \code{EXPNAME}: unique name associated with the main instrument ({\em AMS-01}, {\em KASCADE}, etc.). Most of the data before the 90's come from balloons, and to ensure a unique identifier \crdb{} uses the syntax: {\em Balloon (YYYY)} for a single flight, {\em Balloon (1966,1967,\dots)} for balloons flown several times, and {\em Balloon (1967+1968+\dots)} for experiments that report only the combined analysis of several flights.

   \item \code{TYPE}: in \crdbv{4.0}, we allow for three experiment types (\code{balloon}, \code{ground}, or \code{space}) that represent three broad categories of measurement conditions with different caveats and systematic uncertainties. The experiments in the `Experiments/Data' tab (see Sect.\ref{sec:ExpData-tab}) on the \crdb{} website can be sorted according to date, name, and this type.

   \item \code{DATE} and \code{HTML}: we also keep track of the starting year of the experiment and a link to the official experiment website is provided (if it exists).

 \end{itemize}

\subsubsection{\code{SUBEXP} table}
\label{sec:table-subexp}
The \code{SUBEXP} table contains detailed information on possible sub-detector(s) from which the data were obtained; this includes the data taking conditions and, in some cases, the analysis technique. In \crdbv{4.0}, a new key was added to track the energy-scale uncertainty (i.e. a calibration uncertainty on the energy measurement), which is especially important for ground-based experiments ($\gtrsim$ 10\,\%).
We list all keys in the \code{SUBEXP} table (see Fig.~\ref{fig:mysql}) and then comment on the effect of Solar modulation on multi-GeV GCR data \citep{2013LRSP...10....3P}. In particular, we stress how the last three keys listed below are necessary inputs for Solar modulation calculations.
 \begin{itemize}

   \item \code{ID}: unique identifier for a sub-experiment

   \item \code{EXP\_ID}: \code{ID} of the experiment in the \code{EXP} table to which the sub-exp is attached.

   \item \code{NAME}: concatenation of the experiment (or sub-experiment) name with the sub-detector/specific detection technique used ({\em PAMELA-CALO\dots}), or with the air shower MC generator used ({\em IceTop SIBYLL-2.1\dots}), and always with the data taking period (e.g. {\em IMP7 (1973/05-1973/08)}).

   \item \code{DESCRIPTION}: rough description of the detection techniques and salient features of the detector (in the sub-experiment). If available from the collaboration, an image of the detector is displayed in the user interface (see Sect.~\ref{sec:website}).

   \item \code{INFO}: additional information relevant for the sub-exp. This can be the location of the balloon flight (e.g. {\em 1 flight McMurdo, Antarctica}), some details about the reconstructed data (e.g. {\em Fractions of H,He,O,Fe-like showers in 4-component fit to $\xmax$ data}), indication of times series (e.g. {\em Monthly average 2007/03}), etc.

   \item \code{ESCALE\_RELERR}: if provided in the publication, \crdbv{4.0} stores the relative energy-scale uncertainty, $(\Delta E/E)_{\rm scale}$, which applies to all energies measured by the sub-experiment\footnote{In principle, $(\Delta E/E)_{\rm scale}$ does depend on energy. If future CR publications release energy-dependent $(\Delta E/E)_{\rm scale}$, their implementation in \crdb{} will require modifications of the \code{ESCALE\_RELERR} key.}. We refer the reader to App.~\ref{app:Escale} for more details on the origin of this uncertainty.

   \item \code{DATES}: list of data taking periods for the instrument taken from the publication or, if not reported in old balloon flights, retrieved from the {\sc StratoCat} database\footnote{\url{http://stratocat.com.ar/globos/indexe.html}}.

   \item \code{DISTANCE}: most of the experiments are located on Earth, or are orbiting Earth. However, a few satellites (e.g. {\em Ulysses}, {\em Voyager}, etc.) took data at various locations in the Solar system. We provide in \crdb{} the average distance to the Sun in a.u. (an astronomical unit is the distance Sun-Earth) during the data taking period, as provided in the publications.

   \item \code{SMALL\_PHI}: Accurate Solar modulation models are complex and depend on many parameters. The simplest model, the Force-Field approximation \citep{1967ApJ...149L.115G,1968ApJ...154.1011G,1987A&A...184..119P}, depends on a single parameter $\phi(t)$, and remains useful for many applications despite its limitations \citep{2004JGRA..109.1101C}. Therefore we provide the mean modulation level $\langle\phi_{\rm FF}\rangle$ calculated externally from $\phi$ time series (see below) over the sub-experiment data taking periods.

 \end{itemize}%

\paragraph{\em From interstellar (IS) to top-of-atmosphere (TOA) fluxes}
GCR fluxes are Solar modulated, the modulation level depending on the Solar cycle \citep{2015LRSP...12....4H} and the position in the Solar cavity. IS fluxes refer to fluxes outside of the solar cavity while TOA fluxes correspond to modulated fluxes at Earth. The impact of the Solar modulation decreases with energy and can only be disregarded when it becomes much smaller than the data precision\footnote{For instance, AMS-02 proton data \citep{2015PhRvL.114q1103A} have a $\lesssim2\%$ precision up to a few hundreds of GV, to be compared with a modulation of $\sim 10-20\%$ at 20~GV, $\sim2-5$ at 100~GV, and at sub-percent above several hundreds of GV.}.
Most \code{balloon} and \code{space} experiments in \crdb{} are TOA data from experiments inside the solar cavity. To be compared to them, IS fluxes in GCR propagation models must therefore be Solar modulated. This makes the bookkeeping of the experiment data taking periods and position in the Solar cavity mandatory. These dates and positions can then be fed to dedicated Solar propagation codes---for instance {\tt SolarProp}\footnote{\url{http://www.th.physik.uni-bonn.de/nilles/people/kappl}} \citep{2016CoPhC.207..386K} or {\tt HelMod}\footnote{\url{http://www.helmod.org}} \citep{2018AdSpR..62.2859B}. Alternatively, if \crdb{} users wish to use the simpler Force-Field approximation  \citep{2004JGRA..109.1101C}, they can directly retrieve $\langle\phi_{\rm FF}\rangle$ values provided in \crdb{}.

\paragraph{\em Average Force-Field levels $\langle\phi_{\rm FF}\rangle$ from $\phi(t)$ time series} Modulation time series (in the Force-Field approximation) can be obtained from the analysis \citep{2005JGRA..11012108U,2017JGRA..122.3875U,2017AdSpR..60..833G} of neutron monitor data\footnote{\url{http://www01.nmdb.eu}}. These analyses were pioneered by \citet{2005JGRA..11012108U} in 2005 and updated in \citet{2017JGRA..122.3875U} in 2017. Following a similar approach, and based on improved propagation of error \citep{2015AdSpR..55..363M} and interstellar spectra \citep{2016A&A...591A..94G}, our team independently reconstructed its own $\phi$ time series in \citet{2017AdSpR..60..833G}, roughly finding
\[
     \phi_{\rm [Ghe17]}(t)\approx \phi_{\rm [Uso17]}(t) \approx \phi_{\rm [Uso05]}(t)+ 100~{\rm MV}.
\]
In \crdbv{4.0}, we provide these three values in the \code{SMALL\_PHI} keys (of the \code{SUBEXP} table), tagged as [Uso05], [Uso17], and [Ghe17]. We stress that whereas [Uso05] and [Uso17] are based on monthly-averaged distributed values\footnote{\url{http://cosmicrays.oulu.fi/phi}}, [Ghe17] relies on daily averaged values. Independently of the respective merits of these three sets, daily average $\phi$ values are better suited than monthly ones for short duration experiments. Our values [Ghe17] cover all years from 1950 till now and are automatically updated. These values can also be retrieved from the website (see Sect.~\ref{sec:tab-phi}).

\subsection{\code{PUBLI} and \code{SUBEXP\_PUBLI} tables and meta-data}
\label{sec:tables-publi+subexppubli}

CR data are published by experimental teams and collaborations in scientific publications. To save time and avoid confusion, we have a strong preference to include only final data published in peer-reviewed journals, and most data in \crdb{} come from refereed journals. An important exception is the proceedings of the biyearly International Cosmic-Ray Conference (ICRC, started in 1947): until the 90's, many balloon flight results were published in the ICRC proceedings only, and even nowadays these proceedings still often contain the latest preliminary results and updates of previously published results.

\subsubsection{\code{PUBLI} table}
\label{sec:table-publi}

To store what a publication is, the following keys are used.

\begin{itemize}

   \item \code{ID}: unique identifier for a publication.

   \item \code{HTML}: unique ID for the reference, taken from the Astrophysics Data System (ADS)\footnote{\url{https://ui.adsabs.harvard.edu/}}, e.g. \href{https://ui.adsabs.harvard.edu/abs/2014A&A...569A..32M}{\code{2014A\&A...569A..32M}}.

   \item \code{REF} and \code{BIBTEX}: these two keys provide a full publication reference and \textsc{Bib}\TeX\footnote{\url{www.bibtex.org}} entry, automatically retrieved from the ADS ID thanks to the ADS API (application programming interface)\footnote{\url{https://github.com/adsabs/adsabs-dev-api}}.

   \item \code{SUPERSEDED\_BY}: in some cases, the same data (same quantity from the same sub-exp) are re-analysed and updated in a subsequent publication. This key enables to store the \code{ID} (in the \code{PUBLI} table) of the newer publication analysis, keeping track at the same time of deprecated analyses and references. A single level of superseded data can always be seen from the `Experiments/Data' tab (see Sect.~\ref{sec:webinterface}).

\end{itemize}

\paragraph{\em Deprecated keys in \crdbv{4.0}}
One major modification in the \code{PUBLI} table in this release was to abandon book-keeping Solar modulation informations {\em from the publication}, i.e. Solar modulation levels calculated by the authors for their data.
This feature was introduced in \crdbv{2.1}---requiring many extra keys in the \code{PUBLI} table, see \citet{2014A&A...569A..32M}---in order to highlight the various assumptions behind the modulation levels provided. The motivation was to ease comparisons between modulation levels for data taken at similar periods but provided by different experimental teams. However, picking these informations from the publication was time consuming for very little practical use in retrospect\footnote{Indeed, the IS spectra and the Solar modulation models used are too patchy. For IS fluxes, publications used spectra derived from their data, from other authors, of from inputs of GCR propagation models. For Solar modulation, the models found in the publications were (i) for very old data, the outdated diffusion/convection model and its $\eta$ parameter \citep{1958PhRv..110.1445P,1958PhRv..109.1874P,1972ApJ...171..363L}, (ii) since the 70's, the still widely used Force-Field approximation and its $\phi_{\rm FF}$ parameter, or the more evolved spherically symmetric model and its effective $\phi_{\rm 1D}$ free parameter \citep{1969JGR....74.4973F,1971JGR....76..221F,1993ApJ...413..268B}, (iii) in more recent studies, the modern sign-charge dependent 3D model with parameters for the current sheet, tilt angle, diffusion tensor, etc. \citep{1979ApJ...234..384J,1985ApJ...294..425P,2013LRSP...10....3P} or, most of the time, no modulation value at all (just the data).}.
Moreover, as stressed in the previous section, modern CR propagation model studies either make use of $\phi_{\rm FF}$ or rely on public Solar modulation codes, and the informations stores in the \code{SUBEXP} table allow both these practices (see Sect.~\ref{sec:table-subexp}). A major benefit of not considering the modulation from the publication is that it simplifies the structure of the database, the extraction of data, and the submission of new data.

\subsubsection{\code{SUBEXP\_PUBLI} bridge table}
\label{sec:table-subexppubli}

To avoid duplicates, for cases where different data refer to the same publication or the same sub-experiment, a bridge table (\code{SUBEXP\_PUBLI}, see Fig.~\ref{fig:mysql}) enables to link entries from the \code{SUBEXP} and \code{PUBLI} tables. Its keys are as follows.

\begin{itemize}

   \item \code{ID}: unique identifier for a subexp-publi association.

   \item \code{QUANTITIES}: lists all the quantities (fluxes, ratios, etc.) provided by a given sub-exp/publi association.

   \item \code{DATA\_UPLOAD}: stores the date at which the corresponding data were uploaded.

   \item \code{SUBEXP\_ID}: bridges a sub-experiment/publication result to a `sub-exp' \code{ID} (\code{SUBEXP} table).

   \item \code{PUBLI\_ID}: bridges a sub-experiment/publication result to a `publi' \code{ID} (\code{PUBLI} table).

\end{itemize}

\section{New data in \crdbv{4.0}}
\label{sec:v4.0-newdata}

We now turn to the database content. As much as possible, data are added in \crdb{} on a continuous basis (by one of us); new data can also be submitted (see Sect.~\ref{sec:tab-submitdata}). Thanks to the feedback of many users, typos and mismatches in the data and meta-data are regularly corrected---the list of corrections and people who helped us (see acknowledgements)  are available online (see \crdb{} log file).

In this section, we highlight the main datasets added since the first release \citep{2014A&A...569A..32M}, in particular in the two most recent releases \crdbv{3.1} (\S\ref{sec:v4.0-updates}) and {\tt v4.0}; for the latter we detail the motivation and specifics of UHCR data (\S\ref{sec:v4.0-UHCRs}), antinuclei upper limits (\S\ref{sec:v4.0-UL}), and UHECR data (\S\ref{sec:v4.0-UHECRs}). The complete list of experiments and publications (presently in \crdb{}) is gathered in App.~\ref{app:bibtex}.
%

\subsection{Relevant data updates before \crdbv{4.0}}
\label{sec:v4.0-updates}

Since the first release of \crdb{} seven years ago, two major data updates are worth highlighting. The first, tagged as \crdbv{3.0}, added H and He data from a large number of balloon flights going back to the 50's. These data sets were used to derive Solar modulation levels over sixty years and compared to values derived from neutron monitor data in \citet{2017AdSpR..60..833G}.

The second major upload, tagged \crdbv{3.1}, was made a few weeks before preparing this release. The update is the result of a tremendous concentrated effort to include a wealth of data released in the last two years from various experiments (AMS-02, CALET, NUCLEON, TRACER, Voyager, and a few others); they correspond to some of the most interesting and high-precision CR data ever recorded. In addition, time-dependent fluxes and ratios from AMS-02 \citep{2018PhRvL.121e1101A,2018PhRvL.121e1102A} and PAMELA \citep{2018ApJ...854L...2M} were added.

We emphasise at this point that the \crdb{} is open for data submissions from experiments since its first release, and contributions are very welcome. We further simplified the submission interface, presented in Sect.~\ref{sec:tab-submitdata}, to encourage more data submissions in the future.

\subsection{Very heavy and ultra-heavy CRs}
\label{sec:v4.0-UHCRs}

The theory of stellar nucleosynthesis to explain Solar system abundances goes back to the fifties \citep{1957RvMP...29..547B}. In the following decade, $Z\ge30$ \citep{1967RSPSA.301...39F,1969PhRvL..23..338B} and even a few $Z>90$ (actinides) \citep{1970RSPSA.318....1F,1971PhRvL..26..463O,1971PhRvD...3..815P} CRs were detected. The abundance of these UHCRs are driven by three processes, corresponding to three categories of stable heavy nuclides: the p-, r-, and s-nuclides, respectively located at the neutron-deficient side, neutron-rich side, and bottom of the valley of nuclear stability \citep{2003PhR...384....1A}. How much p- \citep{2003PhR...384....1A,2013RPPh...76f6201R}, r-, and s-processes \citep{1993PhR...227..283C,1994ARA&A..32..153M} contribute to the various UHCRs, and which astrophysical sites are involved remain debated. Moreover, the abundance of UHCRs slightly differs from the Solar system ones, and these differences are likely to be related to specific acceleration \citep[e.g.][]{2019ApJS..245...30L} or transport processes \citep[e.g.][]{2005A&A...435..151C}.

Measuring UHCR data is challenging because the fluxes are low. From $Z=28$ (Fe) to $Z=30$ (Zn) and then $Z=40$ (Zr), the flux is suppressed by a factor $\sim 10^3$ and $\sim 10^5$ \citep{2014ApJ...788...18B}. The rarest CRs, the actinides $Z\geq90$ (Th and U), are about seven orders of magnitude less abundant than Fe \citep{1981Natur.291...45F}. Standard techniques used for measurements of $Z<30$ CRs can still be pushed to cover the $30\leq Z\leq60$ region \citep{2014ApJ...788...18B}, but come slightly short for the heaviest ones \citep{1987ApJ...314..739F,1989ApJ...346..997B}. An alternative consists of passive detectors exposed for long durations (several years). The detection principle is based on the chemical modification made in a solid state nuclear track detector when ionising particles go through it. After long exposure, the detector is etched in a chemical agent and each track left is analysed and its properties can be related to the charge and velocity of the CR.

We added in \crdbv{4.0} most of the CR data found for $Z>30$. Most measurements do not resolve elements, and we needed to create new name holders for groups of charges, for instance \code{Pt-group} ($74 \leq Z\leq 80$), \code{Pb-group} ($81\leq Z\leq 87$), \code{Subactinides} ($74\leq Z\leq 87$), and \code{Actinides} ($Z\geq 88$)---see Table~\ref{tab:cr_quantity}. We note that in the literature different groups sometimes used slightly different charge range in their definition, but we stick to this one for all implemented data---data uncertainties are larger than the error made from using slightly enlarged or reduced charge groups. The data themselves mostly consist of ratios of elements (or of groups of elements) in an unspecified energy band (related to the detector capabilities). In practice, CR fluxes are expected to be maximal at a few GeV/n, and several experiments reported no energy evolution (within the uncertainties) of their measured ratios. For this reason, and for simplification, we set all UHCR data $Z>50$ to a generic energy bin around 1.5 GeV/n.

More than sixty years after their discovery, only very few UHCR data are available and almost no new experiment is running or planned. 
Except for the relatively recent superTIGER experiment in the $Z=30-40$ range \citep{2016ApJ...831..148M}, the experiments in the high charge range date back to the end of the seventies (Ariel6 \citep{1987ApJ...314..739F}, HEAO3-HNE \citep{1981ApJ...247L.115B,1983ICRC....9..115S}), the early eighties (UHCRE-LDEF \citep{1996RadM...26..825D,2003RadM...36..287D,2012ApJ...747...40D}), or at best the nineties (Trek \citep{1998Natur.396...50W,2002ApJ...569..493W}). The most recent results on subactinides and actinides come from indirect measurements of the CRs from their impact on meteorite olivines \citep{2010PhyU...53..805A,2013JETPL..97..708B,2016ApJ...829..120A}, namely the Olimpiya Experiment (not implemented yet in \crdb{}). The solution to the origin of these elements may come from yet another route: recently, optical follow-up observations of a gravitational wave event have shown that binary star mergers could be a major r-process contributor to UHCRs \citep{2017ApJ...846..143K}. More such events are likely to be detected in the future, and they will provide a complementary view to that of CR data.

\subsection{Antimatter CRs}
\label{sec:v4.0-UL}

So far, antiprotons are the only antinuclei found in CRs. They were first detected at the end of the seventies \citep{1979ICRC....1..330B,1981ApJ...248.1179B} and soon interpreted by \citet{1984PhRvL..53..624S} as possible dark matter signatures. The pioneering theoretical aspects and subsequent efforts were acknowledged by the 2019 Cosmology Gruber prize: {\em [...] Silk recognised dark matter's indirect signatures such as antiprotons in cosmic rays and high energy neutrinos from the Sun}. This exploration continues with modern experiments, which have detected several tens of thousands of antiprotons \citep{2016PhRvL.117i1103A}.

The quest for $Z<-1$ CRs is also very much alive but the expected fluxes are very low. Indeed, most if not all CRs antiprotons can be attributed to nuclear production (H and He CRs on the interstellar gas) during CR propagation in the Galaxy. The cost to create and fuse together an extra antiproton or antineutron from colliding protons is a roughly $10^{-4}$ suppression factor \citep[e.g.][]{2020arXiv200210481K}. As a consequence, the ratio of astrophysical CR antiprotons to protons is at most $\sim10^{-4}$ \citep[e.g.][]{2016PhRvL.117i1103A}, and the ratio of astrophysical CR antinuclei (of atomic number $A$) to protons is expected to go very roughly as $10^{-4A}$. So far, only upper limits have been derived on the latter \citep{2002RPPh...65.1243G}. CR antideuterons are particularly difficult to detect because they must be isotopically separated from the $\bar{p}$ tail, and also because charge confusion rejection from $p$ should be at the level of $10^8$. The AMS-02 experiment on the ISS or the GAPS experiment based on a novel detection technique could detect a few events in the near future \citep{2016PhR...618....1A}. Although at a much lower level, antihelium events may have been seen in AMS-02 very preliminary analyses \citep{Ting}. Whether these events are real, and whether this is compatible with no detection of antideuterons and can be accounted for physics scenario is still at the exploratory stage \citep{2019PhRvD..99b3016P,2020arXiv200108749C,2020arXiv200204163V}; a confirmation of these events would have huge consequences.

\begin{figure}[t]
\begin{center}
\frame{\includegraphics[width=0.8\textwidth]{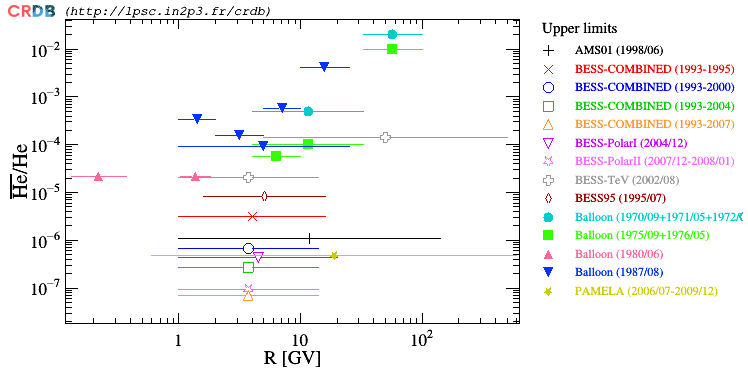}}
\vspace{2mm}\\
\frame{\includegraphics[width=0.8\textwidth]{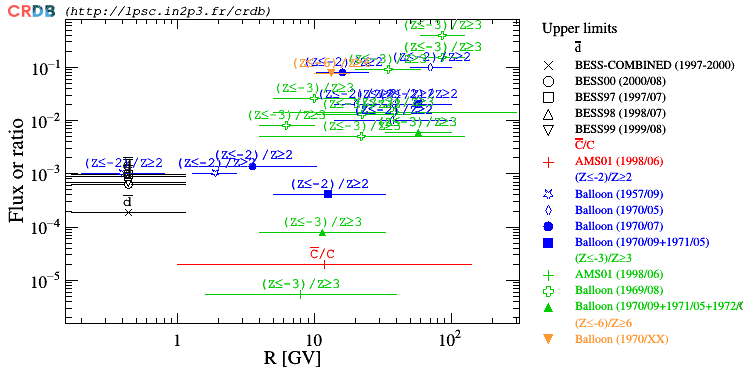}}
\caption{New antinuclear data in \crdbv{4.0}. {\em Top panel:} Upper limits on antihelium-to-helium ratio \citep{1975PhRvL..35..258S,1978Natur.274..137B,1981ApJ...248.1179B,1997ApJ...482L.187O,1997ApJ...479..992G,1998PhLB..422..319S,1999PhLB..461..387A,2002NuPhS.113..202S,2008AdSpR..42..450S,2011JETPL..93..628M,2012PhRvL.108m1301A}. {\em Bottom panel:} Upper limits on antideuteron flux  \cite{2005PhRvL..95h1101F} (black symbols) and various combinations of antielement-to-element ratios \citep{1961PhRv..121.1206A,1971Natur.230..170G,1971ICRC....1..203G,1972Natur.236..335B,1972ApJ...176..797E,1972NPhS..240..135V,1974ApJ...192..747G,1975PhRvL..35..258S,2002NuPhS.113..195C,2005PhRvL..95h1101F}.}
\label{fig:upper-limits}
\end{center}
\end{figure}
To tackle upper limits, a new key in the CR database table was added (see Sect.~\ref{sec:table-data}). Limits on antielements are also usually derived with respect to elements, and we added in \crdbv{4.0} new names to cover all associated data (see Table~\ref{tab:cr_quantity}). The best limits so far come from the BESS balloon flights \citep{2012PhRvL.108m1301A} and PAMELA mission \citep{2014PhR...544..323A}, as shown in Fig.~\ref{fig:upper-limits} with results from other experiments derived over the years (all included in \crdbv{4.0}).

\subsection{Ultra-high energy CRs}
\label{sec:v4.0-UHECRs}

A major extension of the already comprehensive CR database is the inclusion of data from air shower experiments. High-energy CRs are measured indirectly with ground-based experiments. They detect the particle cascades initiated by CRs in Earth's atmosphere (air showers). Since the CR flux drops rapidly with increasing energy, the air-shower technique is the only feasible way to obtain CR data above PeV energies. The air-showers allow for large aperture experiments due to their extensive footprint on the ground in charged particles and Cherenkov light, and their visibility from a distance in ultra-violet light.

The energy and direction of a CR can be well inferred from air showers, but the identity of each individual CR cannot be determined accurately. The identity, more specifically the mass $A$ of the CR, has to be inferred from air shower properties that naturally fluctuate, like the slant depth $\xmax$ of shower maximum measured from the top of the atmosphere or the total number $\nmu$ of muons produced in the shower. These natural fluctuations make it impossible to distinguish species on an event-by-event basis. Therefore, fluxes of individual elements or isotopes cannot be determined.

Experiments most commonly report the all-particle flux. In principle, this also includes stable electromagnetically and weakly interacting particles, but the flux of these components at the PeV scale and above is negligible compared to nuclei~\cite{Kampert:2012mx,Apel:2017ocm}. In addition, some experiments report the fluxes of mass groups by splitting the observed mass range into two groups (proton-helium group and the rest), four groups (proton, helium, oxygen-, and iron-group), or five (as before, but adding a silicon group). These groups span over roughly equal intervals in $\ln A$, since the mass sensitivity is roughly constant in this variable. Beside the usual individual element or isotope names, the list of CR quantity names in \crdbv{4.0} was expanded to handle these groups (e.g. \code{H-He-group}, \code{Fe-group}, \code{AllParticle}, see Table~\ref{tab:cr_quantity}).

We note that these results are obtained by simulating air showers initiated by the leading element in each group only, and by fitting the sum of the air shower response distributions to the observed distribution~\cite{Aab:2014aea,Apel:2012tda}. Representing a whole mass-group by a single element is a great simplification, but unavoidable since the relative fluxes of species within a mass group are unknown. The analysis approach works in practice because air shower fluctuations are large compared to the small differences in the response to individual elements within a group. 

Another peculiarity of high-energy data is that one raw data set often has several interpretations, in which the fluxes of mass groups and, to a lesser degree, also the all-particle flux vary. These parallel interpretations are reported by the experiments due to significant theoretical uncertainties in simulated air showers. Detailed air shower simulations are required to infer the mass $A$ from air shower properties like $\xmax$ and $\nmu$, and these simulations use hadronic interaction models that extrapolate fixed-target and collider measurements of particle interactions using theory and phenomenology. Air showers are dominated by soft QCD interactions, which so far cannot be predicted accurately from first principles. Therefore the interpreted measurements depend on the hadronic interaction model used in air shower simulations, which is listed with the interpreted data.

\begin{figure}[t]
\begin{center}
\frame{\includegraphics[width=0.8\textwidth]{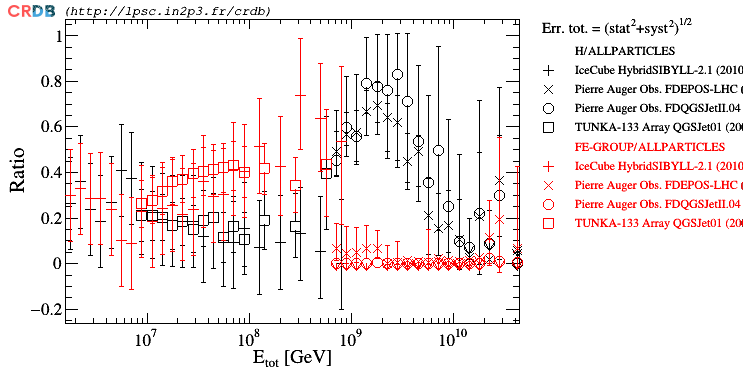}}
\caption{Sample of UHECR data added in \crdbv{4.0}. For illustration, we show the fraction of light (\code{H}) and heavy (\code{Fe-group}) CRs for IceCube \citep{2015ICRC...34..334R}, Pierre Auger Observatory  \citep{2014PhRvD..90l2005A} for two different Monte Carlo event generators, and TUNKA-133 \citep{2014NIMPA.756...94P}.}
\label{fig:uhecrs}
\end{center}
\end{figure}

For this update, we focus on adding flux measurements to the \crdb{}, but we also consider to include measurements of air shower properties like the average of $\xmax$ or $\nmu$ as a function of the shower energy in the future. In the current update, we include air-shower based flux data from the Pierre Auger Observatory, Telescope Array, the IceCube Neutrino Observatory, the KASCADE-Grande experiment, the TUNKA-133 Array, and the H.E.S.S. observatory. An illustration of the evolution with energy of the ratio of two groups of elements is shown in Fig.~\ref{fig:uhecrs}.

\section{User-interface updates and new submission form}
\label{sec:website}

\crdb{} is hosted by the LPSC laboratory website. It runs on a {\sf LAMP} solution, i.e. a stack of free open source softwares: {\sf Linux} (operating system), {\sf Apache HTTP} (server), {\sf MySQL} (database), and \php{} (hypertext pre-processor language). The web interface relies on several third-party libraries (\jquery{}, \jquery{-ui}, \jquery{.cluetip}, and \tsorter{}) that are used to sort and display the database content, as briefly presented in the various sections below. For efficiency and speed, all web pages make use of \ajax{} (Asynchronous {\sf JavaScript} and {\sf XML}) web development techniques.

Accessing the data in \crdb{} can be achieved in two ways, either by accessing the resources via \crdb{} web pages, or via a \rest{} (representational state transfer) interface. While web pages allow to access the fully contextualised data and meta-data (thanks to several sorting, selections, display, and download of data options), the \rest{} interface is useful for users who want to access and retrieve data from command lines (i.e. from their terminal or codes, without going through the \crdb{} web pages).

Below, we discuss separately the various web interfaces. We start with the description of the improvements brought on those already available in the first release (\S\ref{sec:webinterface}). We then describe the interface for extracting $\phi_{\rm FF}$ time series (\S\ref{sec:tab-phi}), the new help page for the \rest{} interface (\S\ref{sec:tab-rest}), and the simplified procedure (and format) to submit new data (\S\ref{sec:tab-submitdata}).

\subsection{Web user interface}
\label{sec:webinterface}

\begin{figure}[t]
\begin{center}
\frame{\includegraphics[width=\textwidth]{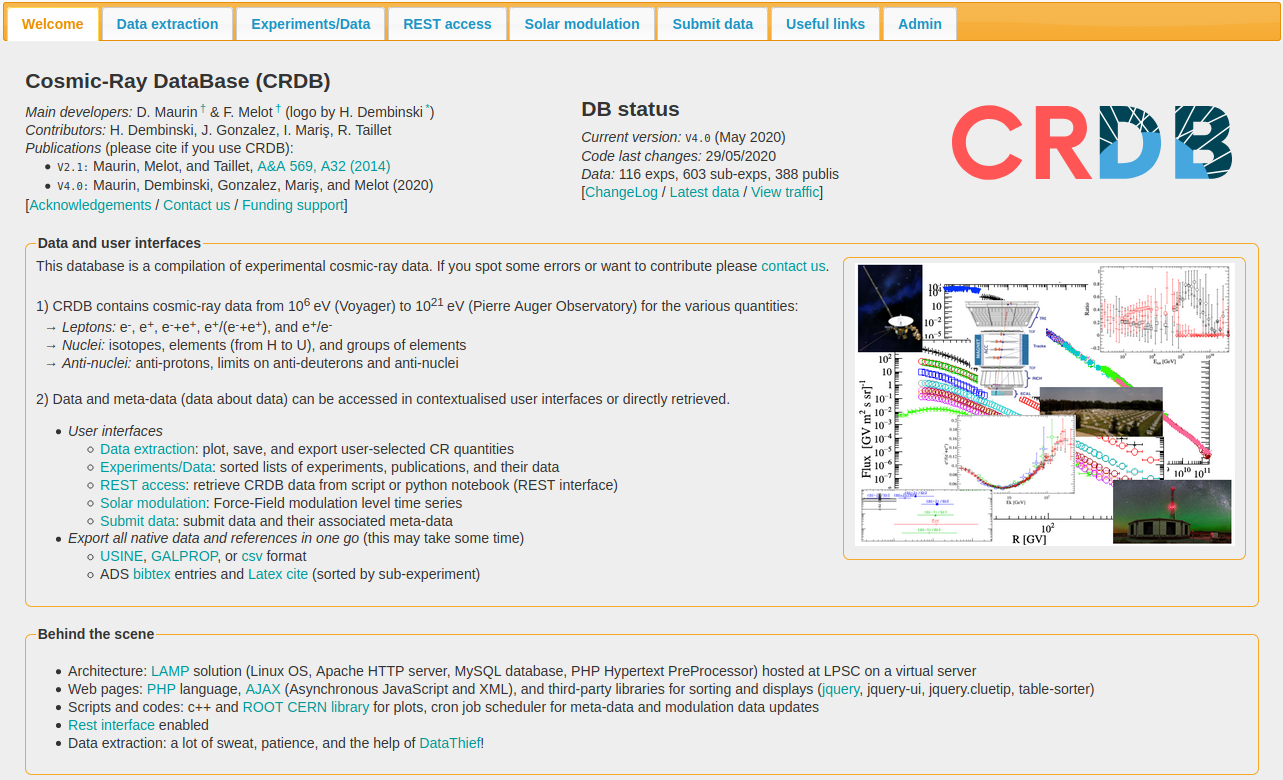}}
\caption{Snapshot of the `Welcome' tab to which users are directed from the \crdb{} webpage \url{https://lpsc.in2p3.fr/crdb}. Clicking on the tabs (shown on top) allows to access the various user-interfaces presenting in different ways the contextualised data and meta-data (see text).}
\label{fig:tab-welcome}
\end{center}
\end{figure}

The address  \url{https://lpsc.in2p3.fr/crdb} leads the user to a choice of various tabs, as illustrated in Fig.~\ref{fig:tab-welcome}. Among them, two tabs provide differently contextualised informations on CR experiments and data, while other tabs provide general information on \crdb{}, external CR resources, etc.
We give below a brief description of these tabs, highlighting the most salient features and novelties of this version, and providing a few snapshots as illustrations.

\subsubsection{`Welcome' tab}
This is the unique entry point of the website (see Fig.~\ref{fig:tab-welcome}). It contains a brief description of the database content and structure, and informations on \crdb{} including versions, log, logo, and publications. In particular, the log file tracks the changes made in the various releases (data corrections and past and ongoing developments).

\subsubsection{`Experiments/Data' tab}
\label{sec:ExpData-tab}
\begin{figure}[t]
\begin{center}
\frame{\includegraphics[width=0.597\columnwidth]{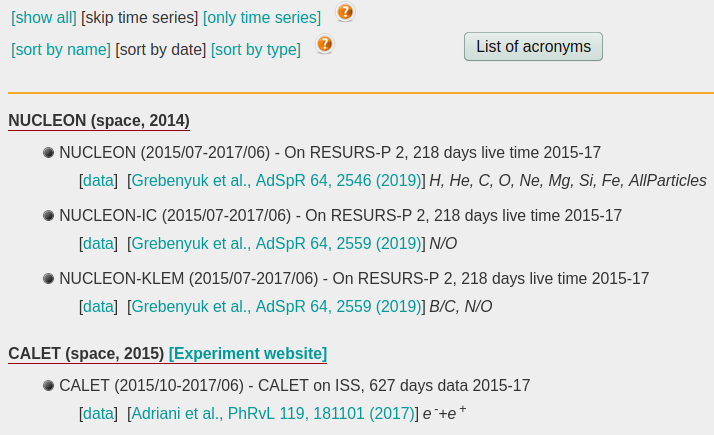}}
\hspace{0.pt}
\frame{\includegraphics[width=0.274\columnwidth]{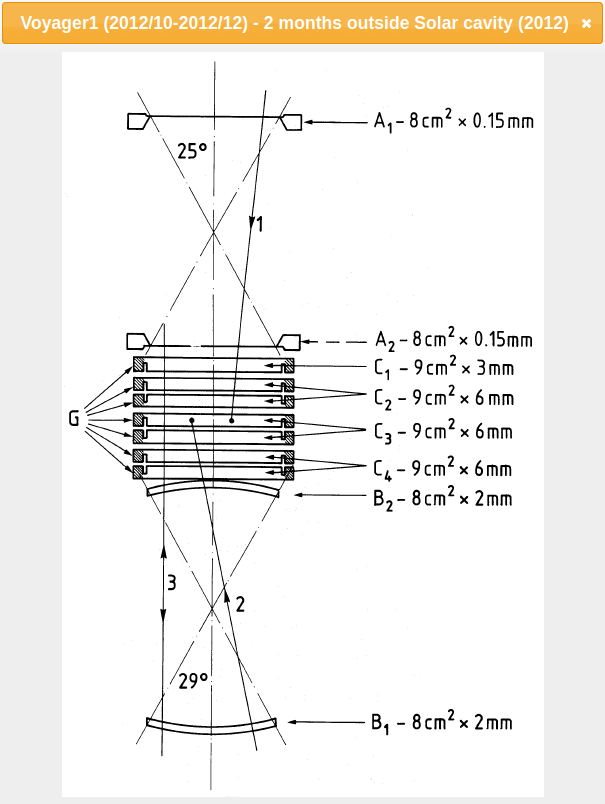}}
\\
[3pt]
\frame{\includegraphics[width=0.7\columnwidth]{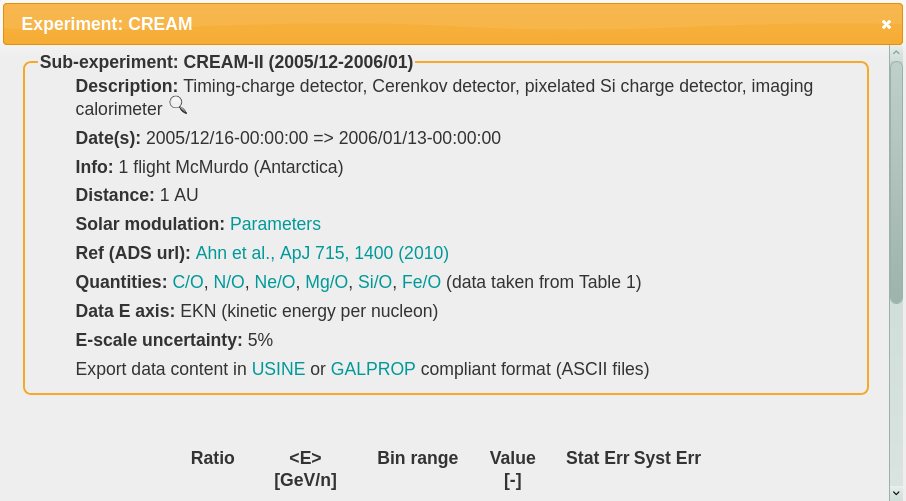}}
\caption{Snapshot from the {\sc Experiments/Data} tab illustrating how experiments, sub-experiments, and publication meta-data are organised in this tab (top left panel). From here, clicking on (i) the `List of acronyms' button provides the meaning of acronyms for all experiments in \crdb{} (not shown), (ii) the magnifying glass icon (seen beside sub-experiment names) pops-up the detector schematics (top right panel) showing Voyager in this example, (iii) the `[data]' link pops-up the full meta-data informations and associated data (bottom panel).}
\label{fig:tab-expdata}
\end{center}
\end{figure}

This is the tab to go for users looking for a specific experiment, its meta-data, and the data it gathered; experiments can be sorted according to their name, starting date, or type. This tab lists the quantities they measured. It also provides links to the experiment web page, the associated publications, and pictures of the sub-detectors (by clicking on the `magnifying glass' icon); see Fig.~\ref{fig:tab-expdata} for an illustration.
Clicking on `[data]' for any given sub-exp gives further information on the CR data origin: data taking period, how data were retrieved (numbers from tables, extracted from figures, or from personal communications), details on the detector, etc.

We stress that the data presented in this tab are `native' data only (see App.~\ref{app:rules}), i.e. data as provided in publications and uploaded in the database without modification (which is not the case in the `Data extraction' tab). Accordingly, the energy axes for the data are as provided in the publications.

\subsubsection{`Data extraction' tab}

\begin{figure}[t]
\begin{center}
\frame{\includegraphics[width=0.8\columnwidth]{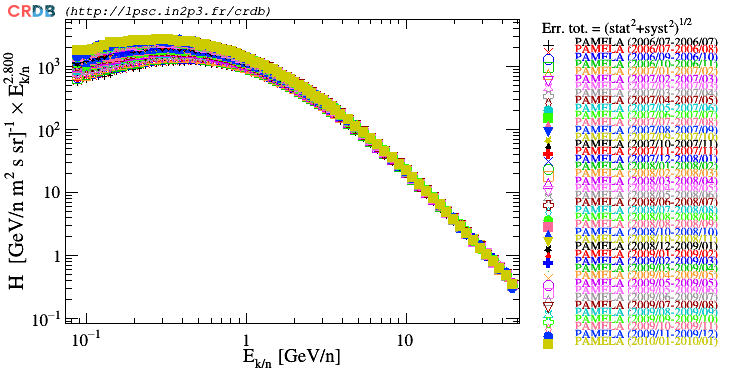}}
\caption{Illustration of the `time series only' option in the query selection criterion on PAMELA data \citep{2013ApJ...765...91A}.}
\label{fig:plot-timeseries}
\end{center}
\end{figure}

This is the main interface to select, extract and plot data, displaying also some meta-data (sub-exp names, links and \textsc{Bib}\TeX\ references for publications matching the selection), with the possibility to save the outputs in different formats (images, \ascii{} files, etc.).
The web page consists of a selection box allowing to choose a CR quantity (numerator and denominator) and to display all native and combined data (see App.~\ref{app:rules})\footnote{Publications cannot provide all combinations of their data: for instance, if $n$ fluxes are measured, $n\times (n-1)/2$ independent ratios can be formed, which can be a large number. For this reason, when extracting data, the query looks first among native data, and if none is found, then for possible combinations leading to the desired quantity.}---for caution about the usage of overlapping data taking periods for data from the same experiment, we refer the reader to App.~\ref{app:t-overlap}.
More selection criteria allow to ask for matching sub-exp (partial or full) names, specific dates and energy range, and flux rescaling with energy. In the context of long time series provided by both the AMS-02 \citep{2018PhRvL.121e1101A,2018PhRvL.121e1102A,2019PhRvL.123r1102A} and PAMELA \citep{2018ApJ...854L...2M} experiments, \crdbv{4.0} allows to select only (or discard) data from time series, in order not to overcrowd the display of data. This is illustrated on a `time-series only' selection in Fig.~\ref{fig:plot-timeseries}, where we show Carrington rotation-averaged PAMELA H data from 2006 to 2010 \citep{2013ApJ...765...91A}. This list triggers on the presence of the word `average' (in yearly, monthly, daily, Bartels rotation\dots averages) in the `subexp-info' key in the \code{SUBEXP} table (see Sect.~\ref{sec:table-subexp}).
A critical choice is that of the energy axis (among $E_{k/n}$, $R$, $E_{\rm tot}$, etc.): if no native data exist for this energy axis, conversion are performed to rescale the data from its native axis to the queried one. However, this conversion is only possible for fluxes (not ratios), and is exact for CR isotopes and leptons (whose $A$, $Z$, and $m$ are identified) but approximate for elements and group of elements (see App.~\ref{app:Econversion}).

Hitting the `Extract Selection' button pops-up a new window displaying: (i) a plot of the data, (ii) list of meta-data (per sub-exp) associated with the data, (iii) a summary of the extraction process (i.e. whether combinations and approximations were made), and (iv) a full listing of all retrieved data points (energy, values, and uncertainties). In \crdbv{4.0}, we improved the browsability between these informations. From the plot, we can export the data in a format compliant with the \usine{} and \galprop{} propagation code, or as a \code{cvs} or tarball of \ascii{} files. We can also directly retrieve images or \rootcern{}\footnote{\url{http://root.cern.ch}}-compliant macros ({\tt .root}, {\tt .C}). A `replot' button allows to further trim the data and modify the plot axes, look, and error bars shown, with the same options (as in the original plot) to export and save the results.

\begin{figure}[t]
\begin{center}
\frame{\includegraphics[width=0.75\columnwidth]{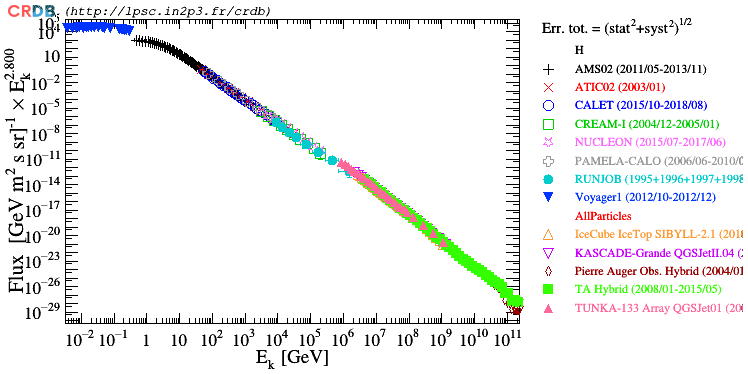}}
\\
[5pt]
\frame{\includegraphics[width=0.75\columnwidth]{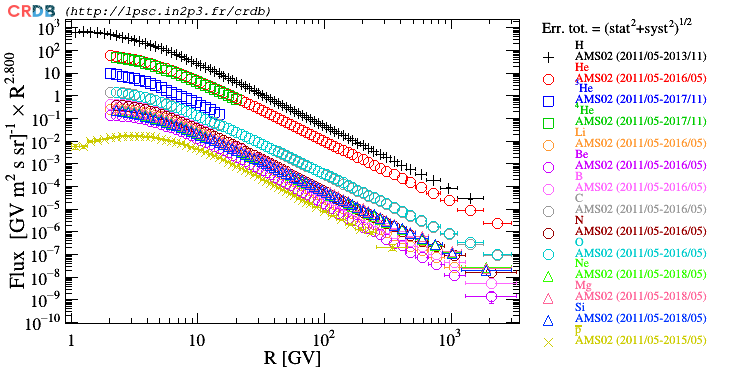}}
\caption{Illustration of the new multi-quantity plotting capability in \crdbv{4.0}: {\em Top panel:} \code{H} flux from IS MeV data 
(Voyager1 \citep{2013Sci...341..150S}) up to PeV data (AMS02 \citep{2015PhRvL.114q1103A}, ATIC02 \citep{2009BRASP..73..564P}, CALET \citep{2019PhRvL.122r1102A}, CREAM-I \citep{2011ApJ...728..122Y}, NUCLEON \citep{2019AdSpR..64.2546G}, RUNJOB \citep{2005ApJ...628L..41D}, PAMELA-CALO \citep{2013AdSpR..51..219A}); and  \code{AllParticle} flux from PeV to EeV energies
(KASCADE-Grande \citep{2015ICRC...34..263S}, IceCube IceTop \citep{2015ICRC...34..334R}, and Pierre Auger Observatory \citep{2015arXiv150903732T}), Telescope Array \citep{2015ICRC...34..349I}, and TUNKA-133 Array \citep{2014NIMPA.756...94P}). {\em Bottom panel:} Nuclear component measured to date by AMS-02 \citep{2015PhRvL.114q1103A,2016PhRvL.117i1103A,2017PhRvL.119y1101A,2018PhRvL.120b1101A,2018PhRvL.121e1103A,2019PhRvL.123r1102A,Aguilar:2020ohx}.}
\label{fig:tab-dataextract}
\end{center}
\end{figure}

To conclude on this tab, we are also pleased to provide in \crdbv{4.0} an additional selection option, that allows to display a comma-separated list of quantities: we stress that all selection criteria (specific energy range, dates, sub-experiments, etc.) are enforced to all quantities in the list. An illustration is provided Fig.~\ref{fig:tab-dataextract}, where we show the full CR spectrum from MeV to EeV energies from a selection of recent experiments (top panel), and a selection of all nuclear fluxes from the AMS-02 experiment (bottom panel).

\subsubsection{`Admin' tabs}
\label{sec:admin-tab}
\begin{figure}[!t]
\begin{center}
\includegraphics[width=0.85\columnwidth]{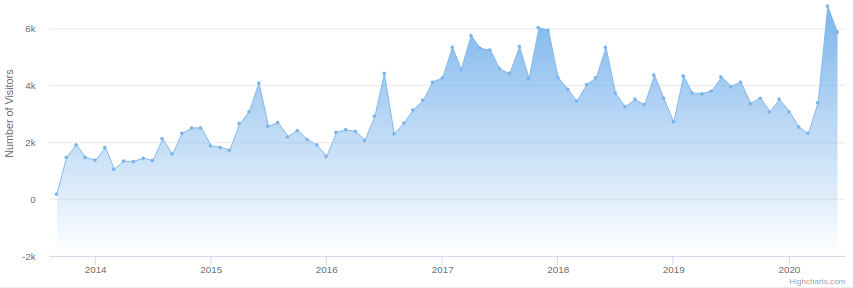}
\caption{Statistics of monthly \crdb{} connections since August 2013. The peak in May 2020 is related to queries made for systematic checks before the release of \crdbv{4.0}; to a lesser extent, a similar peak is visible in August 2016 related to the release of \crdbv{3.0}. In \crdbv{4.0}, this traffic plot is now accessible for any user via a link in the `Welcome' tab.}
\label{fig:stat}
\end{center}
\end{figure}

This tab can be accessed by authenticated users only (\crdb{} maintainers). In this page, various scripts provide internal checks of the database content (missing images for the detectors, orphan ID in the various tables, list of superseded publications, etc.). This page used to handle validation forms for submitted data, but they were removed thanks to the new submission form in \crdbv{4.0} (see Sect.~\ref{sec:tab-submitdata}). We can also monitor from this page the traffic on \crdb{}, stored in a specific \code{LOG\_QUERIES} table (connection date, IP address, and page visited, see Fig.~\ref{fig:mysql}). This traffic since August 2013 is shown in Fig.~\ref{fig:stat}. \crdb{} received more than a quarter million queries, from $\sim20,000$ different IP addresses in over a hundred countries---most queries originate from Germany, USA, Italy, France, China, Switzerland, and Japan, i.e. countries strongly involved in recent experiments and phenomenology.

\subsubsection{`Useful links' tab}
This tab gathers links to many online CR resources. These links include propagation codes for Solar, Galactic and extra-galactic CRs, other useful CR databases, and CR-related websites. If you feel that some important resources are missing, please contact us (\href{mailto:crdb@lpsc.in2p3.fr}{\tt crdb@lpsc.in2p3.fr}), we will be happy to add them.

\subsection{`Solar modulation' tab}
\label{sec:tab-phi}
This tab was added in \crdbv{3.0} and thus not described in \citet{2014A&A...569A..32M}. Its purpose is to provide Solar modulation level $\phi_{\rm FF}$ in the Force-Field approximation \citep[e.g.][]{2004JGRA..109.1101C} for any date between the 50's and today. The calculation is based on neutron monitor (NM) data and the reconstruction algorithm of \citet{2017AdSpR..60..833G}.

\begin{figure}[t]
\begin{center}
\frame{\includegraphics[width=0.8\columnwidth]{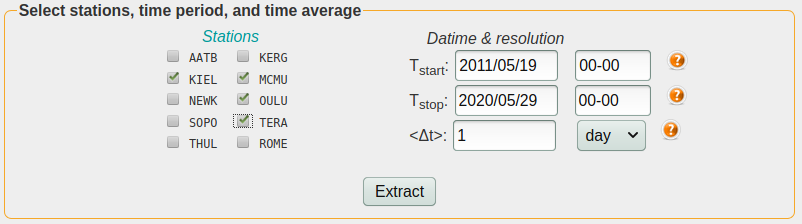}}
\\
[3pt]
\frame{\includegraphics[width=0.8\columnwidth]{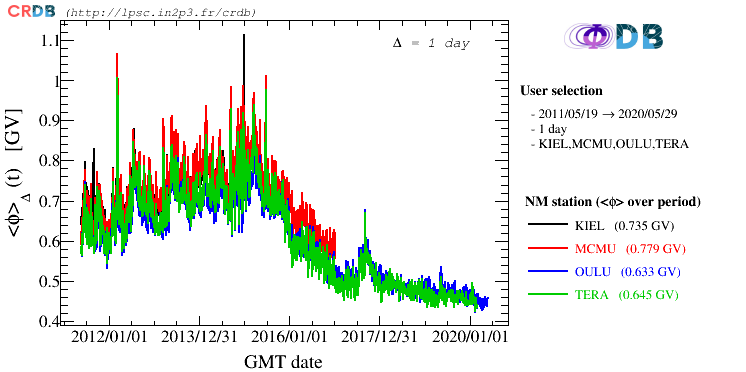}}
\caption{Illustration of $\phi$ time series retrieved from the `$\phi$ from NM' tab, based on the calculation of \citet{2017AdSpR..60..833G}. The top panel shows the user-selection for NM stations (OULU, KIEL, MCMU, and TERA) and time interval 2011/05 to today, corresponding to the time since AMS-02 was installed on the ISS. The bottom panel shows the resulting daily averaged values $\langle\phi\rangle_d\,(t)$ as coloured lines for different stations.}
\label{fig:phi_timesseries}
\end{center}
\end{figure}

Neutron monitors are devices developed in the 50's to monitor Solar activity \citep{2000SSRv...93...11S}. They combine a very good time resolution (a few minutes) and a good stability over decades. For this reason, they are well-suited to get time series over large time periods \citep{2015AdSpR..55..363M}. These values are for instance used to fill $\phi_{FF}$ from any \crdb{} sub-experiment as discussed in Sect.~\ref{sec:table-subexp}. Averaged $\phi(t)$ time series (bottom panel) can also be retrieved, as illustrated in Fig.~\ref{fig:phi_timesseries}, from a selection (top panel) of NM stations, a time period, and a time resolution (from the finer-grain 10~minute to monthly averaged time series).

In practice, scripts on LPSC servers retrieve and process NM data from \href{http://www01.nmdb.eu}{NMDB}. They produce, behind the scene, \ascii{} files for all available stations at a 10~minute time resolution. Then for each \crdb{} query, another layer of scripts reads these files to calculate and display the user selection. To ensure a continuous service and update, a \cron{}\footnote{\url{https://en.wikipedia.org/wiki/Cron}} time-based job scheduler is used on a daily basis.

\subsection{`\rest{} access' tab}
\label{sec:tab-rest}

The database offers a \rest\footnote{\url{https://en.wikipedia.org/wiki/Representational_state_transfer}} interface, which makes it easy to download data sets according to selection criteria from another program. The interface has been available since version \crdbv{1.3}, but the new help page in \crdbv{4.0} makes this feature more prominent.

Data sets can be selected by quantity (element, isotope, electron or positron, or mass group), experiment, energy range, and time range. Users must specify the energy type in their query. As mentioned in previous sections, data sets are stored in their native energy type, which can be kinetic energy, kinetic energy per nucleon, total energy, rigidity. The native type is converted to match the request, and details on the conversion are given in App.~\ref{app:Econversion}. Other parameters of the interface control the output format, and whether ratios should be synthesised from individual flux measurements.

\begin{table}
{
\caption{\rest{} interface parameters, description, and examples values. Only the first three parameters are mandatory. See text for how to use \rest{} in a command line.}
\label{tab:rest}
\centering
\begin{tabular}{l p{5.9cm} p{5.6cm} }
\toprule
\bf Parameter & \bf Description & \bf Allowed values (examples) [default] \\
\midrule

\multicolumn{3}{c}{\em Native data selection (mandatory)}\\
\\[-0.8em]

\code{num}$^a$ & Numerator of quantity to retrieve (element, isotope, lepton, mass group)& \crdb{} name (\code{H}, \code{1H-BAR}, \code{e-}, \code{O-group}) \\

\\[-0.75em]

\code{den}$^a$ & Denominator of quantity to retrieve (only for ratio, leave empty for flux)& \crdb{} name (\code{He}, \code{1H}, \code{e-+e+}, \code{AllParticles})\\

\\[-0.75em]

\code{energy\_type} & Energy axis for returned quantities: kinetic energy per nucleon (GeV/n), kinetic energy (GeV), rigidity (GV), total energy (GeV), or total energy per nucleon (GeV/A) & \crdb{} keyword (\code{EKN}, \code{EK}, \code{R}, \code{ETOT}, \code{ETOTN})\\

\\[-0.1em]
\multicolumn{3}{c}{\em Data combinations and/or modifications (optional)}\\
\\[-0.8em]

\code{combo\_level}$^b$ & Add combinations via ratio or product of native data (from the same sub-exp at the same energy) that match quantities in \code{list} (e.g. compute B/C from native B and C). Three levels of combos are enabled: \code{0} (native data only, no combo), \code{1} (exact combos), or \code{2} (exact and approximate combos): in level \code{1}, the mean energy (or energy bin) of the too quantities must be within 5\%, whereas for level \code{2}, it must be within 20\% & Integer in 0-2 (\code{0}, \code{1}, \code{2}) [\code{1}]\\
\\[-0.75em]
\code{energy\_convert\_level}$^c$\!\!\!\!\!\! & Add data obtained from an exact or approximate \code{energy\_type} conversion (from native to queried). Three levels of conversion are enabled: \code{0} (native data only, no conversion), \code{1} (exact conversion only, which applies to isotopic and leptonic fluxes), and \code{2} (exact and approximate conversions, the latter applying to flux of elements and of groups of elements) & Integer in 0-2 (\code{0}, \code{1}, \code{2}) [\code{1}]\\
\\[-0.75em]
\code{flux\_rescaling} & Multiply fluxes by $\langle E \rangle^\mathtt{flux\_rescaling}$ & Decimal number (1, 2.5) [0.] \\

\\[-0.1em]
\multicolumn{3}{c}{\em Trimming the selection (optional)}\\
\\[-0.8em]

\code{energy\_start} & Lower limit for \code{energy\_type} & Floating point number (2e-1, 1.2e2) [1.e-40] \\
\code{energy\_stop} & Upper limit for \code{energy\_type} & Floating point number (1e5, 5.3e7) [1.e40] \\
\\[-0.75em]
\code{time\_start} & Lower limit for interval selection & Date\,\code{YYYY}\,or\,\code{YYYY/MM}\,(2014,\,2010/06)\,[1950] \\
\code{time\_stop} & Upper limit for interval selection & Date\,\code{YYYY}\,or\,\code{YYYY/MM}\,(2020,\,2019/06)\,[2050] \\
\\[-0.75em]
\code{time\_series} & Whether to discard, select only, or keep time series data in query & \crdb{} keyword (\code{no}, \code{only}, \code{all}) [\code{no}] \\
\\[-0.75em]
\code{exp\_dates} & Comma-separated list (optional time intervals) of sub-experiment partial or full names  & Partial name (\code{AMS}, \code{ACE}), full name (\code{BESSPolarI}), or combinations of names and time intervals (\code{PAMELA(2006:2008)}, \code{BESS(2002/01:2004/12)}) [n/a]\\

\\[-0.1em]
\multicolumn{3}{c}{\em Outputs (optional)}\\
\\[-0.8em]

\code{format} & File output format & \crdb{} keyword\,(\code{csv},\,\code{usine},\,\code{galprop})\,[\code{csv}]\!\!\!\! \\
\code{modulation} & Source of Solar modulation values & \crdb{} keyword\,(\code{USO05},\,\code{USO17},\,\code{GHE17})\,[\code{GHE17}]\!\!\!\!\\

\bottomrule
\end{tabular}
\\[0.5em]

\begin{tabular}{l@{\hspace{0.5em}} p{0.9\textwidth}}
a) & Url syntax requires \code{e\%2B} instead of \code{e+} for a positron. \\
b) & See App.~\ref{app:rules} for a discussion on how to form combos and propagate errors from the native data. \\
c) & Energy type conversions are exact for isotopic and leptonic fluxes, but not for elements, since an elemental flux in general is the sum of several isotope fluxes in often unknown fractions. Usually, isotope fractions have to be assumed (see Sect.~\ref{app:Econversion}). \\
\end{tabular}
}
\end{table}

The available parameters are documented in Table~\ref{tab:rest} (also reproduced on the website). To give an example, the boron-to-carbon flux ratio as a function of the kinetic energy per nucleon can be queried with the command-line utility \texttt{curl} as follows:
\begin{verbatim}
     curl -L 'http://lpsc.in2p3.fr/crdb/rest.php?num=B&den=C&energy_type=EKN' > db.dat
\end{verbatim}
The output in USINE format (table with columns separated by spaces) is stored in the file \code{db.dat}. Queries can be made from any general purpose programming language. For Python users, we provide a simple module to run queries from Python and return the result as a \code{numpy} array. The code and a tutorial with example usage are available on GitHub\footnote{\url{https://github.com/crdb-project/tutorial}} and linked to from the \crdb{} website.

\subsection{`Submit data' tab: new submission form}
\label{sec:tab-submitdata}

In \crdbv{2.1}, the submit data interface was based on a sequential online 4-step procedure. The user had to fill meta-data first (experiment, sub-experiment, and then publication) and the data. Each step had to be completed and validated by one of us (via the admin interface) before being able to go to the next step. In practice, because of the too many steps, probably not documented enough procedure, and delay between the validation steps, this was almost never used.

\begin{table}[t]
\caption{Description of the 23 expected columns in the \code{csv} file prepared by the user to submit data. The table shows the column ID, the associated keyword in the database structure (see Fig.~\ref{fig:mysql}), the expected content of the column, and an example (and associated unit) for this column---entries corresponding to starred keywords$^\star$ can be left empty (i.e. `` '') if the user is unsure. For more details on \crdb{} data and meta-data definitions, see \S\ref{sec:table-data} (data points), \S\ref{sec:table-exp} (experiments), \S\ref{sec:table-subexp} (sub-experiments), and \S\ref{sec:table-publi} (publications).}
\label{tab:submit_data}
\centering
\begin{tabular}{rlll}
\toprule
  \textbf{Col.} & \textbf{(meta-)data}  &  \textbf{Description} & \textbf{`Example''  [Unit]}\\
\midrule
 1  & \code{EXP-NAME}                   &  Experiment name                             & ``AMS02''  \\
 2  & \code{EXP-TYPE}                   &  \code{balloon}, \code{ground}, or \code{space} & ``\code{space}''  \\
 3  & \code{EXP-HTML}$^\star$           &  Experiment official website                 & ``\url{http://www.ams02.org}''  \\
 4  & \code{EXP-STARTYEAR}$^\star$      &  Experiment starting year                    & ``2011''  \\
 5  & \code{SUBEXP-NAME}$^\star$        &  Name specific to analysis (\S\ref{sec:table-subexp}) & ``AMS02 (2011/05-2012/12)''  \\
 6  & \code{SUBEXP-DESCRIPTION}$^\star$ &  Detector information                        & ``9 planes of silicon tracker, TRD, etc''  \\
 7  & \code{SUBEXP-ESCALE\_RELERR}      &  Energy scale relative error (fraction)      & ``0.15''  \\
 8  & \code{SUBEXP-INFO}$^\star$        &  Summary or additional info                  & ``18 months data 2011/05-2012/12''  \\
 9  & \code{SUBEXP-DISTANCE}            &  Sub-exp. average distance to Sun            & ``1'' [AU] \\
 10 & \code{SUBEXP-DATES}               &  Start-stop data taking period               & ``2011/05/19-000000:2012/12/10-000000''  \\
 11 & \code{PUBLI-HTML}                 &  Publication ADS ID or \href{https://www.doi.org/}{doi} name & ``\href{https://ui.adsabs.harvard.edu/abs/2013PhRvL.110n1102A}{2013PhRvL.110n1102A}''  \\
 12 & \code{PUBLI-DATAORIGIN}$^\star$   &  Source for the data in the publication      & ``Table 1, Figs. 2 and 3'' \\
 13 & \code{DATA-QTY}$^\dagger$         &  Measured quantity (\crdb{} names)           & ``B/C''  \\
 14 & \code{DATA-EAXIS}                 &  \code{R} [GV],  \code{EK} [GeV], \code{ETOT} [GeV], &  ``\code{R}'' [GV]\\
    &                                   &  \code{EKN} [GeV/n], or \code{ETOTN} [GeV/A] &  \\
 15 & \code{DATA-E\_M}EAN               &  $\langle E\rangle$ for bin (empty if no value) &  `` '' \\
 16 & \code{DATA-E\_BIN\_L}             &  $E_{\rm lo}$ in unit of [\code{DATA-EAXIS}]  & ``4.88'' [GV] \\
 17 & \code{DATA-E\_BIN\_U}             &  $E_{\rm up}$ in unit of [\code{DATA-EAXIS}]  & ``5.37'' [GV]  \\
 18 & \code{DATA-VAL}                   &  $y$, measured value                   &  ``0.3214'' [m$^{-2}$~s$^{-1}$~sr$^{-1}$~GV$^{-1}$]  \\
 19 & \code{DATA-VAL\_ERRSTAT\_L}       &  $\Delta y_{\rm lo}$, lower stat. err.  &  ``\,-\,0.0011'' [m$^{-2}$~s$^{-1}$~sr$^{-1}$~GV$^{-1}$]  \\
 20 & \code{DATA-VAL\_ERRSTAT\_U}       &  $\Delta y_{\rm up}$, upper stat. err.  &  ``+0.0011'' [m$^{-2}$~s$^{-1}$~sr$^{-1}$~GV$^{-1}$]  \\
 21 & \code{DATA-VAL\_ERRSYST\_L}       &  $\Delta y_{\rm lo}$, lower syst. err.   & ``\,-\,0.075'' [m$^{-2}$~s$^{-1}$~sr$^{-1}$~GV$^{-1}$]  \\
 22 & \code{DATA-VAL\_ERRSYST\_U}       &  $\Delta y_{\rm up}$, upper syst. err.   & ``+0.0089'' [m$^{-2}$~s$^{-1}$~sr$^{-1}$~GV$^{-1}$]  \\
 23 & \code{DATA-ISUPPERLIM}            &  ``0'' (measured) or ``1'' (upper limit)     & ``0'' \\
\bottomrule
\end{tabular}
\\[7pt]
\small{
$^\star$Columns that can be left empty (but not omitted) if the user is unsure about what he should write.
\\$^\dagger$ To help pick the correct denomination for the measured quantities, we list in the webpage all currently defined names. If necessary, quantities not defined yet will be added along with the submitted data.}
\end{table}
In \crdbv{4.0}, we decided to completely change the approach and only ask for a single \code{csv} file, whose expected columns are described in Table~\ref{tab:submit_data} (and also reported on the `Submit data' tab); files must be sent to \href{mailto:crdb@lpsc.in2p3.fr}{\tt crdb@lpsc.in2p3.fr}. One of us then uses dedicated scripts on these files to finalise the upload on \crdb{}.
In this new format, informations on the data and meta-data are still required, but to simplify the submitter task, many entries can be left empty (and will be filled by us), as indicated by stars in Table~\ref{tab:submit_data}.
We hope that this simplified submission procedure and lessened number of meta-data entry to provide will convince experimentalist to make the extra step of submitting the data to \crdb{} after they submit their publication to a journal.

\section{Summary and future developments}
\label{sec:concl}

In this article, we have presented \crdbv{4.0}, a new version of the CR database for charged species, highlighting several improvements on the database structure and content, and on the user interfaces. We summarise the changes made below:
   \begin{itemize}
      \item Database structure: a few tables were simplified and three became deprecated (Solar modulation table from the publication, users and validation tables for the old data submission interface). The pre-defined key values were also modified for the experiment type (now \code{balloon}, \code{ground}, or \code{space}) and energy type (now also includes \code{ETOTN}). Two new keys were added in the in the \code{SUBEX\_PUBLI} and \code{PUBLI} tables respectively, in order to keep track of the data origin (retrieved from from a table or figure in the publication, etc.), and to store a possible energy scale uncertainty in the data (single number for a given sub-experiment).

      \item Database content: this version is associated with a massive upload of CR data. For \crdbv{3.1}, released a couple of weeks before this version, we filled important GCR high-precision data published in the last two years (AMS-02, DAMPE, NUCLEON, TRACER, Voyager, etc.). For \crdbv{4.0}, we extended the CR data content towards higher energies (KASCADE-GRANDE, Pierre Auger Observatory, etc.), higher charges $Z>30$ (LDEF UHCRE, TREK, etc.), and lower charges $Z<-1$ (upper limits  from AMS-01, BESS, PAMELA, etc.). The database now contains more than 100 experiments and 350 publications, covering 14 decades in energy (from $10^6$ to $10^{20}$~eV).

      \item User interfaces: the many improvements we implemented provide users with more options to retrieve data, along with a clearer presentation of the various interfaces. These changes include: (i) more data selection options (e.g. on times series) and plotting/export options (.pdf, .csv), including the possibility to display several CR quantities on the same plot; (ii) new help page and an example python notebook script to retrieve data via the \rest{} interface; (iii) updated interface to retrieve Solar modulation level $\phi_{\rm FF}$ time series (or average) from any time period since 1950 with its dedicated \rest{} interface, (iv) simplified file format to submit new data; (v) extended list of links for online CR resources. We hope in particular that the new submission format, with clearer explanations about the data and meta-data to fill, will help us always keeping \crdb{} up-to-date in the future.

\end{itemize}

\paragraph {\em Final thoughts on \crdb{} evolution}
\crdb{} was thought from the start as a service to the CR community. In an ideal world, we would like experimentalists to submit their data to \crdb{} as soon as their results appear on the arXiv preprint server\footnote{\url{https://arxiv.org/}} or in a peer-review journal.

There are several possible directions along which \crdb{} could be improved. On a technical aspect, we are planning to make the website responsive for a better rendering on diverse electronic devices. On the data content aspect, we will obviously continue to upload new data as they appear, as quickly as we can.
We would also be grateful for any help to extend the datasets to new quantities. Indeed, without any change in the database structure, we could quite easily include (i) more relevant quantities for UHECRs, i.e. the mean logarithmic mass $\langle\ln A\rangle$, the mean depth of shower maximum $\langle X_{\rm max}\rangle$, etc.; (ii) dipole anisotropy spectral data as presented in \citet{2017PrPNP..94..184A}, (iii) upper limits or data on the neutrino CR spectrum discovered by the \citet{2013Sci...342E...1I}. As a first step in this direction, near the completion of this article, a preliminary agreement was reached with the KCDC team \citep{2018EPJC...78..741H} to include their data in a forthcoming \crdb{} release.
Finally, we could also imagine \crdb{} as a portal to gather more CR-related data, for instance hosting the many nuclear cross-section data presented in \citet{2018PhRvC..98c4611G}. These future developments will depend on the workforce available and feedback from the CR community.

To conclude, any help and feedback to further expand the database is welcome. Comments, questions, suggestions, and corrections are to be sent at \href{mailto:crdb@lpsc.in2p3.fr}{\tt crdb@lpsc.in2p3.fr}.

\vspace{6pt}
\authorcontributions{Conceptualisation, D.M., H.D., J.G., and I.M.; methodology, D.M and H.D.; software, F.M. and D.M.; writing--original draft preparation, D.M. and H.D.; writing--review and editing, D.M., H.D., J.G, I.M; supervision, D.M.. All authors have read and agreed to the published version of the manuscript.}

\funding{This work was supported by the Programme National des Hautes Energies of CNRS/INSU with INP and IN2P3, co-funded by CEA and CNES.}

\acknowledgments{We warmly thank L. Baldini, F. Barao, M. Boudaud, B. Coste, C. Deil, T. Delahaye, L. Derome,  F. Donato, M. Duranti, K. Egberts, C. Evoli, H. Gast, J. Gieseler, A. Gil-Swiderska, F. Giovacchini, I. Grenier, S. Haino, G. Jóhannesson, A. Karelin, J. Lavalle, M. Orcinha, M. Paniccia, A. Panov, A. Putze, T. Räihä, B. Shan, A. Strong, L. Sujie, M. Vecchi, A. Vincent, P. von Doetinchem, W. Xu, and S. Zimmer for their feedback that helped correct data in \crdb{}. D.M. thanks C. Combet for a careful reading of the manuscript, and K. Lockhart for her help on the use of the ADS API. This research has made use of NASA’s Astrophysics Data System Bibliographic Services.}

\conflictsofinterest{The authors declare no conflict of interest}

\appendix
\section{Tips and caveats on data and their extraction}
\label{app:tips}

\crdb{} users go to the online `Data extraction' tab or use the \rest{} interface to retrieve data, which is fine. However, they generally overlook how the extraction was made and miss possible caveats with some of the data. Below, we go through a few important items that users should keep in mind. We stress that the \rest{} interface goes through exactly the same steps to extract data from \crdb{} and thus the same cautions apply.

\subsection{Difference between `native' and `combined' data}
\label{app:rules}

We recall the core data of \crdb{} are `native' only, i.e. measurements made available in the publications (in tables or extracted from figures\footnote{In the former case, the data match exactly those provided by the experimental teams, but in the latter case, the manual procedure may lead to data points (energy, central value, and uncertainties) that may deviate from the intended ones. This is especially the case for data shown with large symbols or for fluxes covering several decades in logarithmic scale. However, the precision on the data and uncertainty values obtained via the manual extraction is much better than the size of the data uncertainty (for most data extracted), so that this additional deviation is not an issue. Obviously, if a specific data set obtained by this procedure is crucial for a user's analysis, the user is encouraged to check on the original figure in the publication.} using {\sc DataThief~III}\footnote{\url{http://datathief.org}}). However, many useful combinations can be formed from native data. For instance, starting from C and O native fluxes in \crdb{}, the user can extract C/O. In order to get meaningful results, this procedure has to ensure that C and O are from the same energy range or use the same central energy point. As described in details in App.~A of \citet{2014A&A...569A..32M}, priority rules exist to form such ratios, and tick boxes in the `Data extraction' interface enable to select the energy tolerance for which \crdb{} accepts to form the ratios, in order to have {\em exact} (energy point within 5\%) or {\em approximate} (energy point between [5\%-20\%]) combinations.

\subsection{Data uncertainties in \crdb{}}
\label{app:data_unc}
Most pre-80's experiments were sensitivity-limited, meaning that the error budget was dominated by statistical uncertainties. When retrieved from \crdb{}, these data only show statistical errors and no systematics.

With improved detectors, the practice in the literature shifted towards paying more and more attention to systematic uncertainties. The latter are of various origins (inefficiencies, fragmentation in the detector, etc.). They are either provided as raw text description in the publication (see the example of HEAO-3 data discussed in App.~A.2 of \cite{2014A&A...569A..32M}), or as a single number per data point in tables (separately from the statistical uncertainties). In the latter case, \crdb{} can directly provide statistical and systematics uncertainties as given in the publication. In the former case, we use the text description to estimate the systematic uncertainties, combining quadratically the various sources of systematics. In case in which the systematics description is not clear enough, or when data are indirectly retrieved (i.e. taken from plots), all the uncertainties are assumed to be of statistical origin\footnote{In principle, statistical errors should be symmetric, but when they include sub-dominant systematics or when data are manually retrieved from plots, asymmetric error bars are obtained (which are filled as statistical uncertainties in \crdb{}).}.

With the advent of high-precision experiments (e.g. AMS-02), the situation has become even trickier. It is now routine in publications to have tables with statistical and several systematics separately. Furthermore, as highlighted in \citet{2019A&A...627A.158D}, systematics uncertainties usually are likely to be correlated for nearby energy bins. We have not yet thought of modifications in \crdb{} to handle more than one systematics in the extraction, even less accounting for covariance matrices of systematics. Until further notice, all systematics are combined quadratically  and provided as a single number per data in \crdb{}.

\subsection{Energy-scale uncertainty}
\label{app:Escale}

Energy measurements of CR detectors suffer from two types of distortions, random and systematic. The random distortions differ from event to event and average to zero over many events. In flux measurements, these types of distortions lead to a softening of the measured flux compared to the true flux. Correcting these distortions is possible with unfolding methods, which is the responsibility of each experiment. In the following we discuss the second kind of uncertainty. The second type of distortions are systematic and originate from the residual uncertainty of the energy calibration of the detector. These distortions are the same for each event, do not average out, and lead to shifts in the measured flux compared to the true flux.

In general, the energy distortion itself varies with the (true) energy, but experiments are designed and controlled to keep the relative calibration of energies within the covered energy range to higher accuracy. Therefore, the uncertainty can be usually summarised by a single number, the relative energy-scale uncertainty, for each experiment. Since the CR fluxes are mostly steeply falling power laws $(\propto E^{-3})$, even a small energy-scale uncertainty has a significant impact on the uncertainty on the flux. An uncertainty of 10\,\% in the energy scale translates to a 30\,\% uncertainty on the flux.

In practice, experiments either include this as a systematic uncertainty for each data point, or quote the energy-scale uncertainty separately (and not propagate it to the flux). This is especially common for air shower experiments, which have energy-scale uncertainties of 10\,\% to 24\,\%.
Whenever available in the publications, we fed this number to \crdb{}, so that it can be retrieved by users. However, we stress out that this uncertainty is never accounted for in \crdb{} displays.

\begin{table}[t]
\caption{Conversion between \crdb{} energy axes. The columns are (i) the native energy axis $(E)$, (ii) the queried axis $E^*$ to which to convert, (iii) the relation between $E^*$ and $E$, (iv) the relation between the converted flux $dJ/dE^*$ and the native flux $dJ/dE$, and (v) $\beta=v/c$ expressed as a function of the native energy axis $E$. In the formulae below, $m$ is the CR mass in GeV, and energies are in GeV ($E_k$ and $E_{\rm tot}$), GeV/n ($E_{k/n}$), GV ($R$), or GeV/A ($E_{\rm tot/A}$).}
\label{tab:convert}
\centering
\begin{tabular}{llllc}
\toprule
Native $E$       & Queried $E^*$   & $E^*=f(E,A,m,Z)$  & $\displaystyle \frac{(dJ/dE^*)}{(dJ/dE)}$ & \textbf{$\beta_E$}\\
\midrule
  $E_{k/n}$      & $E_k$           & $E_k=AE_{k/n}$                        & $1/A$                        & \multirow{4}{*}{$\displaystyle\beta_{E_{k/n}}=\frac{\sqrt{AE_{k/n}(AE_{k/n}+2m)}}{(AE_{k/n}+m)}$} \\
  $E_{k/n}$      & $R$             & $R=\sqrt{AE_{k/n}(AE_{k/n}+2m)}/|Z|$ & $|Z|\beta_{E_{k/n}}/A$        &  \\
  $E_{k/n}$      & $E_{\rm tot}$   & $E_{\rm tot}=AE_{k/n}+m$              & $1/A$                        &  \\
  $E_{k/n}$      & $E_{\rm tot/A}$ & $E_{\rm tot/A}=E_{k/n}+ m/A$          & $1$                          &  \\[7pt]
  $E_k$          & $E_{k/n}$       & $E_{k/n}=E_k/A$                       & $A$                          &  \multirow{4}{*}{$\displaystyle\beta_{E_k}=\frac{\sqrt{E_k(E_k+2m)}}{(E_k+m)}$}\\
  $E_k$          & $R$             & $R=\sqrt{E_k(E_k+2m)}/|Z|$            & $|Z|\beta_{E}$               &  \\
  $E_k$          & $E_{\rm tot}$   & $E_{\rm tot}=E_k+m$                   & $1$                          &  \\
  $E_k$          & $E_{\rm tot/A}$ & $E_{\rm tot/A}=(E_k+m)/A$             & $A$                          &  \\[7pt]
  $R$            & $E_{k/n}$       & $E_{k/n}=\sqrt{(RZ)^2+m^2)-m}/A$      & $A/(|Z|\beta_{R}$)           &  \multirow{4}{*}{$\displaystyle\beta_{R}=\frac{R}{\sqrt{R^2+(m/Z)^2}}$}\\
  $R$            & $E_k$           & $E_k=\sqrt{(RZ^2)+m^2}-m$             & $1/(|Z|\beta_{R}$)           &  \\
  $R$            & $E_{\rm tot}$   & $E_{\rm tot}=\sqrt{(RZ)^2+m^2}$       & $1/(|Z|\beta_{R}$)           &  \\
  $R$            & $E_{\rm tot/A}$ & $E_{\rm tot/A}=\sqrt{(RZ)^2+m^2}/A$   & $A/(|Z|\beta_{R}$)           &  \\[7pt]
  $E_{\rm tot}$  & $E_{k/n}$       & $E_{k/n}=(E_{\rm tot}-m)/A$           & $A$                          &  \multirow{4}{*}{$\displaystyle\beta_{E_{\rm tot}}=\frac{\sqrt{E_{\rm tot}^2-m^2}}{E_{\rm tot}}$}\\
  $E_{\rm tot}$  & $E_k$           & $E_k=(E_{\rm tot}-m)$                 & $1$                          &  \\
  $E_{\rm tot}$  & $R$             & $R=\sqrt{E_{\rm tot}^2-m^2}/|Z|$      & $|Z|\beta_{E_{\rm tot}}$     &  \\
  $E_{\rm tot}$  & $E_{\rm tot/A}$ & $E_{\rm tot/A}=E_{\rm tot}/A$         & $A$                          &  \\[7pt]
  $E_{\rm tot/A}$& $E_{k/n}$       & $E_{k/n}=E_{\rm tot/A}(m/A)$          & $1$                          &  \multirow{4}{*}{$\displaystyle\beta_{E_{\rm tot/A}}=\frac{\sqrt{E_{\rm tot/A}^2-(m/A)^2}}{E_{\rm tot/A}}$}\\
  $E_{\rm tot/A}$& $E_k$           & $E_k=AE_{\rm tot/A}-m$                & $1/A$                        &  \\
  $E_{\rm tot/A}$& $R$             & $R=\sqrt{(AE_{\rm tot/A})^2-m^2}/|Z|$ & $|Z|\beta_{E_{\rm tot/A}}/A$ &  \\
  $E_{\rm tot/A}$& $E_{\rm tot}$   & $E_{\rm tot}=AE_{\rm tot/A}$          & $1/A$                        &  \\
\bottomrule
\end{tabular}
\end{table}

\subsection{Energy axes and conversion to other energy axes}
\label{app:Econversion}

Most GCR data in \crdb{} are in kinetic energy per nucleon, as it was the standard for a long time. However, from an instrumental point of view, spectrometers measure rigidities and calorimeters the total energy, etc. With the advent of high-precision experiments, most data are published in their `natural' energy axis, and \crdb{} enables the following energy axes: $R$, $E_k$, $E_{k/n}$, $E_{\rm tot}$, and $E_{\rm tot}/A$.

For isotopic fluxes, the mass, atomic number, and charge are uniquely defined, so that conversions between different energy axes are exact and enabled in \crdb{}---both the energy and the flux are modified. We list in Table~\ref{tab:convert} the formulae used to move from a native energy axis $E$ into a queried axis $E^*$, and how the converted CR flux $dJ/dE^*$ is linked to the original data $dJ/dE$.

For measurements that can only resolve fluxes from elements (or groups of elements), the conversion is no longer possible, because the mandatory (energy-dependent) isotopic content to do so is unknown. Nevertheless, assuming that elements (or groups of elements) are dominated by a single isotope, an approximate conversion can be provided. As this proves useful for several users, this approximate conversion is enabled since \crdbv{1.2}. It takes as the dominant isotope the most abundant one in the Solar system \citep{2003ApJ...591.1220L}, and this is defined, along with the associated values for $A$, $Z$, and $m$, in the \code{ISOTOPE\_PROXY} table of the database (see Fig.~\ref{fig:mysql}). 
We list in Table~\ref{tab:cr_quantity} the proxies enabled for the new \code{CR\_QUANTITY} names introduced in \crdbv{4.0}.

\begin{table}[t]
\caption{List of new \crdb{} names for \code{CR\_QUANTITY} (see Sect.~\ref{sec:table-data}), along with their proxy (i.e. isotope) used to perform approximate energy axis conversions (n/a indicates that no energy conversion at all is enabled).}
\label{tab:cr_quantity}
\centering
\begin{tabular}{lll}
\toprule
\crdb{} name    & Description & Proxy for $E$ conversion\\
\midrule
   \multicolumn{3}{c}{\em For UHCR data (Sect.~\ref{sec:v4.0-UHCRs})}\\[3pt]
   \code{HS-group}      & Heavy secondary ($70\leq Z\leq73$)             &    n/a  \\
   \code{LS-group}      & Light secondary ($62\leq Z\leq69$)             &    n/a  \\
   \code{Pt-group}      & Platinum group of element ($74\leq Z\leq80$)   &    n/a  \\
   \code{Pb-group}      & Lead group of elements ($81\leq Z\leq87$)      &    n/a  \\
   \code{Subactinides}  & Pt+Pb groups ($74\leq Z\leq87$)                &    n/a  \\
   \code{Actinides}     & All the heaviest CRs ($Z\geq88$)               &    n/a  \\
   \code{Zgeq70}        & CRs heavier than charge 70 ($Z\geq70$)         &    n/a  \\[10pt]
   \multicolumn{3}{c}{\em For upper limits on antinuclei (Sect. \ref{sec:v4.0-UL})}\\[3pt]
   \code{Zgeq1}      & Flux of all CR nuclei $Z\geq1$                     &  $^1\rm H$ \\
   \code{Zgeq1-bar}  & Flux of all CR antinuclei $Z\leq-1$               &  $\overline{^1\rm H}$ \\
   \code{Zgeq2}      & Flux of all CR nuclei $Z\geq2$                     &  $^4\rm He$ \\
   \code{Zgeq2-bar}  & Flux of all CR antinuclei $Z\leq-2$               &  $\overline{^4\rm He}$ \\
   \code{H-BAR}      & Antihydrogen flux ($\overline{\rm H}$)            &  $\overline{^1\rm H}$ \\
   \code{1H-BAR}     & Antiproton flux ($\overline{p}$)                  &  n/a \\
   \code{2H-BAR}     & Antideuteron flux ($\overline{^2\rm H}$)          &  n/a \\
   \code{He-BAR}     & Antihelium flux ($\overline{\rm He}$)             &  $\overline{^4\rm He}$ \\
   \dots             & {(\em down to isotopes and elements $Z=-8$)}       &   \\[10pt]
   \multicolumn{3}{c}{\em For UHECR data (Sect.~\ref{sec:v4.0-UHECRs})}\\[3pt]
   \code{H-He-group} & Flux inferred from mixture of H and He in air-shower simulations & n/a \\
   \code{C-Fe-group} & Flux inferred from mixture of C to Fe in air-shower simulations  & n/a \\
   \code{N-group}    & Flux inferred from N in air-shower simulations   & $^{14}$N  \\
   \code{O-group}    & Flux inferred from O in air-shower simulations   & $^{16}$O  \\
   \code{Al-group}   & Flux inferred from Al in air-shower simulations  & $^{27}$Al  \\
   \code{Si-group}   & Flux inferred from Si in air-shower simulations  & $^{28}$Si  \\
   \code{Fe-group}   & Flux inferred from Fe in air-shower simulations  & $^{56}$Fe  \\
   \code{O-Fe-group} & Flux inferred from a mixture of O to Fe in air-shower simulations  & n/a \\
   \code{AllParticles} & Flux of all particles (sum over all isotopes and species)  & n/a \\
\bottomrule
\end{tabular}
\end{table}

\subsection{Data taking period (possible) overlap}
\label{app:t-overlap}

By default, for a given sub-exp, the data extraction tool returns all requested data, regardless of their data taking periods. We caution the user to check for any overlap between dates. Because of the time-dependent Solar activity, datasets from different though overlapping data taking periods can be of interest in the context of Solar modulation studies. However, if one is only interested in the accumulated statistics, one should only consider datasets accumulated on the longest time period.

In any case, datasets from overlapping data taking periods are not independent. The user has to critically decide which data sets are relevant for her/his analysis, and refine the selection criteria accordingly.

\section{List of sub-experiments and publications}
\label{app:bibtex}

In this appendix, we provide several lists of experiments and their associated publications, for several types of CR entries: UHECRs (App.~\ref{app:uhecr}), antinuclei (App.~\ref{app:UL}), antiprotons (App.~\ref{app:antiprotons}), leptons (App.~\ref{app:leptons}), nuclei $Z\leq30$ (App.~\ref{app:nuclei}), and UHCRs (App.~\ref{app:uhcr}). We sort them by experiment type (\code{balloon}, \code{space}, or \code{ground}) and in alphabetical order.

{

\subsection{UHECR~Data}
\label{app:uhecr}

\begin{description}
   \item \underline{Ground}
      \begin{itemize}[leftmargin=5pt]\vspace{-7pt}
         \item Pierre Auger Observatory Hybrid (2004/01-2014/12): \citeauthor{2015arXiv150903732T} \citeyear{2015arXiv150903732T} \cite{2015arXiv150903732T}
         \item Pierre Auger Observatory FDSIBYLL-2.1 (2004/12-2012/12): \citeauthor{2014PhRvD..90l2005A} \citeyear{2014PhRvD..90l2005A} \cite{2014PhRvD..90l2005A}
         \item Pierre Auger Observatory FDQGSJetII.04 (2004/12-2012/12): \citeauthor{2014PhRvD..90l2005A} \citeyear{2014PhRvD..90l2005A} \cite{2014PhRvD..90l2005A}
         \item Pierre Auger Observatory FDEPOS-LHC (2004/12-2012/12): \citeauthor{2014PhRvD..90l2005A} \citeyear{2014PhRvD..90l2005A} \cite{2014PhRvD..90l2005A}
         \item Telescope Array Hybrid (2008/01-2015/05): \citeauthor{2015ICRC...34..349I} \citeyear{2015ICRC...34..349I} \cite{2015ICRC...34..349I}
         \item KASCADE-Grande QGSJetII.04 (2003/01-2009/03): \citeauthor{2015ICRC...34..263S} \citeyear{2015ICRC...34..263S} \cite{2015ICRC...34..263S}
         \item TUNKA-133Array QGSJet01 (2009/10-2012/04): \citeauthor{2014NIMPA.756...94P} \citeyear{2014NIMPA.756...94P} \cite{2014NIMPA.756...94P}
         \item IceCube Hybrid SIBYLL-2.1 (2010/06-2013/05): \citeauthor{2015ICRC...34..334R} \citeyear{2015ICRC...34..334R} \cite{2015ICRC...34..334R}
         \item IceCubeIceTop SIBYLL-2.1 (2010/06-2013/05): \citeauthor{2015ICRC...34..334R} \citeyear{2015ICRC...34..334R} \cite{2015ICRC...34..334R}
         \item H.E.S.S. Average of QGSJet01 and SIBILL-2.1 (2004/01-2006/12): \citeauthor{2007PhRvD..75d2004A} \citeyear{2007PhRvD..75d2004A} \cite{2007PhRvD..75d2004A}
   \end{itemize}
\end{description}

\subsection{Upper Limits on Antinuclei ($Z<-1$ and $A\geq2$)}
\label{app:UL}

\begin{description}
   \item \underline{Balloon}
      \begin{itemize}[leftmargin=5pt]\vspace{-7pt}
         \item Balloon (1957/09): \citeauthor{1961PhRv..121.1206A} \citeyear{1961PhRv..121.1206A} \cite{1961PhRv..121.1206A}
         \item Balloon (1969/08): \citeauthor{1971ICRC....1..203G} \citeyear{1971ICRC....1..203G,1974ApJ...192..747G} \cite{1971ICRC....1..203G,1974ApJ...192..747G}
         \item Balloon (1970/05): \citeauthor{1972NPhS..240..135V} \citeyear{1972NPhS..240..135V} \cite{1972NPhS..240..135V}
         \item Balloon (1970/07): \citeauthor{1972ApJ...176..797E} \citeyear{1972ApJ...176..797E} \cite{1972ApJ...176..797E}
         \item Balloon (1970/XX): \citeauthor{1971Natur.230..170G} \citeyear{1971Natur.230..170G} \cite{1971Natur.230..170G}
         \item Balloon (1970/09+1971/05): \citeauthor{1972Natur.236..335B} \citeyear{1972Natur.236..335B} \cite{1972Natur.236..335B}
         \item Balloon (1970/09+1971/05+1972/09): \citeauthor{1975PhRvL..35..258S} \citeyear{1975PhRvL..35..258S} \cite{1975PhRvL..35..258S}
         \item Balloon (1975/09+1976/05): \citeauthor{1978Natur.274..137B} \citeyear{1978Natur.274..137B} \cite{1978Natur.274..137B}
         \item Balloon (1980/06): \citeauthor{1981ApJ...248.1179B} \citeyear{1981ApJ...248.1179B} \cite{1981ApJ...248.1179B}
         \item Balloon (1987/08): \citeauthor{1997ApJ...479..992G} \citeyear{1997ApJ...479..992G} \cite{1997ApJ...479..992G}
         \item BESS95 (1995/07): \citeauthor{1997ApJ...482L.187O} \citeyear{1997ApJ...482L.187O} \cite{1997ApJ...482L.187O}
         \item BESS97 (1997/07): \citeauthor{2005PhRvL..95h1101F} \citeyear{2005PhRvL..95h1101F} \cite{2005PhRvL..95h1101F}
         \item BESS98 (1998/07): \citeauthor{2005PhRvL..95h1101F} \citeyear{2005PhRvL..95h1101F} \cite{2005PhRvL..95h1101F}
         \item BESS99 (1999/08): \citeauthor{2005PhRvL..95h1101F} \citeyear{2005PhRvL..95h1101F} \cite{2005PhRvL..95h1101F}
         \item BESS00 (2000/08): \citeauthor{2005PhRvL..95h1101F} \citeyear{2005PhRvL..95h1101F} \cite{2005PhRvL..95h1101F}
         \item BESS-TeV (2002/08): \citeauthor{2008AdSpR..42..450S} \citeyear{2008AdSpR..42..450S} \cite{2008AdSpR..42..450S}
         \item BESS-PolarI (2004/12): \citeauthor{2008AdSpR..42..450S} \citeyear{2008AdSpR..42..450S} \cite{2008AdSpR..42..450S}
         \item BESS-PolarII (2007/12-2008/01): \citeauthor{2012PhRvL.108m1301A} \citeyear{2012PhRvL.108m1301A} \cite{2012PhRvL.108m1301A}
         \item BESS-COMBINED (1993-1995): \citeauthor{1998PhLB..422..319S} \citeyear{1998PhLB..422..319S} \cite{1998PhLB..422..319S}
         \item BESS-COMBINED (1993-2000): \citeauthor{2002NuPhS.113..202S} \citeyear{2002NuPhS.113..202S} \cite{2002NuPhS.113..202S}
         \item BESS-COMBINED (1993-2004): \citeauthor{2008AdSpR..42..450S} \citeyear{2008AdSpR..42..450S} \cite{2008AdSpR..42..450S}
         \item BESS-COMBINED (1993-2007): \citeauthor{2012PhRvL.108m1301A} \citeyear{2012PhRvL.108m1301A} \cite{2012PhRvL.108m1301A}
         \item BESS-COMBINED (1997-2000): \citeauthor{2005PhRvL..95h1101F} \citeyear{2005PhRvL..95h1101F} \cite{2005PhRvL..95h1101F}
      \end{itemize}
   \item \underline{Space}
      \begin{itemize}[leftmargin=5pt]\vspace{-7pt}
         \item AMS01 (1998/06): \citeauthor{1999PhLB..461..387A} \citeyear{1999PhLB..461..387A} \cite{1999PhLB..461..387A}, \citeauthor{2002NuPhS.113..195C} \citeyear{2002NuPhS.113..195C} \cite{2002NuPhS.113..195C}
         \item PAMELA (2006/07-2009/12): \citeauthor{2011JETPL..93..628M} \citeyear{2011JETPL..93..628M} \cite{2011JETPL..93..628M}
      \end{itemize}
\end{description}

\subsection{Antiproton~Data}
\label{app:antiprotons}

\begin{description}
   \item \underline{Balloon}
      \begin{itemize}[leftmargin=5pt]\vspace{-7pt}
         \item Balloon (1979/06): \citeauthor{1979PhRvL..43.1196G} \citeyear{1979PhRvL..43.1196G} \cite{1979PhRvL..43.1196G}
         \item Balloon (1979/06): \citeauthor{1984ApL....24...75G} \citeyear{1984ApL....24...75G} \cite{1984ApL....24...75G}
         \item Balloon (1980/06): \citeauthor{1981ApJ...248.1179B} \citeyear{1981ApJ...248.1179B} \cite{1981ApJ...248.1179B}
         \item Balloon (1984/06+1984/07): \citeauthor{1987ICRC....2...72B} \citeyear{1987ICRC....2...72B} \cite{1987ICRC....2...72B}
         \item BESS93 (1993/07): \citeauthor{1997ApJ...474..479M} \citeyear{1997ApJ...474..479M} \cite{1997ApJ...474..479M}
         \item BESS95 (1995/07): \citeauthor{1998PhRvL..81.4052M} \citeyear{1998PhRvL..81.4052M} \cite{1998PhRvL..81.4052M}
         \item BESS97 (1997/07): \citeauthor{2000PhRvL..84.1078O} \citeyear{2000PhRvL..84.1078O} \cite{2000PhRvL..84.1078O}
         \item BESS98 (1998/07): \citeauthor{2001APh....16..121M} \citeyear{2001APh....16..121M} \cite{2001APh....16..121M}
         \item BESS99 (1999/08): \citeauthor{2002PhRvL..88e1101A} \citeyear{2002PhRvL..88e1101A} \cite{2002PhRvL..88e1101A}
         \item BESS00 (2000/08): \citeauthor{2002PhRvL..88e1101A} \citeyear{2002PhRvL..88e1101A} \cite{2002PhRvL..88e1101A}
         \item BESS-TeV (2002/08): \citeauthor{2005ICRC....3...13H} \citeyear{2005ICRC....3...13H} \cite{2005ICRC....3...13H}
         \item BESS-PolarI (2004/12): \citeauthor{2008PhLB..670..103B} \citeyear{2008PhLB..670..103B} \cite{2008PhLB..670..103B}
         \item BESS-PolarII (2007/12-2008/01): \citeauthor{2012PhRvL.108e1102A} \citeyear{2012PhRvL.108e1102A} \cite{2012PhRvL.108e1102A}
         \item CAPRICE94 (1994/08): \citeauthor{1997ApJ...487..415B} \citeyear{1997ApJ...487..415B} \cite{1997ApJ...487..415B}
         \item CAPRICE98 (1998/05): \citeauthor{2001ApJ...561..787B} \citeyear{2001ApJ...561..787B} \cite{2001ApJ...561..787B}
         \item HEAT-pbar (2000/06): \citeauthor{2001PhRvL..87A1101B} \citeyear{2001PhRvL..87A1101B} \cite{2001PhRvL..87A1101B}
         \item IMAX92 (1992/07): \citeauthor{1996PhRvL..76.3057M} \citeyear{1996PhRvL..76.3057M} \cite{1996PhRvL..76.3057M}
         \item MASS91 (1991/09): \citeauthor{1999ICRC....3...77B} \citeyear{1999ICRC....3...77B} \cite{1999ICRC....3...77B}
         \item MASS91 (1991/09): \citeauthor{1996ApJ...467L..33H} \citeyear{1996ApJ...467L..33H} \cite{1996ApJ...467L..33H}
      \end{itemize}
   \item \underline{Space}
      \begin{itemize}[leftmargin=5pt]\vspace{-7pt}
         \item AMS01 (1998/06): \citeauthor{2002PhR...366..331A} \citeyear{2002PhR...366..331A} \cite{2002PhR...366..331A}
         \item AMS02 (2011/05-2015/05): \citeauthor{2016PhRvL.117i1103A} \citeyear{2016PhRvL.117i1103A} \cite{2016PhRvL.117i1103A}
         \item PAMELA (2006/07-2008/06): \citeauthor{2009PhRvL.102e1101A} \citeyear{2009PhRvL.102e1101A} \cite{2009PhRvL.102e1101A}
         \item PAMELA (2006/07-2008/12): \citeauthor{2010PhRvL.105l1101A} \citeyear{2010PhRvL.105l1101A} \cite{2010PhRvL.105l1101A}
         \item PAMELA (2006/07-2009/12): \citeauthor{2013JETPL..96..621A} \citeyear{2013JETPL..96..621A} \cite{2013JETPL..96..621A}
      \end{itemize}
\end{description}

\subsection{Lepton~Data}
\label{app:leptons}

\begin{description}
   \item \underline{Balloon}
      \begin{itemize}[leftmargin=5pt]\vspace{-7pt}
           \item Balloon (1963/04): \citeauthor{1965PhRvL..15..769D} \citeyear{1965PhRvL..15..769D} \cite{1965PhRvL..15..769D}
           \item Balloon (1963/07): \citeauthor{1965JGR....70.5753F} \citeyear{1965JGR....70.5753F} \cite{1965JGR....70.5753F}
           \item Balloon (1964/07): \citeauthor{1967ApJ...148..399L} \citeyear{1967ApJ...148..399L} \cite{1967ApJ...148..399L}
           \item Balloon (1965/06+1965/07): \citeauthor{1967ApJ...148..399L} \citeyear{1967ApJ...148..399L} \cite{1967ApJ...148..399L}
           \item Balloon (1965/07): \citeauthor{1973ICRC....2..760W} \citeyear{1973ICRC....2..760W} \cite{1973ICRC....2..760W}
           \item Balloon (1965/07): \citeauthor{1965ICRC....1..327B} \citeyear{1965ICRC....1..327B} \cite{1965ICRC....1..327B}
           \item Balloon (1965/07+1965/08+1966/06: \citeauthor{1968ApJ...152..783F} \citeyear{1968ApJ...152..783F} \cite{1968ApJ...152..783F}
           \item Balloon (1965/07+1965/08+1966/06: \citeauthor{1969ApJ...158..771F} \citeyear{1969ApJ...158..771F} \cite{1969ApJ...158..771F}
           \item Balloon (1965/07+1966/07): \citeauthor{1968CaJPS..46.1014B} \citeyear{1968CaJPS..46.1014B} \cite{1968CaJPS..46.1014B}
           \item Balloon (1966/03): \citeauthor{1968PhRvL..20..764A} \citeyear{1968PhRvL..20..764A} \cite{1968PhRvL..20..764A}
           \item Balloon (1966/06): \citeauthor{1968CaJPS..46..892L} \citeyear{1968CaJPS..46..892L} \cite{1968CaJPS..46..892L}
           \item Balloon (1966/07): \citeauthor{1973ICRC....2..760W} \citeyear{1973ICRC....2..760W} \cite{1973ICRC....2..760W}
           \item Balloon (1966/07): \citeauthor{1972JGR....77.1087E} \citeyear{1972JGR....77.1087E} \cite{1972JGR....77.1087E}
           \item Balloon (1966/08): \citeauthor{1970ICRC....1..209B} \citeyear{1970ICRC....1..209B} \cite{1970ICRC....1..209B}
           \item Balloon (1966/09): \citeauthor{1972JGR....77.1087E} \citeyear{1972JGR....77.1087E} \cite{1972JGR....77.1087E}
           \item Balloon (1966/09): \citeauthor{1968CaJPh..46..530D} \citeyear{1968CaJPh..46..530D} \cite{1968CaJPh..46..530D}
           \item Balloon (1967/06+1967/07): \citeauthor{1968PhRvL..20.1053I} \citeyear{1968PhRvL..20.1053I} \cite{1968PhRvL..20.1053I}
           \item Balloon (1967/07): \citeauthor{1969NCimL...1...53A} \citeyear{1969NCimL...1...53A} \cite{1969NCimL...1...53A}
           \item Balloon (1967/07): \citeauthor{1972JGR....77.1087E} \citeyear{1972JGR....77.1087E} \cite{1972JGR....77.1087E}
           \item Balloon (1967/08): \citeauthor{1970ICRC....1..209B} \citeyear{1970ICRC....1..209B} \cite{1970ICRC....1..209B}
           \item Balloon (1967/11+1968/06): \citeauthor{1971Ap&SS..14..301F} \citeyear{1971Ap&SS..14..301F} \cite{1971Ap&SS..14..301F}
           \item Balloon (1968+\dots+2001): \citeauthor{2012ApJ...760..146K} \citeyear{2012ApJ...760..146K} \cite{2012ApJ...760..146K}
           \item Balloon (1968/05): \citeauthor{1973ICRC....1..355A} \citeyear{1973ICRC....1..355A} \cite{1973ICRC....1..355A}
           \item Balloon (1968/06): \citeauthor{1971A&A....11...53S} \citeyear{1971A&A....11...53S} \cite{1971A&A....11...53S}
           \item Balloon (1968/06+1968/07): \citeauthor{1975JGR....80.1701F} \citeyear{1975JGR....80.1701F} \cite{1975JGR....80.1701F}
           \item Balloon (1968/07): \citeauthor{1969PhRvL..22..412B} \citeyear{1969PhRvL..22..412B} \cite{1969PhRvL..22..412B}
           \item Balloon (1968/07): \citeauthor{1969PhRvL..22..412B} \citeyear{1969PhRvL..22..412B} \cite{1969PhRvL..22..412B}
           \item Balloon (1968/07): \citeauthor{1973ICRC....2..760W} \citeyear{1973ICRC....2..760W} \cite{1973ICRC....2..760W}
           \item Balloon (1968/07): \citeauthor{1970ICRC....1..209B} \citeyear{1970ICRC....1..209B} \cite{1970ICRC....1..209B}
           \item Balloon (1969/04+1970/11): \citeauthor{1973JGR....78.7165S} \citeyear{1973JGR....78.7165S} \cite{1973JGR....78.7165S}
           \item Balloon (1969/06+1969/07): \citeauthor{1975JGR....80.1701F} \citeyear{1975JGR....80.1701F} \cite{1975JGR....80.1701F}
           \item Balloon (1969/07): \citeauthor{1973ICRC....2..760W} \citeyear{1973ICRC....2..760W} \cite{1973ICRC....2..760W}
           \item Balloon (1969/09+1973/05): \citeauthor{1975ApJ...197..219M} \citeyear{1975ApJ...197..219M} \cite{1975ApJ...197..219M}
           \item Balloon (1970/05+1970/09): \citeauthor{1973ApJ...186..841M} \citeyear{1973ApJ...186..841M} \cite{1973ApJ...186..841M}
           \item Balloon (1970/06+1970/07): \citeauthor{1975JGR....80.1701F} \citeyear{1975JGR....80.1701F} \cite{1975JGR....80.1701F}
           \item Balloon (1970/11): \citeauthor{1976JGR....81.3944S} \citeyear{1976JGR....81.3944S} \cite{1976JGR....81.3944S}
           \item Balloon (1971/06+1971/07): \citeauthor{1975JGR....80.1701F} \citeyear{1975JGR....80.1701F} \cite{1975JGR....80.1701F}
           \item Balloon (1971/07): \citeauthor{1973ICRC....2..760W} \citeyear{1973ICRC....2..760W} \cite{1973ICRC....2..760W}
           \item Balloon (1971/07+1972/07): \citeauthor{1973ICRC....2..760W} \citeyear{1973ICRC....2..760W} \cite{1973ICRC....2..760W}
           \item Balloon (1972/07): \citeauthor{1975ApJ...198..493D} \citeyear{1975ApJ...198..493D} \cite{1975ApJ...198..493D}
           \item Balloon (1972/07): \citeauthor{1973ICRC....2..760W} \citeyear{1973ICRC....2..760W} \cite{1973ICRC....2..760W}
           \item Balloon (1972/07): \citeauthor{1975JGR....80.1701F} \citeyear{1975JGR....80.1701F} \cite{1975JGR....80.1701F}
           \item Balloon (1972/10): \citeauthor{1977ApJ...213..588F} \citeyear{1977ApJ...213..588F} \cite{1977ApJ...213..588F}
           \item Balloon (1972/10): \citeauthor{1976JGR....81.3944S} \citeyear{1976JGR....81.3944S} \cite{1976JGR....81.3944S}
           \item Balloon (1972/11+1973/05): \citeauthor{1975ApJ...199..669B} \citeyear{1975ApJ...199..669B} \cite{1975ApJ...199..669B}
           \item Balloon (1973/06): \citeauthor{1973ICRC....5.3073I} \citeyear{1973ICRC....5.3073I} \cite{1973ICRC....5.3073I}
           \item Balloon (1973/07): \citeauthor{1975ICRC....3.1000C} \citeyear{1975ICRC....3.1000C} \cite{1975ICRC....3.1000C}
           \item Balloon (1974/07): \citeauthor{1977ICRC...11..203C} \citeyear{1977ICRC...11..203C} \cite{1977ICRC...11..203C}
           \item Balloon (1974/07+1974/08): \citeauthor{1976ApJ...204..927H} \citeyear{1976ApJ...204..927H} \cite{1976ApJ...204..927H}
           \item Balloon (1975/07): \citeauthor{1977ICRC...11..203C} \citeyear{1977ICRC...11..203C} \cite{1977ICRC...11..203C}
           \item Balloon (1975/10): \citeauthor{1979ApJ...227..676P} \citeyear{1979ApJ...227..676P} \cite{1979ApJ...227..676P}
           \item Balloon (1976/05): \citeauthor{1984ApJ...287..622G} \citeyear{1984ApJ...287..622G} \cite{1984ApJ...287..622G}
           \item Balloon (1976/05): \citeauthor{1987A&A...188..145G} \citeyear{1987A&A...188..145G} \cite{1987A&A...188..145G}
           \item Balloon (1977/07): \citeauthor{1979ICRC....1..462E} \citeyear{1979ICRC....1..462E} \cite{1979ICRC....1..462E}
           \item Balloon (1979/08): \citeauthor{1984JGR....89.2647E} \citeyear{1984JGR....89.2647E} \cite{1984JGR....89.2647E}
           \item Balloon (1980/10): \citeauthor{1984ApJ...278..881T} \citeyear{1984ApJ...278..881T} \cite{1984ApJ...278..881T}
           \item Balloon (1982/08): \citeauthor{1984JGR....89.2647E} \citeyear{1984JGR....89.2647E} \cite{1984JGR....89.2647E}
           \item Balloon (1982/10): \citeauthor{1986JGR....91.2858G} \citeyear{1986JGR....91.2858G} \cite{1986JGR....91.2858G}
           \item Balloon (1984/04): \citeauthor{1987ApJ...312..183M} \citeyear{1987ApJ...312..183M} \cite{1987ApJ...312..183M}
           \item Balloon (1984/09): \citeauthor{1986JGR....91.2858G} \citeyear{1986JGR....91.2858G} \cite{1986JGR....91.2858G}
           \item Balloon (1987/08): \citeauthor{1995JGR...100.7873E} \citeyear{1995JGR...100.7873E} \cite{1995JGR...100.7873E}
           \item Balloon (1990/08): \citeauthor{1995JGR...100.7873E} \citeyear{1995JGR...100.7873E} \cite{1995JGR...100.7873E}
           \item Balloon (1992/08): \citeauthor{1995JGR...100.7873E} \citeyear{1995JGR...100.7873E} \cite{1995JGR...100.7873E}
           \item Balloon (1994/08): \citeauthor{1995JGR...100.7873E} \citeyear{1995JGR...100.7873E} \cite{1995JGR...100.7873E}
           \item AESOP94 (1994/08): \citeauthor{1996ApJ...464..507C} \citeyear{1996ApJ...464..507C} \cite{1996ApJ...464..507C}
           \item AESOP97+98 (1997/09+1998/08): \citeauthor{2000JGR...10523099C} \citeyear{2000JGR...10523099C} \cite{2000JGR...10523099C}
           \item AESOP99 (1999/08): \citeauthor{2002ApJ...568..216C} \citeyear{2002ApJ...568..216C} \cite{2002ApJ...568..216C}
           \item AESOP00 (2000/08): \citeauthor{2002ApJ...568..216C} \citeyear{2002ApJ...568..216C} \cite{2002ApJ...568..216C}
           \item AESOP02 (2002/08): \citeauthor{2004JGRA..109.7107C} \citeyear{2004JGRA..109.7107C} \cite{2004JGRA..109.7107C}
           \item AESOP06 (2006/08): \citeauthor{2009JGRA..11410108C} \citeyear{2009JGRA..11410108C} \cite{2009JGRA..11410108C}
           \item ATIC01\&02 (2001/01+2003/01): \citeauthor{2008Natur.456..362C} \citeyear{2008Natur.456..362C} \cite{2008Natur.456..362C}
           \item BETS04 (2004/01): \citeauthor{2008AdSpR..42.1670Y} \citeyear{2008AdSpR..42.1670Y} \cite{2008AdSpR..42.1670Y}
           \item BETS97\&98 (1997/06+1998/05): \citeauthor{2001ApJ...559..973T} \citeyear{2001ApJ...559..973T} \cite{2001ApJ...559..973T}
           \item CAPRICE94 (1994/08): \citeauthor{2000ApJ...532..653B} \citeyear{2000ApJ...532..653B} \cite{2000ApJ...532..653B}
           \item CAPRICE98 (1998/05): \citeauthor{2001AdSpR..27..669B} \citeyear{2001AdSpR..27..669B} \cite{2001AdSpR..27..669B}
           \item HEAT-pbar (2000/06): \citeauthor{2004PhRvL..93x1102B} \citeyear{2004PhRvL..93x1102B} \cite{2004PhRvL..93x1102B}
           \item HEAT94 (1994/05): \citeauthor{1998ApJ...498..779B} \citeyear{1998ApJ...498..779B} \cite{1998ApJ...498..779B}
           \item HEAT94 (1994/05): \citeauthor{1997ApJ...482L.191B} \citeyear{1997ApJ...482L.191B} \cite{1997ApJ...482L.191B}
           \item HEAT94\&95 (1994/05+1995/08): \citeauthor{2001ApJ...559..296D} \citeyear{2001ApJ...559..296D} \cite{2001ApJ...559..296D}
           \item HEAT94\&95 (1994/05+1995/08): \citeauthor{1997ApJ...482L.191B} \citeyear{1997ApJ...482L.191B} \cite{1997ApJ...482L.191B}
           \item HEAT95 (1995/08): \citeauthor{2001ApJ...559..296D} \citeyear{2001ApJ...559..296D} \cite{2001ApJ...559..296D}
           \item HEAT95 (1995/08): \citeauthor{1997ApJ...482L.191B} \citeyear{1997ApJ...482L.191B} \cite{1997ApJ...482L.191B}
           \item MASS89 (1989/09): \citeauthor{1994ApJ...436..769G} \citeyear{1994ApJ...436..769G} \cite{1994ApJ...436..769G}
           \item MASS91 (1991/09): \citeauthor{2002A&A...392..287G} \citeyear{2002A&A...392..287G} \cite{2002A&A...392..287G}
           \item TS93 (1993/09): \citeauthor{1996ApJ...457L.103G} \citeyear{1996ApJ...457L.103G} \cite{1996ApJ...457L.103G}
      \end{itemize}
   \item \underline{Space}
      \begin{itemize}[leftmargin=5pt]\vspace{-7pt}
           \item AMS01-BremsstrahlungPhotons (1998/06): \citeauthor{2007PhLB..646..145A} \citeyear{2007PhLB..646..145A} \cite{2007PhLB..646..145A}
           \item AMS01-singleTrack (1998/06): \citeauthor{2000PhLB..484...10A} \citeyear{2000PhLB..484...10A} \cite{2000PhLB..484...10A}
           \item AMS02 (2011/05-2012/12): \citeauthor{2013PhRvL.110n1102A} \citeyear{2013PhRvL.110n1102A} \cite{2013PhRvL.110n1102A}
           \item AMS02 (2011/05-2013/11): \citeauthor{2014PhRvL.113v1102A} \citeyear{2014PhRvL.113v1102A} \cite{2014PhRvL.113v1102A}
           \item AMS02 (2011/05-2015/05): \citeauthor{2016PhRvL.117i1103A} \citeyear{2016PhRvL.117i1103A} \cite{2016PhRvL.117i1103A}
           \item AMS02 (2011/05-2017/11): \citeauthor{2019PhRvL.122j1101A} \citeyear{2019PhRvL.122j1101A} \cite{2019PhRvL.122j1101A}
           \item AMS02 (2011/05-2018/05): \citeauthor{Aguilar:2020ohx} \citeyear{Aguilar:2020ohx} \cite{Aguilar:2020ohx}
          \item AMS02 time series: \citeauthor{2018PhRvL.121e1102A} \citeyear{2018PhRvL.121e1102A} \cite{2018PhRvL.121e1102A}
           \item CALET (2015/10-2017/06): \citeauthor{2017PhRvL.119r1101A} \citeyear{2017PhRvL.119r1101A} \cite{2017PhRvL.119r1101A}
           \item CALET (2015/10-2017/11): \citeauthor{2018PhRvL.120z1102A} \citeyear{2018PhRvL.120z1102A} \cite{2018PhRvL.120z1102A}
           \item DAMPE (2015/12-2017/06): \citeauthor{2017Natur.552...63D} \citeyear{2017Natur.552...63D} \cite{2017Natur.552...63D}
           \item Fermi-LAT-LE (2008/06-2009/06): \citeauthor{2010PhRvD..82i2004A} \citeyear{2010PhRvD..82i2004A} \cite{2010PhRvD..82i2004A}
           \item Fermi-LAT-HE (2008/06-2009/06): \citeauthor{2010PhRvD..82i2004A} \citeyear{2010PhRvD..82i2004A} \cite{2010PhRvD..82i2004A}
           \item Fermi-LAT-LE (2008/08-2015/06): \citeauthor{2017PhRvD..95h2007A} \citeyear{2017PhRvD..95h2007A} \cite{2017PhRvD..95h2007A}
           \item Fermi-LAT-HE (2008/08-2015/06): \citeauthor{2017PhRvD..95h2007A} \citeyear{2017PhRvD..95h2007A} \cite{2017PhRvD..95h2007A}
           \item IMP1 (1963/11-1964/05): \citeauthor{1964PhRvL..13..786C} \citeyear{1964PhRvL..13..786C} \cite{1964PhRvL..13..786C}
           \item IMP3 (1965/07-1966/03): \citeauthor{1968ApJ...151..737F} \citeyear{1968ApJ...151..737F} \cite{1968ApJ...151..737F}
           \item ISEE3-MEH (1978/08-1979/02): \citeauthor{1979ICRC....1..462E} \citeyear{1979ICRC....1..462E} \cite{1979ICRC....1..462E}
           \item ISEE3-MEH (1979/02-1980/03): \citeauthor{1981ICRC...10...77E} \citeyear{1981ICRC...10...77E} \cite{1981ICRC...10...77E}
           \item ISEE3-MEH (1980/03-1981/03): \citeauthor{1981ICRC...10...77E} \citeyear{1981ICRC...10...77E} \cite{1981ICRC...10...77E}
           \item OGO5 (1968/04-1968/05): \citeauthor{1974JGR....79.1533B} \citeyear{1974JGR....79.1533B} \cite{1974JGR....79.1533B}
           \item OGO5 (1968/04-1969/04): \citeauthor{1972ApJ...171..363L} \citeyear{1972ApJ...171..363L} \cite{1972ApJ...171..363L}
           \item OGO5 (1968/06-1968/10): \citeauthor{1975JGR....80.1701F} \citeyear{1975JGR....80.1701F} \cite{1975JGR....80.1701F}
           \item OGO5 (1969/06-1969/07): \citeauthor{1975JGR....80.1701F} \citeyear{1975JGR....80.1701F} \cite{1975JGR....80.1701F}
           \item OGO5 (1969/06-1969/07): \citeauthor{1973JGR....78..292B} \citeyear{1973JGR....78..292B} \cite{1973JGR....78..292B}
           \item OGO5 (1970/06-1970/07): \citeauthor{1975JGR....80.1701F} \citeyear{1975JGR....80.1701F} \cite{1975JGR....80.1701F}
           \item OGO5 (1970/06-1970/07): \citeauthor{1973JGR....78..292B} \citeyear{1973JGR....78..292B} \cite{1973JGR....78..292B}
           \item OGO5 (1971/05-1971/08): \citeauthor{1975JGR....80.1701F} \citeyear{1975JGR....80.1701F} \cite{1975JGR....80.1701F}
           \item OGO5 (1971/07-1971/08): \citeauthor{1973JGR....78..292B} \citeyear{1973JGR....78..292B} \cite{1973JGR....78..292B}
           \item OGO5 (1972/06): \citeauthor{1975JGR....80.1701F} \citeyear{1975JGR....80.1701F} \cite{1975JGR....80.1701F}
           \item OGO5 (1972/06-1972/07): \citeauthor{1974JGR....79.1533B} \citeyear{1974JGR....79.1533B} \cite{1974JGR....79.1533B}
           \item PAMELA (2006/07-2006/11): \citeauthor{2015ApJ...810..142A} \citeyear{2015ApJ...810..142A} \cite{2015ApJ...810..142A}
           \item PAMELA (2006/07-2008/02): \citeauthor{2009Natur.458..607A} \citeyear{2009Natur.458..607A} \cite{2009Natur.458..607A}
           \item PAMELA (2006/07-2008/12): \citeauthor{2010APh....34....1A} \citeyear{2010APh....34....1A} \cite{2010APh....34....1A}
           \item PAMELA (2006/07-2009/12): \citeauthor{2013PhRvL.111h1102A} \citeyear{2013PhRvL.111h1102A} \cite{2013PhRvL.111h1102A}
           \item PAMELA (2006/07-2010/01): \citeauthor{2011PhRvL.106t1101A} \citeyear{2011PhRvL.106t1101A} \cite{2011PhRvL.106t1101A}
           \item PAMELA time series: \citeauthor{2015ApJ...810..142A} \citeyear{2015ApJ...810..142A} \cite{2015ApJ...810..142A}
           \item Pioneer8\&9 (1967/12-1969/04): \citeauthor{1973JGR....78.1487W} \citeyear{1973JGR....78.1487W} \cite{1973JGR....78.1487W}
           \item Ulysses-KET (1990/10-1991/02): \citeauthor{1996A&A...307..981R} \citeyear{1996A&A...307..981R} \cite{1996A&A...307..981R}
           \item Ulysses-KET (1992/07-1992/10): \citeauthor{1996A&A...307..981R} \citeyear{1996A&A...307..981R} \cite{1996A&A...307..981R}
           \item Ulysses-KET (1993/07-1993/10): \citeauthor{1996A&A...307..981R} \citeyear{1996A&A...307..981R} \cite{1996A&A...307..981R}
           \item Ulysses-KET (1994/07-1994/10): \citeauthor{1996A&A...307..981R} \citeyear{1996A&A...307..981R} \cite{1996A&A...307..981R}
           \item Voyager1-HET (2010/01-2010/03): \citeauthor{2010ApJ...725..121C} \citeyear{2010ApJ...725..121C} \cite{2010ApJ...725..121C}
           \item Voyager2-HET (2010/01-2010/03): \citeauthor{2010ApJ...725..121C} \citeyear{2010ApJ...725..121C} \cite{2010ApJ...725..121C}
           \item Voyager1-BSe (2012/09-2012/12): \citeauthor{2019NatAs...3.1013S} \citeyear{2019NatAs...3.1013S} \cite{2019NatAs...3.1013S}
           \item Voyager1-HET-Bse (2012/12-2015/06): \citeauthor{2016ApJ...831...18C} \citeyear{2016ApJ...831...18C} \cite{2016ApJ...831...18C}
           \item Voyager1-TET (2012/09-2012/12): \citeauthor{2019NatAs...3.1013S} \citeyear{2019NatAs...3.1013S} \cite{2019NatAs...3.1013S}
           \item Voyager1-TET (2012/12-2015/06): \citeauthor{2016ApJ...831...18C} \citeyear{2016ApJ...831...18C} \cite{2016ApJ...831...18C}
           \item Voyager2-BSe (2019/03-2019/07): \citeauthor{2019NatAs...3.1013S} \citeyear{2019NatAs...3.1013S} \cite{2019NatAs...3.1013S}
           \item Voyager2-TET (2019/03-2019/07): \citeauthor{2019NatAs...3.1013S} \citeyear{2019NatAs...3.1013S} \cite{2019NatAs...3.1013S}
      \end{itemize}
   \item \underline{Ground}
      \begin{itemize}[leftmargin=5pt]\vspace{-7pt}
         \item H.E.S.S. (2004/10-2007/08): \citeauthor{2008PhRvL.101z1104A} \citeyear{2008PhRvL.101z1104A} \cite{2008PhRvL.101z1104A}
         \item H.E.S.S.-LEAnalysis (2004/10-2005/12): \citeauthor{2009A&A...508..561A} \citeyear{2009A&A...508..561A} \cite{2009A&A...508..561A}
      \end{itemize}

\end{description}

\subsection{Data for Nuclei ($Z\leq30$)}
\label{app:nuclei}

These data are the most numerous in terms of experiments and publications. For~readability, we~further group them by their experiment name~below.\\

\noindent\underline{Balloon}

\begin{itemize}[leftmargin=15pt]\vspace{-8pt}
   \item Balloon (1950,1954,1957)
      \begin{itemize}[leftmargin=15pt]\vspace{-8pt}
         \item Balloon (1950/10): \citeauthor{1957PMag....2..157F} \citeyear{1957PMag....2..157F} \cite{1957PMag....2..157F}
         \item Balloon (1954/06): \citeauthor{1958Natur.181.1319F} \citeyear{1958Natur.181.1319F} \cite{1958Natur.181.1319F}
         \item Balloon (1957/05): \citeauthor{1957PMag....2..157F} \citeyear{1957PMag....2..157F} \cite{1957PMag....2..157F}
         \item Balloon (1957/06): \citeauthor{1961PhRv..121.1206A} \citeyear{1961PhRv..121.1206A} \cite{1961PhRv..121.1206A}
         \item Balloon (1957/09): \citeauthor{1961PhRv..121.1206A} \citeyear{1961PhRv..121.1206A} \cite{1961PhRv..121.1206A}
      \end{itemize}
   \item Balloon (1955,1956,1958,1959)
      \begin{itemize}[leftmargin=15pt]\vspace{-8pt}
         \item Balloon (1955/07): \citeauthor{1956PhRv..104.1723M} \citeyear{1956PhRv..104.1723M} \cite{1956PhRv..104.1723M}
         \item Balloon (1956/03): \citeauthor{1959PhRv..115..194M} \citeyear{1959PhRv..115..194M} \cite{1959PhRv..115..194M}
         \item Balloon (1956/08): \citeauthor{1957PhRv..107.1386M} \citeyear{1957PhRv..107.1386M} \cite{1957PhRv..107.1386M}
         \item Balloon (1956/09): \citeauthor{1964JGR....69.3097W} \citeyear{1964JGR....69.3097W} \cite{1964JGR....69.3097W}
         \item Balloon (1958/02): \citeauthor{1959PhRv..116..462M} \citeyear{1959PhRv..116..462M} \cite{1959PhRv..116..462M}
         \item Balloon (1958/07): \citeauthor{1960JGR....65..767M} \citeyear{1960JGR....65..767M} \cite{1960JGR....65..767M}
         \item Balloon (1959/05): \citeauthor{1964JGR....69.3097W} \citeyear{1964JGR....69.3097W} \cite{1964JGR....69.3097W}
         \item Balloon (1959/06): \citeauthor{1964JGR....69.3097W} \citeyear{1964JGR....69.3097W} \cite{1964JGR....69.3097W}
      \end{itemize}
   \item Balloon (1957,1958)
      \begin{itemize}[leftmargin=15pt]\vspace{-8pt}
         \item Balloon (1957/07): \citeauthor{1961NCim...19.1090E} \citeyear{1961NCim...19.1090E} \cite{1961NCim...19.1090E}
         \item Balloon (1958/08): \citeauthor{1961NCim...20.1157E} \citeyear{1961NCim...20.1157E} \cite{1961NCim...20.1157E}
      \end{itemize}
   \item Balloon (1960,1961)
      \begin{itemize}[leftmargin=15pt]\vspace{-8pt}
         \item Balloon (1960/09): \citeauthor{1963PhRv..129.2275M} \citeyear{1963PhRv..129.2275M} \cite{1963PhRv..129.2275M}
         \item Balloon (1961/08): \citeauthor{1963PhRv..129.2275M} \citeyear{1963PhRv..129.2275M} \cite{1963PhRv..129.2275M}
      \end{itemize}
   \item Balloon (1960,1963,1964,1965,1968)
      \begin{itemize}[leftmargin=15pt]\vspace{-8pt}
         \item Balloon (1960/08): \citeauthor{1965JGR....70.2111F} \citeyear{1965JGR....70.2111F} \cite{1965JGR....70.2111F}
         \item Balloon (1964/07): \citeauthor{1968JGR....73.4261F} \citeyear{1968JGR....73.4261F} \cite{1968JGR....73.4261F}
      \end{itemize}
   \item Balloon (1961,1962,1963,1964)
      \begin{itemize}[leftmargin=15pt]\vspace{-8pt}
         \item Balloon (1961/07): \citeauthor{1964PhRv..133..818F} \citeyear{1964PhRv..133..818F} \cite{1964PhRv..133..818F}
         \item Balloon (1962/07): \citeauthor{1964JGR....69.3293F} \citeyear{1964JGR....69.3293F} \cite{1964JGR....69.3293F}
         \item Balloon (1963/06): \citeauthor{1965ICRC....1..412O} \citeyear{1965ICRC....1..412O} \cite{1965ICRC....1..412O}
         \item Balloon (1964/06): \citeauthor{1967JGR....72.2765D} \citeyear{1967JGR....72.2765D} \cite{1967JGR....72.2765D}
      \end{itemize}
   \item Balloon (1962/05): \citeauthor{1967NCimA..47..189F} \citeyear{1967NCimA..47..189F} \cite{1967NCimA..47..189F}
   \item Balloon (1963/04): \citeauthor{1968CaJPS..46..652A} \citeyear{1968CaJPS..46..652A} \cite{1968CaJPS..46..652A}
   \item Balloon (1963,1964,1967)
      \begin{itemize}[leftmargin=15pt]\vspace{-8pt}
         \item Balloon (1963/06): \citeauthor{1977JGR....82.2419B} \citeyear{1977JGR....82.2419B} \cite{1977JGR....82.2419B}
         \item Balloon (1964/06): \citeauthor{1977JGR....82.2419B} \citeyear{1977JGR....82.2419B} \cite{1977JGR....82.2419B}
         \item Balloon (1967/07): \citeauthor{1977JGR....82.2419B} \citeyear{1977JGR....82.2419B} \cite{1977JGR....82.2419B}
      \end{itemize}
   \item Balloon (1963,1965)
      \begin{itemize}[leftmargin=15pt]\vspace{-8pt}
         \item Balloon (1963/06): \citeauthor{1964JGR....69.3289B} \citeyear{1964JGR....69.3289B} \cite{1964JGR....69.3289B}
         \item Balloon (1965/06): \citeauthor{1966JGR....71.1771B} \citeyear{1966JGR....71.1771B} \cite{1966JGR....71.1771B}
      \end{itemize}
   \item Balloon (1963,1965,1966,1967)
      \begin{itemize}[leftmargin=15pt]\vspace{-8pt}
         \item Balloon (1963/01): \citeauthor{1964PhRvL..13..106O} \citeyear{1964PhRvL..13..106O} \cite{1964PhRvL..13..106O}
         \item Balloon (1963/04): \citeauthor{1964PhRvL..13..106O} \citeyear{1964PhRvL..13..106O} \cite{1964PhRvL..13..106O}
         \item Balloon (1965/06): \citeauthor{1968JGR....73.4231O} \citeyear{1968JGR....73.4231O} \cite{1968JGR....73.4231O}
         \item Balloon (1963/11): \citeauthor{1964PhRvL..13..106O} \citeyear{1964PhRvL..13..106O} \cite{1964PhRvL..13..106O}
         \item Balloon (1966/07): \citeauthor{1968JGR....73.4231O} \citeyear{1968JGR....73.4231O} \cite{1968JGR....73.4231O}
         \item Balloon (1967/06): \citeauthor{1968JGR....73.4231O} \citeyear{1968JGR....73.4231O} \cite{1968JGR....73.4231O}
         \item Balloon (64/03-65/7): \citeauthor{1967JGR....72.5957W} \citeyear{1967JGR....72.5957W} \cite{1967JGR....72.5957W}
         \item Balloon (66/07+67/05-06-07): \citeauthor{1969Ap&SS...3...80V} \citeyear{1969Ap&SS...3...80V} \cite{1969Ap&SS...3...80V}
      \end{itemize}
   \item Balloon (1964/09): \citeauthor{1966P&SS...14..503C} \citeyear{1966P&SS...14..503C} \cite{1966P&SS...14..503C}
   \item Balloon (1965/05): \citeauthor{1966PhRvL..16..109H} \citeyear{1966PhRvL..16..109H,1967P&SS...15..715H} \cite{1966PhRvL..16..109H,1967P&SS...15..715H}
   \item Balloon (1965/07+1965/08+1966/06): \citeauthor{1967ApJ...150..371H} \citeyear{1967ApJ...150..371H} \cite{1967ApJ...150..371H}
   \item Balloon (1965,1966,1967,1968,1969,1970,1971,1972)
      \begin{itemize}[leftmargin=15pt]\vspace{-8pt}
         \item Balloon (1965/08): \citeauthor{1971JGR....76.7445R} \citeyear{1971JGR....76.7445R} \cite{1971JGR....76.7445R}
         \item Balloon (1966/08): \citeauthor{1971JGR....76.7445R} \citeyear{1971JGR....76.7445R} \cite{1971JGR....76.7445R}
         \item Balloon (1967/07): \citeauthor{1971JGR....76.7445R} \citeyear{1971JGR....76.7445R} \cite{1971JGR....76.7445R}
         \item Balloon (1968/07): \citeauthor{1971JGR....76.7445R} \citeyear{1971JGR....76.7445R} \cite{1971JGR....76.7445R}
         \item Balloon (1969/07): \citeauthor{1971JGR....76.7445R} \citeyear{1971JGR....76.7445R} \cite{1971JGR....76.7445R}
         \item Balloon (1970/07): \citeauthor{1974JGR....79.4127R} \citeyear{1974JGR....79.4127R} \cite{1974JGR....79.4127R}
         \item Balloon (1971/07): \citeauthor{1974JGR....79.4127R} \citeyear{1974JGR....79.4127R} \cite{1974JGR....79.4127R}
         \item Balloon (1972/07): \citeauthor{1974JGR....79.4127R} \citeyear{1974JGR....79.4127R} \cite{1974JGR....79.4127R}
      \end{itemize}
   \item Balloon (1965,1966,1968,1969,1970)
      \begin{itemize}[leftmargin=15pt]\vspace{-8pt}
         \item Balloon (1965/06+1966/06): \citeauthor{1973ICRC....2..732G} \citeyear{1973ICRC....2..732G} \cite{1973ICRC....2..732G}
         \item Balloon (1969/06): \citeauthor{1973ICRC....2..732G} \citeyear{1973ICRC....2..732G} \cite{1973ICRC....2..732G}
         \item Balloon (1970/06): \citeauthor{1973ICRC....2..732G} \citeyear{1973ICRC....2..732G} \cite{1973ICRC....2..732G}
      \end{itemize}
   \item Balloon (1965,1966,1968,1969,1971+1972)
      \begin{itemize}[leftmargin=15pt]\vspace{-8pt}
         \item Balloon (1965/07): \citeauthor{1967JGR....72.2783W} \citeyear{1967JGR....72.2783W} \cite{1967JGR....72.2783W}
         \item Balloon (1971/07+1972/07): \citeauthor{1973ICRC....2..760W} \citeyear{1973ICRC....2..760W} \cite{1973ICRC....2..760W}
      \end{itemize}
   \item Balloon (1966/07+1966/08): \citeauthor{1967PhRv..163.1327B} \citeyear{1967PhRv..163.1327B} \cite{1967PhRv..163.1327B}
   \item Balloon (1970/07+1970/06): \citeauthor{1976A&A....52..327B} \citeyear{1976A&A....52..327B} \cite{1976A&A....52..327B}
   \item Balloon (1968)
      \begin{itemize}[leftmargin=15pt]\vspace{-8pt}
         \item Balloon (1968/05): \citeauthor{1971A&A....11...53S} \citeyear{1971A&A....11...53S} \cite{1971A&A....11...53S}
         \item Balloon (1968/07): \citeauthor{1979Ap&SS..62..465S} \citeyear{1979Ap&SS..62..465S} \cite{1979Ap&SS..62..465S}
      \end{itemize}
   \item Ballon (1968\dots 1994)
      \begin{itemize}[leftmargin=15pt]\vspace{-8pt}
         \item Balloon (1973/07): \citeauthor{1975ICRC....3.1000C} \citeyear{1975ICRC....3.1000C} \cite{1975ICRC....3.1000C}
         \item Balloon (1974/07): \citeauthor{1977ICRC...11..203C} \citeyear{1977ICRC...11..203C} \cite{1977ICRC...11..203C}
         \item Balloon (1975/07): \citeauthor{1977ICRC...11..203C} \citeyear{1977ICRC...11..203C} \cite{1977ICRC...11..203C}
         \item Balloon (1977/07): \citeauthor{1979ICRC....1..462E} \citeyear{1979ICRC....1..462E} \cite{1979ICRC....1..462E}
         \item Balloon (1979/08): \citeauthor{1984JGR....89.2647E} \citeyear{1984JGR....89.2647E} \cite{1984JGR....89.2647E}
         \item Balloon (1982/08): \citeauthor{1984JGR....89.2647E} \citeyear{1984JGR....89.2647E} \cite{1984JGR....89.2647E}
         \item Balloon (1982/10): \citeauthor{1986JGR....91.2858G} \citeyear{1986JGR....91.2858G} \cite{1986JGR....91.2858G}
         \item Balloon (1984/09): \citeauthor{1986JGR....91.2858G} \citeyear{1986JGR....91.2858G} \cite{1986JGR....91.2858G}
         \item Balloon (1987/08): \citeauthor{1995JGR...100.7873E} \citeyear{1995JGR...100.7873E} \cite{1995JGR...100.7873E}
         \item Balloon (1990/08): \citeauthor{1995JGR...100.7873E} \citeyear{1995JGR...100.7873E} \cite{1995JGR...100.7873E}
         \item Balloon (1992/08): \citeauthor{1995JGR...100.7873E} \citeyear{1995JGR...100.7873E} \cite{1995JGR...100.7873E}
         \item Balloon (1994/08): \citeauthor{1995JGR...100.7873E} \citeyear{1995JGR...100.7873E} \cite{1995JGR...100.7873E}
      \end{itemize}
   \item Balloon (1970)
      \begin{itemize}[leftmargin=15pt]\vspace{-8pt}
         \item Balloon (1970/05): \citeauthor{1972NPhS..240..135V} \citeyear{1972NPhS..240..135V} \cite{1972NPhS..240..135V}
         \item Balloon (1970/06): \citeauthor{1979ZPhyA.291..383B} \citeyear{1979ZPhyA.291..383B} \cite{1979ZPhyA.291..383B}
         \item Balloon (1970/07): \citeauthor{1972Ap&SS..15..245W} \citeyear{1972Ap&SS..15..245W} \cite{1972Ap&SS..15..245W}
         \item Balloon (1970/11): \citeauthor{1972PhRvL..28..985R} \citeyear{1972PhRvL..28..985R} \cite{1972PhRvL..28..985R}
      \end{itemize}
   \item Balloon (1970/09+1971/05): \citeauthor{1973ApJ...180..987S} \citeyear{1973ApJ...180..987S} \cite{1973ApJ...180..987S}
   \item Balloon (1971/05): \citeauthor{1973ICRC....1..126A} \citeyear{1973ICRC....1..126A} \cite{1973ICRC....1..126A}
   \item Balloon (1971/09+1972/10): \citeauthor{1974ApJ...191..331J} \citeyear{1974ApJ...191..331J} \cite{1974ApJ...191..331J}
   \item Balloon (1973/08): 
   \citeauthor{1976ApJ...205..938F} \citeyear{1976ApJ...205..938F} \cite{1976ApJ...205..938F}, \citeauthor{1977ApJ...212..262H} \citeyear{1977ApJ...212..262H} \cite{1977ApJ...212..262H}, \citeauthor{1977Ap&SS..47..163M} \citeyear{1977Ap&SS..47..163M} \cite{1977Ap&SS..47..163M}, \citeauthor{1978ApJ...221.1110L} \citeyear{1978ApJ...221.1110L} \cite{1978ApJ...221.1110L}
   \item Balloon (1973+1974+1975)
      \begin{itemize}[leftmargin=15pt]\vspace{-8pt}
         \item Balloon (1973/09+1974/05): \citeauthor{1978ApJ...224..691D} \citeyear{1978ApJ...224..691D} \cite{1978ApJ...224..691D}
         \item Balloon (1973/09+1974/05+\dots): \citeauthor{1987ApJ...322..981D} \citeyear{1987ApJ...322..981D} \cite{1987ApJ...322..981D}
         \item Balloon (1975/09+1975/10): \citeauthor{1981ApJ...248..847M} \citeyear{1981ApJ...248..847M} \cite{1981ApJ...248..847M}
      \end{itemize}
   \item Balloon (1974/10): \citeauthor{1978Ap&SS..59..301S} \citeyear{1978Ap&SS..59..301S} \cite{1978Ap&SS..59..301S}
   \item Balloon (1974/07+/08+1976/09): \citeauthor{1978ApJ...223..676L} \citeyear{1978ApJ...223..676L} \cite{1978ApJ...223..676L}
   \item Balloon (1975/12): \citeauthor{1979ICRC....1..330B} \citeyear{1979ICRC....1..330B} \cite{1979ICRC....1..330B}
   \item Balloon (1976,1979)
      \begin{itemize}[leftmargin=15pt]\vspace{-8pt}
         \item Balloon (1976/09): \citeauthor{1992IJRAI..20..415D} \citeyear{1992IJRAI..20..415D} \cite{1992IJRAI..20..415D}
         \item Balloon (1976/10): \citeauthor{1980ApJ...239..712S} \citeyear{1980ApJ...239..712S} \cite{1980ApJ...239..712S}
         \item Balloon (1979/06): \citeauthor{1987ICRC....1..325W} \citeyear{1987ICRC....1..325W} \cite{1987ICRC....1..325W}
      \end{itemize}
   \item Balloon (1977)
      \begin{itemize}[leftmargin=15pt]\vspace{-8pt}
         \item Balloon (1977/05): \citeauthor{1978ApJ...226..355B} \citeyear{1978ApJ...226..355B} \cite{1978ApJ...226..355B} \cite{1978ApJ...226..355B}
         \item Balloon (1977/07): \citeauthor{1983ApJ...275..391W}  \citeyear{1983ApJ...275..391W} \cite{1983ApJ...275..391W}, \citeauthor{1987ICRC....1..325W} \citeyear{1987ICRC....1..325W} \cite{1987ICRC....1..325W}
         \item Balloon (1977/09): \citeauthor{1979ICRC....1..389W} \citeyear{1979ICRC....1..389W} \cite{1979ICRC....1..389W}, \citeauthor{1982ApJ...252..386W} \citeyear{1981ICRC....2...80W}, \citeyear{1982ApJ...252..386W} \cite{1981ICRC....2...80W,1982ApJ...252..386W}, \citeauthor{1979ICRC...12...51W} \citeyear{1979ICRC...12...51W} \cite{1979ICRC...12...51W}
      \end{itemize}
   \item Balloon (1979/06): \citeauthor{1979PhRvL..43.1196G} \citeyear{1979PhRvL..43.1196G}, \citeyear{1984ApL....24...75G}  \cite{1979PhRvL..43.1196G,1984ApL....24...75G}
   \item Balloon (1980/06): \citeauthor{1981ApJ...248.1179B} \citeyear{1981ApJ...248.1179B} \cite{1981ApJ...248.1179B}
   \item Balloon (1981)
      \begin{itemize}[leftmargin=15pt]\vspace{-8pt}
         \item Balloon (1981/04): \citeauthor{1985ApJ...291..207J} \citeyear{1985ApJ...291..207J} \cite{1985ApJ...291..207J}
         \item Balloon (1981/09): \citeauthor{1985ICRC....2...16W} \citeyear{1985ICRC....2...16W} \cite{1985ICRC....2...16W,1985ICRC....2...88W}
      \end{itemize}
   \item Balloon (1984/06+1984/07): \citeauthor{1987ICRC....2...72B} \citeyear{1987ICRC....2...72B} \cite{1987ICRC....2...72B}
   \item Balloon (1987/05+1988/05+1989/05+1991/05): \citeauthor{1993PhRvD..48.1949I} \citeyear{1993PhRvD..48.1949I} \cite{1993PhRvD..48.1949I}
   \item Balloon (1989/09): \citeauthor{1995PhRvD..52.6219H} \citeyear{1995PhRvD..52.6219H} \cite{1995PhRvD..52.6219H}
   \item Balloon (1989+1991)
      \begin{itemize}[leftmargin=15pt]\vspace{-8pt}
         \item Balloon-EWAsym (1989/05+1991/05): \citeauthor{1997APh.....6..155K} \citeyear{1997APh.....6..155K} \cite{1997APh.....6..155K}
         \item Balloon-OpeningAngle (1989/05+1991/05): \citeauthor{1997APh.....6..155K} \citeyear{1997APh.....6..155K} \cite{1997APh.....6..155K}
      \end{itemize}
   \item Balloon (1990/07): \citeauthor{1995ICRC....2..598B} \citeyear{1995ICRC....2..598B} \cite{1995ICRC....2..598B}
   \item Balloon (1991/09): \citeauthor{1994ApJ...429..736B} \citeyear{1994ApJ...429..736B} \cite{1994ApJ...429..736B}
   \item~ALICE
      \begin{itemize}[leftmargin=15pt]\vspace{-8pt}
         \item ALICE (1987/08): \citeauthor{1996A&A...314..785H} \citeyear{1996A&A...314..785H} \cite{1996A&A...314..785H}
         \item ALICE (1987/08+1987/08): \citeauthor{1992APh.....1...33E} \citeyear{1992APh.....1...33E} \cite{1992APh.....1...33E}
     \end{itemize}
   \item ATIC02 (2003/01): \citeauthor{2008ICRC....2....3P} \citeyear{2008ICRC....2....3P}, \citeyear{2009BRASP..73..564P} \cite{2008ICRC....2....3P,2009BRASP..73..564P}
   \item~BESS
      \begin{itemize}[leftmargin=15pt]\vspace{-8pt}
         \item BESS93 (1993/07): \citeauthor{1997ApJ...474..479M} \citeyear{1997ApJ...474..479M} \cite{1997ApJ...474..479M}, \citeauthor{2002ApJ...564..244W} \citeyear{2002ApJ...564..244W} \cite{2002ApJ...564..244W}
         \item BESS94 (1994/07): \citeauthor{2001AdSpR..26.1831S} \citeyear{2001AdSpR..26.1831S} \cite{2001AdSpR..26.1831S}, \citeauthor{2003ICRC....4.1805M} \citeyear{2003ICRC....4.1805M}, \citeyear{2005AdSpR..35..151M} \cite{2003ICRC....4.1805M,2005AdSpR..35..151M}, \citeauthor{2013AdSpR..51..234K} \citeyear{2013AdSpR..51..234K} \cite{2013AdSpR..51..234K}
         \item BESS95 (1995/07): 
         \citeauthor{1998PhRvL..81.4052M} \citeyear{1998PhRvL..81.4052M} \cite{1998PhRvL..81.4052M}, \citeauthor{2001AdSpR..26.1831S} \citeyear{2001AdSpR..26.1831S} \cite{2001AdSpR..26.1831S}, \citeauthor{2003ICRC....4.1805M} \citeyear{2003ICRC....4.1805M}, \citeyear{2005AdSpR..35..151M} \cite{2003ICRC....4.1805M,2005AdSpR..35..151M}, \citeauthor{2013AdSpR..51..234K}  \citeyear{2013AdSpR..51..234K} \cite{2013AdSpR..51..234K}
         \item BESS97 (1997/07): \citeauthor{2000PhRvL..84.1078O} \citeyear{2000PhRvL..84.1078O} \cite{2000PhRvL..84.1078O}, \citeauthor{2003ICRC....4.1805M} \citeyear{2003ICRC....4.1805M}, \citeyear{2005AdSpR..35..151M} \cite{2003ICRC....4.1805M,2005AdSpR..35..151M}, \citeauthor{2007APh....28..154S} \citeyear{2007APh....28..154S} \cite{2007APh....28..154S}, \citeauthor{2013AdSpR..51..234K} \citeyear{2013AdSpR..51..234K} \cite{2013AdSpR..51..234K}
         \item BESS98 (1998/07): \citeauthor{2000ApJ...545.1135S} \citeyear{2000ApJ...545.1135S} \cite{2000ApJ...545.1135S}, \citeauthor{2001APh....16..121M} \citeyear{2001APh....16..121M} \cite{2001APh....16..121M}, \citeauthor{2003ICRC....4.1805M} \citeyear{2003ICRC....4.1805M}, \citeyear{2005AdSpR..35..151M} \cite{2003ICRC....4.1805M,2005AdSpR..35..151M}, \citeauthor{2007APh....28..154S} \citeyear{2007APh....28..154S} \cite{2007APh....28..154S}
         \item BESS99 (1999/08): \citeauthor{2002PhRvL..88e1101A} \citeyear{2002PhRvL..88e1101A} \cite{2002PhRvL..88e1101A}, \citeauthor{2007APh....28..154S} \citeyear{2007APh....28..154S} \cite{2007APh....28..154S}
         \item BESS00 (2000/08): \citeauthor{2002PhRvL..88e1101A} \citeyear{2002PhRvL..88e1101A} \cite{2002PhRvL..88e1101A}, \citeauthor{2007APh....28..154S} \citeyear{2007APh....28..154S} \cite{2007APh....28..154S}, \citeauthor{2013AdSpR..51..234K} \citeyear{2013AdSpR..51..234K} \cite{2013AdSpR..51..234K}
         \item BESS-TeV (2002/08): \citeauthor{2004PhLB..594...35H} \citeyear{2004PhLB..594...35H} \cite{2004PhLB..594...35H}, \citeauthor{2005ICRC....3...13H}  \citeyear{2005ICRC....3...13H} \cite{2005ICRC....3...13H}, \citeauthor{2007APh....28..154S} \citeyear{2007APh....28..154S} \cite{2007APh....28..154S}
         \item BESS-PolarI (2004/12): \citeauthor{2008PhLB..670..103B} \citeyear{2008PhLB..670..103B}, \citeyear{2015arXiv150601267A} \cite{2008PhLB..670..103B,2015arXiv150601267A}
         \item BESS-PolarII (2007/12-2008/01): \citeauthor{2012PhRvL.108e1102A} \citeyear{2012PhRvL.108e1102A}, \citeyear{2015arXiv150601267A} \cite{2012PhRvL.108e1102A,2015arXiv150601267A}
      \end{itemize}
   \item~CAPRICE
      \begin{itemize}[leftmargin=15pt]\vspace{-8pt}
         \item CAPRICE94 (1994/08): \citeauthor{1997ApJ...487..415B} \citeyear{1997ApJ...487..415B}, \citeyear{1999ApJ...518..457B} \cite{1997ApJ...487..415B,1999ApJ...518..457B}
         \item CAPRICE98 (1998/05): \citeauthor{2001ApJ...561..787B} \citeyear{2001ApJ...561..787B}, \citeyear{2003APh....19..583B} \cite{2001ApJ...561..787B,2003APh....19..583B}, \citeauthor{2003ICRC....4.1809M} \citeyear{2003ICRC....4.1809M} \cite{2003ICRC....4.1809M}, \citeauthor{2004ApJ...615..259P}  \citeyear{2004ApJ...615..259P} \cite{2004ApJ...615..259P}
      \end{itemize}
   \item~CREAM
      \begin{itemize}[leftmargin=15pt]\vspace{-8pt}
         \item CREAM-I (2004/12-2005/01): \citeauthor{2008APh....30..133A} \citeyear{2008APh....30..133A} \cite{2008APh....30..133A}, \citeauthor{2011ApJ...728..122Y} \citeyear{2011ApJ...728..122Y} \cite{2011ApJ...728..122Y}
         \item CREAM-I+III (2004+2007): \citeauthor{2017ApJ...839....5Y} \citeyear{2017ApJ...839....5Y} \cite{2017ApJ...839....5Y}
         \item CREAM-II (2005/12-2006/01): \citeauthor{2009ApJ...707..593A} \citeyear{2009ApJ...707..593A}, \citeyear{2010ApJ...715.1400A} \cite{2009ApJ...707..593A,2010ApJ...715.1400A}
         \item CREAM-III (2007/12-2008/01): \citeauthor{2017ApJ...839....5Y} \citeyear{2017ApJ...839....5Y} \cite{2017ApJ...839....5Y}
      \end{itemize}
   \item CRISIS (1977/05): \citeauthor{1980ApJ...240L..53F} \citeyear{1980ApJ...240L..53F} \cite{1980ApJ...240L..53F}, \citeauthor{1981ApJ...246.1014Y} \citeyear{1981ApJ...246.1014Y} \cite{1981ApJ...246.1014Y}
   \item HEAT-pbar (2000/06): \citeauthor{2001PhRvL..87A1101B} \citeyear{2001PhRvL..87A1101B} \cite{2001PhRvL..87A1101B}
   \item HEIST (1988/08): \citeauthor{1992ApJ...391L..89G} \citeyear{1992ApJ...391L..89G} \cite{1992ApJ...391L..89G}
   \item IMAX92 (1992/07): \citeauthor{1996PhRvL..76.3057M} \citeyear{1996PhRvL..76.3057M} \cite{1996PhRvL..76.3057M}, \citeauthor{1998ApJ...496..490R}  \citeyear{1998ApJ...496..490R} \cite{1998ApJ...496..490R}, \citeauthor{2000ApJ...533..281M}  \citeyear{2000ApJ...533..281M} \cite{2000ApJ...533..281M}, \citeauthor{2000AIPC..528..425D} \citeyear{2000AIPC..528..425D} \cite{2000AIPC..528..425D}
   \item IRIS (1976/09): \citeauthor{1979ApJ...230..607T} \citeyear{1979ApJ...230..607T} \cite{1979ApJ...230..607T,1979ApJ...232L.161T}
   \item ISOMAX (1998/08): \citeauthor{2004ApJ...611..892H} \citeyear{2004ApJ...611..892H} \cite{2004ApJ...611..892H}
   \item JACEE (1979+\dots+1995): \citeauthor{1998ApJ...502..278A} \citeyear{1998ApJ...502..278A} \cite{1998ApJ...502..278A}
   \item LEAP (1987/08): \citeauthor{1991ApJ...378..763S} \citeyear{1991ApJ...378..763S} \cite{1991ApJ...378..763S}
   \item~MASS
      \begin{itemize}[leftmargin=15pt]\vspace{-8pt}
         \item MASS89 (1989/09): \citeauthor{1991ApJ...380..230W} \citeyear{1991ApJ...380..230W} \cite{1991ApJ...380..230W}
         \item MASS91 (1991/09): \citeauthor{1996ApJ...467L..33H} \citeyear{1996ApJ...467L..33H} \cite{1996ApJ...467L..33H}, \citeauthor{1999PhRvD..60e2002B} \citeyear{1999PhRvD..60e2002B} \cite{1999PhRvD..60e2002B}, \citeauthor{1999ICRC....3...77B} \citeyear{1999ICRC....3...77B} \cite{1999ICRC....3...77B}
      \end{itemize}
   \item MUBEE (1975/09+1978/08+1986/07+1987/07): \citeauthor{1993ICRC....2...13Z} \citeyear{1993ICRC....2...13Z} \cite{1993ICRC....2...13Z}
   \item RICH-II (1997/10): \citeauthor{2003APh....18..487D} \citeyear{2003APh....18..487D} \cite{2003APh....18..487D}
   \item RUNJOB (1995+1996+1997+1998+1999): \citeauthor{2005ApJ...628L..41D} \citeyear{2005ApJ...628L..41D} \cite{2005ApJ...628L..41D}
   \item~SMILI
      \begin{itemize}[leftmargin=15pt]\vspace{-8pt}
         \item SMILI-I (1989/09): \citeauthor{1993ApJ...413..268B} \citeyear{1993ApJ...413..268B} \cite{1993ApJ...413..268B}
         \item SMILI-II (1991/07): \citeauthor{2000ApJ...534..757A} \citeyear{2000ApJ...534..757A} \cite{2000ApJ...534..757A}, \citeauthor{1995ICRC....2..630W} \citeyear{1995ICRC....2..630W} \cite{1995ICRC....2..630W}
      \end{itemize}
   \item~TRACER
      \begin{itemize}[leftmargin=15pt]\vspace{-8pt}
         \item TRACER99: \citeauthor{2004ApJ...607..333G} \citeyear{2004ApJ...607..333G} \cite{2004ApJ...607..333G}
         \item TRACER03 (2003/12): \citeauthor{2008ApJ...678..262A} \citeyear{2008ApJ...678..262A} \cite{2008ApJ...678..262A}
         \item TRACER06 (2006/07): \citeauthor{2011ApJ...742...14O} \citeyear{2011ApJ...742...14O} \cite{2011ApJ...742...14O}
      \end{itemize}
\noindent\underline{Space}\vspace{5pt}
   \item~ACE
      \begin{itemize}[leftmargin=15pt]\vspace{-8pt}
         \item ACE-CRIS (1997/08-1998/04): \citeauthor{2009ApJ...698.1666G} \citeyear{2009ApJ...698.1666G} \cite{2009ApJ...698.1666G}, \citeauthor{2013ApJ...770..117L} \citeyear{2013ApJ...770..117L} \cite{2013ApJ...770..117L}
         \item ACE-CRIS (1997/08-1998/12): \citeauthor{1999ApJ...523L..61W} \citeyear{1999ApJ...523L..61W} \cite{1999ApJ...523L..61W}
         \item ACE-CRIS (1997/08-1999/07): \citeauthor{2001ApJ...563..768Y} \citeyear{2001ApJ...563..768Y} \cite{2001ApJ...563..768Y}, \citeauthor{2009ApJ...695..666O} \citeyear{2009ApJ...695..666O} \cite{2009ApJ...695..666O}
         \item ACE-CRIS (1997/12-1999/09): \citeauthor{2005ApJ...634..351B} \citeyear{2005ApJ...634..351B} \cite{2005ApJ...634..351B}
         \item ACE-CRIS (1997/12-2000/07): \citeauthor{2001AdSpR..27..767B} \citeyear{2001AdSpR..27..767B} \cite{2001AdSpR..27..767B}
         \item ACE-CRIS (1998/01-1999/01): \citeauthor{2006AdSpR..38.1558D} \citeyear{2006AdSpR..38.1558D} \cite{2006AdSpR..38.1558D}
         \item ACE-CRIS (2001/05-2003/09):  \citeauthor{2009ApJ...698.1666G} \citeyear{2009ApJ...698.1666G} \cite{2009ApJ...698.1666G}, \citeauthor{2013ApJ...770..117L} \citeyear{2013ApJ...770..117L} \cite{2013ApJ...770..117L}
         \item ACE-CRIS (2009/03-2010/01): \citeauthor{2013ApJ...770..117L} \citeyear{2013ApJ...770..117L} \cite{2013ApJ...770..117L}
         \item ACE-SIS (1997/08-1999/07): \citeauthor{2001ApJ...563..768Y} \citeyear{2001ApJ...563..768Y} \cite{2001ApJ...563..768Y}
      \end{itemize}
   \item~AMS01
      \begin{itemize}[leftmargin=15pt]\vspace{-8pt}
         \item AMS01 (1998/06): \citeauthor{2000PhLB..490...27A} \citeyear{2000PhLB..490...27A} \cite{2000PhLB..490...27A,2000PhLB..494..193A}, \citeauthor{2003JHEP...11..048X} \citeyear{2003JHEP...11..048X} \cite{2003JHEP...11..048X}, \citeauthor{2002PhR...366..331A} \citeyear{2002PhR...366..331A}, \citeyear{2010ApJ...724..329A}, \citeyear{2011ApJ...736..105A} \cite{2002PhR...366..331A,2010ApJ...724..329A,2011ApJ...736..105A}
      \end{itemize}
   \item~AMS02
      \begin{itemize}[leftmargin=15pt]\vspace{-8pt}
         \item AMS02 (2011/05-2011/06): \citeauthor{2018PhRvL.121e1102A} \citeyear{2018PhRvL.121e1102A} \cite{2018PhRvL.121e1102A,2018PhRvL.121e1101A}
         \item AMS02 (2011/05-2011/08): \citeauthor{2019PhRvL.123r1102A} \citeyear{2019PhRvL.123r1102A} \cite{2019PhRvL.123r1102A}
         \item AMS02 (2011/05-2013/11): \citeauthor{2014PDU.....4....6B}  \citeyear{2014PDU.....4....6B} \cite{2014PDU.....4....6B}, \citeauthor{2014PhRvL.113v1102A} \citeyear{2014PhRvL.113v1102A}, \citeyear{2015PhRvL.114q1103A} \cite{2014PhRvL.113v1102A,2015PhRvL.114q1103A,2015PhRvL.115u1101A}
         \item AMS02 (2011/05-2016/05): \citeauthor{2016PhRvL.117w1102A} \citeyear{2016PhRvL.117w1102A}, \citeyear{2017PhRvL.119y1101A}, \citeyear{2018PhRvL.121e1103A} \cite{2016PhRvL.117w1102A,2017PhRvL.119y1101A,2018PhRvL.120b1101A,2018PhRvL.121e1103A}
         \item AMS02 (2011/05-2017/05): \citeauthor{2018PhRvL.121e1102A} \citeyear{2018PhRvL.121e1102A} \cite{2018PhRvL.121e1102A}
         \item AMS02 (2011/05-2017/11): \citeauthor{2019PhRvL.122d1102A} \citeyear{2019PhRvL.122d1102A} \cite{2019PhRvL.122d1102A,2019PhRvL.123r1102A}
         \item AMS02 time series: \citeauthor{2018PhRvL.121e1101A} \citeyear{2018PhRvL.121e1101A}, \citeyear{2019PhRvL.123r1102A} \cite{2018PhRvL.121e1101A,2019PhRvL.123r1102A}
      \end{itemize}
   \item CALET (2015/10-2018/08): \citeauthor{2019PhRvL.122r1102A} \citeyear{2019PhRvL.122r1102A} \cite{2019PhRvL.122r1102A}
   \item CRN-Spacelab2 (1985/07-1985/08): \citeauthor{1990ApJ...349..625S} \citeyear{1990ApJ...349..625S} \cite{1990ApJ...349..625S}, \citeauthor{1991ApJ...374..356M} \citeyear{1991ApJ...374..356M} \cite{1991ApJ...374..356M}
   \item~CRRES
      \begin{itemize}[leftmargin=15pt]\vspace{-8pt}
      \item CRRES (1990/07-1992/10): \citeauthor{1996ApJ...466..457D} \citeyear{1996ApJ...466..457D} \cite{1996ApJ...466..457D}
      \item CRRES (1990/12-1991/03): \citeauthor{2000SoPh..195..175C} \citeyear{2000SoPh..195..175C} \cite{2000SoPh..195..175C}
      \end{itemize}
   \item Discoverer36 (1961/12): \citeauthor{1964JGR....69.3939S} \citeyear{1964JGR....69.3939S} \cite{1964JGR....69.3939S}
   \item E6 (1974/12-1977/12): \citeauthor{2019A&A...625A.153M} \citeyear{2019A&A...625A.153M} \cite{2019A&A...625A.153M}
   \item EPHIN time series (1995-2014): \citeauthor{2016SoPh..291..965K} \citeyear{2016SoPh..291..965K} \cite{2016SoPh..291..965K}
   \item Explorer12 (1961/08): \citeauthor{1962JGR....67.4983B} \citeyear{1962JGR....67.4983B} \cite{1962JGR....67.4983B}
   \item Gemini11 (1966/08): \citeauthor{1970PhRvD...1.1021D} \citeyear{1970PhRvD...1.1021D} \cite{1970PhRvD...1.1021D}
   \item HEAO3-C2 (1979/10-1980/06): \citeauthor{1988A&A...193...69F} \citeyear{1988A&A...193...69F} \cite{1988A&A...193...69F}, \citeauthor{1990A&A...233...96E} \citeyear{1990A&A...233...96E} \cite{1990A&A...233...96E}
   \item~IMP
      \begin{itemize}[leftmargin=15pt]\vspace{-8pt}
         \item IMP1 (1963/11-1964/02): \citeauthor{1965JGR....70.5333G} \citeyear{1965JGR....70.5333G} \cite{1965JGR....70.5333G}
         \item IMP1 (1963/11-1964/05): \citeauthor{1965JGR....70.3515F} \citeyear{1965JGR....70.3515F} \cite{1965JGR....70.3515F}, \citeauthor{1964PhRvL..13..783M} \citeyear{1964PhRvL..13..783M} \cite{1964PhRvL..13..783M}
         \item IMP1 (1964/02-1964/05): \citeauthor{1965JGR....70.5333G} \citeyear{1965JGR....70.5333G} \cite{1965JGR....70.5333G}
         \item IMP3 (1965/05-1965/06): \citeauthor{1966JGR....71.1771B} \citeyear{1966JGR....71.1771B} \cite{1966JGR....71.1771B}
         \item IMP3 (1965/05-1965/09): \citeauthor{1966PhRvL..16..813F} \citeyear{1966PhRvL..16..813F} \cite{1966PhRvL..16..813F,1966PhRvL..17..329F}, \citeauthor{1970ApJ...159...61H} \citeyear{1970ApJ...159...61H} \cite{1970ApJ...159...61H}
         \item IMP3 (1965/06-1965/12): \citeauthor{1971ApJ...166..221H} \citeyear{1971ApJ...166..221H} \cite{1971ApJ...166..221H}
         \item IMP4 (1967/06-1967/10): \citeauthor{1971ApJ...166..221H} \citeyear{1971ApJ...166..221H} \cite{1971ApJ...166..221H}
         \item IMP4 (1967/07-1967/10): \citeauthor{1970ApJ...159...61H} \citeyear{1970ApJ...159...61H} \cite{1970ApJ...159...61H}
         \item IMP5 (1969/06-1969/09): \citeauthor{1971ApJ...166..221H} \citeyear{1971ApJ...166..221H} \cite{1971ApJ...166..221H}, \citeauthor{1972ApJ...171..139M} \citeyear{1972ApJ...171..139M} \cite{1972ApJ...171..139M}
         \item IMP5 (1969/06-1970/06): \citeauthor{1972ApJ...171..139M} \citeyear{1972ApJ...171..139M} \cite{1972ApJ...171..139M}
         \item IMP5 (1970/04-1970/10): \citeauthor{1975ApJ...202..265G} \citeyear{1975ApJ...202..265G} \cite{1975ApJ...202..265G}
         \item IMP5 (1971/06-1971/09): \citeauthor{1975ApJ...202..265G} \citeyear{1975ApJ...202..265G} \cite{1975ApJ...202..265G}
         \item IMP5 (1972/06-1972/09): \citeauthor{1975ApJ...202..265G} \citeyear{1975ApJ...202..265G} \cite{1975ApJ...202..265G}
         \item IMP7 (1972/09-1972/12): \citeauthor{1975ApJ...202..815T} \citeyear{1975ApJ...202..815T} \cite{1975ApJ...202..815T}
         \item IMP7 (1972/10-1976/10): \citeauthor{1979ApJ...232L..95G} \citeyear{1979ApJ...232L..95G} \cite{1979ApJ...232L..95G}
         \item IMP7 (1973/01-1974-10): \citeauthor{1976ApJ...206..616M} \citeyear{1976ApJ...206..616M} \cite{1976ApJ...206..616M}
         \item IMP7 (1973/05-1973/07): \citeauthor{1975ApJ...202..265G} \citeyear{1975ApJ...202..265G} \cite{1975ApJ...202..265G}
         \item IMP7 (1973/05-1973/08): \citeauthor{1975ApJ...202..265G} \citeyear{1975ApJ...202..265G} \cite{1975ApJ...202..265G}
         \item IMP7 (1973/05-1973/12): \citeauthor{1975ApJ...202..265G} \citeyear{1975ApJ...202..265G} \cite{1975ApJ...202..265G}
         \item IMP7\&8 (1972/09-1975/09): \citeauthor{1977ApJ...217..859G} \citeyear{1977ApJ...217..859G} \cite{1977ApJ...217..859G,1977ICRC....1..301G}
         \item IMP7\&8 (1973/02-1977/09): \citeauthor{1981ApJ...244..695G} \citeyear{1981ApJ...244..695G} \cite{1981ApJ...244..695G}
         \item IMP7\&8 (1974/01-1980/05): \citeauthor{1981ICRC....2...72G} \citeyear{1981ICRC....2...72G} \cite{1981ICRC....2...72G}
         \item IMP8 (1974/01-1977/11): \citeauthor{1985ApJ...294..455B} \citeyear{1985ApJ...294..455B} \cite{1985ApJ...294..455B}
         \item IMP8 (1974/01-1978/10): \citeauthor{1987ApJS...64..269G} \citeyear{1987ApJS...64..269G} \cite{1987ApJS...64..269G}
         \item IMP8 (1974/07-1974/09): \citeauthor{1975ApJ...202..265G} \citeyear{1975ApJ...202..265G} \cite{1975ApJ...202..265G}
         \item IMP8 (1975/02-1976/05): \citeauthor{1985ApJ...294..455B} \citeyear{1985ApJ...294..455B} \cite{1985ApJ...294..455B}
         \item IMP8 (1976/06-1977/04): \citeauthor{1985ApJ...294..455B} \citeyear{1985ApJ...294..455B} \cite{1985ApJ...294..455B}
         \item IMP8 (1977/01-1979/12): \citeauthor{1983ApJ...275L..15E} \citeyear{1983ApJ...275L..15E} \cite{1983ApJ...275L..15E}
      \end{itemize}
   \item~ISEE
      \begin{itemize}[leftmargin=15pt]\vspace{-8pt}
         \item ISEE3-HIST (1978/08-1978/12): \citeauthor{1980ApJ...236L.121M} \citeyear{1980ApJ...236L.121M}, \citeyear{1981ApJ...251L..27M}, \citeyear{1986ApJ...311..979M} \cite{1980ApJ...236L.121M,1980ApJ...235L..95M,1981ApJ...251L..27M,1986ApJ...311..979M}
         \item ISEE3-MEH (1978/08-1978/12): \citeauthor{1986ApJ...303..816K} \citeyear{1986ApJ...303..816K} \cite{1986ApJ...303..816K}
         \item ISEE3-HKH (1978/08-1979/08): \citeauthor{1980ApJ...239L.139W} \citeyear{1980ApJ...239L.139W} \cite{1980ApJ...239L.139W}
         \item ISEE3-HKH (1978/08-1980/05): \citeauthor{1981ApJ...247L.119W} \citeyear{1981ApJ...247L.119W} \cite{1981ApJ...247L.119W,1981PhRvL..46..682W}
         \item ISEE3-HKH (1978/08-1981/04): \citeauthor{1983ICRC....9..147W} \citeyear{1983ICRC....9..147W} \cite{1983ICRC....9..147W}, \citeauthor{1988ApJ...328..940K} \citeyear{1988ApJ...328..940K} \cite{1988ApJ...328..940K}, \citeauthor{1993ApJ...405..567L} \citeyear{1993ApJ...405..567L} \cite{1993ApJ...405..567L}
         \item ISEE3-MEH (1978/08-1984/04): \citeauthor{1986ApJ...303..816K} \citeyear{1986ApJ...303..816K} \cite{1986ApJ...303..816K}
         \item ISEE3-MEH (1979/01-1979/12): \citeauthor{1986ApJ...303..816K} \citeyear{1986ApJ...303..816K} \cite{1986ApJ...303..816K}
         \item ISEE3-MEH (1980/01-1980/12): \citeauthor{1986ApJ...303..816K} \citeyear{1986ApJ...303..816K} \cite{1986ApJ...303..816K}
         \item ISEE3-MEH (1981/01-1981/12): \citeauthor{1986ApJ...303..816K} \citeyear{1986ApJ...303..816K} \cite{1986ApJ...303..816K}
         \item ISEE3-MEH (1982/01-1982/12): \citeauthor{1986ApJ...303..816K} \citeyear{1986ApJ...303..816K} \cite{1986ApJ...303..816K}
         \item ISEE3-MEH (1983/01-1983/12): \citeauthor{1986ApJ...303..816K} \citeyear{1986ApJ...303..816K} \cite{1986ApJ...303..816K}
         \item ISEE3-MEH (1984/01-1984/12): \citeauthor{1986ApJ...303..816K} \citeyear{1986ApJ...303..816K} \cite{1986ApJ...303..816K}
      \end{itemize}
   \item~NUCLEON
      \begin{itemize}[leftmargin=15pt]\vspace{-8pt}
         \item NUCLEON (2015/07-2017/06): \citeauthor{2019AdSpR..64.2546G} \citeyear{2019AdSpR..64.2546G} \cite{2019AdSpR..64.2546G}
         \item NUCLEON-IC (2015/07-2017/06): \citeauthor{2019AdSpR..64.2559G} \citeyear{2019AdSpR..64.2559G} \cite{2019AdSpR..64.2559G}
         \item NUCLEON-KLEM (2015/07-2017/06): \citeauthor{2019AdSpR..64.2559G} \citeyear{2019AdSpR..64.2559G} \cite{2019AdSpR..64.2559G}
      \end{itemize}
   \item~OGO
      \begin{itemize}[leftmargin=15pt]\vspace{-8pt}
         \item OGO1 (1964/10-1965/11): \citeauthor{1969ApJ...155..609C} \citeyear{1969ApJ...155..609C} \cite{1969ApJ...155..609C}
         \item OGO1 (1965/03-1965/06): \citeauthor{1966JGR....71.1771B} \citeyear{1966JGR....71.1771B} \cite{1966JGR....71.1771B}
         \item OGO5 (1968/03-1968/10): \citeauthor{1970ICRC....1..345T} \citeyear{1970ICRC....1..345T} \cite{1970ICRC....1..345T}
      \end{itemize}
   \item~PAMELA
      \begin{itemize}[leftmargin=15pt]\vspace{-8pt}
         \item PAMELA (2006/07-2006/07): \citeauthor{2013ApJ...765...91A} \citeyear{2013ApJ...765...91A} \cite{2013ApJ...765...91A}
         \item PAMELA (2006/07-2006/08): \citeauthor{2013ApJ...765...91A} \citeyear{2013ApJ...765...91A} \cite{2013ApJ...765...91A}
         \item PAMELA (2006/07-2008/03): \citeauthor{2014ApJ...791...93A} \citeyear{2014ApJ...791...93A} \cite{2014ApJ...791...93A}
         \item PAMELA (2006/07-2008/06): \citeauthor{2009PhRvL.102e1101A} \citeyear{2009PhRvL.102e1101A} \cite{2009PhRvL.102e1101A}
         \item PAMELA (2006/07-2008/12): \citeauthor{2010PhRvL.105l1101A} \citeyear{2010PhRvL.105l1101A}, \citeyear{2011Sci...332...69A} \cite{2010PhRvL.105l1101A,2011Sci...332...69A}
         \item PAMELA (2006/07-2009/12): \citeauthor{2013JETPL..96..621A} \citeyear{2013JETPL..96..621A} \cite{2013JETPL..96..621A}
         \item PAMELA (2006/08-2006/09): \citeauthor{2013ApJ...765...91A} \citeyear{2013ApJ...765...91A} \cite{2013ApJ...765...91A}
         \item PAMELA-CALO (2006/06-2010/01): \citeauthor{2013AdSpR..51..219A} \citeyear{2013AdSpR..51..219A} \cite{2013AdSpR..51..219A}
         \item PAMELA-CALO (2006/07-2007/12): \citeauthor{2016ApJ...818...68A} \citeyear{2016ApJ...818...68A} \cite{2016ApJ...818...68A}
         \item PAMELA-CALO (2006/07-2014/09): \citeauthor{2018ApJ...862..141M} \citeyear{2018ApJ...862..141M} \cite{2018ApJ...862..141M}
         \item PAMELA-TOF (2006/07-2007/12): \citeauthor{2013ApJ...770....2A} \citeyear{2013ApJ...770....2A}, \citeyear{2016ApJ...818...68A} \cite{2013ApJ...770....2A,2016ApJ...818...68A}
         \item PAMELA-TOF (2006/07-2014/09): \citeauthor{2018ApJ...862..141M} \citeyear{2018ApJ...862..141M} \cite{2018ApJ...862..141M}
         \item PAMELA time series: \citeauthor{2013ApJ...765...91A} \citeyear{2013ApJ...765...91A}, \citeyear{2015ApJ...810..142A} \cite{2013ApJ...765...91A,2015ApJ...810..142A}, \citeauthor{2018ApJ...854L...2M} \citeyear{2018ApJ...854L...2M} \cite{2018ApJ...854L...2M}
      \end{itemize}
   \item~Pioneer
      \begin{itemize}[leftmargin=15pt]\vspace{-8pt}
         \item Pioneer8 (1968/04): \citeauthor{1971JGR....76.1605L} \citeyear{1971JGR....76.1605L} \cite{1971JGR....76.1605L}
         \item Pioneer10-HET (1972/03-1973/03): \citeauthor{1975ApJ...202..815T} \citeyear{1975ApJ...202..815T} \cite{1975ApJ...202..815T}
         \item Pioneer10-HET (1985/04-1988/11): \citeauthor{1994ApJ...435..464W} \citeyear{1994ApJ...435..464W} \cite{1994ApJ...435..464W}
      \end{itemize}
   \item SOKOL (1984/03-1986/01): \citeauthor{1993ICRC....2...17I} \citeyear{1993ICRC....2...17I} \cite{1993ICRC....2...17I}, \citeauthor{2017AdSpR..59..496T} \citeyear{2017AdSpR..59..496T} \cite{2017AdSpR..59..496T}
   \item Trek-MIR (1991/06-1992/12): \citeauthor{1996ApJ...468..679W} \citeyear{1996ApJ...468..679W} \cite{1996ApJ...468..679W}
   \item~Ulysses
      \begin{itemize}[leftmargin=15pt]\vspace{-8pt}
         \item Ulysses-HET (1990/10-1995/07): \citeauthor{1996ApJ...465..982D} \citeyear{1996ApJ...465..982D} \cite{1996ApJ...465..982D}, \citeauthor{1996A&A...316..555D} \citeyear{1996A&A...316..555D} \cite{1996A&A...316..555D}, \citeauthor{1997ApJ...475L..61C} \citeyear{1997ApJ...475L..61C} \cite{1997ApJ...475L..61C}, \citeauthor{1997ApJ...481..241D} \citeyear{1997ApJ...481..241D} \cite{1997ApJ...481..241D}, \citeauthor{1997ApJ...482..792T} \citeyear{1997ApJ...482..792T} \cite{1997ApJ...482..792T}
         \item Ulysses-HET (1990/10-1996/12): \citeauthor{1998ApJ...497L..85S}  \citeyear{1998ApJ...497L..85S} \cite{1998ApJ...497L..85S}, \citeauthor{1997ICRC....3..381C} \citeyear{1997ICRC....3..381C} \cite{1997ICRC....3..381C}
         \item Ulysses-HET (1990/10-1997/12): \citeauthor{1998ApJ...501L..59C} \citeyear{1998ApJ...501L..59C}
          \cite{1998ApJ...501L..59C,1998ApJ...509L..97C}, \citeauthor{1999ICRC....3...33C} \citeyear{1999ICRC....3...33C} \cite{1999ICRC....3...33C}, \citeauthor{2005ApJ...634..351B} \citeyear{2005ApJ...634..351B} \cite{2005ApJ...634..351B}
      \end{itemize}
   \item~Voyager
      \begin{itemize}[leftmargin=15pt]\vspace{-8pt}
         \item Voyager1 (2008/05): \citeauthor{2009JGRA..11402103W} \citeyear{2009JGRA..11402103W} \cite{2009JGRA..11402103W}
         \item Voyager1 (2012/09-2012/12): \citeauthor{2013Sci...341..150S} \citeyear{2013Sci...341..150S} \cite{2013Sci...341..150S}
         \item Voyager1 (2012/10-2012/12): \citeauthor{2013Sci...341..150S} \citeyear{2013Sci...341..150S} \cite{2013Sci...341..150S}
         \item Voyager1\&2 (1977/01-1991/12): \citeauthor{1994ApJ...423..426L} \citeyear{1994ApJ...423..426L} \cite{1994ApJ...423..426L}
         \item Voyager1\&2 (1977/01-1993/12): \citeauthor{1994ApJ...430L..69L}  \citeyear{1994ApJ...430L..69L} \cite{1994ApJ...430L..69L}, \citeauthor{1996ApJ...457..435W} \citeyear{1996ApJ...457..435W} \cite{1996ApJ...457..435W}
         \item Voyager1\&2 (1977/01-1996/12): \citeauthor{1997ApJ...488..454L} \citeyear{1997ApJ...488..454L} \cite{1997ApJ...488..454L,1997ICRC....3..389L}, \citeauthor{1997ApJ...476..766W} \citeyear{1997ApJ...476..766W} \cite{1997ApJ...476..766W}
         \item Voyager1\&2 (1977/01-1998/12): \citeauthor{1999ICRC....3...41L} \citeyear{1999ICRC....3...41L} \cite{1999ICRC....3...41L}
         \item Voyager1\&2 (1986/01-1989/12): \citeauthor{1994ApJ...426..366L} \citeyear{1994ApJ...426..366L} \cite{1994ApJ...426..366L}
         \item Voyager1-HET (1977/10-1977/11): \citeauthor{1983ApJ...275..391W} \citeyear{1983ApJ...275..391W} \cite{1983ApJ...275..391W}
         \item Voyager1-HET (1994/01-1994/09): \citeauthor{1995ApJ...451L..33S} \citeyear{1995ApJ...451L..33S} \cite{1995ApJ...451L..33S}
         \item Voyager1-HET (2012/12-2015/06): \citeauthor{2016ApJ...831...18C} \citeyear{2016ApJ...831...18C} \cite{2016ApJ...831...18C}
         \item Voyager1-HET-Aend (1977/10-1977/11): \citeauthor{1983ApJ...275..391W} \citeyear{1983ApJ...275..391W} \cite{1983ApJ...275..391W}
         \item Voyager1-HET-Aend (2012/12-2015/06): \citeauthor{2016ApJ...831...18C} \citeyear{2016ApJ...831...18C} \cite{2016ApJ...831...18C}
         \item Voyager1-HET-Bend (2012/12-2014/12): \citeauthor{2016ApJ...831...18C} \citeyear{2016ApJ...831...18C} \cite{2016ApJ...831...18C}
         \item Voyager1-LET (2012/12-2015/06): \citeauthor{2016ApJ...831...18C} \citeyear{2016ApJ...831...18C} \cite{2016ApJ...831...18C}
         \item Voyager2 (2019/03-2019/07): \citeauthor{2019NatAs...3.1013S} \citeyear{2019NatAs...3.1013S} \cite{2019NatAs...3.1013S}
         \item Voyager2-HET (1986/01-1987/12): \citeauthor{1991A&A...247..163F} \citeyear{1991A&A...247..163F} \cite{1991A&A...247..163F}
         \item Voyager2-HET (1987/01-1987/12): \citeauthor{1994ApJ...432..656S} \citeyear{1994ApJ...432..656S} \cite{1994ApJ...432..656S}
      \end{itemize}
   \noindent\underline{Ground}\vspace{5pt}
      \item H.E.S.S.-LEAnalysis (2004/10-2005/12): \citeauthor{2007PhRvD..75d2004A} \citeyear{2007PhRvD..75d2004A} \cite{2007PhRvD..75d2004A}
      \item~VERITAS
      \begin{itemize}[leftmargin=15pt]\vspace{-8pt}
         \item VERITAS (2007/09-2009/05): \citeauthor{2010PhDT........37W} \citeyear{2010PhDT........37W} \cite{2010PhDT........37W}
         \item VERITAS (2009/09-2012/05): \citeauthor{2018PhRvD..98b2009A} \citeyear{2018PhRvD..98b2009A} \cite{2018PhRvD..98b2009A}
      \end{itemize}

\end{itemize}

\subsection{Data for UHCRs ($Z>30$)}
\label{app:uhcr}

\begin{description}
   \item \underline{Balloon}
      \begin{itemize}[leftmargin=5pt]\vspace{-7pt}
         \item TIGER (2001/12+2003/12): \citeauthor{2009ApJ...697.2083R} \citeyear{2009ApJ...697.2083R} \cite{2009ApJ...697.2083R}
         \item SuperTIGER (2012/12-2013/03):   \citeauthor{2016ApJ...831..148M} \citeyear{2016ApJ...831..148M} \cite{2016ApJ...831..148M}
      \end{itemize}
   \item \underline{Space}
      \begin{itemize}[leftmargin=5pt]\vspace{-7pt}
         \item Ariel6 (1979/06-1982/02):   \citeauthor{1987ApJ...314..739F} \citeyear{1987ApJ...314..739F} \cite{1987ApJ...314..739F}
         \item HEAO3-HNE (1979/10-1980/06):   \citeauthor{1981ApJ...247L.115B} \citeyear{1981ApJ...247L.115B} \cite{1981ApJ...247L.115B}
         \item HEAO3-HNE (1979/10-1981/01):   \citeauthor{1982ApJ...261L.117B} \citeyear{1982ApJ...261L.117B} \cite{1982ApJ...261L.117B}, \citeauthor{1983ICRC....9..115S} \citeyear{1983ICRC....9..115S} \cite{1983ICRC....9..115S}
         \item HEAO3-HNE (1979/10-1981/01):   \citeauthor{1983ApJ...267L..93B} \citeyear{1983ApJ...267L..93B} \cite{1983ApJ...267L..93B}
         \item HEAO3-HNE (1979/10-1981/06):   \citeauthor{1985ApJ...297..111B} \citeyear{1985ApJ...297..111B} \cite{1985ApJ...297..111B}
         \item HEAO3-HNE (1979/10-1981/18):   \citeauthor{1989ApJ...346..997B} \citeyear{1989ApJ...346..997B} \cite{1989ApJ...346..997B}
         \item Trek (1991/06-1994/01):   \citeauthor{1998Natur.396...50W} \citeyear{1998Natur.396...50W} \cite{1998Natur.396...50W}
         \item Trek-Extended (1991/06-1995/11):   \citeauthor{2002ApJ...569..493W} \citeyear{2002ApJ...569..493W} \cite{2002ApJ...569..493W}
         \item UHCRE-LDEF (1984/01-1990/01):   \citeauthor{1996RadM...26..825D} \citeyear{1996RadM...26..825D} \cite{1996RadM...26..825D}, \citeauthor{2003RadM...36..287D} \citeyear{2003RadM...36..287D} \cite{2003RadM...36..287D}, \citeauthor{2012ApJ...747...40D}          \citeyear{2012ApJ...747...40D} \cite{2012ApJ...747...40D}
      \end{itemize}
\end{description}
}

\bibliography{crdb}

\begin{thebibliography}{-------}
\providecommand{\natexlab}[1]{#1}

\bibitem[{Ghelfi} \em{et~al.}(2017){Ghelfi}, {Maurin}, {Cheminet}, {Derome},
  {Hubert}, and {Melot}]{2017AdSpR..60..833G}
{Ghelfi}, A.; {Maurin}, D.; {Cheminet}, A.; {Derome}, L.; {Hubert}, G.;
  {Melot}, F.
\newblock {Neutron monitors and muon detectors for solar modulation studies: 2.
  phi time series}.
\newblock {\em AdSR} {\bf 2017}, {\em 60},~833--847,
  \href{http://xxx.lanl.gov/abs/1607.01976}{{\normalfont
  [arXiv:astro-ph.HE/1607.01976]}}.
\newblock
  doi:{\changeurlcolor{black}\href{https://doi.org/10.1016/j.asr.2016.06.027}{\detokenize{10.1016/j.asr.2016.06.027}}}.

\bibitem[{Shen} \em{et~al.}(2019){Shen}, {Qin}, {Zuo}, and
  {Wei}]{2019ApJ...887..132S}
{Shen}, Z.N.; {Qin}, G.; {Zuo}, P.; {Wei}, F.
\newblock {Modulation of Galactic Cosmic Rays from Helium to Nickel in the
  Inner Heliosphere}.
\newblock {\em \apj} {\bf 2019}, {\em 887},~132.
\newblock
  doi:{\changeurlcolor{black}\href{https://doi.org/10.3847/1538-4357/ab5520}{\detokenize{10.3847/1538-4357/ab5520}}}.

\bibitem[{Donnelly} \em{et~al.}(2012){Donnelly}, {Thompson}, {O\'Sullivan},
  {Daly}, {Drury}, {Domingo}, and {Wenzel}]{2012ApJ...747...40D}
{Donnelly}, J.; {Thompson}, A.; {O\'Sullivan}, D.; {Daly}, J.; {Drury}, L.;
  {Domingo}, V.; {Wenzel}, K.P.
\newblock {Actinide and Ultra-Heavy Abundances in the Local Galactic Cosmic
  Rays: An Analysis of the Results from the LDEF Ultra-Heavy Cosmic-Ray
  Experiment}.
\newblock {\em \\apj} {\bf 2012}, {\em 747},~40.
\newblock
  doi:{\changeurlcolor{black}\href{https://doi.org/10.1088/0004-637X/747/1/40}{\detokenize{10.1088/0004-637X/747/1/40}}}.

\bibitem[{Binns} \em{et~al.}(2014){Binns}, {Bose}, {Braun}, {Brandt},
  {Daniels}, {Dowkontt}, {Fitzsimmons}, {Hahne}, {Hams}, {Israel}, and
  et~al]{2014ApJ...788...18B}
{Binns}, W.R.; {Bose}, R.G.; {Braun}, D.L.; {Brandt}, T.J.; {Daniels}, W.M.;
  {Dowkontt}, P.F.; {Fitzsimmons}, S.P.; {Hahne}, D.J.; {Hams}, T.; {Israel},
  M.H.; et~al.
\newblock {The SUPERTIGER Instrument: Measurement of Elemental Abundances of
  Ultra-Heavy Galactic Cosmic Rays}.
\newblock {\em \apj} {\bf 2014}, {\em 788},~18.
\newblock
  doi:{\changeurlcolor{black}\href{https://doi.org/10.1088/0004-637X/788/1/18}{\detokenize{10.1088/0004-637X/788/1/18}}}.

\bibitem[{Abbott} \em{et~al.}(2017){Abbott}, {Abbott}, {Abbott}, {Acernese},
  {Ackley}, {Adams}, {Adams}, {Addesso}, others, {LIGO Scientific
  Collaboration}, and {Virgo Collaboration}]{2017PhRvL.119p1101A}
{Abbott}, B.P.; {Abbott}, R.; {Abbott}, T.D.; {Acernese}, F.; {Ackley}, K.;
  {Adams}, C.; {Adams}, T.; {Addesso}, P.; others.; {LIGO Scientific
  Collaboration}.; {Virgo Collaboration}.
\newblock {GW170817: Observation of Gravitational Waves from a Binary Neutron
  Star Inspiral}.
\newblock {\em \prl} {\bf 2017}, {\em 119},~161101,
  \href{http://xxx.lanl.gov/abs/1710.05832}{{\normalfont
  [arXiv:gr-qc/1710.05832]}}.
\newblock
  doi:{\changeurlcolor{black}\href{https://doi.org/10.1103/PhysRevLett.119.161101}{\detokenize{10.1103/PhysRevLett.119.161101}}}.

\bibitem[{Nicholl} \em{et~al.}(2017){Nicholl}, {Berger}, {Kasen}, {Metzger},
  {Elias}, {Brice{\~n}o}, {Alexander}, {Blanchard}, {Chornock},
  {Cowperthwaite}, and et~al]{2017ApJ...848L..18N}
{Nicholl}, M.; {Berger}, E.; {Kasen}, D.; {Metzger}, B.D.; {Elias}, J.;
  {Brice{\~n}o}, C.; {Alexander}, K.D.; {Blanchard}, P.K.; {Chornock}, R.;
  {Cowperthwaite}, P.S.; et~al.
\newblock {The Electromagnetic Counterpart of the Binary Neutron Star Merger
  LIGO/Virgo GW170817. III. Optical and UV Spectra of a Blue Kilonova from Fast
  Polar Ejecta}.
\newblock {\em \apjl} {\bf 2017}, {\em 848},~L18,
  \href{http://xxx.lanl.gov/abs/1710.05456}{{\normalfont
  [arXiv:astro-ph.HE/1710.05456]}}.
\newblock
  doi:{\changeurlcolor{black}\href{https://doi.org/10.3847/2041-8213/aa9029}{\detokenize{10.3847/2041-8213/aa9029}}}.

\bibitem[{Chornock} \em{et~al.}(2017){Chornock}, {Berger}, {Kasen},
  {Cowperthwaite}, {Nicholl}, {Villar}, {Alexand er}, {Blanchard}, {Eftekhari},
  {Fong}, and et~al]{2017ApJ...848L..19C}
{Chornock}, R.; {Berger}, E.; {Kasen}, D.; {Cowperthwaite}, P.S.; {Nicholl},
  M.; {Villar}, V.A.; {Alexand er}, K.D.; {Blanchard}, P.K.; {Eftekhari}, T.;
  {Fong}, W.; et~al.
\newblock {The Electromagnetic Counterpart of the Binary Neutron Star Merger
  LIGO/Virgo GW170817. IV. Detection of Near-infrared Signatures of r-process
  Nucleosynthesis with Gemini-South}.
\newblock {\em \apjl} {\bf 2017}, {\em 848},~L19,
  \href{http://xxx.lanl.gov/abs/1710.05454}{{\normalfont
  [arXiv:astro-ph.HE/1710.05454]}}.
\newblock
  doi:{\changeurlcolor{black}\href{https://doi.org/10.3847/2041-8213/aa905c}{\detokenize{10.3847/2041-8213/aa905c}}}.

\bibitem[{Grenier} \em{et~al.}(2015){Grenier}, {Black}, and
  {Strong}]{2015ARA&A..53..199G}
{Grenier}, I.A.; {Black}, J.H.; {Strong}, A.W.
\newblock {The Nine Lives of Cosmic Rays in Galaxies}.
\newblock {\em \araa} {\bf 2015}, {\em 53},~199--246.
\newblock
  doi:{\changeurlcolor{black}\href{https://doi.org/10.1146/annurev-astro-082214-122457}{\detokenize{10.1146/annurev-astro-082214-122457}}}.

\bibitem[{Lavalle} and {Salati}(2012)]{2012CRPhy..13..740L}
{Lavalle}, J.; {Salati}, P.
\newblock {Dark matter indirect signatures}.
\newblock {\em Comptes Rendus Physique} {\bf 2012}, {\em 13},~740--782,
  \href{http://xxx.lanl.gov/abs/1205.1004}{{\normalfont
  [arXiv:astro-ph.HE/1205.1004]}}.
\newblock
  doi:{\changeurlcolor{black}\href{https://doi.org/10.1016/j.crhy.2012.05.001}{\detokenize{10.1016/j.crhy.2012.05.001}}}.

\bibitem[{Abe} \em{et~al.}(2012){Abe}, {Fuke}, {Haino}, {Hams}, {Hasegawa},
  {Horikoshi}, {Itazaki}, {Kim}, {Kumazawa}, {Kusumoto}, {Lee}, {Makida},
  {Matsuda}, {Matsukawa}, {Matsumoto}, {Mitchell}, {Myers}, {Nishimura},
  {Nozaki}, {Orito}, {Ormes}, {Sakai}, {Sasaki}, {Seo}, {Shikaze}, {Shinoda},
  {Streitmatter}, {Suzuki}, {Takasugi}, {Takeuchi}, {Tanaka}, {Thakur},
  {Yamagami}, {Yamamoto}, {Yoshida}, and {Yoshimura}]{2012PhRvL.108m1301A}
{Abe}, K.; {Fuke}, H.; {Haino}, S.; {Hams}, T.; {Hasegawa}, M.; {Horikoshi},
  A.; {Itazaki}, A.; {Kim}, K.C.; {Kumazawa}, T.; {Kusumoto}, A.; {Lee}, M.H.;
  {Makida}, Y.; {Matsuda}, S.; {Matsukawa}, Y.; {Matsumoto}, K.; {Mitchell},
  J.W.; {Myers}, Z.; {Nishimura}, J.; {Nozaki}, M.; {Orito}, R.; {Ormes}, J.F.;
  {Sakai}, K.; {Sasaki}, M.; {Seo}, E.S.; {Shikaze}, Y.; {Shinoda}, R.;
  {Streitmatter}, R.E.; {Suzuki}, J.; {Takasugi}, Y.; {Takeuchi}, K.; {Tanaka},
  K.; {Thakur}, N.; {Yamagami}, T.; {Yamamoto}, A.; {Yoshida}, T.; {Yoshimura},
  K.
\newblock {Search for Antihelium with the BESS-Polar Spectrometer}.
\newblock {\em \\prl} {\bf 2012}, {\em 108},~131301,
  \href{http://xxx.lanl.gov/abs/1201.2967}{{\normalfont
  [arXiv:astro-ph.CO/1201.2967]}}.
\newblock
  doi:{\changeurlcolor{black}\href{https://doi.org/10.1103/PhysRevLett.108.131301}{\detokenize{10.1103/PhysRevLett.108.131301}}}.

\bibitem[{Aramaki} \em{et~al.}(2016){Aramaki}, {Boggs}, {Bufalino}, {Dal}, {von
  Doetinchem}, {Donato}, {Fornengo}, {Fuke}, {Grefe}, {Hailey}, and
  et~al]{2016PhR...618....1A}
{Aramaki}, T.; {Boggs}, S.; {Bufalino}, S.; {Dal}, L.; {von Doetinchem}, P.;
  {Donato}, F.; {Fornengo}, N.; {Fuke}, H.; {Grefe}, M.; {Hailey}, C.; et~al.
\newblock {Review of the theoretical and experimental status of dark matter
  identification with cosmic-ray antideuterons}.
\newblock {\em \physrep} {\bf 2016}, {\em 618},~1--37,
  \href{http://xxx.lanl.gov/abs/1505.07785}{{\normalfont
  [arXiv:hep-ph/1505.07785]}}.
\newblock
  doi:{\changeurlcolor{black}\href{https://doi.org/10.1016/j.physrep.2016.01.002}{\detokenize{10.1016/j.physrep.2016.01.002}}}.

\bibitem[Ting(2016)]{Ting}
Ting, S.
\newblock The First Five Years of the Alpha Magnetic Spectrometer on the ISS.
\newblock CERN Colloquium,  2016.

\bibitem[{Aguilar} \em{et~al.}(2015a){Aguilar}, {Aisa}, {Alpat}, {Alvino},
  {Ambrosi}, {Andeen}, {Arruda}, {Attig}, {Azzarello}, {Bachlechner}, and
  et~al]{2015PhRvL.114q1103A}
{Aguilar}, M.; {Aisa}, D.; {Alpat}, B.; {Alvino}, A.; {Ambrosi}, G.; {Andeen},
  K.; {Arruda}, L.; {Attig}, N.; {Azzarello}, P.; {Bachlechner}, A.; et~al.
\newblock {Precision Measurement of the Proton Flux in Primary Cosmic Rays from
  Rigidity 1 GV to 1.8 TV with the Alpha Magnetic Spectrometer on the
  International Space Station}.
\newblock {\em \prl} {\bf 2015a}, {\em 114},~171103.
\newblock
  doi:{\changeurlcolor{black}\href{https://doi.org/10.1103/PhysRevLett.114.171103}{\detokenize{10.1103/PhysRevLett.114.171103}}}.

\bibitem[{Aguilar} \em{et~al.}(2015b){Aguilar}, {Aisa}, {Alpat}, {Alvino},
  {Ambrosi}, {Andeen}, {Arruda}, {Attig}, {Azzarello}, {Bachlechner}, and
  et~al]{2015PhRvL.115u1101A}
{Aguilar}, M.; {Aisa}, D.; {Alpat}, B.; {Alvino}, A.; {Ambrosi}, G.; {Andeen},
  K.; {Arruda}, L.; {Attig}, N.; {Azzarello}, P.; {Bachlechner}, A.; et~al.
\newblock {Precision Measurement of the Helium Flux in Primary Cosmic Rays of
  Rigidities 1.9 GV to 3 TV with the Alpha Magnetic Spectrometer on the
  International Space Station}.
\newblock {\em \prl} {\bf 2015b}, {\em 115},~211101.
\newblock
  doi:{\changeurlcolor{black}\href{https://doi.org/10.1103/PhysRevLett.115.211101}{\detokenize{10.1103/PhysRevLett.115.211101}}}.

\bibitem[{Aguilar} \em{et~al.}(2018a){Aguilar}, {Ali Cavasonza}, {Ambrosi},
  {Arruda}, {Attig}, {Aupetit}, {Azzarello}, {Bachlechner}, {Barao}, {Barrau},
  and et~al]{2018PhRvL.120b1101A}
{Aguilar}, M.; {Ali Cavasonza}, L.; {Ambrosi}, G.; {Arruda}, L.; {Attig}, N.;
  {Aupetit}, S.; {Azzarello}, P.; {Bachlechner}, A.; {Barao}, F.; {Barrau}, A.;
  et~al.
\newblock {Observation of New Properties of Secondary Cosmic Rays Lithium,
  Beryllium, and Boron by the Alpha Magnetic Spectrometer on the International
  Space Station}.
\newblock {\em \prl} {\bf 2018a}, {\em 120},~021101.
\newblock
  doi:{\changeurlcolor{black}\href{https://doi.org/10.1103/PhysRevLett.120.021101}{\detokenize{10.1103/PhysRevLett.120.021101}}}.

\bibitem[{Aartsen} \em{et~al.}(2019){Aartsen}, {Ackermann}, {Adams}, {Aguilar},
  {Ahlers}, {Ahrens}, {Alispach}, {Andeen}, {Anderson}, {Ansseau}, and
  et~al]{2019PhRvD.100h2002A}
{Aartsen}, M.G.; {Ackermann}, M.; {Adams}, J.; {Aguilar}, J.A.; {Ahlers}, M.;
  {Ahrens}, M.; {Alispach}, C.; {Andeen}, K.; {Anderson}, T.; {Ansseau}, I.;
  et~al.
\newblock {Cosmic ray spectrum and composition from PeV to EeV using 3 years of
  data from IceTop and IceCube}.
\newblock {\em \prd} {\bf 2019}, {\em 100},~082002,
  \href{http://xxx.lanl.gov/abs/1906.04317}{{\normalfont
  [arXiv:astro-ph.HE/1906.04317]}}.
\newblock
  doi:{\changeurlcolor{black}\href{https://doi.org/10.1103/PhysRevD.100.082002}{\detokenize{10.1103/PhysRevD.100.082002}}}.

\bibitem[{Globus} \em{et~al.}(2015){Globus}, {Allard}, and
  {Parizot}]{2015PhRvD..92b1302G}
{Globus}, N.; {Allard}, D.; {Parizot}, E.
\newblock {A complete model of the cosmic ray spectrum and composition across
  the Galactic to extragalactic transition}.
\newblock {\em \prd} {\bf 2015}, {\em 92},~021302,
  \href{http://xxx.lanl.gov/abs/1505.01377}{{\normalfont
  [arXiv:astro-ph.HE/1505.01377]}}.
\newblock
  doi:{\changeurlcolor{black}\href{https://doi.org/10.1103/PhysRevD.92.021302}{\detokenize{10.1103/PhysRevD.92.021302}}}.

\bibitem[{Thoudam} \em{et~al.}(2016){Thoudam}, {Rachen}, {van Vliet},
  {Achterberg}, {Buitink}, {Falcke}, and {H{\"o}rand el}]{2016A&A...595A..33T}
{Thoudam}, S.; {Rachen}, J.P.; {van Vliet}, A.; {Achterberg}, A.; {Buitink},
  S.; {Falcke}, H.; {H{\"o}rand el}, J.R.
\newblock {Cosmic-ray energy spectrum and composition up to the ankle: the case
  for a second Galactic component}.
\newblock {\em \aap} {\bf 2016}, {\em 595},~A33,
  \href{http://xxx.lanl.gov/abs/1605.03111}{{\normalfont
  [arXiv:astro-ph.HE/1605.03111]}}.
\newblock
  doi:{\changeurlcolor{black}\href{https://doi.org/10.1051/0004-6361/201628894}{\detokenize{10.1051/0004-6361/201628894}}}.

\bibitem[{Kotera} and {Olinto}(2011)]{2011ARA&A..49..119K}
{Kotera}, K.; {Olinto}, A.V.
\newblock {The Astrophysics of Ultrahigh-Energy Cosmic Rays}.
\newblock {\em \araa} {\bf 2011}, {\em 49},~119--153,
  \href{http://xxx.lanl.gov/abs/1101.4256}{{\normalfont
  [arXiv:astro-ph.HE/1101.4256]}}.
\newblock
  doi:{\changeurlcolor{black}\href{https://doi.org/10.1146/annurev-astro-081710-102620}{\detokenize{10.1146/annurev-astro-081710-102620}}}.

\bibitem[{Strong} and {Moskalenko}(2009)]{2009arXiv0907.0565S}
{Strong}, A.W.; {Moskalenko}, I.V.
\newblock {A Galactic Cosmic-Ray Database}.
\newblock {\em ArXiv:0907.0565} {\bf 2009},
  \href{http://xxx.lanl.gov/abs/0907.0565}{{\normalfont
  [arXiv:astro-ph.HE/0907.0565]}}.

\bibitem[{Maurin}(2020)]{2020CoPhC.24706942M}
{Maurin}, D.
\newblock {USINE: Semi-analytical models for Galactic cosmic-ray propagation}.
\newblock {\em Computer Physics Communications} {\bf 2020}, {\em 247},~106942.
\newblock
  doi:{\changeurlcolor{black}\href{https://doi.org/10.1016/j.cpc.2019.106942}{\detokenize{10.1016/j.cpc.2019.106942}}}.

\bibitem[{Maurin} \em{et~al.}(2014){Maurin}, {Melot}, and
  {Taillet}]{2014A&A...569A..32M}
{Maurin}, D.; {Melot}, F.; {Taillet}, R.
\newblock {A database of charged cosmic rays}.
\newblock {\em \aap} {\bf 2014}, {\em 569},~A32,
  \href{http://xxx.lanl.gov/abs/1302.5525}{{\normalfont
  [arXiv:astro-ph.HE/1302.5525]}}.
\newblock
  doi:{\changeurlcolor{black}\href{https://doi.org/10.1051/0004-6361/201321344}{\detokenize{10.1051/0004-6361/201321344}}}.

\bibitem[{Di Felice} \em{et~al.}(2017){Di Felice}, {Pizzolotto}, {D'Urso},
  {Dari}, {Navarra}, {Primavera}, and {Bertucci}]{2017ICRC...35.1073D}
{Di Felice}, V.; {Pizzolotto}, C.; {D'Urso}, D.; {Dari}, S.; {Navarra}, D.;
  {Primavera}, R.; {Bertucci}, B.
\newblock {Looking for cosmic ray data? The ASI Cosmic Ray Database}.
\newblock  35th International Cosmic Ray Conference (ICRC2017),  2017, Vol.
  301, {\em International Cosmic Ray Conference}, p. 1073.

\bibitem[{Haungs} \em{et~al.}(2018){Haungs}, {Kang}, {Schoo}, {Wochele},
  {Wochele}, {Apel}, {Arteaga-Vel{\'a}zquez}, {Bekk}, {Bertaina}, {Bl{\"u}mer},
  and et~al]{2018EPJC...78..741H}
{Haungs}, A.; {Kang}, D.; {Schoo}, S.; {Wochele}, D.; {Wochele}, J.; {Apel},
  W.D.; {Arteaga-Vel{\'a}zquez}, J.C.; {Bekk}, K.; {Bertaina}, M.;
  {Bl{\"u}mer}, J.; et~al.
\newblock {The KASCADE Cosmic-ray Data Centre KCDC: granting open access to
  astroparticle physics research data}.
\newblock {\em European Physical Journal C} {\bf 2018}, {\em 78},~741,
  \href{http://xxx.lanl.gov/abs/1806.05493}{{\normalfont
  [arXiv:astro-ph.IM/1806.05493]}}.
\newblock
  doi:{\changeurlcolor{black}\href{https://doi.org/10.1140/epjc/s10052-018-6221-2}{\detokenize{10.1140/epjc/s10052-018-6221-2}}}.

\bibitem[{Lafferty} and {Wyatt}(1995)]{1995NIMPA.355..541L}
{Lafferty}, G.D.; {Wyatt}, T.R.
\newblock {Where to stick your data points: The treatment of measurements
  within wide bins}.
\newblock {\em Nuclear Instruments and Methods in Physics Research A} {\bf
  1995}, {\em 355},~541--547.
\newblock
  doi:{\changeurlcolor{black}\href{https://doi.org/10.1016/0168-9002(94)01112-5}{\detokenize{10.1016/0168-9002(94)01112-5}}}.

\bibitem[{Derome} \em{et~al.}(2019){Derome}, {Maurin}, {Salati}, {Boudaud},
  {G{\'e}nolini}, and {Kunz{\'e}}]{2019A&A...627A.158D}
{Derome}, L.; {Maurin}, D.; {Salati}, P.; {Boudaud}, M.; {G{\'e}nolini}, Y.;
  {Kunz{\'e}}, P.
\newblock {Fitting B/C cosmic-ray data in the AMS-02 era: a cookbook. Model
  numerical precision, data covariance matrix of errors, cross-section nuisance
  parameters, and mock data}.
\newblock {\em \aap} {\bf 2019}, {\em 627},~A158.
\newblock
  doi:{\changeurlcolor{black}\href{https://doi.org/10.1051/0004-6361/201935717}{\detokenize{10.1051/0004-6361/201935717}}}.

\bibitem[{Potgieter}(2013)]{2013LRSP...10....3P}
{Potgieter}, M.
\newblock {Solar Modulation of Cosmic Rays}.
\newblock {\em Living Reviews in Solar Physics} {\bf 2013}, {\em 10},~3,
  \href{http://xxx.lanl.gov/abs/1306.4421}{{\normalfont
  [arXiv:physics.space-ph/1306.4421]}}.
\newblock
  doi:{\changeurlcolor{black}\href{https://doi.org/10.12942/lrsp-2013-3}{\detokenize{10.12942/lrsp-2013-3}}}.

\bibitem[{Gleeson} and {Axford}(1967)]{1967ApJ...149L.115G}
{Gleeson}, L.J.; {Axford}, W.I.
\newblock {Cosmic Rays in the Interplanetary Medium}.
\newblock {\em \apjl} {\bf 1967}, {\em 149},~L115.
\newblock
  doi:{\changeurlcolor{black}\href{https://doi.org/10.1086/180070}{\detokenize{10.1086/180070}}}.

\bibitem[{Gleeson} and {Axford}(1968)]{1968ApJ...154.1011G}
{Gleeson}, L.J.; {Axford}, W.I.
\newblock {Solar Modulation of Galactic Cosmic Rays}.
\newblock {\em \apj} {\bf 1968}, {\em 154},~1011.
\newblock
  doi:{\changeurlcolor{black}\href{https://doi.org/10.1086/149822}{\detokenize{10.1086/149822}}}.

\bibitem[{Perko}(1987)]{1987A&A...184..119P}
{Perko}, J.S.
\newblock {Solar modulation of galactic antiprotons}.
\newblock {\em \aap} {\bf 1987}, {\em 184},~119--121.

\bibitem[{Caballero-Lopez} and {Moraal}(2004)]{2004JGRA..109.1101C}
{Caballero-Lopez}, R.A.; {Moraal}, H.
\newblock {Limitations of the force field equation to describe cosmic ray
  modulation}.
\newblock {\em Journal of Geophysical Research (Space Physics)} {\bf 2004},
  {\em 109},~1101.
\newblock
  doi:{\changeurlcolor{black}\href{https://doi.org/10.1029/2003JA010098}{\detokenize{10.1029/2003JA010098}}}.

\bibitem[{Hathaway}(2015)]{2015LRSP...12....4H}
{Hathaway}, D.H.
\newblock {The Solar Cycle}.
\newblock {\em Living Reviews in Solar Physics} {\bf 2015}, {\em 12},~4,
  \href{http://xxx.lanl.gov/abs/1502.07020}{{\normalfont
  [arXiv:astro-ph.SR/1502.07020]}}.
\newblock
  doi:{\changeurlcolor{black}\href{https://doi.org/10.1007/lrsp-2015-4}{\detokenize{10.1007/lrsp-2015-4}}}.

\bibitem[{Kappl}(2016)]{2016CoPhC.207..386K}
{Kappl}, R.
\newblock {SOLARPROP: Charge-sign dependent solar modulation for everyone}.
\newblock {\em Computer Physics Communications} {\bf 2016}, {\em
  207},~386--399,  \href{http://xxx.lanl.gov/abs/1511.07875}{{\normalfont
  [arXiv:astro-ph.SR/1511.07875]}}.
\newblock
  doi:{\changeurlcolor{black}\href{https://doi.org/10.1016/j.cpc.2016.05.025}{\detokenize{10.1016/j.cpc.2016.05.025}}}.

\bibitem[{Boschini} \em{et~al.}(2018){Boschini}, {Della Torre}, {Gervasi}, {La
  Vacca}, and {Rancoita}]{2018AdSpR..62.2859B}
{Boschini}, M.J.; {Della Torre}, S.; {Gervasi}, M.; {La Vacca}, G.; {Rancoita},
  P.G.
\newblock {Propagation of cosmic rays in heliosphere: The HELMOD model}.
\newblock {\em Advances in Space Research} {\bf 2018}, {\em 62},~2859--2879,
  \href{http://xxx.lanl.gov/abs/1704.03733}{{\normalfont
  [arXiv:astro-ph.SR/1704.03733]}}.
\newblock
  doi:{\changeurlcolor{black}\href{https://doi.org/10.1016/j.asr.2017.04.017}{\detokenize{10.1016/j.asr.2017.04.017}}}.

\bibitem[{Usoskin} \em{et~al.}(2005){Usoskin}, {Alanko-Huotari}, {Kovaltsov},
  and {Mursula}]{2005JGRA..11012108U}
{Usoskin}, I.G.; {Alanko-Huotari}, K.; {Kovaltsov}, G.A.; {Mursula}, K.
\newblock {Heliospheric modulation of cosmic rays: Monthly reconstruction for
  1951-2004}.
\newblock {\em Journal of Geophysical Research (Space Physics)} {\bf 2005},
  {\em 110},~12108.
\newblock
  doi:{\changeurlcolor{black}\href{https://doi.org/10.1029/2005JA011250}{\detokenize{10.1029/2005JA011250}}}.

\bibitem[{Usoskin} \em{et~al.}(2017){Usoskin}, {Gil}, {Kovaltsov}, {Mishev},
  and {Mikhailov}]{2017JGRA..122.3875U}
{Usoskin}, I.G.; {Gil}, A.; {Kovaltsov}, G.A.; {Mishev}, A.L.; {Mikhailov},
  V.V.
\newblock {Heliospheric modulation of cosmic rays during the neutron monitor
  era: Calibration using PAMELA data for 2006-2010}.
\newblock {\em Journal of Geophysical Research (Space Physics)} {\bf 2017},
  {\em 122},~3875--3887,
  \href{http://xxx.lanl.gov/abs/1705.07197}{{\normalfont
  [arXiv:physics.space-ph/1705.07197]}}.
\newblock
  doi:{\changeurlcolor{black}\href{https://doi.org/10.1002/2016JA023819}{\detokenize{10.1002/2016JA023819}}}.

\bibitem[{Maurin} \em{et~al.}(2015){Maurin}, {Cheminet}, {Derome}, {Ghelfi},
  and {Hubert}]{2015AdSpR..55..363M}
{Maurin}, D.; {Cheminet}, A.; {Derome}, L.; {Ghelfi}, A.; {Hubert}, G.
\newblock {Neutron monitors and muon detectors for solar modulation studies:
  Interstellar flux, yield function, and assessment of critical parameters in
  count rate calculations}.
\newblock {\em AdSR} {\bf 2015}, {\em 55},~363--389,
  \href{http://xxx.lanl.gov/abs/1403.1612}{{\normalfont
  [arXiv:astro-ph.EP/1403.1612]}}.
\newblock
  doi:{\changeurlcolor{black}\href{https://doi.org/10.1016/j.asr.2014.06.021}{\detokenize{10.1016/j.asr.2014.06.021}}}.

\bibitem[{Ghelfi} \em{et~al.}(2016){Ghelfi}, {Barao}, {Derome}, and
  {Maurin}]{2016A&A...591A..94G}
{Ghelfi}, A.; {Barao}, F.; {Derome}, L.; {Maurin}, D.
\newblock {Non-parametric determination of H and He interstellar fluxes from
  cosmic-ray data}.
\newblock {\em \aap} {\bf 2016}, {\em 591},~A94,
  \href{http://xxx.lanl.gov/abs/1511.08650}{{\normalfont
  [arXiv:astro-ph.HE/1511.08650]}}.
\newblock
  doi:{\changeurlcolor{black}\href{https://doi.org/10.1051/0004-6361/201527852}{\detokenize{10.1051/0004-6361/201527852}}}.

\bibitem[{Parker}(1958{\natexlab{a}})]{1958PhRv..110.1445P}
{Parker}, E.N.
\newblock {Cosmic-Ray Modulation by Solar Wind}.
\newblock {\em Physical Review} {\bf 1958}, {\em 110},~1445--1449.
\newblock
  doi:{\changeurlcolor{black}\href{https://doi.org/10.1103/PhysRev.110.1445}{\detokenize{10.1103/PhysRev.110.1445}}}.

\bibitem[{Parker}(1958{\natexlab{b}})]{1958PhRv..109.1874P}
{Parker}, E.N.
\newblock {Dynamical Instability in an Anisotropic Ionized Gas of Low Density}.
\newblock {\em Physical Review} {\bf 1958}, {\em 109},~1874--1876.
\newblock
  doi:{\changeurlcolor{black}\href{https://doi.org/10.1103/PhysRev.109.1874}{\detokenize{10.1103/PhysRev.109.1874}}}.

\bibitem[{L'Heureux} \em{et~al.}(1972){L'Heureux}, {Fan}, and
  {Meyer}]{1972ApJ...171..363L}
{L'Heureux}, J.; {Fan}, C.Y.; {Meyer}, P.
\newblock {The Quiet-Tiem Spectra of Cosmic-Ray Electrons of Energies Between
  10 and 200 MeV Observed on OGO-5}.
\newblock {\em \apj} {\bf 1972}, {\em 171},~363.
\newblock
  doi:{\changeurlcolor{black}\href{https://doi.org/10.1086/151287}{\detokenize{10.1086/151287}}}.

\bibitem[{Fisk} and {Axford}(1969)]{1969JGR....74.4973F}
{Fisk}, L.A.; {Axford}, W.I.
\newblock {Solar modulation of galactic cosmic rays, 1}.
\newblock {\em \jgr} {\bf 1969}, {\em 74},~4973.
\newblock
  doi:{\changeurlcolor{black}\href{https://doi.org/10.1029/JA074i021p04973}{\detokenize{10.1029/JA074i021p04973}}}.

\bibitem[{Fisk}(1971)]{1971JGR....76..221F}
{Fisk}, L.A.
\newblock {Solar modulation of galactic cosmic rays, 2}.
\newblock {\em \jgr} {\bf 1971}, {\em 76},~221.
\newblock
  doi:{\changeurlcolor{black}\href{https://doi.org/10.1029/JA076i001p00221}{\detokenize{10.1029/JA076i001p00221}}}.

\bibitem[{Beatty} \em{et~al.}(1993){Beatty}, {Ficenec}, {Tobias}, {Mitchell},
  {McKee}, {Nutter}, {Tarle}, {Tomasch}, {Clem}, {Guzik}, and
  et~al]{1993ApJ...413..268B}
{Beatty}, J.J.; {Ficenec}, D.J.; {Tobias}, S.; {Mitchell}, J.W.; {McKee}, S.;
  {Nutter}, S.; {Tarle}, G.; {Tomasch}, A.; {Clem}, J.; {Guzik}, T.G.; et~al.
\newblock {The cosmic-ray He-3/He-4 ratio from 100 to 1600 MeV/amu}.
\newblock {\em \apj} {\bf 1993}, {\em 413},~268--280.
\newblock
  doi:{\changeurlcolor{black}\href{https://doi.org/10.1086/172994}{\detokenize{10.1086/172994}}}.

\bibitem[{Jokipii} and {Kopriva}(1979)]{1979ApJ...234..384J}
{Jokipii}, J.R.; {Kopriva}, D.A.
\newblock {Effects of particle drift on the transport of cosmic rays. III -
  Numerical models of galactic cosmic-ray modulation}.
\newblock {\em \apj} {\bf 1979}, {\em 234},~384--392.
\newblock
  doi:{\changeurlcolor{black}\href{https://doi.org/10.1086/157506}{\detokenize{10.1086/157506}}}.

\bibitem[{Potgieter} and {Moraal}(1985)]{1985ApJ...294..425P}
{Potgieter}, M.S.; {Moraal}, H.
\newblock {A drift model for the modulation of galactic cosmic rays}.
\newblock {\em \apj} {\bf 1985}, {\em 294},~425--440.
\newblock
  doi:{\changeurlcolor{black}\href{https://doi.org/10.1086/163309}{\detokenize{10.1086/163309}}}.

\bibitem[{Aguilar} \em{et~al.}(2018{\natexlab{a}}){Aguilar}, {Ali Cavasonza},
  {Alpat}, {Ambrosi}, {Arruda}, {Attig}, {Aupetit}, {Azzarello}, {Bachlechner},
  {Barao}, and et~al]{2018PhRvL.121e1101A}
{Aguilar}, M.; {Ali Cavasonza}, L.; {Alpat}, B.; {Ambrosi}, G.; {Arruda}, L.;
  {Attig}, N.; {Aupetit}, S.; {Azzarello}, P.; {Bachlechner}, A.; {Barao}, F.;
  et~al.
\newblock {Observation of Fine Time Structures in the Cosmic Proton and Helium
  Fluxes with the Alpha Magnetic Spectrometer on the International Space
  Station}.
\newblock {\em \prl} {\bf 2018}, {\em 121},~051101.
\newblock
  doi:{\changeurlcolor{black}\href{https://doi.org/10.1103/PhysRevLett.121.051101}{\detokenize{10.1103/PhysRevLett.121.051101}}}.

\bibitem[{Aguilar} \em{et~al.}(2018{\natexlab{b}}){Aguilar}, {Cavasonza},
  {Ambrosi}, {Arruda}, {Attig}, {Aupetit}, {Azzarello}, {Bachlechner}, {Barao},
  {Barrau}, and et~al]{2018PhRvL.121e1102A}
{Aguilar}, M.; {Cavasonza}, L.A.; {Ambrosi}, G.; {Arruda}, L.; {Attig}, N.;
  {Aupetit}, S.; {Azzarello}, P.; {Bachlechner}, A.; {Barao}, F.; {Barrau}, A.;
  et~al.
\newblock {Observation of Complex Time Structures in the Cosmic-Ray Electron
  and Positron Fluxes with the Alpha Magnetic Spectrometer on the International
  Space Station}.
\newblock {\em \prl} {\bf 2018}, {\em 121},~051102.
\newblock
  doi:{\changeurlcolor{black}\href{https://doi.org/10.1103/PhysRevLett.121.051102}{\detokenize{10.1103/PhysRevLett.121.051102}}}.

\bibitem[{Martucci} \em{et~al.}(2018){Martucci}, {Munini}, {Boezio}, {Di
  Felice}, {Adriani}, {Barbarino}, {Bazilevskaya}, {Bellotti}, {Bongi},
  {Bonvicini}, , and et~al]{2018ApJ...854L...2M}
{Martucci}, M.; {Munini}, R.; {Boezio}, M.; {Di Felice}, V.; {Adriani}, O.;
  {Barbarino}, G.C.; {Bazilevskaya}, G.A.; {Bellotti}, R.; {Bongi}, M.;
  {Bonvicini}, V.; .; et~al.
\newblock {Proton Fluxes Measured by the PAMELA Experiment from the Minimum to
  the Maximum Solar Activity for Solar Cycle 24}.
\newblock {\em \apjl} {\bf 2018}, {\em 854},~L2,
  \href{http://xxx.lanl.gov/abs/1801.07112}{{\normalfont
  [arXiv:physics.space-ph/1801.07112]}}.
\newblock
  doi:{\changeurlcolor{black}\href{https://doi.org/10.3847/2041-8213/aaa9b2}{\detokenize{10.3847/2041-8213/aaa9b2}}}.

\bibitem[{Burbidge} \em{et~al.}(1957){Burbidge}, {Burbidge}, {Fowler}, and
  {Hoyle}]{1957RvMP...29..547B}
{Burbidge}, E.M.; {Burbidge}, G.R.; {Fowler}, W.A.; {Hoyle}, F.
\newblock {Synthesis of the Elements in Stars}.
\newblock {\em Reviews of Modern Physics} {\bf 1957}, {\em 29},~547--650.
\newblock
  doi:{\changeurlcolor{black}\href{https://doi.org/10.1103/RevModPhys.29.547}{\detokenize{10.1103/RevModPhys.29.547}}}.

\bibitem[{Fowler} \em{et~al.}(1967){Fowler}, {Adams}, {Cowen}, and
  {Kidd}]{1967RSPSA.301...39F}
{Fowler}, P.H.; {Adams}, R.A.; {Cowen}, V.G.; {Kidd}, J.M.
\newblock {The Charge Spectrum of Very Heavy Cosmic Ray Nuclei}.
\newblock {\em Royal Society of London Proceedings Series A} {\bf 1967}, {\em
  301},~39--45.
\newblock
  doi:{\changeurlcolor{black}\href{https://doi.org/10.1098/rspa.1967.0188}{\detokenize{10.1098/rspa.1967.0188}}}.

\bibitem[{Blanford} \em{et~al.}(1969){Blanford}, {Friedlander}, {Klarmann},
  {Walker}, {Wefel}, {Wells}, {Fleischer}, {Nichols}, and
  {Price}]{1969PhRvL..23..338B}
{Blanford}, G.E.; {Friedlander}, M.W.; {Klarmann}, J.; {Walker}, R.M.; {Wefel},
  J.P.; {Wells}, W.C.; {Fleischer}, R.L.; {Nichols}, G.E.; {Price}, P.B.
\newblock {Observation of Trans-Iron Nuclei in the Primary Cosmic Radiation}.
\newblock {\em Physical Review Letters} {\bf 1969}, {\em 23},~338--342.
\newblock
  doi:{\changeurlcolor{black}\href{https://doi.org/10.1103/PhysRevLett.23.338}{\detokenize{10.1103/PhysRevLett.23.338}}}.

\bibitem[{Fowler} \em{et~al.}(1970){Fowler}, {Clapham}, {Cowen}, {Kidd}, and
  {Moses}]{1970RSPSA.318....1F}
{Fowler}, P.H.; {Clapham}, V.M.; {Cowen}, V.G.; {Kidd}, J.M.; {Moses}, R.T.
\newblock {The Charge Spectrum of Very Heavy Cosmic Ray Nuclei}.
\newblock {\em Royal Society of London Proceedings Series A} {\bf 1970}, {\em
  318},~1--43.
\newblock
  doi:{\changeurlcolor{black}\href{https://doi.org/10.1098/rspa.1970.0132}{\detokenize{10.1098/rspa.1970.0132}}}.

\bibitem[{O'Sullivan} \em{et~al.}(1971){O'Sullivan}, {Price}, {Shirk},
  {Fowler}, {Kidd}, {Kobetich}, and {Thorne}]{1971PhRvL..26..463O}
{O'Sullivan}, D.; {Price}, P.B.; {Shirk}, E.K.; {Fowler}, P.H.; {Kidd}, J.M.;
  {Kobetich}, E.J.; {Thorne}, R.
\newblock {High-Resolution Measurements of Slowing Cosmic Rays from Fe to U}.
\newblock {\em Physical Review Letters} {\bf 1971}, {\em 26},~463--466.
\newblock
  doi:{\changeurlcolor{black}\href{https://doi.org/10.1103/PhysRevLett.26.463}{\detokenize{10.1103/PhysRevLett.26.463}}}.

\bibitem[{Price} \em{et~al.}(1971){Price}, {Fowler}, {Kidd}, {Kobetich},
  {Fleischer}, and {Nichols}]{1971PhRvD...3..815P}
{Price}, P.B.; {Fowler}, P.H.; {Kidd}, J.M.; {Kobetich}, E.J.; {Fleischer},
  R.L.; {Nichols}, G.E.
\newblock {Study of the Charge Spectrum of Extremely Heavy Cosmic Rays Using
  Combined Plastic Detectors and Nuclear Emulsions}.
\newblock {\em \prd} {\bf 1971}, {\em 3},~815--823.
\newblock
  doi:{\changeurlcolor{black}\href{https://doi.org/10.1103/PhysRevD.3.815}{\detokenize{10.1103/PhysRevD.3.815}}}.

\bibitem[{Arnould} and {Goriely}(2003)]{2003PhR...384....1A}
{Arnould}, M.; {Goriely}, S.
\newblock {The p-process of stellar nucleosynthesis: astrophysics and nuclear
  physics status}.
\newblock {\em \physrep} {\bf 2003}, {\em 384},~1--84.
\newblock
  doi:{\changeurlcolor{black}\href{https://doi.org/10.1016/S0370-1573(03)00242-4}{\detokenize{10.1016/S0370-1573(03)00242-4}}}.

\bibitem[{Rauscher} \em{et~al.}(2013){Rauscher}, {Dauphas}, {Dillmann},
  {Fr{\"o}hlich}, {F{\"u}l{\"o}p}, and {Gy{\"u}rky}]{2013RPPh...76f6201R}
{Rauscher}, T.; {Dauphas}, N.; {Dillmann}, I.; {Fr{\"o}hlich}, C.;
  {F{\"u}l{\"o}p}, Z.; {Gy{\"u}rky}, G.
\newblock {Constraining the astrophysical origin of the p-nuclei through
  nuclear physics and meteoritic data}.
\newblock {\em Reports on Progress in Physics} {\bf 2013}, {\em 76},~066201,
  \href{http://xxx.lanl.gov/abs/1303.2666}{{\normalfont
  [arXiv:astro-ph.SR/1303.2666]}}.
\newblock
  doi:{\changeurlcolor{black}\href{https://doi.org/10.1088/0034-4885/76/6/066201}{\detokenize{10.1088/0034-4885/76/6/066201}}}.

\bibitem[{Cameron} \em{et~al.}(1993){Cameron}, {Thielemann}, and
  {Cowan}]{1993PhR...227..283C}
{Cameron}, A.G.W.; {Thielemann}, F.K.; {Cowan}, J.J.
\newblock {s- and r-process contributions to extinct radioactivities}.
\newblock {\em \physrep} {\bf 1993}, {\em 227},~283--291.
\newblock
  doi:{\changeurlcolor{black}\href{https://doi.org/10.1016/0370-1573(93)90073-M}{\detokenize{10.1016/0370-1573(93)90073-M}}}.

\bibitem[{Meyer}(1994)]{1994ARA&A..32..153M}
{Meyer}, B.S.
\newblock {The r-, s-, and p-Processes in Nucleosynthesis}.
\newblock {\em \araa} {\bf 1994}, {\em 32},~153--190.
\newblock
  doi:{\changeurlcolor{black}\href{https://doi.org/10.1146/annurev.aa.32.090194.001101}{\detokenize{10.1146/annurev.aa.32.090194.001101}}}.

\bibitem[{Lingenfelter}(2019)]{2019ApJS..245...30L}
{Lingenfelter}, R.E.
\newblock {The Origin of Cosmic Rays: How Their Composition Defines Their
  Sources and Sites and the Processes of Their Mixing, Injection, and
  Acceleration}.
\newblock {\em \apjs} {\bf 2019}, {\em 245},~30,
  \href{http://xxx.lanl.gov/abs/1903.06330}{{\normalfont
  [arXiv:astro-ph.HE/1903.06330]}}.
\newblock
  doi:{\changeurlcolor{black}\href{https://doi.org/10.3847/1538-4365/ab4b58}{\detokenize{10.3847/1538-4365/ab4b58}}}.

\bibitem[{Combet} \em{et~al.}(2005){Combet}, {Maurin}, {Donnelly}, {O'C.
  Drury}, and {Vangioni-Flam}]{2005A&A...435..151C}
{Combet}, C.; {Maurin}, D.; {Donnelly}, J.; {O'C. Drury}, L.; {Vangioni-Flam},
  E.
\newblock {Spallation-dominated propagation of heavy cosmic rays and the Local
  Interstellar Medium (LISM)}.
\newblock {\em \aap} {\bf 2005}, {\em 435},~151--160,
  \href{http://xxx.lanl.gov/abs/astro-ph/0412015}{{\normalfont
  [arXiv:astro-ph/astro-ph/0412015]}}.
\newblock
  doi:{\changeurlcolor{black}\href{https://doi.org/10.1051/0004-6361:20042459}{\detokenize{10.1051/0004-6361:20042459}}}.

\bibitem[{Fowler} \em{et~al.}(1981){Fowler}, {Walker}, {Masheder}, {Moses}, and
  {Worley}]{1981Natur.291...45F}
{Fowler}, P.H.; {Walker}, R.N.F.; {Masheder}, M.R.W.; {Moses}, R.T.; {Worley},
  A.
\newblock {Ultra-heavy cosmic ray studies with Ariel VI}.
\newblock {\em \nat} {\bf 1981}, {\em 291},~45--47.
\newblock
  doi:{\changeurlcolor{black}\href{https://doi.org/10.1038/291045a0}{\detokenize{10.1038/291045a0}}}.

\bibitem[{Fowler} \em{et~al.}(1987){Fowler}, {Walker}, {Masheder}, {Moses},
  {Worley}, and {Gay}]{1987ApJ...314..739F}
{Fowler}, P.H.; {Walker}, R.N.F.; {Masheder}, M.R.W.; {Moses}, R.T.; {Worley},
  A.; {Gay}, A.M.
\newblock {Ariel 6 Measurements of the Fluxes of Ultra-heavy Cosmic Rays}.
\newblock {\em \apj} {\bf 1987}, {\em 314},~739.
\newblock
  doi:{\changeurlcolor{black}\href{https://doi.org/10.1086/165101}{\detokenize{10.1086/165101}}}.

\bibitem[{Binns} \em{et~al.}(1989){Binns}, {Garrard}, {Gibner}, {Israel},
  {Kertzman}, {Klarmann}, {Newport}, {Stone}, and
  {Waddington}]{1989ApJ...346..997B}
{Binns}, W.R.; {Garrard}, T.L.; {Gibner}, P.S.; {Israel}, M.H.; {Kertzman},
  M.P.; {Klarmann}, J.; {Newport}, B.J.; {Stone}, E.C.; {Waddington}, C.J.
\newblock {Abundances of Ultraheavy Elements in the Cosmic Radiation: Results
  from HEAO 3}.
\newblock {\em \apj} {\bf 1989}, {\em 346},~997.
\newblock
  doi:{\changeurlcolor{black}\href{https://doi.org/10.1086/168082}{\detokenize{10.1086/168082}}}.

\bibitem[{Murphy} \em{et~al.}(2016){Murphy}, {Sasaki}, {Binns}, {Brandt},
  {Hams}, {Israel}, {Labrador}, {Link}, {Mewaldt}, {Mitchell}, {Rauch},
  {Sakai}, {Stone}, {Waddington}, {Walsh}, {Ward}, and
  {Wiedenbeck}]{2016ApJ...831..148M}
{Murphy}, R.P.; {Sasaki}, M.; {Binns}, W.R.; {Brandt}, T.J.; {Hams}, T.;
  {Israel}, M.H.; {Labrador}, A.W.; {Link}, J.T.; {Mewaldt}, R.A.; {Mitchell},
  J.W.; {Rauch}, B.F.; {Sakai}, K.; {Stone}, E.C.; {Waddington}, C.J.; {Walsh},
  N.E.; {Ward}, J.E.; {Wiedenbeck}, M.E.
\newblock {Galactic Cosmic Ray Origins and OB Associations: Evidence from
  SuperTIGER Observations of Elements $_{26}$Fe through $_{40}$Zr}.
\newblock {\em \\apj} {\bf 2016}, {\em 831},~148,
  \href{http://xxx.lanl.gov/abs/1608.08183}{{\normalfont
  [arXiv:astro-ph.HE/1608.08183]}}.
\newblock
  doi:{\changeurlcolor{black}\href{https://doi.org/10.3847/0004-637X/831/2/148}{\detokenize{10.3847/0004-637X/831/2/148}}}.

\bibitem[{Binns} \em{et~al.}(1981){Binns}, {Fickle}, {Waddington}, {Garrard},
  {Stone}, {Israel}, and {Klarmann}]{1981ApJ...247L.115B}
{Binns}, W.R.; {Fickle}, R.K.; {Waddington}, C.J.; {Garrard}, T.L.; {Stone},
  E.C.; {Israel}, M.H.; {Klarmann}, J.
\newblock {Cosmic-ray abundances of elements with atomic number 26 less than or
  equal to 40 measured on HEAO 3}.
\newblock {\em \\apjl} {\bf 1981}, {\em 247},~L115--L118.
\newblock
  doi:{\changeurlcolor{black}\href{https://doi.org/10.1086/183602}{\detokenize{10.1086/183602}}}.

\bibitem[{Stone} \em{et~al.}(1983){Stone}, {Garrard}, {Krombel}, {Binns},
  {Israel}, {Klarmann}, {Brewster}, {Fickle}, and
  {Waddington}]{1983ICRC....9..115S}
{Stone}, E.C.; {Garrard}, T.L.; {Krombel}, K.E.; {Binns}, W.R.; {Israel}, M.H.;
  {Klarmann}, J.; {Brewster}, N.R.; {Fickle}, R.K.; {Waddington}, C.J.
\newblock {Cosmic-ray abundances of the even charge elements from $_{50}$Snto
  $_{58}$Ce measured on HEAO-3.}
\newblock  International Cosmic Ray Conference,  1983, Vol.~9, {\em
  International Cosmic Ray Conference}, pp. 115--118.

\bibitem[{Domingo} \em{et~al.}(1996){Domingo}, {Font}, {Baixeras}, and
  {Fern{\\\'a}ndez}]{1996RadM...26..825D}
{Domingo}, C.; {Font}, J.; {Baixeras}, C.; {Fern{\\\'a}ndez}, F.
\newblock {Source abundances of ultra heavy elements derived from UHCRE
  measurements}.
\newblock {\em Radiation Measurements} {\bf 1996}, {\em 26},~825--832.
\newblock
  doi:{\changeurlcolor{black}\href{https://doi.org/10.1016/S1350-4487(96)00090-X}{\detokenize{10.1016/S1350-4487(96)00090-X}}}.

\bibitem[{Dutta} \em{et~al.}(2003){Dutta}, {Batra}, and
  {Biswas}]{2003RadM...36..287D}
{Dutta}, A.; {Batra}, V.; {Biswas}, S.
\newblock {Abundance of actinides in cosmic radiation}.
\newblock {\em Radiation Measurements} {\bf 2003}, {\em 36},~287--290.
\newblock
  doi:{\changeurlcolor{black}\href{https://doi.org/10.1016/S1350-4487(03)00137-9}{\detokenize{10.1016/S1350-4487(03)00137-9}}}.

\bibitem[{Westphal} \em{et~al.}(1998){Westphal}, {Price}, {Weaver}, and
  {Afanasiev}]{1998Natur.396...50W}
{Westphal}, A.J.; {Price}, P.B.; {Weaver}, B.A.; {Afanasiev}, V.G.
\newblock {Evidence against stellar chromospheric origin of Galactic cosmic
  rays}.
\newblock {\em \ at} {\bf 1998}, {\em 396},~50--52.
\newblock
  doi:{\changeurlcolor{black}\href{https://doi.org/10.1038/23887}{\detokenize{10.1038/23887}}}.

\bibitem[{Weaver} and {Westphal}(2002)]{2002ApJ...569..493W}
{Weaver}, B.A.; {Westphal}, A.J.
\newblock {Extended Analysis of the Trek Ultraheavy Collector}.
\newblock {\em \\apj} {\bf 2002}, {\em 569},~493--500.
\newblock
  doi:{\changeurlcolor{black}\href{https://doi.org/10.1086/339265}{\detokenize{10.1086/339265}}}.

\bibitem[{Aleksandrov} \em{et~al.}(2010){Aleksandrov}, {Bagulya}, {Vladimirov},
  {Goncharova}, {Ivliev}, {Kalinina}, {Kashkarov}, {Konovalova}, {Okat'eva},
  {Polukhina}, {Rusetskii}, and {Starkov}]{2010PhyU...53..805A}
{Aleksandrov}, A.B.; {Bagulya}, A.V.; {Vladimirov}, M.S.; {Goncharova}, L.A.;
  {Ivliev}, A.I.; {Kalinina}, G.V.; {Kashkarov}, L.L.; {Konovalova}, N.S.;
  {Okat'eva}, N.M.; {Polukhina}, N.G.; {Rusetskii}, A.S.; {Starkov}, N.I.
\newblock {INSTRUMENTS AND METHODS OF INVESTIGATION Charge spectrum of galactic
  cosmic ray nuclei as measured in meteorite olivines}.
\newblock {\em Physics Uspekhi} {\bf 2010}, {\em 53},~805--808.
\newblock
  doi:{\changeurlcolor{black}\href{https://doi.org/10.3367/UFNe.0180.201008c.0839}{\detokenize{10.3367/UFNe.0180.201008c.0839}}}.

\bibitem[{Bagulya} \em{et~al.}(2013){Bagulya}, {Kashkarov}, {Konovalova},
  {Okat'eva}, {Polukhina}, and {Starkov}]{2013JETPL..97..708B}
{Bagulya}, A.V.; {Kashkarov}, L.L.; {Konovalova}, N.S.; {Okat'eva}, N.M.;
  {Polukhina}, N.G.; {Starkov}, N.I.
\newblock {Search for superheavy elements in galactic cosmic rays}.
\newblock {\em Soviet Journal of Experimental and Theoretical Physics Letters}
  {\bf 2013}, {\em 97},~708--719.
\newblock
  doi:{\changeurlcolor{black}\href{https://doi.org/10.1134/S0021364013120047}{\detokenize{10.1134/S0021364013120047}}}.

\bibitem[{Alexeev} \em{et~al.}(2016){Alexeev}, {Bagulya}, {Chernyavsky},
  {Gippius}, {Goncharova}, {Gorbunov}, {Gorshenkov}, {Kalinina}, {Konovalova},
  {Liu}, {Zhai}, {Okatyeva}, {Pavlova}, {Polukhina}, {Starkov}, {Naing Soe},
  {Trautmann}, {Savchenko}, {Shchedrina}, {Vasiliev}, and
  {Volkov}]{2016ApJ...829..120A}
{Alexeev}, V.; {Bagulya}, A.; {Chernyavsky}, M.; {Gippius}, A.; {Goncharova},
  L.; {Gorbunov}, S.; {Gorshenkov}, M.; {Kalinina}, G.; {Konovalova}, N.;
  {Liu}, J.; {Zhai}, P.; {Okatyeva}, N.; {Pavlova}, T.; {Polukhina}, N.;
  {Starkov}, N.; {Naing Soe}, T.; {Trautmann}, C.; {Savchenko}, E.;
  {Shchedrina}, T.; {Vasiliev}, A.; {Volkov}, A.
\newblock {Charge Spectrum of Heavy and Superheavy Components of Galactic
  Cosmic Rays: Results of the Olimpiya Experiment}.
\newblock {\em \apj} {\bf 2016}, {\em 829},~120.
\newblock
  doi:{\changeurlcolor{black}\href{https://doi.org/10.3847/0004-637X/829/2/120}{\detokenize{10.3847/0004-637X/829/2/120}}}.

\bibitem[{Komiya} and {Shigeyama}(2017)]{2017ApJ...846..143K}
{Komiya}, Y.; {Shigeyama}, T.
\newblock {R-process Element Cosmic Rays from Neutron Star Mergers}.
\newblock {\em \apj} {\bf 2017}, {\em 846},~143,
  \href{http://xxx.lanl.gov/abs/1708.05638}{{\normalfont
  [arXiv:astro-ph.HE/1708.05638]}}.
\newblock
  doi:{\changeurlcolor{black}\href{https://doi.org/10.3847/1538-4357/aa86b3}{\detokenize{10.3847/1538-4357/aa86b3}}}.

\bibitem[{Bogomolov} \em{et~al.}(1979){Bogomolov}, {Lubyanaya}, {Romanov},
  {Stepanov}, and {Shulakova}]{1979ICRC....1..330B}
{Bogomolov}, E.A.; {Lubyanaya}, N.D.; {Romanov}, V.A.; {Stepanov}, S.V.;
  {Shulakova}, M.S.
\newblock {a Stratospheric Magnetic Spectrometer Investigation of the Singly
  Charged Component Spectra and Composition of the Primary and Secondary Cosmic
  Radiation}.
\newblock  International Cosmic Ray Conference,  1979, Vol.~1, {\em
  International Cosmic Ray Conference}, p. 330.

\bibitem[{Buffington} \em{et~al.}(1981){Buffington}, {Schindler}, and
  {Pennypacker}]{1981ApJ...248.1179B}
{Buffington}, A.; {Schindler}, S.M.; {Pennypacker}, C.R.
\newblock {A measurement of the cosmic-ray antiproton flux and a search for
  antihelium}.
\newblock {\em \apj} {\bf 1981}, {\em 248},~1179--1193.
\newblock
  doi:{\changeurlcolor{black}\href{https://doi.org/10.1086/159247}{\detokenize{10.1086/159247}}}.

\bibitem[{Silk} and {Srednicki}(1984)]{1984PhRvL..53..624S}
{Silk}, J.; {Srednicki}, M.
\newblock {Cosmic-Ray Antiprotons as a Probe of a Photino-Dominated Universe}.
\newblock {\em \prl} {\bf 1984}, {\em 53},~624--627.
\newblock
  doi:{\changeurlcolor{black}\href{https://doi.org/10.1103/PhysRevLett.53.624}{\detokenize{10.1103/PhysRevLett.53.624}}}.

\bibitem[{Aguilar} \em{et~al.}(2016){Aguilar}, {Ali Cavasonza}, {Alpat},
  {Ambrosi}, {Arruda}, {Attig}, {Aupetit}, {Azzarello}, {Bachlechner}, {Barao},
  and et~al]{2016PhRvL.117i1103A}
{Aguilar}, M.; {Ali Cavasonza}, L.; {Alpat}, B.; {Ambrosi}, G.; {Arruda}, L.;
  {Attig}, N.; {Aupetit}, S.; {Azzarello}, P.; {Bachlechner}, A.; {Barao}, F.;
  et~al.
\newblock {Antiproton Flux, Antiproton-to-Proton Flux Ratio, and Properties of
  Elementary Particle Fluxes in Primary Cosmic Rays Measured with the Alpha
  Magnetic Spectrometer on the International Space Station}.
\newblock {\em Physical Review Letters} {\bf 2016}, {\em 117},~091103.
\newblock
  doi:{\changeurlcolor{black}\href{https://doi.org/10.1103/PhysRevLett.117.091103}{\detokenize{10.1103/PhysRevLett.117.091103}}}.

\bibitem[{Kachelriess} \em{et~al.}(2020){Kachelriess}, {Ostapchenko}, and
  {Tjemsland}]{2020arXiv200210481K}
{Kachelriess}, M.; {Ostapchenko}, S.; {Tjemsland}, J.
\newblock {Revisiting cosmic ray antinuclei fluxes with a new coalescence
  mode}.
\newblock {\em arXiv e-prints} {\bf 2020}, p. arXiv:2002.10481,
  \href{http://xxx.lanl.gov/abs/2002.10481}{{\normalfont
  [arXiv:hep-ph/2002.10481]}}.

\bibitem[{Galaktionov}(2002)]{2002RPPh...65.1243G}
{Galaktionov}, Y.V.
\newblock {Antimatter in cosmic rays}.
\newblock {\em Reports on Progress in Physics} {\bf 2002}, {\em
  65},~1243--1270.
\newblock
  doi:{\changeurlcolor{black}\href{https://doi.org/10.1088/0034-4885/65/9/201}{\detokenize{10.1088/0034-4885/65/9/201}}}.

\bibitem[{Poulin} \em{et~al.}(2019){Poulin}, {Salati}, {Cholis},
  {Kamionkowski}, and {Silk}]{2019PhRvD..99b3016P}
{Poulin}, V.; {Salati}, P.; {Cholis}, I.; {Kamionkowski}, M.; {Silk}, J.
\newblock {Where do the AMS-02 antihelium events come from?}
\newblock {\em \prd} {\bf 2019}, {\em 99},~023016,
  \href{http://xxx.lanl.gov/abs/1808.08961}{{\normalfont
  [arXiv:astro-ph.HE/1808.08961]}}.
\newblock
  doi:{\changeurlcolor{black}\href{https://doi.org/10.1103/PhysRevD.99.023016}{\detokenize{10.1103/PhysRevD.99.023016}}}.

\bibitem[{Cholis} \em{et~al.}(2020){Cholis}, {Linden}, and
  {Hooper}]{2020arXiv200108749C}
{Cholis}, I.; {Linden}, T.; {Hooper}, D.
\newblock {Anti-Deuterons and Anti-Helium Nuclei from Annihilating Dark
  Matter}.
\newblock {\em arXiv e-prints} {\bf 2020}, p. arXiv:2001.08749,
  \href{http://xxx.lanl.gov/abs/2001.08749}{{\normalfont
  [arXiv:astro-ph.HE/2001.08749]}}.

\bibitem[{von Doetinchem} \em{et~al.}(2020){von Doetinchem}, {Perez},
  {Aramaki}, {Baker}, {Barwick}, {Bird}, {Boezio}, {Boggs}, {Cui}, {Datta}, and
  et~al]{2020arXiv200204163V}
{von Doetinchem}, P.; {Perez}, K.; {Aramaki}, T.; {Baker}, S.; {Barwick}, S.;
  {Bird}, R.; {Boezio}, M.; {Boggs}, S.E.; {Cui}, M.; {Datta}, A.; et~al.
\newblock {Cosmic-ray Antinuclei as Messengers of New Physics: Status and
  Outlook for the New Decade}.
\newblock {\em arXiv e-prints} {\bf 2020}, p. arXiv:2002.04163,
  \href{http://xxx.lanl.gov/abs/2002.04163}{{\normalfont
  [arXiv:astro-ph.HE/2002.04163]}}.

\bibitem[{Smoot} \em{et~al.}(1975){Smoot}, {Buffington}, and
  {Orth}]{1975PhRvL..35..258S}
{Smoot}, G.F.; {Buffington}, A.; {Orth}, C.D.
\newblock {Search for Cosmic-Ray Antimatter}.
\newblock {\em \\prl} {\bf 1975}, {\em 35},~258--261.
\newblock
  doi:{\changeurlcolor{black}\href{https://doi.org/10.1103/PhysRevLett.35.258}{\detokenize{10.1103/PhysRevLett.35.258}}}.

\bibitem[{Badhwar} \em{et~al.}(1978){Badhwar}, {Golden}, {Lacy}, {Zipse},
  {Daniel}, and {Stephens}]{1978Natur.274..137B}
{Badhwar}, G.D.; {Golden}, R.L.; {Lacy}, J.L.; {Zipse}, J.E.; {Daniel}, R.R.;
  {Stephens}, S.A.
\newblock {Relative abundance of antiprotons and antihelium in the primary
  cosmic radiation}.
\newblock {\em \ at} {\bf 1978}, {\em 274},~137--139.
\newblock
  doi:{\changeurlcolor{black}\href{https://doi.org/10.1038/274137b0}{\detokenize{10.1038/274137b0}}}.

\bibitem[{Ormes} \em{et~al.}(1997){Ormes}, {Moiseev}, {Saeki}, {Anraku},
  {Orito}, {Golden}, {Imori}, {Inaba}, {Kimbell}, {Kimura}, {Makida},
  {Matsumoto}, {Matsunaga}, {Mitchell}, {Motoki}, {Nishimura}, {Nozaki},
  {Streitmatter}, {Suzuki}, {Tanaka}, {Ueda}, {Yajima}, {Yamagami}, {Yamamoto},
  {Yoshida}, {Yoshimura}, and {BESS Collaboration}]{1997ApJ...482L.187O}
{Ormes}, J.F.; {Moiseev}, A.A.; {Saeki}, T.; {Anraku}, K.; {Orito}, S.;
  {Golden}, R.L.; {Imori}, M.; {Inaba}, S.; {Kimbell}, B.L.; {Kimura}, N.;
  {Makida}, Y.; {Matsumoto}, H.; {Matsunaga}, H.; {Mitchell}, J.W.; {Motoki},
  M.; {Nishimura}, J.; {Nozaki}, M.; {Streitmatter}, R.E.; {Suzuki}, J.;
  {Tanaka}, K.; {Ueda}, I.; {Yajima}, N.; {Yamagami}, T.; {Yamamoto}, A.;
  {Yoshida}, T.; {Yoshimura}, K.; {BESS Collaboration}.
\newblock {Antihelium in Cosmic Rays: A New Upper Limit and Its Significance}.
\newblock {\em \\apjl} {\bf 1997}, {\em 482},~L187--L190.
\newblock
  doi:{\changeurlcolor{black}\href{https://doi.org/10.1086/310700}{\detokenize{10.1086/310700}}}.

\bibitem[{Golden} \em{et~al.}(1997){Golden}, {Stochaj}, {Stephens}, {Moiseev},
  {Ormes}, {Streitmatter}, {Bowen}, {Moats}, and
  {Lloyd-Evans}]{1997ApJ...479..992G}
{Golden}, R.L.; {Stochaj}, S.J.; {Stephens}, S.A.; {Moiseev}, A.A.; {Ormes},
  J.F.; {Streitmatter}, R.E.; {Bowen}, T.; {Moats}, A.; {Lloyd-Evans}, J.
\newblock {Search for Antihelium in the Cosmic Rays}.
\newblock {\em \\apj} {\bf 1997}, {\em 479},~992--996.
\newblock
  doi:{\changeurlcolor{black}\href{https://doi.org/10.1086/303886}{\detokenize{10.1086/303886}}}.

\bibitem[{Saeki} \em{et~al.}(1998){Saeki}, {Anraku}, {Orito}, {Ormes}, {Imori},
  {Kimbell}, {Makida}, {Matsumoto}, {Matsunaga}, {Mitchell}, {Motoki},
  {Nishimura}, {Nozaki}, {Otoba}, {Sanuki}, {Streitmatter}, {Suzuki}, {Tanaka},
  {Ueda}, {Yajima}, {Yamagami}, {Yamamoto}, {Yoshida}, and
  {Yoshimura}]{1998PhLB..422..319S}
{Saeki}, T.; {Anraku}, K.; {Orito}, S.; {Ormes}, J.; {Imori}, M.; {Kimbell},
  B.; {Makida}, Y.; {Matsumoto}, H.; {Matsunaga}, H.; {Mitchell}, J.; {Motoki},
  M.; {Nishimura}, J.; {Nozaki}, M.; {Otoba}, M.; {Sanuki}, T.; {Streitmatter},
  R.; {Suzuki}, J.; {Tanaka}, K.; {Ueda}, I.; {Yajima}, N.; {Yamagami}, T.;
  {Yamamoto}, A.; {Yoshida}, T.; {Yoshimura}, K.
\newblock {A new limit on the flux of cosmic antihelium}.
\newblock {\em Physics Letters B} {\bf 1998}, {\em 422},~319--324,
  \href{http://xxx.lanl.gov/abs/astro-ph/9710228}{{\normalfont
  [arXiv:astro-ph/astro-ph/9710228]}}.
\newblock
  doi:{\changeurlcolor{black}\href{https://doi.org/10.1016/S0370-2693(98)00131-2}{\detokenize{10.1016/S0370-2693(98)00131-2}}}.

\bibitem[{Alcaraz} \em{et~al.}(1999){Alcaraz}, {Alvisi}, {Alpat}, {Ambrosi},
  {Anderhub}, {Ao}, {Arefiev}, {Azzarello}, {Babucci}, {Baldini}, and
  et~al]{1999PhLB..461..387A}
{Alcaraz}, J.; {Alvisi}, D.; {Alpat}, B.; {Ambrosi}, G.; {Anderhub}, H.; {Ao},
  L.; {Arefiev}, A.; {Azzarello}, P.; {Babucci}, E.; {Baldini}, L.; et~al.
\newblock {Search for antihelium in cosmic rays.}
\newblock {\em Physics Letters B} {\bf 1999}, {\em 461},~387--396,
  \href{http://xxx.lanl.gov/abs/hep-ex/0002048}{{\normalfont
  [arXiv:hep-ex/hep-ex/0002048]}}.
\newblock
  doi:{\changeurlcolor{black}\href{https://doi.org/10.1016/S0370-2693(99)00874-6}{\detokenize{10.1016/S0370-2693(99)00874-6}}}.

\bibitem[{Sasaki} \em{et~al.}(2002){Sasaki}, {Matsumoto}, {Nozaki}, {Saeki},
  {Abe}, {Anraku}, {Asaoka}, {Fujikawa}, {Fuke}, {Imori}, {Haino}, {Izumi},
  {Maeno}, {Makida}, {Matsuda}, {Matsui}, {Matsukawa}, {Matsunaga}, {Mitchell},
  {Mitsui}, {Moiseev}, {Motoki}, {Nishimura}, {Orito}, {Ormes}, {Sanuki},
  {Shikaze}, {Seo}, {Sonoda}, {Streitmatter}, {Suzuki}, {Tanaka}, {Tanizaki},
  {Ueda}, {Wang}, {Yajima}, {Yamagami}, {Yamamoto}, {Yamamoto}, {Yamato},
  {Yoshida}, and {Yoshimura}]{2002NuPhS.113..202S}
{Sasaki}, M.; {Matsumoto}, H.; {Nozaki}, M.; {Saeki}, T.; {Abe}, K.; {Anraku},
  K.; {Asaoka}, Y.; {Fujikawa}, M.; {Fuke}, H.; {Imori}, M.; {Haino}, S.;
  {Izumi}, K.; {Maeno}, T.; {Makida}, Y.; {Matsuda}, S.; {Matsui}, N.;
  {Matsukawa}, T.; {Matsunaga}, H.; {Mitchell}, J.W.; {Mitsui}, T.; {Moiseev},
  A.; {Motoki}, M.; {Nishimura}, J.; {Orito}, S.; {Ormes}, J.F.; {Sanuki}, T.;
  {Shikaze}, Y.; {Seo}, E.S.; {Sonoda}, T.; {Streitmatter}, R.; {Suzuki}, J.;
  {Tanaka}, K.; {Tanizaki}, K.; {Ueda}, I.; {Wang}, J.Z.; {Yajima}, Y.;
  {Yamagami}, Y.; {Yamamoto}, A.; {Yamamoto}, Y.; {Yamato}, K.; {Yoshida}, T.;
  {Yoshimura}, K.
\newblock {Progress in search for antihelium with BESS}.
\newblock {\em Nuclear Physics B Proceedings Supplements} {\bf 2002}, {\em
  113},~202--207.
\newblock
  doi:{\changeurlcolor{black}\href{https://doi.org/10.1016/S0920-5632(02)01842-X}{\detokenize{10.1016/S0920-5632(02)01842-X}}}.

\bibitem[{Sasaki} \em{et~al.}(2008){Sasaki}, {Haino}, {Abe}, {Fuke}, {Hams},
  {Kim}, {Lee}, {Makida}, {Matsuda}, {Mitchell}, {Moiseev}, {Nishimura},
  {Nozaki}, {Orito}, {Ormes}, {Sanuki}, {Seo}, {Shikaze}, {Streitmatter},
  {Suzuki}, {Tanaka}, {Yamagami}, {Yamamoto}, {Yoshida}, and
  {Yoshimura}]{2008AdSpR..42..450S}
{Sasaki}, M.; {Haino}, S.; {Abe}, K.; {Fuke}, H.; {Hams}, T.; {Kim}, K.C.;
  {Lee}, M.H.; {Makida}, Y.; {Matsuda}, S.; {Mitchell}, J.W.; {Moiseev}, A.A.;
  {Nishimura}, J.; {Nozaki}, M.; {Orito}, S.; {Ormes}, J.F.; {Sanuki}, T.;
  {Seo}, E.S.; {Shikaze}, Y.; {Streitmatter}, R.E.; {Suzuki}, J.; {Tanaka}, K.;
  {Yamagami}, T.; {Yamamoto}, A.; {Yoshida}, T.; {Yoshimura}, K.
\newblock {Search for antihelium: Progress with BESS}.
\newblock {\em Advances in Space Research} {\bf 2008}, {\em 42},~450--454.
\newblock
  doi:{\changeurlcolor{black}\href{https://doi.org/10.1016/j.asr.2007.09.012}{\detokenize{10.1016/j.asr.2007.09.012}}}.

\bibitem[{Mayorov} \em{et~al.}(2011){Mayorov}, {Galper}, {Adriani},
  {Bazilevskaya}, {Barbarino}, {Bellotti}, {Boezio}, {Bogomolov}, {Bonvicini},
  {Bongi}, {Bonechi}, {Borisov}, {Bottai}, {Bruno}, {Vacci}, {Vannuccini},
  {Vasiliev}, {Voronov}, {Wu}, {Danilchenko}, {Gillard}, {Jerse}, {Zampa},
  {Zampa}, {Zverev}, {Casolino}, {Campana}, {Carbone}, {Karelin}, {Carlson},
  {Castellini}, {Cafagna}, {Kvashnin}, {Koldashov}, {Koldobsky}, {Krutkov},
  {Leonov}, {Malakhov}, {Malvezzi}, {Marcelli}, {Menn}, {Mikhailov},
  {Mocchiutti}, {Mori}, {Nikonov}, {Osteria}, {Pizzolotto}, {Papini}, {De
  Pascale}, {Picozza}, {Pearce}, {Ricci}, {Ricciarini}, {Runtso}, {Simon}, {De
  Simone}, {Sparvoli}, {Spillantini}, {Stozhkov}, {Di Felice}, and
  {Yurkin}]{2011JETPL..93..628M}
{Mayorov}, A.G.; {Galper}, A.M.; {Adriani}, O.; {Bazilevskaya}, G.A.;
  {Barbarino}, G.; {Bellotti}, R.; {Boezio}, M.; {Bogomolov}, E.A.;
  {Bonvicini}, V.; {Bongi}, M.; {Bonechi}, L.; {Borisov}, S.V.; {Bottai}, S.;
  {Bruno}, A.; {Vacci}, S.; {Vannuccini}, E.; {Vasiliev}, G.I.; {Voronov},
  S.A.; {Wu}, Y.; {Danilchenko}, I.A.; {Gillard}, W.; {Jerse}, G.; {Zampa}, G.;
  {Zampa}, N.; {Zverev}, V.G.; {Casolino}, M.; {Campana}, D.; {Carbone}, R.;
  {Karelin}, A.V.; {Carlson}, P.; {Castellini}, G.; {Cafagna}, F.; {Kvashnin},
  A.N.; {Koldashov}, S.V.; {Koldobsky}, S.A.; {Krutkov}, S.Y.; {Leonov}, A.A.;
  {Malakhov}, V.V.; {Malvezzi}, V.; {Marcelli}, L.; {Menn}, W.; {Mikhailov},
  V.V.; {Mocchiutti}, E.; {Mori}, N.; {Nikonov}, N.V.; {Osteria}, G.;
  {Pizzolotto}, S.; {Papini}, P.; {De Pascale}, M.P.; {Picozza}, P.; {Pearce},
  M.; {Ricci}, M.; {Ricciarini}, S.; {Runtso}, M.F.; {Simon}, M.; {De Simone},
  N.; {Sparvoli}, R.; {Spillantini}, P.; {Stozhkov}, Y.I.; {Di Felice}, V.;
  {Yurkin}, Y.T.
\newblock {Upper limit on the antihelium flux in primary cosmic rays}.
\newblock {\em Soviet Journal of Experimental and Theoretical Physics Letters}
  {\bf 2011}, {\em 93},~628--631.
\newblock
  doi:{\changeurlcolor{black}\href{https://doi.org/10.1134/S0021364011110087}{\detokenize{10.1134/S0021364011110087}}}.

\bibitem[{Fuke} \em{et~al.}(2005){Fuke}, {Maeno}, {Abe}, {Haino}, {Makida},
  {Matsuda}, {Matsumoto}, {Mitchell}, {Moiseev}, {Nishimura}, {Nozaki},
  {Orito}, {Ormes}, {Sasaki}, {Seo}, {Shikaze}, {Streitmatter}, {Suzuki},
  {Tanaka}, {Tanizaki}, {Yamagami}, {Yamamoto}, {Yamamoto}, {Yamato},
  {Yoshida}, and {Yoshimura}]{2005PhRvL..95h1101F}
{Fuke}, H.; {Maeno}, T.; {Abe}, K.; {Haino}, S.; {Makida}, Y.; {Matsuda}, S.;
  {Matsumoto}, H.; {Mitchell}, J.W.; {Moiseev}, A.A.; {Nishimura}, J.;
  {Nozaki}, M.; {Orito}, S.; {Ormes}, J.F.; {Sasaki}, M.; {Seo}, E.S.;
  {Shikaze}, Y.; {Streitmatter}, R.E.; {Suzuki}, J.; {Tanaka}, K.; {Tanizaki},
  K.; {Yamagami}, T.; {Yamamoto}, A.; {Yamamoto}, Y.; {Yamato}, K.; {Yoshida},
  T.; {Yoshimura}, K.
\newblock {Search for Cosmic-Ray Antideuterons}.
\newblock {\em \\prl} {\bf 2005}, {\em 95},~081101,
  \href{http://xxx.lanl.gov/abs/astro-ph/0504361}{{\normalfont
  [arXiv:astro-ph/astro-ph/0504361]}}.
\newblock
  doi:{\changeurlcolor{black}\href{https://doi.org/10.1103/PhysRevLett.95.081101}{\detokenize{10.1103/PhysRevLett.95.081101}}}.

\bibitem[{Aizu} \em{et~al.}(1961){Aizu}, {Fujimoto}, {Hasegawa}, {Koshiba},
  {Mito}, {Nishimura}, {Yokoi}, and {Schein}]{1961PhRv..121.1206A}
{Aizu}, H.; {Fujimoto}, Y.; {Hasegawa}, S.; {Koshiba}, M.; {Mito}, I.;
  {Nishimura}, J.; {Yokoi}, K.; {Schein}, M.
\newblock {Heavy Nuclei in the Primary Cosmic Radiation at Prince Albert,
  Canada. II}.
\newblock {\em Physical Review} {\bf 1961}, {\em 121},~1206--1218.
\newblock
  doi:{\changeurlcolor{black}\href{https://doi.org/10.1103/PhysRev.121.1206}{\detokenize{10.1103/PhysRev.121.1206}}}.

\bibitem[{Greenhill} \em{et~al.}(1971){Greenhill}, {Clarke}, and
  {Elliot}]{1971Natur.230..170G}
{Greenhill}, J.G.; {Clarke}, A.R.; {Elliot}, H.
\newblock {Search for Anti-matter in Primary Cosmic Rays}.
\newblock {\em \ at} {\bf 1971}, {\em 230},~170--172.
\newblock
  doi:{\changeurlcolor{black}\href{https://doi.org/10.1038/230170a0}{\detokenize{10.1038/230170a0}}}.

\bibitem[{Golden} \em{et~al.}(1971){Golden}, {Adams}, {Boykin}, {Denny},
  {Marar}, {Heckman}, and {Lindstrom}]{1971ICRC....1..203G}
{Golden}, R.L.; {Adams}, J.H.; {Boykin}, W.R.; {Denny}, C.L.; {Marar}, T.M.K.;
  {Heckman}, H.H.; {Lindstrom}, P.L.
\newblock {The Rigidity Spectrum of Z <= 3 Nuclei from 5 GV to over 300 GV.}
\newblock  12th International Cosmic Ray Conference (ICRC12), Volume 1,  1971,
  Vol.~1, {\em International Cosmic Ray Conference}, p. 203.

\bibitem[{Buffington} \em{et~al.}(1972){Buffington}, {Smith}, {Smoot}, and
  {Alvarez}]{1972Natur.236..335B}
{Buffington}, A.; {Smith}, L.H.; {Smoot}, G.F.; {Alvarez}, L.W.
\newblock {Search for Antimatter in Primary Cosmic Rays}.
\newblock {\em \ at} {\bf 1972}, {\em 236},~335--338.
\newblock
  doi:{\changeurlcolor{black}\href{https://doi.org/10.1038/236335a0}{\detokenize{10.1038/236335a0}}}.

\bibitem[{Evenson}(1972)]{1972ApJ...176..797E}
{Evenson}, P.
\newblock {A Search for Antihelium in Primary Cosmic Radiation}.
\newblock {\em \\apj} {\bf 1972}, {\em 176},~797.
\newblock
  doi:{\changeurlcolor{black}\href{https://doi.org/10.1086/151678}{\detokenize{10.1086/151678}}}.

\bibitem[{Verma} \em{et~al.}(1972){Verma}, {Rengarajan}, {Tandon}, {Damle}, and
  {Pal}]{1972NPhS..240..135V}
{Verma}, R.P.; {Rengarajan}, T.N.; {Tandon}, S.N.; {Damle}, S.V.; {Pal}, Y.
\newblock {Rigidity Spectrum of Helium Nuclei above 17 GV and a Search for High
  Energy Anti-nuclei in Primary Cosmic Rays}.
\newblock {\em Nature Physical Science} {\bf 1972}, {\em 240},~135--137.
\newblock
  doi:{\changeurlcolor{black}\href{https://doi.org/10.1038/physci240135a0}{\detokenize{10.1038/physci240135a0}}}.

\bibitem[{Golden} \em{et~al.}(1974){Golden}, {Adams}, {Deney}, {Badhwar},
  {Marar}, {Heckman}, and {Lindstrom}]{1974ApJ...192..747G}
{Golden}, R.L.; {Adams}, J.~H., J.; {Deney}, C.L.; {Badhwar}, G.D.; {Marar},
  T.M.K.; {Heckman}, H.H.; {Lindstrom}, P.J.
\newblock {Rigidity spectrum of Z {\\ensuremath{\\geq}} 3 cosmic-ray nuclei in
  the range 4 - 285 GV and a search for cosmic antimatter.}
\newblock {\em \\apj} {\bf 1974}, {\em 192},~747--751.
\newblock
  doi:{\changeurlcolor{black}\href{https://doi.org/10.1086/153112}{\detokenize{10.1086/153112}}}.

\bibitem[{Cristinziani}(2002)]{2002NuPhS.113..195C}
{Cristinziani}, M.
\newblock {Antimatter searches with AMS}.
\newblock {\em Nuclear Physics B Proceedings Supplements} {\bf 2002}, {\em
  113},~195--201.
\newblock
  doi:{\changeurlcolor{black}\href{https://doi.org/10.1016/S0920-5632(02)01841-8}{\detokenize{10.1016/S0920-5632(02)01841-8}}}.

\bibitem[{Adriani} \em{et~al.}(2014){Adriani}, {Barbarino}, {Bazilevskaya},
  {Bellotti}, {Boezio}, {Bogomolov}, {Bongi}, {Bonvicini}, {Bottai}, {Bruno},
  and et~al]{2014PhR...544..323A}
{Adriani}, O.; {Barbarino}, G.C.; {Bazilevskaya}, G.A.; {Bellotti}, R.;
  {Boezio}, M.; {Bogomolov}, E.A.; {Bongi}, M.; {Bonvicini}, V.; {Bottai}, S.;
  {Bruno}, A.; et~al.
\newblock {The PAMELA Mission: Heralding a new era in precision cosmic ray
  physics}.
\newblock {\em \physrep} {\bf 2014}, {\em 544},~323--370.
\newblock
  doi:{\changeurlcolor{black}\href{https://doi.org/10.1016/j.physrep.2014.06.003}{\detokenize{10.1016/j.physrep.2014.06.003}}}.

\bibitem[Kampert and Unger(2012)]{Kampert:2012mx}
Kampert, K.H.; Unger, M.
\newblock {Measurements of the Cosmic Ray Composition with Air Shower
  Experiments}.
\newblock {\em Astropart. Phys.} {\bf 2012}, {\em 35},~660--678,
  \href{http://xxx.lanl.gov/abs/1201.0018}{{\normalfont
  [arXiv:astro-ph.HE/1201.0018]}}.
\newblock
  doi:{\changeurlcolor{black}\href{https://doi.org/10.1016/j.astropartphys.2012.02.004}{\detokenize{10.1016/j.astropartphys.2012.02.004}}}.

\bibitem[Apel \em{et~al.}(2017)Apel et~al.]{Apel:2017ocm}
Apel, W.; others.
\newblock {KASCADE-Grande Limits on the Isotropic Diffuse Gamma-Ray Flux
  between 100 TeV and 1 EeV}.
\newblock {\em Astrophys. J.} {\bf 2017}, {\em 848},~1,
  \href{http://xxx.lanl.gov/abs/1710.02889}{{\normalfont
  [arXiv:astro-ph.HE/1710.02889]}}.
\newblock
  doi:{\changeurlcolor{black}\href{https://doi.org/10.3847/1538-4357/aa8bb7}{\detokenize{10.3847/1538-4357/aa8bb7}}}.

\bibitem[Aab \em{et~al.}(2014)Aab et~al.]{Aab:2014aea}
Aab, A.; others.
\newblock {Depth of maximum of air-shower profiles at the Pierre Auger
  Observatory. II. Composition implications}.
\newblock {\em Phys. Rev. D} {\bf 2014}, {\em 90},~122006,
  \href{http://xxx.lanl.gov/abs/1409.5083}{{\normalfont
  [arXiv:astro-ph.HE/1409.5083]}}.
\newblock
  doi:{\changeurlcolor{black}\href{https://doi.org/10.1103/PhysRevD.90.122006}{\detokenize{10.1103/PhysRevD.90.122006}}}.

\bibitem[Apel \em{et~al.}(2012)Apel et~al.]{Apel:2012tda}
Apel, W.; others.
\newblock {The spectrum of high-energy cosmic rays measured with
  KASCADE-Grande}.
\newblock {\em Astropart. Phys.} {\bf 2012}, {\em 36},~183--194.
\newblock
  doi:{\changeurlcolor{black}\href{https://doi.org/10.1016/j.astropartphys.2012.05.023}{\detokenize{10.1016/j.astropartphys.2012.05.023}}}.

\bibitem[{Rawlins} and {IceCube Collaboration}(2015)]{2015ICRC...34..334R}
{Rawlins}, K.; {IceCube Collaboration}.
\newblock {Latest Results on Cosmic Ray Spectrum and Composition from Three
  Years of IceTop and IceCube}.
\newblock  34th International Cosmic Ray Conference (ICRC2015),  2015, Vol.~34,
  {\em International Cosmic Ray Conference}, p. 334.

\bibitem[{Aab} \em{et~al.}(2014){Aab}, {Abreu}, {Aglietta}, {Ahn}, {Al
  Samarai}, {Albuquerque}, {Allekotte}, {Allen}, {Allison}, {Almela}, and
  et~al]{2014PhRvD..90l2005A}
{Aab}, A.; {Abreu}, P.; {Aglietta}, M.; {Ahn}, E.J.; {Al Samarai}, I.;
  {Albuquerque}, I.F.M.; {Allekotte}, I.; {Allen}, J.; {Allison}, P.; {Almela},
  A.; et~al.
\newblock {Depth of maximum of air-shower profiles at the Pierre Auger
  Observatory. I. Measurements at energies above 1 0$^{17.8}$ eV}.
\newblock {\em \\prd} {\bf 2014}, {\em 90},~122005,
  \href{http://xxx.lanl.gov/abs/1409.4809}{{\normalfont
  [arXiv:astro-ph.HE/1409.4809]}}.
\newblock
  doi:{\changeurlcolor{black}\href{https://doi.org/10.1103/PhysRevD.90.122005}{\detokenize{10.1103/PhysRevD.90.122005}}}.

\bibitem[{Prosin} \em{et~al.}(2014){Prosin}, {Berezhnev}, {Budnev},
  {Chiavassa}, {Chvalaev}, {Gress}, {Dyachok}, {Epimakhov}, {Karpov},
  {Kalmykov}, and et~al]{2014NIMPA.756...94P}
{Prosin}, V.V.; {Berezhnev}, S.F.; {Budnev}, N.M.; {Chiavassa}, A.; {Chvalaev},
  O.A.; {Gress}, O.A.; {Dyachok}, A.N.; {Epimakhov}, S.N.; {Karpov}, N.I.;
  {Kalmykov}, N.N.; et~al.
\newblock {Tunka-133: Results of 3 year operation}.
\newblock {\em Nuclear Instruments and Methods in Physics Research A} {\bf
  2014}, {\em 756},~94--101.
\newblock
  doi:{\changeurlcolor{black}\href{https://doi.org/10.1016/j.nima.2013.09.018}{\detokenize{10.1016/j.nima.2013.09.018}}}.

\bibitem[{Aguilar} \em{et~al.}(2019){Aguilar}, {Ali Cavasonza}, {Ambrosi},
  {Arruda}, {Attig}, {Bachlechner}, {Barao}, {Barrau}, {Barrin}, {Bartoloni},
  and et~al]{2019PhRvL.123r1102A}
{Aguilar}, M.; {Ali Cavasonza}, L.; {Ambrosi}, G.; {Arruda}, L.; {Attig}, N.;
  {Bachlechner}, A.; {Barao}, F.; {Barrau}, A.; {Barrin}, L.; {Bartoloni}, A.;
  et~al.
\newblock {Properties of Cosmic Helium Isotopes Measured by the Alpha Magnetic
  Spectrometer}.
\newblock {\em \prl} {\bf 2019}, {\em 123},~181102.
\newblock
  doi:{\changeurlcolor{black}\href{https://doi.org/10.1103/PhysRevLett.123.181102}{\detokenize{10.1103/PhysRevLett.123.181102}}}.

\bibitem[{Adriani} \em{et~al.}(2013){Adriani}, {Barbarino}, {Bazilevskaya},
  {Bellotti}, {Boezio}, {Bogomolov}, {Bongi}, {Bonvicini}, {Borisov}, {Bottai},
  and et~al]{2013ApJ...765...91A}
{Adriani}, O.; {Barbarino}, G.C.; {Bazilevskaya}, G.A.; {Bellotti}, R.;
  {Boezio}, M.; {Bogomolov}, E.A.; {Bongi}, M.; {Bonvicini}, V.; {Borisov}, S.;
  {Bottai}, S.; et~al.
\newblock {Time Dependence of the Proton Flux Measured by PAMELA during the
  2006 July-2009 December Solar Minimum}.
\newblock {\em \apj} {\bf 2013}, {\em 765},~91,
  \href{http://xxx.lanl.gov/abs/1301.4108}{{\normalfont
  [arXiv:astro-ph.HE/1301.4108]}}.
\newblock
  doi:{\changeurlcolor{black}\href{https://doi.org/10.1088/0004-637X/765/2/91}{\detokenize{10.1088/0004-637X/765/2/91}}}.

\bibitem[{Stone} \em{et~al.}(2013){Stone}, {Cummings}, {McDonald}, {Heikkila},
  {Lal}, and {Webber}]{2013Sci...341..150S}
{Stone}, E.C.; {Cummings}, A.C.; {McDonald}, F.B.; {Heikkila}, B.C.; {Lal}, N.;
  {Webber}, W.R.
\newblock {Voyager 1 Observes Low-Energy Galactic Cosmic Rays in a Region
  Depleted of Heliospheric Ions}.
\newblock {\em Science} {\bf 2013}, {\em 341},~150--153.
\newblock
  doi:{\changeurlcolor{black}\href{https://doi.org/10.1126/science.1236408}{\detokenize{10.1126/science.1236408}}}.

\bibitem[{Panov} \em{et~al.}(2009){Panov}, {Adams}, {Ahn}, {Bashinzhagyan},
  {Watts}, {Wefel}, {Wu}, {Ganel}, {Guzik}, {Zatsepin}, and
  et~al]{2009BRASP..73..564P}
{Panov}, A.D.; {Adams}, J.H.; {Ahn}, H.S.; {Bashinzhagyan}, G.L.; {Watts},
  J.W.; {Wefel}, J.P.; {Wu}, J.; {Ganel}, O.; {Guzik}, T.G.; {Zatsepin}, V.I.;
  et~al.
\newblock {Energy spectra of abundant nuclei of primary cosmic rays from the
  data of ATIC-2 experiment: Final results}.
\newblock {\em Bulletin of the Russian Academy of Science, Phys.} {\bf 2009},
  {\em 73},~564--567,  \href{http://xxx.lanl.gov/abs/1101.3246}{{\normalfont
  [arXiv:astro-ph.HE/1101.3246]}}.
\newblock
  doi:{\changeurlcolor{black}\href{https://doi.org/10.3103/S1062873809050098}{\detokenize{10.3103/S1062873809050098}}}.

\bibitem[{Adriani} \em{et~al.}(2019){Adriani}, {Akaike}, {Asano}, {Asaoka},
  {Bagliesi}, {Berti}, {Bigongiari}, {Binns}, {Bonechi}, {Bongi}, and
  et~al]{2019PhRvL.122r1102A}
{Adriani}, O.; {Akaike}, Y.; {Asano}, K.; {Asaoka}, Y.; {Bagliesi}, M.G.;
  {Berti}, E.; {Bigongiari}, G.; {Binns}, W.R.; {Bonechi}, S.; {Bongi}, M.;
  et~al.
\newblock {Direct Measurement of the Cosmic-Ray Proton Spectrum from 50 GeV to
  10 TeV with the Calorimetric Electron Telescope on the International Space
  Station}.
\newblock {\em \prl} {\bf 2019}, {\em 122},~181102,
  \href{http://xxx.lanl.gov/abs/1905.04229}{{\normalfont
  [arXiv:astro-ph.HE/1905.04229]}}.
\newblock
  doi:{\changeurlcolor{black}\href{https://doi.org/10.1103/PhysRevLett.122.181102}{\detokenize{10.1103/PhysRevLett.122.181102}}}.

\bibitem[{Yoon} \em{et~al.}(2011){Yoon}, {Ahn}, {Allison}, {Bagliesi},
  {Beatty}, {Bigongiari}, {Boyle}, {Childers}, {Conklin}, {Coutu}, , and
  et~al]{2011ApJ...728..122Y}
{Yoon}, Y.S.; {Ahn}, H.S.; {Allison}, P.S.; {Bagliesi}, M.G.; {Beatty}, J.J.;
  {Bigongiari}, G.; {Boyle}, P.J.; {Childers}, J.T.; {Conklin}, N.B.; {Coutu},
  S.; .; et~al.
\newblock {Cosmic-ray Proton and Helium Spectra from the First CREAM Flight}.
\newblock {\em \apj} {\bf 2011}, {\em 728},~122,
  \href{http://xxx.lanl.gov/abs/1102.2575}{{\normalfont
  [arXiv:astro-ph.HE/1102.2575]}}.
\newblock
  doi:{\changeurlcolor{black}\href{https://doi.org/10.1088/0004-637X/728/2/122}{\detokenize{10.1088/0004-637X/728/2/122}}}.

\bibitem[{Grebenyuk} \em{et~al.}(2019){Grebenyuk}, {Karmanov}, {Kovalev},
  {Kudryashov}, {Kurganov}, {Panov}, {Podorozhny}, {Tkachenko}, {Tkachev},
  {Turundaevskiy}, {Vasiliev}, and {Voronin}]{2019AdSpR..64.2546G}
{Grebenyuk}, V.; {Karmanov}, D.; {Kovalev}, I.; {Kudryashov}, I.; {Kurganov},
  A.; {Panov}, A.; {Podorozhny}, D.; {Tkachenko}, A.; {Tkachev}, L.;
  {Turundaevskiy}, A.; {Vasiliev}, O.; {Voronin}, A.
\newblock {Energy spectra of abundant cosmic-ray nuclei in the NUCLEON
  experiment}.
\newblock {\em Advances in Space Research} {\bf 2019}, {\em 64},~2546--2558.
\newblock
  doi:{\changeurlcolor{black}\href{https://doi.org/10.1016/j.asr.2019.10.004}{\detokenize{10.1016/j.asr.2019.10.004}}}.

\bibitem[{Derbina} \em{et~al.}(2005){Derbina}, {Galkin}, {Hareyama},
  {Hirakawa}, {Horiuchi}, {Ichimura}, {Inoue}, {Kamioka}, {Kobayashi},
  {Kopenkin}, and et~al]{2005ApJ...628L..41D}
{Derbina}, V.A.; {Galkin}, V.I.; {Hareyama}, M.; {Hirakawa}, Y.; {Horiuchi},
  Y.; {Ichimura}, M.; {Inoue}, N.; {Kamioka}, E.; {Kobayashi}, T.; {Kopenkin},
  V.V.; et~al.
\newblock {Cosmic-Ray Spectra and Composition in the Energy Range of 10-1000
  TeV per Particle Obtained by the RUNJOB Experiment}.
\newblock {\em \apjl} {\bf 2005}, {\em 628},~L41--L44.
\newblock
  doi:{\changeurlcolor{black}\href{https://doi.org/10.1086/432715}{\detokenize{10.1086/432715}}}.

\bibitem[{Adriani} \em{et~al.}(2013){Adriani}, {Barbarino}, {Bazilevskaya},
  {Boezio}, {Bogomolov}, {Bonechi}, {Bongi}, {Bonvicini}, {Borisov}, {Bottai},
  and et~al]{2013AdSpR..51..219A}
{Adriani}, O.; {Barbarino}, G.C.; {Bazilevskaya}, G.A.; {Boezio}, M.;
  {Bogomolov}, E.A.; {Bonechi}, L.; {Bongi}, M.; {Bonvicini}, V.; {Borisov},
  S.V.; {Bottai}, S.; et~al.
\newblock {Measurements of cosmic-ray proton and helium spectra with the PAMELA
  calorimeter}.
\newblock {\em Advances in Space Research} {\bf 2013}, {\em 51},~219--226.
\newblock
  doi:{\changeurlcolor{black}\href{https://doi.org/10.1016/j.asr.2012.09.029}{\detokenize{10.1016/j.asr.2012.09.029}}}.

\bibitem[{Schoo} \em{et~al.}(2015){Schoo}, {Apel}, {Arteaga-Vel{\'a}zquez},
  {Bekk}, {Bertaina}, {Bl{\"u}mer}, {Bozdog}, {Brancus}, {Cantoni},
  {Chiavassa}, and et~al]{2015ICRC...34..263S}
{Schoo}, S.; {Apel}, W.D.; {Arteaga-Vel{\'a}zquez}, J.C.; {Bekk}, K.;
  {Bertaina}, M.; {Bl{\"u}mer}, J.; {Bozdog}, H.; {Brancus}, I.M.; {Cantoni},
  E.; {Chiavassa}, A.; et~al.
\newblock {The energy spectrum of cosmic rays in the range from 10\^\{14\} to
  10\^\{18\}eV}.
\newblock  34th International Cosmic Ray Conference (ICRC2015),  2015, Vol.~34,
  {\em International Cosmic Ray Conference}, p. 263.

\bibitem[{Aab} \em{et~al.}(2015){Aab}, {Abreu}, {Aglietta}, {Ahn}, {Samarai},
  {Albuquerque}, {Allekotte}, {Allison}, {Almela}, and
  et~al]{2015arXiv150903732T}
{Aab}, A.; {Abreu}, P.; {Aglietta}, M.; {Ahn}, E.J.; {Samarai}, I.A.;
  {Albuquerque}, I.F.M.; {Allekotte}, I.; {Allison}, P.; {Almela}, A.; et~al.
\newblock {The Pierre Auger Observatory: Contributions to the 34th
  International Cosmic Ray Conference (ICRC 2015)}.
\newblock {\em arXiv e-prints} {\bf 2015}, p. arXiv:1509.03732,
  \href{http://xxx.lanl.gov/abs/1509.03732}{{\normalfont
  [arXiv:astro-ph.HE/1509.03732]}}.

\bibitem[{Ivanov}(2015)]{2015ICRC...34..349I}
{Ivanov}, D.
\newblock {TA Spectrum Summary}.
\newblock  34th International Cosmic Ray Conference (ICRC2015),  2015, Vol.~34,
  {\em International Cosmic Ray Conference}, p. 349.

\bibitem[{Aguilar} \em{et~al.}(2017){Aguilar}, {Ali Cavasonza}, {Alpat},
  {Ambrosi}, {Arruda}, {Attig}, {Aupetit}, {Azzarello}, {Bachlechner}, {Barao},
  and et~al]{2017PhRvL.119y1101A}
{Aguilar}, M.; {Ali Cavasonza}, L.; {Alpat}, B.; {Ambrosi}, G.; {Arruda}, L.;
  {Attig}, N.; {Aupetit}, S.; {Azzarello}, P.; {Bachlechner}, A.; {Barao}, F.;
  et~al.
\newblock {Observation of the Identical Rigidity Dependence of He, C, and O
  Cosmic Rays at High Rigidities by the Alpha Magnetic Spectrometer on the
  International Space Station}.
\newblock {\em \prl} {\bf 2017}, {\em 119},~251101.
\newblock
  doi:{\changeurlcolor{black}\href{https://doi.org/10.1103/PhysRevLett.119.251101}{\detokenize{10.1103/PhysRevLett.119.251101}}}.

\bibitem[{Aguilar} \em{et~al.}(2018b){Aguilar}, {Ali Cavasonza}, {Alpat},
  {Ambrosi}, {Arruda}, {Attig}, {Aupetit}, {Azzarello}, {Bachlechner}, {Barao},
  and et~al]{2018PhRvL.121e1103A}
{Aguilar}, M.; {Ali Cavasonza}, L.; {Alpat}, B.; {Ambrosi}, G.; {Arruda}, L.;
  {Attig}, N.; {Aupetit}, S.; {Azzarello}, P.; {Bachlechner}, A.; {Barao}, F.;
  et~al.
\newblock {Precision Measurement of Cosmic-Ray Nitrogen and its Primary and
  Secondary Components with the Alpha Magnetic Spectrometer on the
  International Space Station}.
\newblock {\em \prl} {\bf 2018b}, {\em 121},~051103.
\newblock
  doi:{\changeurlcolor{black}\href{https://doi.org/10.1103/PhysRevLett.121.051103}{\detokenize{10.1103/PhysRevLett.121.051103}}}.

\bibitem[Aguilar \em{et~al.}(2020)Aguilar, Ali~Cavasonza, Ambrosi, Arruda,
  Attig, Barao, Barrin, Bartoloni, Başeğmez-du Pree, Battiston, and
  et~al]{Aguilar:2020ohx}
Aguilar, M.; Ali~Cavasonza, L.; Ambrosi, G.; Arruda, L.; Attig, N.; Barao, F.;
  Barrin, L.; Bartoloni, A.; Başeğmez-du Pree, S.; Battiston, R.; et~al.
\newblock {Properties of Neon, Magnesium, and Silicon Primary Cosmic Rays
  Results from the Alpha Magnetic Spectrometer}.
\newblock {\em Phys. Rev. Lett.} {\bf 2020}, {\em 124},~211102.
\newblock
  doi:{\changeurlcolor{black}\href{https://doi.org/10.1103/PhysRevLett.124.211102}{\detokenize{10.1103/PhysRevLett.124.211102}}}.

\bibitem[{Simpson}(2000)]{2000SSRv...93...11S}
{Simpson}, J.A.
\newblock {The Cosmic Ray Nucleonic Component: The Invention and Scientific
  Uses of the Neutron Monitor - (Keynote Lecture)}.
\newblock {\em \ssr} {\bf 2000}, {\em 93},~11--32.
\newblock
  doi:{\changeurlcolor{black}\href{https://doi.org/10.1023/A:1026567706183}{\detokenize{10.1023/A:1026567706183}}}.

\bibitem[{Ahlers} and {Mertsch}(2017)]{2017PrPNP..94..184A}
{Ahlers}, M.; {Mertsch}, P.
\newblock {Origin of small-scale anisotropies in Galactic cosmic rays}.
\newblock {\em Progress in Particle and Nuclear Physics} {\bf 2017}, {\em
  94},~184--216,  \href{http://xxx.lanl.gov/abs/1612.01873}{{\normalfont
  [arXiv:astro-ph.HE/1612.01873]}}.
\newblock
  doi:{\changeurlcolor{black}\href{https://doi.org/10.1016/j.ppnp.2017.01.004}{\detokenize{10.1016/j.ppnp.2017.01.004}}}.

\bibitem[{IceCube Collaboration}(2013)]{2013Sci...342E...1I}
{IceCube Collaboration}.
\newblock {Evidence for High-Energy Extraterrestrial Neutrinos at the IceCube
  Detector}.
\newblock {\em Science} {\bf 2013}, {\em 342},~1242856,
  \href{http://xxx.lanl.gov/abs/1311.5238}{{\normalfont
  [arXiv:astro-ph.HE/1311.5238]}}.
\newblock
  doi:{\changeurlcolor{black}\href{https://doi.org/10.1126/science.1242856}{\detokenize{10.1126/science.1242856}}}.

\bibitem[{G{\'e}nolini} \em{et~al.}(2018){G{\'e}nolini}, {Maurin},
  {Moskalenko}, and {Unger}]{2018PhRvC..98c4611G}
{G{\'e}nolini}, Y.; {Maurin}, D.; {Moskalenko}, I.V.; {Unger}, M.
\newblock {Current status and desired precision of the isotopic production
  cross sections relevant to astrophysics of cosmic rays: Li, Be, B, C, and N}.
\newblock {\em \prc} {\bf 2018}, {\em 98},~034611,
  \href{http://xxx.lanl.gov/abs/1803.04686}{{\normalfont
  [arXiv:astro-ph.HE/1803.04686]}}.
\newblock
  doi:{\changeurlcolor{black}\href{https://doi.org/10.1103/PhysRevC.98.034611}{\detokenize{10.1103/PhysRevC.98.034611}}}.

\bibitem[{Lodders}(2003)]{2003ApJ...591.1220L}
{Lodders}, K.
\newblock {Solar System Abundances and Condensation Temperatures of the
  Elements}.
\newblock {\em \apj} {\bf 2003}, {\em 591},~1220--1247.
\newblock
  doi:{\changeurlcolor{black}\href{https://doi.org/10.1086/375492}{\detokenize{10.1086/375492}}}.

\bibitem[{Aharonian} \em{et~al.}(2007){Aharonian}, {Akhperjanian},
  {Bazer-Bachi}, {Beilicke}, {Benbow}, {Berge}, {Bernl{\"o}hr}, {Boisson},
  {Bolz}, {Borrel}, and et~al]{2007PhRvD..75d2004A}
{Aharonian}, F.; {Akhperjanian}, A.G.; {Bazer-Bachi}, A.R.; {Beilicke}, M.;
  {Benbow}, W.; {Berge}, D.; {Bernl{\"o}hr}, K.; {Boisson}, C.; {Bolz}, O.;
  {Borrel}, V.; et~al.
\newblock {First ground-based measurement of atmospheric Cherenkov light from
  cosmic rays}.
\newblock {\em \prd} {\bf 2007}, {\em 75},~042004,
  \href{http://xxx.lanl.gov/abs/astro-ph/0701766}{{\normalfont
  [astro-ph/0701766]}}.
\newblock
  doi:{\changeurlcolor{black}\href{https://doi.org/10.1103/PhysRevD.75.042004}{\detokenize{10.1103/PhysRevD.75.042004}}}.

\bibitem[{Golden} \em{et~al.}(1979){Golden}, {Horan}, {Mauger}, {Badhwar},
  {Lacy}, {Stephens}, {Daniel}, and {Zipse}]{1979PhRvL..43.1196G}
{Golden}, R.L.; {Horan}, S.; {Mauger}, B.G.; {Badhwar}, G.D.; {Lacy}, J.L.;
  {Stephens}, S.A.; {Daniel}, R.R.; {Zipse}, J.E.
\newblock {Evidence for the existence of cosmic-ray antiprotons}.
\newblock {\em Physical Review Letters} {\bf 1979}, {\em 43},~1196--1199.
\newblock
  doi:{\changeurlcolor{black}\href{https://doi.org/10.1103/PhysRevLett.43.1196}{\detokenize{10.1103/PhysRevLett.43.1196}}}.

\bibitem[{Golden} \em{et~al.}(1984){Golden}, {Mauger}, {Nunn}, and
  {Horan}]{1984ApL....24...75G}
{Golden}, R.L.; {Mauger}, B.G.; {Nunn}, S.; {Horan}, S.
\newblock {Energy Dependence of the P/P Ratio in Cosmic Rays}.
\newblock {\em \\aplett} {\bf 1984}, {\em 24},~75.

\bibitem[{Bogomolov} \em{et~al.}(1987){Bogomolov}, {Krut\'kov}, {Lubyanaya},
  {Romanov}, {Shulakova}, {Stepanov}, and {Vasilyev}]{1987ICRC....2...72B}
{Bogomolov}, E.A.; {Krut\'kov}, S.Y.; {Lubyanaya}, N.D.; {Romanov}, V.A.;
  {Shulakova}, M.S.; {Stepanov}, S.V.; {Vasilyev}, G.I.
\newblock {Galactic Antiproton Spectrum in the 0.2-5 GEV Range}.
\newblock  International Cosmic Ray Conference,  1987, Vol.~2, {\em
  International Cosmic Ray Conference}, p.~72.

\bibitem[{Moiseev} \em{et~al.}(1997){Moiseev}, {Yoshimura}, {Ueda}, {Anraku},
  {Golden}, {Imori}, {Inaba}, {Kimball}, {Kimura}, {Makida}, and
  et~al]{1997ApJ...474..479M}
{Moiseev}, A.; {Yoshimura}, K.; {Ueda}, I.; {Anraku}, K.; {Golden}, R.;
  {Imori}, M.; {Inaba}, S.; {Kimball}, B.; {Kimura}, N.; {Makida}, Y.; et~al.
\newblock {Cosmic-Ray Antiproton Flux in the Energy Range from 200 to 600 MeV}.
\newblock {\em \apj} {\bf 1997}, {\em 474},~479.
\newblock
  doi:{\changeurlcolor{black}\href{https://doi.org/10.1086/303463}{\detokenize{10.1086/303463}}}.

\bibitem[{Matsunaga} \em{et~al.}(1998){Matsunaga}, {Orito}, {Matsumoto},
  {Yoshimura}, {Moiseev}, {Anraku}, {Golden}, {Imori}, {Makida}, {Mitchell}, ,
  and et~al]{1998PhRvL..81.4052M}
{Matsunaga}, H.; {Orito}, S.; {Matsumoto}, H.; {Yoshimura}, K.; {Moiseev}, A.;
  {Anraku}, K.; {Golden}, R.; {Imori}, M.; {Makida}, Y.; {Mitchell}, J.; .;
  et~al.
\newblock {Measurement of Low-Energy Cosmic-Ray Antiprotons at Solar Minimum}.
\newblock {\em Physical Review Letters} {\bf 1998}, {\em 81},~4052--4055,
  \href{http://xxx.lanl.gov/abs/arXiv:astro-ph/9809326}{{\normalfont
  [arXiv:astro-ph/9809326]}}.
\newblock
  doi:{\changeurlcolor{black}\href{https://doi.org/10.1103/PhysRevLett.81.4052}{\detokenize{10.1103/PhysRevLett.81.4052}}}.

\bibitem[{Orito} \em{et~al.}(2000){Orito}, {Maeno}, {Matsunaga}, {Abe},
  {Anraku}, {Asaoka}, {Fujikawa}, {Imori}, {Ishino}, {Makida}, and
  et~al]{2000PhRvL..84.1078O}
{Orito}, S.; {Maeno}, T.; {Matsunaga}, H.; {Abe}, K.; {Anraku}, K.; {Asaoka},
  Y.; {Fujikawa}, M.; {Imori}, M.; {Ishino}, M.; {Makida}, Y.; et~al.
\newblock {Precision Measurement of Cosmic-Ray Antiproton Spectrum}.
\newblock {\em Physical Review Letters} {\bf 2000}, {\em 84},~1078--1081,
  \href{http://xxx.lanl.gov/abs/arXiv:astro-ph/9906426}{{\normalfont
  [arXiv:astro-ph/9906426]}}.
\newblock
  doi:{\changeurlcolor{black}\href{https://doi.org/10.1103/PhysRevLett.84.1078}{\detokenize{10.1103/PhysRevLett.84.1078}}}.

\bibitem[{Maeno} \em{et~al.}(2001){Maeno}, {Orito}, {Matsunaga}, {Abe},
  {Anraku}, {Asaoka}, {Fujikawa}, {Imori}, {Makida}, {Matsui}, and
  et~al]{2001APh....16..121M}
{Maeno}, T.; {Orito}, S.; {Matsunaga}, H.; {Abe}, K.; {Anraku}, K.; {Asaoka},
  Y.; {Fujikawa}, M.; {Imori}, M.; {Makida}, Y.; {Matsui}, N.; et~al.
\newblock {Successive measurements of cosmic-ray antiproton spectrum in a
  positive phase of the solar cycle}.
\newblock {\em Astroparticle Physics} {\bf 2001}, {\em 16},~121--128,
  \href{http://xxx.lanl.gov/abs/arXiv:astro-ph/0010381}{{\normalfont
  [arXiv:astro-ph/0010381]}}.
\newblock
  doi:{\changeurlcolor{black}\href{https://doi.org/10.1016/S0927-6505(01)00107-4}{\detokenize{10.1016/S0927-6505(01)00107-4}}}.

\bibitem[{Asaoka} \em{et~al.}(2002){Asaoka}, {Shikaze}, {Abe}, {Anraku},
  {Fujikawa}, {Fuke}, {Haino}, {Imori}, {Izumi}, {Maeno}, and
  et~al]{2002PhRvL..88e1101A}
{Asaoka}, Y.; {Shikaze}, Y.; {Abe}, K.; {Anraku}, K.; {Fujikawa}, M.; {Fuke},
  H.; {Haino}, S.; {Imori}, M.; {Izumi}, K.; {Maeno}, T.; et~al.
\newblock {Measurements of Cosmic-Ray Low-Energy Antiproton and Proton Spectra
  in a Transient Period of Solar Field Reversal}.
\newblock {\em Physical Review Letters} {\bf 2002}, {\em 88},~051101,
  \href{http://xxx.lanl.gov/abs/arXiv:astro-ph/0109007}{{\normalfont
  [arXiv:astro-ph/0109007]}}.
\newblock
  doi:{\changeurlcolor{black}\href{https://doi.org/10.1103/PhysRevLett.88.051101}{\detokenize{10.1103/PhysRevLett.88.051101}}}.

\bibitem[{Haino} \em{et~al.}(2005){Haino}, {Abe}, {Fuke}, {Maeno}, {Makida},
  {Matsumoto}, {Mitchell}, {Moiseev}, {Nishimura}, {Nozaki}, and
  et~al]{2005ICRC....3...13H}
{Haino}, S.; {Abe}, K.; {Fuke}, H.; {Maeno}, T.; {Makida}, Y.; {Matsumoto}, H.;
  {Mitchell}, J.W.; {Moiseev}, A.A.; {Nishimura}, J.; {Nozaki}, M.; et~al.
\newblock {Measurement of cosmic-ray antiproton spectrum with BESS-2002}.
\newblock  International Cosmic Ray Conference,  2005, Vol.~3, {\em
  International Cosmic Ray Conference}, p.~13.

\bibitem[{Abe} \em{et~al.}(2008){Abe}, {Fuke}, {Haino}, {Hams}, {Itazaki},
  {Kim}, {Kumazawa}, {Lee}, {Makida}, {Matsuda}, and
  et~al]{2008PhLB..670..103B}
{Abe}, K.; {Fuke}, H.; {Haino}, S.; {Hams}, T.; {Itazaki}, A.; {Kim}, K.C.;
  {Kumazawa}, T.; {Lee}, M.H.; {Makida}, Y.; {Matsuda}, S.; et~al.
\newblock {Measurement of the cosmic-ray low-energy antiproton spectrum with
  the first BESS-Polar Antarctic flight}.
\newblock {\em Physics Letters B} {\bf 2008}, {\em 670},~103--108,
  \href{http://xxx.lanl.gov/abs/0805.1754}{{\normalfont [0805.1754]}}.
\newblock
  doi:{\changeurlcolor{black}\href{https://doi.org/10.1016/j.physletb.2008.10.053}{\detokenize{10.1016/j.physletb.2008.10.053}}}.

\bibitem[{Abe} \em{et~al.}(2012){Abe}, {Fuke}, {Haino}, {Hams}, {Hasegawa},
  {Horikoshi}, {Kim}, {Kusumoto}, {Lee}, {Makida}, and
  et~al]{2012PhRvL.108e1102A}
{Abe}, K.; {Fuke}, H.; {Haino}, S.; {Hams}, T.; {Hasegawa}, M.; {Horikoshi},
  A.; {Kim}, K.C.; {Kusumoto}, A.; {Lee}, M.H.; {Makida}, Y.; et~al.
\newblock {Measurement of the Cosmic-Ray Antiproton Spectrum at Solar Minimum
  with a Long-Duration Balloon Flight over Antarctica}.
\newblock {\em Physical Review Letters} {\bf 2012}, {\em 108},~051102,
  \href{http://xxx.lanl.gov/abs/1107.6000}{{\normalfont
  [arXiv:astro-ph.HE/1107.6000]}}.
\newblock
  doi:{\changeurlcolor{black}\href{https://doi.org/10.1103/PhysRevLett.108.051102}{\detokenize{10.1103/PhysRevLett.108.051102}}}.

\bibitem[{Boezio} \em{et~al.}(1997){Boezio}, {Carlson}, {Francke}, {Weber},
  {Suffert}, {Hof}, {Menn}, {Simon}, {Stephens}, {Bellotti}, and
  et~al]{1997ApJ...487..415B}
{Boezio}, M.; {Carlson}, P.; {Francke}, T.; {Weber}, N.; {Suffert}, M.; {Hof},
  M.; {Menn}, W.; {Simon}, M.; {Stephens}, S.A.; {Bellotti}, R.; et~al.
\newblock {The Cosmic-Ray Antiproton Flux between 0.62 and 3.19 G eV Measured
  Near Solar Minimum Activity}.
\newblock {\em \apj} {\bf 1997}, {\em 487},~415.
\newblock
  doi:{\changeurlcolor{black}\href{https://doi.org/10.1086/304593}{\detokenize{10.1086/304593}}}.

\bibitem[{Boezio} \em{et~al.}(2001){Boezio}, {Bonvicini}, {Schiavon}, {Vacchi},
  {Zampa}, {Bergstr{\"o}m}, {Carlson}, {Francke}, {Grinstein}, {Suffert}, and
  et~al]{2001ApJ...561..787B}
{Boezio}, M.; {Bonvicini}, V.; {Schiavon}, P.; {Vacchi}, A.; {Zampa}, N.;
  {Bergstr{\"o}m}, D.; {Carlson}, P.; {Francke}, T.; {Grinstein}, S.;
  {Suffert}, M.; et~al.
\newblock {The Cosmic-Ray Antiproton Flux between 3 and 49 GeV}.
\newblock {\em \apj} {\bf 2001}, {\em 561},~787--799,
  \href{http://xxx.lanl.gov/abs/arXiv:astro-ph/0103513}{{\normalfont
  [arXiv:astro-ph/0103513]}}.
\newblock
  doi:{\changeurlcolor{black}\href{https://doi.org/10.1086/323366}{\detokenize{10.1086/323366}}}.

\bibitem[{Beach} \em{et~al.}(2001){Beach}, {Beatty}, {Bhattacharyya}, {Bower},
  {Coutu}, {Duvernois}, {Labrador}, {McKee}, {Minnick}, {M{\"u}ller}, and
  et~al]{2001PhRvL..87A1101B}
{Beach}, A.S.; {Beatty}, J.J.; {Bhattacharyya}, A.; {Bower}, C.; {Coutu}, S.;
  {Duvernois}, M.A.; {Labrador}, A.W.; {McKee}, S.; {Minnick}, S.A.;
  {M{\"u}ller}, D.; et~al.
\newblock {Measurement of the Cosmic-Ray Antiproton-to-Proton Abundance Ratio
  between 4 and 50 GeV}.
\newblock {\em Physical Review Letters} {\bf 2001}, {\em 87},~A261101,
  \href{http://xxx.lanl.gov/abs/arXiv:astro-ph/0111094}{{\normalfont
  [arXiv:astro-ph/0111094]}}.
\newblock
  doi:{\changeurlcolor{black}\href{https://doi.org/10.1103/PhysRevLett.87.271101}{\detokenize{10.1103/PhysRevLett.87.271101}}}.

\bibitem[{Mitchell} \em{et~al.}(1996){Mitchell}, {Barbier}, {Christian},
  {Krizmanic}, {Krombel}, {Ormes}, {Streitmatter}, {Labrador}, {Davis},
  {Mewaldt}, and et~al]{1996PhRvL..76.3057M}
{Mitchell}, J.W.; {Barbier}, L.M.; {Christian}, E.R.; {Krizmanic}, J.F.;
  {Krombel}, K.; {Ormes}, J.F.; {Streitmatter}, R.E.; {Labrador}, A.W.;
  {Davis}, A.J.; {Mewaldt}, R.A.; et~al.
\newblock {Measurement of 0.25-3.2 GeV Antiprotons in the Cosmic Radiation}.
\newblock {\em Physical Review Letters} {\bf 1996}, {\em 76},~3057--3060.
\newblock
  doi:{\changeurlcolor{black}\href{https://doi.org/10.1103/PhysRevLett.76.3057}{\detokenize{10.1103/PhysRevLett.76.3057}}}.

\bibitem[{Basini}(1999)]{1999ICRC....3...77B}
{Basini}, G.
\newblock {The Flux of Cosmic Ray Antiprotons from 3.7 to 24 GeV}.
\newblock  International Cosmic Ray Conference,  1999, Vol.~3, {\em
  International Cosmic Ray Conference}, p.~77.

\bibitem[{Hof} \em{et~al.}(1996){Hof}, {Menn}, {Pfeifer}, {Simon}, {Golden},
  {Stochaj}, {Stephens}, {Basini}, {Ricci}, {Brancaccio}, and
  et~al]{1996ApJ...467L..33H}
{Hof}, M.; {Menn}, W.; {Pfeifer}, C.; {Simon}, M.; {Golden}, R.L.; {Stochaj},
  S.J.; {Stephens}, S.A.; {Basini}, G.; {Ricci}, M.; {Brancaccio}, F.M.; et~al.
\newblock {Measurement of Cosmic-Ray Antiprotons from 3.7 to 19 GeV}.
\newblock {\em \apjl} {\bf 1996}, {\em 467},~L33.
\newblock
  doi:{\changeurlcolor{black}\href{https://doi.org/10.1086/310185}{\detokenize{10.1086/310185}}}.

\bibitem[{Aguilar} \em{et~al.}(2002){Aguilar}, {Alcaraz}, {Allaby}, {Alpat},
  {Ambrosi}, {Anderhub}, {Ao}, {Arefiev}, {Azzarello}, and
  et~al]{2002PhR...366..331A}
{Aguilar}, M.; {Alcaraz}, J.; {Allaby}, J.; {Alpat}, B.; {Ambrosi}, G.;
  {Anderhub}, H.; {Ao}, L.; {Arefiev}, A.; {Azzarello}, P.; et~al.
\newblock {The Alpha Magnetic Spectrometer (AMS) on the International Space
  Station: Part I - results from the test flight on the space shuttle}.
\newblock {\em \physrep} {\bf 2002}, {\em 366},~331--405.
\newblock
  doi:{\changeurlcolor{black}\href{https://doi.org/10.1016/S0370-1573(02)00013-3}{\detokenize{10.1016/S0370-1573(02)00013-3}}}.

\bibitem[{Adriani} \em{et~al.}(2009){Adriani}, {Barbarino}, {Bazilevskaya},
  {Bellotti}, {Boezio}, {Bogomolov}, {Bonechi}, {Bongi}, {Bonvicini}, {Bottai},
  and et~al]{2009PhRvL.102e1101A}
{Adriani}, O.; {Barbarino}, G.C.; {Bazilevskaya}, G.A.; {Bellotti}, R.;
  {Boezio}, M.; {Bogomolov}, E.A.; {Bonechi}, L.; {Bongi}, M.; {Bonvicini}, V.;
  {Bottai}, S.; et~al.
\newblock {New Measurement of the Antiproton-to-Proton Flux Ratio up to 100 GeV
  in the Cosmic Radiation}.
\newblock {\em Physical Review Letters} {\bf 2009}, {\em 102},~051101,
  \href{http://xxx.lanl.gov/abs/0810.4994}{{\normalfont [0810.4994]}}.
\newblock
  doi:{\changeurlcolor{black}\href{https://doi.org/10.1103/PhysRevLett.102.051101}{\detokenize{10.1103/PhysRevLett.102.051101}}}.

\bibitem[{Adriani} \em{et~al.}(2010){Adriani}, {Barbarino}, {Bazilevskaya},
  {Bellotti}, {Boezio}, {Bogomolov}, {Bonechi}, {Bongi}, {Bonvicini},
  {Borisov}, and et~al]{2010PhRvL.105l1101A}
{Adriani}, O.; {Barbarino}, G.C.; {Bazilevskaya}, G.A.; {Bellotti}, R.;
  {Boezio}, M.; {Bogomolov}, E.A.; {Bonechi}, L.; {Bongi}, M.; {Bonvicini}, V.;
  {Borisov}, S.; et~al.
\newblock {PAMELA Results on the Cosmic-Ray Antiproton Flux from 60 MeV to 180
  GeV in Kinetic Energy}.
\newblock {\em Physical Review Letters} {\bf 2010}, {\em 105},~121101,
  \href{http://xxx.lanl.gov/abs/1007.0821}{{\normalfont
  [arXiv:astro-ph.HE/1007.0821]}}.
\newblock
  doi:{\changeurlcolor{black}\href{https://doi.org/10.1103/PhysRevLett.105.121101}{\detokenize{10.1103/PhysRevLett.105.121101}}}.

\bibitem[{Adriani} \em{et~al.}(2013){Adriani}, {Bazilevskaya}, {Barbarino},
  {Bellotti}, {Boezio}, {Bogomolov}, {Bonvicini}, {Bongi}, {Bonechi},
  {Borisov}, and et~al]{2013JETPL..96..621A}
{Adriani}, O.; {Bazilevskaya}, G.A.; {Barbarino}, G.C.; {Bellotti}, R.;
  {Boezio}, M.; {Bogomolov}, E.A.; {Bonvicini}, V.; {Bongi}, M.; {Bonechi}, L.;
  {Borisov}, S.V.; et~al.
\newblock {Measurement of the flux of primary cosmic ray antiprotons with
  energies of 60 MeV to 350 GeV in the PAMELA experiment}.
\newblock {\em Soviet Journal of Experimental and Theoretical Physics Letters}
  {\bf 2013}, {\em 96},~621--627.
\newblock
  doi:{\changeurlcolor{black}\href{https://doi.org/10.1134/S002136401222002X}{\detokenize{10.1134/S002136401222002X}}}.

\bibitem[{Daniel} and {Stephens}(1965)]{1965PhRvL..15..769D}
{Daniel}, R.R.; {Stephens}, S.A.
\newblock {Electron Component of the Primary Cosmic Radiation at Energies >=15
  GeV}.
\newblock {\em Physical Review Letters} {\bf 1965}, {\em 15},~769--772.
\newblock
  doi:{\changeurlcolor{black}\href{https://doi.org/10.1103/PhysRevLett.15.769}{\detokenize{10.1103/PhysRevLett.15.769}}}.

\bibitem[{Freir} and {Waddington}(1965)]{1965JGR....70.5753F}
{Freir}, P.S.; {Waddington}, C.J.
\newblock {Electron, Hydrogen Nuclei, and Helium Nuclei Observed in Primary
  Cosmic Radiation during 1963}.
\newblock {\em \jgr} {\bf 1965}, {\em 70},~5753--5768.
\newblock
  doi:{\changeurlcolor{black}\href{https://doi.org/10.1029/JZ070i023p05753}{\detokenize{10.1029/JZ070i023p05753}}}.

\bibitem[{L'Heureux}(1967)]{1967ApJ...148..399L}
{L'Heureux}, J.
\newblock {The Primary Cosmic-Ray Electron Spectrum Near Solar Minimum}.
\newblock {\em \apj} {\bf 1967}, {\em 148},~399.
\newblock
  doi:{\changeurlcolor{black}\href{https://doi.org/10.1086/149162}{\detokenize{10.1086/149162}}}.

\bibitem[{Webber} \em{et~al.}(1973){Webber}, {Kish}, and
  {Rockstroh}]{1973ICRC....2..760W}
{Webber}, W.R.; {Kish}, J.; {Rockstroh}, J.M.
\newblock {Measurements of the Primary Cosmic Ray Electron Spectrum from 1965
  to 1972}.
\newblock  International Cosmic Ray Conference,  1973, Vol.~2, {\em
  International Cosmic Ray Conference}, p. 760.

\bibitem[{Bleeker} \em{et~al.}(1965){Bleeker}, {Burger}, {Scheepmaker},
  {Swanenburg}, and {Tanaka}]{1965ICRC....1..327B}
{Bleeker}, J.A.M.; {Burger}, J.J.; {Scheepmaker}, A.; {Swanenburg}, B.N.;
  {Tanaka}, Y.
\newblock {A balloon observation of high energy electrons.}
\newblock  International Cosmic Ray Conference,  1965, Vol.~1, {\em
  International Cosmic Ray Conference}, p. 327.

\bibitem[{Fanselow}(1968)]{1968ApJ...152..783F}
{Fanselow}, J.L.
\newblock {The Primary Cosmic-Ray Electron Spectrum Between 0.09 and 8.4 BeV in
  1965}.
\newblock {\em \apj} {\bf 1968}, {\em 152},~783.
\newblock
  doi:{\changeurlcolor{black}\href{https://doi.org/10.1086/149595}{\detokenize{10.1086/149595}}}.

\bibitem[{Fanselow} \em{et~al.}(1969){Fanselow}, {Hartman}, {Hildebrad}, and
  {Meyer}]{1969ApJ...158..771F}
{Fanselow}, J.L.; {Hartman}, R.C.; {Hildebrad}, R.H.; {Meyer}, P.
\newblock {Charge Composition and Energy Spectrum of Primary Cosmic-Ray
  Electrons}.
\newblock {\em \apj} {\bf 1969}, {\em 158},~771.
\newblock
  doi:{\changeurlcolor{black}\href{https://doi.org/10.1086/150236}{\detokenize{10.1086/150236}}}.

\bibitem[{Beedle} and {Webber}(1968)]{1968CaJPS..46.1014B}
{Beedle}, R.E.; {Webber}, W.R.
\newblock {Measurements of cosmic-ray electrons in the energy range 4 MeV to 6
  BeV at 2 g/cm$^{2}$ atmospheric depth at Ft. Churchill}.
\newblock {\em Canadian Journal of Physics Supplement} {\bf 1968}, {\em
  46},~1014.

\bibitem[{Anand} \em{et~al.}(1968){Anand}, {Daniel}, and
  {Stephens}]{1968PhRvL..20..764A}
{Anand}, K.C.; {Daniel}, R.R.; {Stephens}, S.A.
\newblock {Cosmic-Ray Electron Spectrum above 50 BeV and Its Implications for
  Cosmic-Ray Confinement}.
\newblock {\em Physical Review Letters} {\bf 1968}, {\em 20},~764--768.
\newblock
  doi:{\changeurlcolor{black}\href{https://doi.org/10.1103/PhysRevLett.20.764}{\detokenize{10.1103/PhysRevLett.20.764}}}.

\bibitem[{L'Heureux} and {Meyer}(1968)]{1968CaJPS..46..892L}
{L'Heureux}, J.; {Meyer}, P.
\newblock {The primary cosmic-ray electron spectrum in the energy range from
  300 MeV to 4 BeV from 1964 to 1966}.
\newblock {\em Canadian Journal of Physics Supplement} {\bf 1968}, {\em
  46},~892.

\bibitem[{Earl} \em{et~al.}(1972){Earl}, {Neely}, and
  {Rygg}]{1972JGR....77.1087E}
{Earl}, J.A.; {Neely}, D.E.; {Rygg}, T.A.
\newblock {Balloon measurements of the energy spectrum of cosmic electrons
  between 1 and 25 Gev}.
\newblock {\em \jgr} {\bf 1972}, {\em 77},~1087.
\newblock
  doi:{\changeurlcolor{black}\href{https://doi.org/10.1029/JA077i007p01087}{\detokenize{10.1029/JA077i007p01087}}}.

\bibitem[{Bleeker} \em{et~al.}(1970){Bleeker}, {Burger}, {Deerenberg}, {van de
  Hulst}, {Scheepmaker}, {Swanenburg}, and {Tanaka}]{1970ICRC....1..209B}
{Bleeker}, J.A.M.; {Burger}, J.J.; {Deerenberg}, A.J.M.; {van de Hulst}, H.C.;
  {Scheepmaker}, A.; {Swanenburg}, B.N.; {Tanaka}, Y.
\newblock {Time variations in the cosmic ray electron spectrum above 500 MeV}.
\newblock  International Cosmic Ray Conference,  1970, Vol.~1, {\em
  International Cosmic Ray Conference}, p. 209.

\bibitem[{Danjo} \em{et~al.}(1968){Danjo}, {Hayakawa}, {Makino}, and
  {Tanaka}]{1968CaJPh..46..530D}
{Danjo}, A.; {Hayakawa}, S.; {Makino}, F.; {Tanaka}, Y.
\newblock {Electron component in primary cosmic rays. I. Experiment.}
\newblock {\em Canadian Journal of Physics} {\bf 1968}, {\em 46},~530.

\bibitem[{Israel} and {Vogt}(1968)]{1968PhRvL..20.1053I}
{Israel}, M.H.; {Vogt}, R.E.
\newblock {Flux of Cosmic-Ray Electrons Between 17 and 63 MeV}.
\newblock {\em Physical Review Letters} {\bf 1968}, {\em 20},~1053--1056.
\newblock
  doi:{\changeurlcolor{black}\href{https://doi.org/10.1103/PhysRevLett.20.1053}{\detokenize{10.1103/PhysRevLett.20.1053}}}.

\bibitem[{Agrinier} \em{et~al.}(1969){Agrinier}, {Koechlin}, {Parlier}, {Paul},
  {Vasseur}, {Boella}, {Dilworth}, {Scarsi}, {Sironi}, and
  {Russo}]{1969NCimL...1...53A}
{Agrinier}, B.; {Koechlin}, Y.; {Parlier}, B.; {Paul}, J.; {Vasseur}, J.;
  {Boella}, G.; {Dilworth}, C.; {Scarsi}, L.; {Sironi}, G.; {Russo}, A.
\newblock {East-West asymmetry and charge sign ratio of primary cosmic-ray
  electrons at 8.3 GV rigidity cut-off.}
\newblock {\em Nuovo Cimento Lettere} {\bf 1969}, {\em 1},~53--56.

\bibitem[{Fanselow} \em{et~al.}(1971){Fanselow}, {Hartman}, {Meyer}, and
  {Schmidt}]{1971Ap&SS..14..301F}
{Fanselow}, J.L.; {Hartman}, R.C.; {Meyer}, P.; {Schmidt}, P.J.
\newblock {The Energy Spectrum of Primary Cosmic Ray Electrons from 2 GeV to
  200 GeV}.
\newblock {\em \apss} {\bf 1971}, {\em 14},~301--313.
\newblock
  doi:{\changeurlcolor{black}\href{https://doi.org/10.1007/BF00653319}{\detokenize{10.1007/BF00653319}}}.

\bibitem[{Kobayashi} \em{et~al.}(2012){Kobayashi}, {Komori}, {Yoshida},
  {Yanagisawa}, {Nishimura}, {Yamagami}, {Saito}, {Tateyama}, {Yuda}, and
  {Wilkes}]{2012ApJ...760..146K}
{Kobayashi}, T.; {Komori}, Y.; {Yoshida}, K.; {Yanagisawa}, K.; {Nishimura},
  J.; {Yamagami}, T.; {Saito}, Y.; {Tateyama}, N.; {Yuda}, T.; {Wilkes}, R.J.
\newblock {Observations of High-energy Cosmic-Ray Electrons from 30 GeV to 3
  TeV with Emulsion Chambers}.
\newblock {\em \apj} {\bf 2012}, {\em 760},~146,
  \href{http://xxx.lanl.gov/abs/1210.2813}{{\normalfont
  [arXiv:astro-ph.HE/1210.2813]}}.
\newblock
  doi:{\changeurlcolor{black}\href{https://doi.org/10.1088/0004-637X/760/2/146}{\detokenize{10.1088/0004-637X/760/2/146}}}.

\bibitem[{Anand} \em{et~al.}(1973){Anand}, {Daniel}, and
  {Stephens}]{1973ICRC....1..355A}
{Anand}, K.C.; {Daniel}, R.R.; {Stephens}, S.A.
\newblock {Final Results on the TIFR Cosmic Ray Electron Spectrum in the Region
  10 to 800 GeV}.
\newblock  International Cosmic Ray Conference,  1973, Vol.~1, {\em
  International Cosmic Ray Conference}, p. 355.

\bibitem[{Scheepmaker} and {Tanaka}(1971)]{1971A&A....11...53S}
{Scheepmaker}, A.; {Tanaka}, Y.
\newblock {Primary Cosmic-Ray Electron Spectrum between 5 and \~{}300GeV in
  1968}.
\newblock {\em \aap} {\bf 1971}, {\em 11},~53.

\bibitem[{Fulks}(1975)]{1975JGR....80.1701F}
{Fulks}, G.J.
\newblock {Solar modulation of galactic cosmic ray electrons, protons, and
  alphas}.
\newblock {\em \jgr} {\bf 1975}, {\em 80},~1701--1714.
\newblock
  doi:{\changeurlcolor{black}\href{https://doi.org/10.1029/JA080i013p01701}{\detokenize{10.1029/JA080i013p01701}}}.

\bibitem[{Beuermann} \em{et~al.}(1969){Beuermann}, {Rice}, {Stone}, and
  {Vogt}]{1969PhRvL..22..412B}
{Beuermann}, K.P.; {Rice}, C.J.; {Stone}, E.C.; {Vogt}, R.E.
\newblock {Cosmic-Ray Negatron and Positron Spectra Between 12 and 220 MeV}.
\newblock {\em Physical Review Letters} {\bf 1969}, {\em 22},~412--415.
\newblock
  doi:{\changeurlcolor{black}\href{https://doi.org/10.1103/PhysRevLett.22.412}{\detokenize{10.1103/PhysRevLett.22.412}}}.

\bibitem[{Silverberg} \em{et~al.}(1973){Silverberg}, {Ormes}, and
  {Balasubrahmanyan}]{1973JGR....78.7165S}
{Silverberg}, R.F.; {Ormes}, J.F.; {Balasubrahmanyan}, V.K.
\newblock {Primary cosmic ray electrons above 10 Gev: Measurements using a
  thick detector}.
\newblock {\em \jgr} {\bf 1973}, {\em 78},~7165.
\newblock
  doi:{\changeurlcolor{black}\href{https://doi.org/10.1029/JA078i031p07165}{\detokenize{10.1029/JA078i031p07165}}}.

\bibitem[{Meegan} and {Earl}(1975)]{1975ApJ...197..219M}
{Meegan}, C.A.; {Earl}, J.A.
\newblock {The spectrum of cosmic electrons with energies between 6 and 100
  GeV}.
\newblock {\em \apj} {\bf 1975}, {\em 197},~219--233.
\newblock
  doi:{\changeurlcolor{black}\href{https://doi.org/10.1086/153505}{\detokenize{10.1086/153505}}}.

\bibitem[{Muller} and {Meyer}(1973)]{1973ApJ...186..841M}
{Muller}, D.; {Meyer}, P.
\newblock {The Spectrum of Galactic Electrons with Energies Between 10 and 900
  GEV}.
\newblock {\em \apj} {\bf 1973}, {\em 186},~841--858.
\newblock
  doi:{\changeurlcolor{black}\href{https://doi.org/10.1086/152551}{\detokenize{10.1086/152551}}}.

\bibitem[{Silverberg}(1976)]{1976JGR....81.3944S}
{Silverberg}, R.F.
\newblock {Measurement of the primary cosmic electron spectrum from 10 to about
  250 GeV}.
\newblock {\em \jgr} {\bf 1976}, {\em 81},~3944--3952.
\newblock
  doi:{\changeurlcolor{black}\href{https://doi.org/10.1029/JA081i022p03944}{\detokenize{10.1029/JA081i022p03944}}}.

\bibitem[{Daugherty} \em{et~al.}(1975){Daugherty}, {Hartman}, and
  {Schmidt}]{1975ApJ...198..493D}
{Daugherty}, J.K.; {Hartman}, R.C.; {Schmidt}, P.J.
\newblock {A measurement of cosmic-ray positron and negatron spectra between 50
  and 800 MV}.
\newblock {\em \apj} {\bf 1975}, {\em 198},~493--505.
\newblock
  doi:{\changeurlcolor{black}\href{https://doi.org/10.1086/153626}{\detokenize{10.1086/153626}}}.

\bibitem[{Freier} \em{et~al.}(1977){Freier}, {Gilman}, and
  {Waddington}]{1977ApJ...213..588F}
{Freier}, P.; {Gilman}, C.; {Waddington}, C.J.
\newblock {Intensity of primary cosmic-ray electrons of energy exceeding 8
  GeV}.
\newblock {\em \apj} {\bf 1977}, {\em 213},~588--598.
\newblock
  doi:{\changeurlcolor{black}\href{https://doi.org/10.1086/155190}{\detokenize{10.1086/155190}}}.

\bibitem[{Buffington} \em{et~al.}(1975){Buffington}, {Orth}, and
  {Smoot}]{1975ApJ...199..669B}
{Buffington}, A.; {Orth}, C.D.; {Smoot}, G.F.
\newblock {Measurement of primary cosmic-ray electrons and positrons from 4 to
  50 GeV}.
\newblock {\em \apj} {\bf 1975}, {\em 199},~669--679.
\newblock
  doi:{\changeurlcolor{black}\href{https://doi.org/10.1086/153736}{\detokenize{10.1086/153736}}}.

\bibitem[{Ishii} \em{et~al.}(1973){Ishii}, {Kobayashi}, {Shigihara}, {Yokoi},
  {Matsuo}, {Nishimura}, {Taira}, and {Niu}]{1973ICRC....5.3073I}
{Ishii}, C.; {Kobayashi}, T.; {Shigihara}, N.; {Yokoi}, K.; {Matsuo}, M.;
  {Nishimura}, J.; {Taira}, T.; {Niu}, K.
\newblock {Observation of High Energy Primary Electrons with Emulsion Chamber}.
\newblock  International Cosmic Ray Conference,  1973, Vol.~5, {\em
  International Cosmic Ray Conference}, p. 3073.

\bibitem[{Caldwell} \em{et~al.}(1975){Caldwell}, {Evenson}, {Jordan}, and
  {Meyer}]{1975ICRC....3.1000C}
{Caldwell}, J.; {Evenson}, P.; {Jordan}, S.; {Meyer}, P.
\newblock {The Cosmic Ray Electron Spectrum in 1973 and 1974}.
\newblock  International Cosmic Ray Conference,  1975, Vol.~3, {\em
  International Cosmic Ray Conference}, p. 1000.

\bibitem[{Caldwell} \em{et~al.}(1977){Caldwell}, {Evenson}, {Jordan}, and
  {Meyer}]{1977ICRC...11..203C}
{Caldwell}, J.H.; {Evenson}, P.; {Jordan}, S.; {Meyer}, P.
\newblock {The Cosmic Ray Electron Spectra in 1974 and 1975 and the
  Implications for Solar Modulation}.
\newblock  International Cosmic Ray Conference,  1977, Vol.~11, {\em
  International Cosmic Ray Conference}, p. 203.

\bibitem[{Hartman} and {Pellerin}(1976)]{1976ApJ...204..927H}
{Hartman}, R.C.; {Pellerin}, C.J.
\newblock {Cosmic-ray positron and negatron spectra between 20 and 800 MeV
  measured in 1974}.
\newblock {\em \apj} {\bf 1976}, {\em 204},~927--933.
\newblock
  doi:{\changeurlcolor{black}\href{https://doi.org/10.1086/154241}{\detokenize{10.1086/154241}}}.

\bibitem[{Prince}(1979)]{1979ApJ...227..676P}
{Prince}, T.A.
\newblock {The energy spectrum of cosmic ray electrons between 9 and 300 GeV}.
\newblock {\em \apj} {\bf 1979}, {\em 227},~676--693.
\newblock
  doi:{\changeurlcolor{black}\href{https://doi.org/10.1086/156778}{\detokenize{10.1086/156778}}}.

\bibitem[{Golden} \em{et~al.}(1984){Golden}, {Mauger}, {Badhwar}, {Daniel},
  {Lacy}, {Stephens}, and {Zipse}]{1984ApJ...287..622G}
{Golden}, R.L.; {Mauger}, B.G.; {Badhwar}, G.D.; {Daniel}, R.R.; {Lacy}, J.L.;
  {Stephens}, S.A.; {Zipse}, J.E.
\newblock {A measurement of the absolute flux of cosmic-ray electrons}.
\newblock {\em \apj} {\bf 1984}, {\em 287},~622--632.
\newblock
  doi:{\changeurlcolor{black}\href{https://doi.org/10.1086/162720}{\detokenize{10.1086/162720}}}.

\bibitem[{Golden} \em{et~al.}(1987){Golden}, {Mauger}, {Horan}, {Stephens},
  {Daniel}, {Badhwar}, {Lacy}, and {Zipse}]{1987A&A...188..145G}
{Golden}, R.L.; {Mauger}, B.G.; {Horan}, S.; {Stephens}, S.A.; {Daniel}, R.R.;
  {Badhwar}, G.D.; {Lacy}, J.L.; {Zipse}, J.E.
\newblock {Observation of cosmic ray positrons in the region from 5 to 50 GeV}.
\newblock {\em \aap} {\bf 1987}, {\em 188},~145--154.

\bibitem[{Evenson} \em{et~al.}(1979){Evenson}, {Meyer}, and
  {Nandkumar}]{1979ICRC....1..462E}
{Evenson}, P.; {Meyer}, P.; {Nandkumar}, R.
\newblock {The Energy Spectrum of Cosmic Ray Electrons, 5-150 Mev in Late 1978
  and Early 1979}.
\newblock  International Cosmic Ray Conference,  1979, Vol.~1, {\em
  International Cosmic Ray Conference}, p. 462.

\bibitem[{Evenson} and {Meyer}(1984)]{1984JGR....89.2647E}
{Evenson}, P.; {Meyer}, P.
\newblock {Solar modulation of cosmic ray electrons 1978-1983}.
\newblock {\em \jgr} {\bf 1984}, {\em 89},~2647--2654.
\newblock
  doi:{\changeurlcolor{black}\href{https://doi.org/10.1029/JA089iA05p02647}{\detokenize{10.1029/JA089iA05p02647}}}.

\bibitem[{Tang}(1984)]{1984ApJ...278..881T}
{Tang}, K.K.
\newblock {The energy spectrum of electrons and cosmic-ray confinement A new
  measurement and its interpretation}.
\newblock {\em \apj} {\bf 1984}, {\em 278},~881--892.
\newblock
  doi:{\changeurlcolor{black}\href{https://doi.org/10.1086/161857}{\detokenize{10.1086/161857}}}.

\bibitem[{Garcia-Munoz} \em{et~al.}(1986){Garcia-Munoz}, {Meyer}, {Pyle},
  {Simpson}, and {Evenson}]{1986JGR....91.2858G}
{Garcia-Munoz}, M.; {Meyer}, P.; {Pyle}, K.R.; {Simpson}, J.A.; {Evenson}, P.
\newblock {The dependence of solar modulation on the sign of the cosmic ray
  particle charge}.
\newblock {\em \jgr} {\bf 1986}, {\em 91},~2858--2866.
\newblock
  doi:{\changeurlcolor{black}\href{https://doi.org/10.1029/JA091iA03p02858}{\detokenize{10.1029/JA091iA03p02858}}}.

\bibitem[{Mueller} and {Tang}(1987)]{1987ApJ...312..183M}
{Mueller}, D.; {Tang}, K.K.
\newblock {Cosmic-ray positrons from 10 to 20 GeV - A balloon-borne measurement
  using the geomagnetic east-west asymmetry}.
\newblock {\em \apj} {\bf 1987}, {\em 312},~183--194.
\newblock
  doi:{\changeurlcolor{black}\href{https://doi.org/10.1086/164859}{\detokenize{10.1086/164859}}}.

\bibitem[{Evenson} \em{et~al.}(1995){Evenson}, {Huber}, {Patterson},
  {Esposito}, {Clements}, and {Clem}]{1995JGR...100.7873E}
{Evenson}, P.; {Huber}, D.; {Patterson}, E.T.; {Esposito}, J.; {Clements}, D.;
  {Clem}, J.
\newblock {Cosmic electron spectra 1987-1994}.
\newblock {\em \jgr} {\bf 1995}, {\em 100},~7873--7875.
\newblock
  doi:{\changeurlcolor{black}\href{https://doi.org/10.1029/95JA00484}{\detokenize{10.1029/95JA00484}}}.

\bibitem[{Clem} \em{et~al.}(1996){Clem}, {Clements}, {Esposito}, {Evenson},
  {Huber}, {L'Heureux}, {Meyer}, and {Constantin}]{1996ApJ...464..507C}
{Clem}, J.M.; {Clements}, D.P.; {Esposito}, J.; {Evenson}, P.; {Huber}, D.;
  {L'Heureux}, J.; {Meyer}, P.; {Constantin}, C.
\newblock {Solar Modulation of Cosmic Electrons}.
\newblock {\em \apj} {\bf 1996}, {\em 464},~507.
\newblock
  doi:{\changeurlcolor{black}\href{https://doi.org/10.1086/177340}{\detokenize{10.1086/177340}}}.

\bibitem[{Clem} \em{et~al.}(2000){Clem}, {Evenson}, {Huber}, {Pyle}, {Lopate},
  and {Simpson}]{2000JGR...10523099C}
{Clem}, J.M.; {Evenson}, P.; {Huber}, D.; {Pyle}, R.; {Lopate}, C.; {Simpson},
  J.A.
\newblock {Charge sign dependence of cosmic ray modulation near a rigidity of 1
  GV}.
\newblock {\em \jgr} {\bf 2000}, {\em 105},~23099--23106.
\newblock
  doi:{\changeurlcolor{black}\href{https://doi.org/10.1029/2000JA000097}{\detokenize{10.1029/2000JA000097}}}.

\bibitem[{Clem} and {Evenson}(2002)]{2002ApJ...568..216C}
{Clem}, J.M.; {Evenson}, P.A.
\newblock {Positron Abundance in Galactic Cosmic Rays}.
\newblock {\em \apj} {\bf 2002}, {\em 568},~216--219.
\newblock
  doi:{\changeurlcolor{black}\href{https://doi.org/10.1086/338841}{\detokenize{10.1086/338841}}}.

\bibitem[{Clem} and {Evenson}(2004)]{2004JGRA..109.7107C}
{Clem}, J.; {Evenson}, P.
\newblock {Observations of cosmic ray electrons and positrons during the early
  stages of the A- magnetic polarity epoch}.
\newblock {\em Journal of Geophysical Research (Space Physics)} {\bf 2004},
  {\em 109},~7107.
\newblock
  doi:{\changeurlcolor{black}\href{https://doi.org/10.1029/2003JA010361}{\detokenize{10.1029/2003JA010361}}}.

\bibitem[{Clem} and {Evenson}(2009)]{2009JGRA..11410108C}
{Clem}, J.; {Evenson}, P.
\newblock {Balloon-borne observations of the galactic positron fraction during
  solar minimum negative polarity}.
\newblock {\em Journal of Geophysical Research (Space Physics)} {\bf 2009},
  {\em 114},~10108.
\newblock
  doi:{\changeurlcolor{black}\href{https://doi.org/10.1029/2009JA014225}{\detokenize{10.1029/2009JA014225}}}.

\bibitem[{Chang} \em{et~al.}(2008){Chang}, {Adams}, {Ahn}, {Bashindzhagyan},
  {Christl}, {Ganel}, {Guzik}, {Isbert}, {Kim}, {Kuznetsov}, and
  et~al]{2008Natur.456..362C}
{Chang}, J.; {Adams}, J.H.; {Ahn}, H.S.; {Bashindzhagyan}, G.L.; {Christl}, M.;
  {Ganel}, O.; {Guzik}, T.G.; {Isbert}, J.; {Kim}, K.C.; {Kuznetsov}, E.N.;
  et~al.
\newblock {An excess of cosmic ray electrons at energies of 300-800GeV}.
\newblock {\em \nat} {\bf 2008}, {\em 456},~362--365.
\newblock
  doi:{\changeurlcolor{black}\href{https://doi.org/10.1038/nature07477}{\detokenize{10.1038/nature07477}}}.

\bibitem[{Yoshida} \em{et~al.}(2008){Yoshida}, {Torii}, {Yamagami}, {Tamura},
  {Kitamura}, {Chang}, {Iijima}, {Kadokura}, {Kasahara}, and
  {Katayose}]{2008AdSpR..42.1670Y}
{Yoshida}, K.; {Torii}, S.; {Yamagami}, T.; {Tamura}, T.; {Kitamura}, H.;
  {Chang}, J.; {Iijima}, I.; {Kadokura}, A.; {Kasahara}, K.; {Katayose},
  Y.a.e.a.
\newblock {Cosmic-ray electron spectrum above 100 GeV from PPB-BETS experiment
  in Antarctica}.
\newblock {\em Advances in Space Research} {\bf 2008}, {\em 42},~1670--1675.
\newblock
  doi:{\changeurlcolor{black}\href{https://doi.org/10.1016/j.asr.2007.04.043}{\detokenize{10.1016/j.asr.2007.04.043}}}.

\bibitem[{Torii} \em{et~al.}(2001){Torii}, {Tamura}, {Tateyama}, {Yoshida},
  {Nishimura}, {Yamagami}, {Murakami}, {Kobayashi}, {Komori}, {Kasahara}, and
  {Yuda}]{2001ApJ...559..973T}
{Torii}, S.; {Tamura}, T.; {Tateyama}, N.; {Yoshida}, K.; {Nishimura}, J.;
  {Yamagami}, T.; {Murakami}, H.; {Kobayashi}, T.; {Komori}, Y.; {Kasahara},
  K.; {Yuda}, T.
\newblock {The Energy Spectrum of Cosmic-Ray Electrons from 10 to 100 GeV
  Observed with a Highly Granulated Imaging Calorimeter}.
\newblock {\em \apj} {\bf 2001}, {\em 559},~973--984.
\newblock
  doi:{\changeurlcolor{black}\href{https://doi.org/10.1086/322274}{\detokenize{10.1086/322274}}}.

\bibitem[{Boezio} \em{et~al.}(2000){Boezio}, {Carlson}, {Francke}, {Weber},
  {Suffert}, {Hof}, {Menn}, {Simon}, {Stephens}, {Bellotti}, and
  et~al]{2000ApJ...532..653B}
{Boezio}, M.; {Carlson}, P.; {Francke}, T.; {Weber}, N.; {Suffert}, M.; {Hof},
  M.; {Menn}, W.; {Simon}, M.; {Stephens}, S.A.; {Bellotti}, R.; et~al.
\newblock {The Cosmic-Ray Electron and Positron Spectra Measured at 1 AU during
  Solar Minimum Activity}.
\newblock {\em \apj} {\bf 2000}, {\em 532},~653--669.
\newblock
  doi:{\changeurlcolor{black}\href{https://doi.org/10.1086/308545}{\detokenize{10.1086/308545}}}.

\bibitem[{Boezio} \em{et~al.}(2001){Boezio}, {Barbiellini}, {Bonvicini},
  {Schiavon}, {Vacchi}, {Zampa}, {Bergstr{\"o}m}, {Carlson}, {Francke},
  {Grinstein}, and et~al]{2001AdSpR..27..669B}
{Boezio}, M.; {Barbiellini}, G.; {Bonvicini}, V.; {Schiavon}, P.; {Vacchi}, A.;
  {Zampa}, N.; {Bergstr{\"o}m}, D.; {Carlson}, P.; {Francke}, T.; {Grinstein},
  S.; et~al.
\newblock {Measurements of cosmic-ray electrons and positrons by the
  Wizard/CAPRICE collaboration}.
\newblock {\em Advances in Space Research} {\bf 2001}, {\em 27},~669--674.
\newblock
  doi:{\changeurlcolor{black}\href{https://doi.org/10.1016/S0273-1177(01)00108-9}{\detokenize{10.1016/S0273-1177(01)00108-9}}}.

\bibitem[{Beatty} \em{et~al.}(2004){Beatty}, {Bhattacharyya}, {Bower}, {Coutu},
  {Duvernois}, {McKee}, {Minnick}, {M{\"u}ller}, {Musser}, {Nutter}, and
  et~al]{2004PhRvL..93x1102B}
{Beatty}, J.J.; {Bhattacharyya}, A.; {Bower}, C.; {Coutu}, S.; {Duvernois},
  M.A.; {McKee}, S.; {Minnick}, S.A.; {M{\"u}ller}, D.; {Musser}, J.; {Nutter},
  S.; et~al.
\newblock {New Measurement of the Cosmic-Ray Positron Fraction from 5 to
  15GeV}.
\newblock {\em Physical Review Letters} {\bf 2004}, {\em 93},~241102,
  \href{http://xxx.lanl.gov/abs/arXiv:astro-ph/0412230}{{\normalfont
  [arXiv:astro-ph/0412230]}}.
\newblock
  doi:{\changeurlcolor{black}\href{https://doi.org/10.1103/PhysRevLett.93.241102}{\detokenize{10.1103/PhysRevLett.93.241102}}}.

\bibitem[{Barwick} \em{et~al.}(1998){Barwick}, {Beatty}, {Bower}, {Chaput},
  {Coutu}, {de Nolfo}, {Duvernois}, {Ellithorpe}, {Ficenec}, {Knapp}, and
  et~al]{1998ApJ...498..779B}
{Barwick}, S.W.; {Beatty}, J.J.; {Bower}, C.R.; {Chaput}, C.J.; {Coutu}, S.;
  {de Nolfo}, G.A.; {Duvernois}, M.A.; {Ellithorpe}, D.; {Ficenec}, D.;
  {Knapp}, J.; et~al.
\newblock {The Energy Spectra and Relative Abundances of Electrons and
  Positrons in the Galactic Cosmic Radiation}.
\newblock {\em \apj} {\bf 1998}, {\em 498},~779,
  \href{http://xxx.lanl.gov/abs/arXiv:astro-ph/9712324}{{\normalfont
  [arXiv:astro-ph/9712324]}}.
\newblock
  doi:{\changeurlcolor{black}\href{https://doi.org/10.1086/305573}{\detokenize{10.1086/305573}}}.

\bibitem[{Barwick} \em{et~al.}(1997){Barwick}, {Beatty}, {Bhattacharyya},
  {Bower}, {Chaput}, {Coutu}, {de Nolfo}, {Knapp}, {Lowder}, {McKee}, and
  et~al]{1997ApJ...482L.191B}
{Barwick}, S.W.; {Beatty}, J.J.; {Bhattacharyya}, A.; {Bower}, C.R.; {Chaput},
  C.J.; {Coutu}, S.; {de Nolfo}, G.A.; {Knapp}, J.; {Lowder}, D.M.; {McKee},
  S.; et~al.
\newblock {Measurements of the Cosmic-Ray Positron Fraction from 1 to 50 GeV}.
\newblock {\em \apjl} {\bf 1997}, {\em 482},~L191,
  \href{http://xxx.lanl.gov/abs/arXiv:astro-ph/9703192}{{\normalfont
  [arXiv:astro-ph/9703192]}}.
\newblock
  doi:{\changeurlcolor{black}\href{https://doi.org/10.1086/310706}{\detokenize{10.1086/310706}}}.

\bibitem[{DuVernois} \em{et~al.}(2001){DuVernois}, {Barwick}, {Beatty},
  {Bhattacharyya}, {Bower}, {Chaput}, {Coutu}, {de Nolfo}, {Lowder}, {McKee},
  and et~al]{2001ApJ...559..296D}
{DuVernois}, M.A.; {Barwick}, S.W.; {Beatty}, J.J.; {Bhattacharyya}, A.;
  {Bower}, C.R.; {Chaput}, C.J.; {Coutu}, S.; {de Nolfo}, G.A.; {Lowder}, D.M.;
  {McKee}, S.; et~al.
\newblock {Cosmic-Ray Electrons and Positrons from 1 to 100 GeV: Measurements
  with HEAT and Their Interpretation}.
\newblock {\em \apj} {\bf 2001}, {\em 559},~296--303.
\newblock
  doi:{\changeurlcolor{black}\href{https://doi.org/10.1086/322324}{\detokenize{10.1086/322324}}}.

\bibitem[{Golden} \em{et~al.}(1994){Golden}, {Grimani}, {Kimbell}, {Stephens},
  {Stochaj}, {Webber}, {Basini}, {Bongiorno}, {Massimo Brancaccio}, {Ricci},
  and et~al]{1994ApJ...436..769G}
{Golden}, R.L.; {Grimani}, C.; {Kimbell}, B.L.; {Stephens}, S.A.; {Stochaj},
  S.J.; {Webber}, W.R.; {Basini}, G.; {Bongiorno}, F.; {Massimo Brancaccio},
  F.; {Ricci}, M.; et~al.
\newblock {Observations of cosmic-ray electrons and positrons using an imaging
  calorimeter}.
\newblock {\em \apj} {\bf 1994}, {\em 436},~769--775.
\newblock
  doi:{\changeurlcolor{black}\href{https://doi.org/10.1086/174951}{\detokenize{10.1086/174951}}}.

\bibitem[{Grimani} \em{et~al.}(2002){Grimani}, {Stephens}, {Cafagna}, {Basini},
  {Bellotti}, {Brunetti}, {Circella}, {Codino}, {De Marzo}, {De Pascale}, and
  et~al]{2002A&A...392..287G}
{Grimani}, C.; {Stephens}, S.A.; {Cafagna}, F.S.; {Basini}, G.; {Bellotti}, R.;
  {Brunetti}, M.T.; {Circella}, M.; {Codino}, A.; {De Marzo}, C.; {De Pascale},
  M.P.; et~al.
\newblock {Measurements of the absolute energy spectra of cosmic-ray positrons
  and electrons above 7 GeV}.
\newblock {\em \aap} {\bf 2002}, {\em 392},~287--294.
\newblock
  doi:{\changeurlcolor{black}\href{https://doi.org/10.1051/0004-6361:20020845}{\detokenize{10.1051/0004-6361:20020845}}}.

\bibitem[{Golden} \em{et~al.}(1996){Golden}, {Stochaj}, {Stephens}, {Aversa},
  {Barbiellini}, {Boezio}, {Bravar}, {Colavita}, {Fratnik}, {Schiavon}, and
  et~al]{1996ApJ...457L.103G}
{Golden}, R.L.; {Stochaj}, S.J.; {Stephens}, S.A.; {Aversa}, F.; {Barbiellini},
  G.; {Boezio}, M.; {Bravar}, U.; {Colavita}, A.; {Fratnik}, F.; {Schiavon},
  P.; et~al.
\newblock {Measurement of the Positron to Electron Ratio in Cosmic Rays above 5
  GeV}.
\newblock {\em \apjl} {\bf 1996}, {\em 457},~L103.
\newblock
  doi:{\changeurlcolor{black}\href{https://doi.org/10.1086/309896}{\detokenize{10.1086/309896}}}.

\bibitem[{Aguilar} \em{et~al.}(2007){Aguilar}, {Alcaraz}, {Allaby}, {Alpat},
  {Ambrosi}, {Anderhub}, {Ao}, {Arefiev}, {Azzarello}, {Baldini}, and
  et~al]{2007PhLB..646..145A}
{Aguilar}, M.; {Alcaraz}, J.; {Allaby}, J.; {Alpat}, B.; {Ambrosi}, G.;
  {Anderhub}, H.; {Ao}, L.; {Arefiev}, A.; {Azzarello}, P.; {Baldini}, L.;
  et~al.
\newblock {Cosmic-ray positron fraction measurement from 1 to 30 GeV with
  AMS-01}.
\newblock {\em Physics Letters B} {\bf 2007}, {\em 646},~145--154,
  \href{http://xxx.lanl.gov/abs/arXiv:astro-ph/0703154}{{\normalfont
  [arXiv:astro-ph/0703154]}}.
\newblock
  doi:{\changeurlcolor{black}\href{https://doi.org/10.1016/j.physletb.2007.01.024}{\detokenize{10.1016/j.physletb.2007.01.024}}}.

\bibitem[{Alcaraz} \em{et~al.}(2000){Alcaraz}, {Alpat}, {Ambrosi}, {Anderhub},
  {Ao}, {Arefiev}, {Azzarello}, {Babucci}, {Baldini}, {Basile}, , and
  et~al]{2000PhLB..484...10A}
{Alcaraz}, J.; {Alpat}, B.; {Ambrosi}, G.; {Anderhub}, H.; {Ao}, L.; {Arefiev},
  A.; {Azzarello}, P.; {Babucci}, E.; {Baldini}, L.; {Basile}, M.; .; et~al.
\newblock {Leptons in near earth orbit}.
\newblock {\em Physics Letters B} {\bf 2000}, {\em 484},~10--22.
\newblock
  doi:{\changeurlcolor{black}\href{https://doi.org/10.1016/S0370-2693(00)00588-8}{\detokenize{10.1016/S0370-2693(00)00588-8}}}.

\bibitem[{Aguilar} \em{et~al.}(2013){Aguilar}, {Alberti}, {Alpat}, {Alvino},
  {Ambrosi}, {Andeen}, {Anderhub}, {Arruda}, {Azzarello}, {Bachlechner}, and
  et~al]{2013PhRvL.110n1102A}
{Aguilar}, M.; {Alberti}, G.; {Alpat}, B.; {Alvino}, A.; {Ambrosi}, G.;
  {Andeen}, K.; {Anderhub}, H.; {Arruda}, L.; {Azzarello}, P.; {Bachlechner},
  A.; et~al.
\newblock {First Result from the Alpha Magnetic Spectrometer on the
  International Space Station: Precision Measurement of the Positron Fraction
  in Primary Cosmic Rays of 0.5-350 GeV}.
\newblock {\em Physical Review Letters} {\bf 2013}, {\em 110},~141102.
\newblock
  doi:{\changeurlcolor{black}\href{https://doi.org/10.1103/PhysRevLett.110.141102}{\detokenize{10.1103/PhysRevLett.110.141102}}}.

\bibitem[{Aguilar} \em{et~al.}(2014){Aguilar}, {Aisa}, {Alpat}, {Alvino},
  {Ambrosi}, {Andeen}, {Arruda}, {Attig}, {Azzarello}, {Bachlechner}, and
  et~al]{2014PhRvL.113v1102A}
{Aguilar}, M.; {Aisa}, D.; {Alpat}, B.; {Alvino}, A.; {Ambrosi}, G.; {Andeen},
  K.; {Arruda}, L.; {Attig}, N.; {Azzarello}, P.; {Bachlechner}, A.; et~al.
\newblock {Precision Measurement of the (e$^{+}$+e$^{-}$) Flux in Primary
  Cosmic Rays from 0.5 GeV to 1 TeV with the Alpha Magnetic Spectrometer on the
  International Space Station}.
\newblock {\em \prl} {\bf 2014}, {\em 113},~221102.
\newblock
  doi:{\changeurlcolor{black}\href{https://doi.org/10.1103/PhysRevLett.113.221102}{\detokenize{10.1103/PhysRevLett.113.221102}}}.

\bibitem[{Aguilar} \em{et~al.}(2019){Aguilar}, {Ali Cavasonza}, {Alpat},
  {Ambrosi}, {Arruda}, {Attig}, {Azzarello}, {Bachlechner}, {Barao}, {Barrau},
  and et~al]{2019PhRvL.122j1101A}
{Aguilar}, M.; {Ali Cavasonza}, L.; {Alpat}, B.; {Ambrosi}, G.; {Arruda}, L.;
  {Attig}, N.; {Azzarello}, P.; {Bachlechner}, A.; {Barao}, F.; {Barrau}, A.;
  et~al.
\newblock {Towards Understanding the Origin of Cosmic-Ray Electrons}.
\newblock {\em \prl} {\bf 2019}, {\em 122},~101101.
\newblock
  doi:{\changeurlcolor{black}\href{https://doi.org/10.1103/PhysRevLett.122.101101}{\detokenize{10.1103/PhysRevLett.122.101101}}}.

\bibitem[{Adriani} \em{et~al.}(2017){Adriani}, {Akaike}, {Asano}, {Asaoka},
  {Bagliesi}, {Bigongiari}, {Binns}, {Bonechi}, {Bongi}, {Brogi}, and
  et~al]{2017PhRvL.119r1101A}
{Adriani}, O.; {Akaike}, Y.; {Asano}, K.; {Asaoka}, Y.; {Bagliesi}, M.G.;
  {Bigongiari}, G.; {Binns}, W.R.; {Bonechi}, S.; {Bongi}, M.; {Brogi}, P.;
  et~al.
\newblock {Energy Spectrum of Cosmic-Ray Electron and Positron from 10 GeV to 3
  TeV Observed with the Calorimetric Electron Telescope on the International
  Space Station}.
\newblock {\em \prl} {\bf 2017}, {\em 119},~181101,
  \href{http://xxx.lanl.gov/abs/1712.01711}{{\normalfont
  [arXiv:astro-ph.HE/1712.01711]}}.
\newblock
  doi:{\changeurlcolor{black}\href{https://doi.org/10.1103/PhysRevLett.119.181101}{\detokenize{10.1103/PhysRevLett.119.181101}}}.

\bibitem[{Adriani} \em{et~al.}(2018){Adriani}, {Akaike}, {Asano}, {Asaoka},
  {Bagliesi}, {Berti}, {Bigongiari}, {Binns}, {Bonechi}, {Bongi}, and
  et~al]{2018PhRvL.120z1102A}
{Adriani}, O.; {Akaike}, Y.; {Asano}, K.; {Asaoka}, Y.; {Bagliesi}, M.G.;
  {Berti}, E.; {Bigongiari}, G.; {Binns}, W.R.; {Bonechi}, S.; {Bongi}, M.;
  et~al.
\newblock {Extended Measurement of the Cosmic-Ray Electron and Positron
  Spectrum from 11 GeV to 4.8 TeV with the Calorimetric Electron Telescope on
  the International Space Station}.
\newblock {\em \prl} {\bf 2018}, {\em 120},~261102,
  \href{http://xxx.lanl.gov/abs/1806.09728}{{\normalfont
  [arXiv:astro-ph.HE/1806.09728]}}.
\newblock
  doi:{\changeurlcolor{black}\href{https://doi.org/10.1103/PhysRevLett.120.261102}{\detokenize{10.1103/PhysRevLett.120.261102}}}.

\bibitem[{Ambrosi} \em{et~al.}(2017){Ambrosi}, {An}, {Asfand iyarov},
  {Azzarello}, {Bernardini}, {Bertucci}, {Cai}, {Chang}, {Chen}, and
  {Chen}]{2017Natur.552...63D}
{Ambrosi}, G.; {An}, Q.; {Asfand iyarov}, R.; {Azzarello}, P.; {Bernardini},
  P.; {Bertucci}, B.; {Cai}, M.S.; {Chang}, J.; {Chen}, D.Y.; {Chen}, H.F.a.a.
\newblock {Direct detection of a break in the teraelectronvolt cosmic-ray
  spectrum of electrons and positrons}.
\newblock {\em \nat} {\bf 2017}, {\em 552},~63--66,
  \href{http://xxx.lanl.gov/abs/1711.10981}{{\normalfont
  [arXiv:astro-ph.HE/1711.10981]}}.
\newblock
  doi:{\changeurlcolor{black}\href{https://doi.org/10.1038/nature24475}{\detokenize{10.1038/nature24475}}}.

\bibitem[{Ackermann} \em{et~al.}(2010){Ackermann}, {Ajello}, {Atwood},
  {Baldini}, {Ballet}, {Barbiellini}, {Bastieri}, {Baughman}, {Bechtol},
  {Bellardi}, and et~al]{2010PhRvD..82i2004A}
{Ackermann}, M.; {Ajello}, M.; {Atwood}, W.B.; {Baldini}, L.; {Ballet}, J.;
  {Barbiellini}, G.; {Bastieri}, D.; {Baughman}, B.M.; {Bechtol}, K.;
  {Bellardi}, F.; et~al.
\newblock {Fermi LAT observations of cosmic-ray electrons from 7 GeV to 1 TeV}.
\newblock {\em \prd} {\bf 2010}, {\em 82},~092004,
  \href{http://xxx.lanl.gov/abs/1008.3999}{{\normalfont
  [arXiv:astro-ph.HE/1008.3999]}}.
\newblock
  doi:{\changeurlcolor{black}\href{https://doi.org/10.1103/PhysRevD.82.092004}{\detokenize{10.1103/PhysRevD.82.092004}}}.

\bibitem[{Abdollahi} \em{et~al.}(2017){Abdollahi}, {Ackermann}, {Ajello},
  {Atwood}, {Baldini}, {Barbiellini}, {Bastieri}, {Bellazzini}, {Bloom},
  {Bonino}, and et~al]{2017PhRvD..95h2007A}
{Abdollahi}, S.; {Ackermann}, M.; {Ajello}, M.; {Atwood}, W.B.; {Baldini}, L.;
  {Barbiellini}, G.; {Bastieri}, D.; {Bellazzini}, R.; {Bloom}, E.D.; {Bonino},
  R.; et~al.
\newblock {Cosmic-ray electron-positron spectrum from 7 GeV to 2 TeV with the
  Fermi Large Area Telescope}.
\newblock {\em \prd} {\bf 2017}, {\em 95},~082007.
\newblock
  doi:{\changeurlcolor{black}\href{https://doi.org/10.1103/PhysRevD.95.082007}{\detokenize{10.1103/PhysRevD.95.082007}}}.

\bibitem[{Cline} \em{et~al.}(1964){Cline}, {Ludwig}, and
  {McDonald}]{1964PhRvL..13..786C}
{Cline}, T.L.; {Ludwig}, G.H.; {McDonald}, F.B.
\newblock {Detection of Interplanetary 3- to 12-MeV Electrons}.
\newblock {\em Physical Review Letters} {\bf 1964}, {\em 13},~786--789.
\newblock
  doi:{\changeurlcolor{black}\href{https://doi.org/10.1103/PhysRevLett.13.786}{\detokenize{10.1103/PhysRevLett.13.786}}}.

\bibitem[{Fan} \em{et~al.}(1968){Fan}, {Gloeckler}, {Simpson}, and
  {Verma}]{1968ApJ...151..737F}
{Fan}, C.Y.; {Gloeckler}, G.; {Simpson}, J.A.; {Verma}, S.D.
\newblock {The Primary Cosmic-Ray Electron Energy Spectrum from 10 TO 40 Mev}.
\newblock {\em \apj} {\bf 1968}, {\em 151},~737.
\newblock
  doi:{\changeurlcolor{black}\href{https://doi.org/10.1086/149471}{\detokenize{10.1086/149471}}}.

\bibitem[{Evenson} \em{et~al.}(1981){Evenson}, {Krawczyk}, {Moses}, and
  {Meyer}]{1981ICRC...10...77E}
{Evenson}, P.; {Krawczyk}, L.; {Moses}, D.; {Meyer}, P.
\newblock {The primary cosmic ray electron spectrum 1978-1980}.
\newblock  International Cosmic Ray Conference,  1981, Vol.~10, {\em
  International Cosmic Ray Conference}, pp. 77--80.

\bibitem[{Burger} and {Swanenburg}(1974)]{1974JGR....79.1533B}
{Burger}, J.J.; {Swanenburg}, B.N.
\newblock {The 1972 cosmic ray electron spectrum above 0.5 GeV}.
\newblock {\em \jgr} {\bf 1974}, {\em 79},~1533.
\newblock
  doi:{\changeurlcolor{black}\href{https://doi.org/10.1029/JA079i010p01533}{\detokenize{10.1029/JA079i010p01533}}}.

\bibitem[{Burger} and {Swanenburg}(1973)]{1973JGR....78..292B}
{Burger}, J.J.; {Swanenburg}, B.N.
\newblock {Energy dependent time lag in the long-term modulation of cosmic
  rays}.
\newblock {\em \jgr} {\bf 1973}, {\em 78},~292.
\newblock
  doi:{\changeurlcolor{black}\href{https://doi.org/10.1029/JA078i001p00292}{\detokenize{10.1029/JA078i001p00292}}}.

\bibitem[{Adriani} \em{et~al.}(2015){Adriani}, {Barbarino}, {Bazilevskaya},
  {Bellotti}, {Boezio}, {Bogomolov}, {Bongi}, {Bonvicini}, {Bottai}, {Bruno},
  and et~al]{2015ApJ...810..142A}
{Adriani}, O.; {Barbarino}, G.C.; {Bazilevskaya}, G.A.; {Bellotti}, R.;
  {Boezio}, M.; {Bogomolov}, E.A.; {Bongi}, M.; {Bonvicini}, V.; {Bottai}, S.;
  {Bruno}, A.; et~al.
\newblock {Time Dependence of the e$^{-}$ Flux Measured by PAMELA during the
  July 2006-December 2009 Solar Minimum.}
\newblock {\em \apj} {\bf 2015}, {\em 810},~142,
  \href{http://xxx.lanl.gov/abs/1512.01079}{{\normalfont
  [arXiv:astro-ph.SR/1512.01079]}}.
\newblock
  doi:{\changeurlcolor{black}\href{https://doi.org/10.1088/0004-637X/810/2/142}{\detokenize{10.1088/0004-637X/810/2/142}}}.

\bibitem[{Adriani} \em{et~al.}(2009){Adriani}, {Barbarino}, {Bazilevskaya},
  {Bellotti}, {Boezio}, {Bogomolov}, {Bonechi}, {Bongi}, {Bonvicini}, {Bottai},
  and et~al]{2009Natur.458..607A}
{Adriani}, O.; {Barbarino}, G.C.; {Bazilevskaya}, G.A.; {Bellotti}, R.;
  {Boezio}, M.; {Bogomolov}, E.A.; {Bonechi}, L.; {Bongi}, M.; {Bonvicini}, V.;
  {Bottai}, S.; et~al.
\newblock {An anomalous positron abundance in cosmic rays with energies
  1.5-100GeV}.
\newblock {\em \nat} {\bf 2009}, {\em 458},~607--609,
  \href{http://xxx.lanl.gov/abs/0810.4995}{{\normalfont [0810.4995]}}.
\newblock
  doi:{\changeurlcolor{black}\href{https://doi.org/10.1038/nature07942}{\detokenize{10.1038/nature07942}}}.

\bibitem[{Adriani} \em{et~al.}(2010){Adriani}, {Barbarino}, {Bazilevskaya},
  {Bellotti}, {Boezio}, {Bogomolov}, {Bonechi}, {Bongi}, {Bonvicini},
  {Borisov}, and et~al]{2010APh....34....1A}
{Adriani}, O.; {Barbarino}, G.C.; {Bazilevskaya}, G.A.; {Bellotti}, R.;
  {Boezio}, M.; {Bogomolov}, E.A.; {Bonechi}, L.; {Bongi}, M.; {Bonvicini}, V.;
  {Borisov}, S.; et~al.
\newblock {A statistical procedure for the identification of positrons in the
  PAMELA experiment}.
\newblock {\em Astroparticle Physics} {\bf 2010}, {\em 34},~1--11,
  \href{http://xxx.lanl.gov/abs/1001.3522}{{\normalfont
  [arXiv:astro-ph.HE/1001.3522]}}.
\newblock
  doi:{\changeurlcolor{black}\href{https://doi.org/10.1016/j.astropartphys.2010.04.007}{\detokenize{10.1016/j.astropartphys.2010.04.007}}}.

\bibitem[{Adriani} \em{et~al.}(2013){Adriani}, {Barbarino}, {Bazilevskaya},
  {Bellotti}, {Bianco}, {Boezio}, {Bogomolov}, {Bongi}, {Bonvicini}, {Bottai},
  and et~al]{2013PhRvL.111h1102A}
{Adriani}, O.; {Barbarino}, G.C.; {Bazilevskaya}, G.A.; {Bellotti}, R.;
  {Bianco}, A.; {Boezio}, M.; {Bogomolov}, E.A.; {Bongi}, M.; {Bonvicini}, V.;
  {Bottai}, S.; et~al.
\newblock {Cosmic-Ray Positron Energy Spectrum Measured by PAMELA}.
\newblock {\em Physical Review Letters} {\bf 2013}, {\em 111},~081102,
  \href{http://xxx.lanl.gov/abs/1308.0133}{{\normalfont
  [arXiv:astro-ph.HE/1308.0133]}}.
\newblock
  doi:{\changeurlcolor{black}\href{https://doi.org/10.1103/PhysRevLett.111.081102}{\detokenize{10.1103/PhysRevLett.111.081102}}}.

\bibitem[{Adriani} \em{et~al.}(2011){Adriani}, {Barbarino}, {Bazilevskaya},
  {Bellotti}, {Boezio}, {Bogomolov}, {Bongi}, {Bonvicini}, {Borisov}, {Bottai},
  and et~al]{2011PhRvL.106t1101A}
{Adriani}, O.; {Barbarino}, G.C.; {Bazilevskaya}, G.A.; {Bellotti}, R.;
  {Boezio}, M.; {Bogomolov}, E.A.; {Bongi}, M.; {Bonvicini}, V.; {Borisov}, S.;
  {Bottai}, S.; et~al.
\newblock {Cosmic-Ray Electron Flux Measured by the PAMELA Experiment between 1
  and 625 GeV}.
\newblock {\em Physical Review Letters} {\bf 2011}, {\em 106},~201101,
  \href{http://xxx.lanl.gov/abs/1103.2880}{{\normalfont
  [arXiv:astro-ph.HE/1103.2880]}}.
\newblock
  doi:{\changeurlcolor{black}\href{https://doi.org/10.1103/PhysRevLett.106.201101}{\detokenize{10.1103/PhysRevLett.106.201101}}}.

\bibitem[{Webber} \em{et~al.}(1973){Webber}, {Lezniak}, and
  {Damle}]{1973JGR....78.1487W}
{Webber}, W.R.; {Lezniak}, J.A.; {Damle}, S.V.
\newblock {Cosmic ray electrons from 0.2 to 8 Mev: Pioneer 8 and 9 measurements
  of their spectrum, time variations, and interplanetary radial gradient}.
\newblock {\em \jgr} {\bf 1973}, {\em 78},~1487.
\newblock
  doi:{\changeurlcolor{black}\href{https://doi.org/10.1029/JA078i010p01487}{\detokenize{10.1029/JA078i010p01487}}}.

\bibitem[{Rastoin} \em{et~al.}(1996){Rastoin}, {Ferrando}, {Raviart}, {Ducros},
  {Petrucci}, {Paizis}, {Kunow}, {Mueller-Mellin}, {Sierks}, and
  {Wibberenz}]{1996A&A...307..981R}
{Rastoin}, C.; {Ferrando}, P.; {Raviart}, A.; {Ducros}, R.; {Petrucci}, P.O.;
  {Paizis}, C.; {Kunow}, H.; {Mueller-Mellin}, R.; {Sierks}, H.; {Wibberenz},
  G.
\newblock {Time and space variations of the Galactic cosmic ray electron
  spectrum in the 3-D heliosphere explored by Ulysses.}
\newblock {\em \aap} {\bf 1996}, {\em 307},~981--995.

\bibitem[{Caballero-Lopez} \em{et~al.}(2010){Caballero-Lopez}, {Moraal}, and
  {McDonald}]{2010ApJ...725..121C}
{Caballero-Lopez}, R.A.; {Moraal}, H.; {McDonald}, F.B.
\newblock {The Modulation of Galactic Cosmic-ray Electrons in the Heliosheath}.
\newblock {\em \apj} {\bf 2010}, {\em 725},~121--127.
\newblock
  doi:{\changeurlcolor{black}\href{https://doi.org/10.1088/0004-637X/725/1/121}{\detokenize{10.1088/0004-637X/725/1/121}}}.

\bibitem[{Stone} \em{et~al.}(2019){Stone}, {Cummings}, {Heikkila}, and
  {Lal}]{2019NatAs...3.1013S}
{Stone}, E.C.; {Cummings}, A.C.; {Heikkila}, B.C.; {Lal}, N.
\newblock {Cosmic ray measurements from Voyager 2 as it crossed into
  interstellar space}.
\newblock {\em Nature Astronomy} {\bf 2019}, {\em 3},~1013--1018.
\newblock
  doi:{\changeurlcolor{black}\href{https://doi.org/10.1038/s41550-019-0928-3}{\detokenize{10.1038/s41550-019-0928-3}}}.

\bibitem[{Cummings} \em{et~al.}(2016){Cummings}, {Stone}, {Heikkila}, {Lal},
  {Webber}, {J{\'o}hannesson}, {Moskalenko}, {Orlando}, and
  {Porter}]{2016ApJ...831...18C}
{Cummings}, A.C.; {Stone}, E.C.; {Heikkila}, B.C.; {Lal}, N.; {Webber}, W.R.;
  {J{\'o}hannesson}, G.; {Moskalenko}, I.V.; {Orlando}, E.; {Porter}, T.A.
\newblock {Galactic Cosmic Rays in the Local Interstellar Medium: Voyager 1
  Observations and Model Results}.
\newblock {\em \apj} {\bf 2016}, {\em 831},~18.
\newblock
  doi:{\changeurlcolor{black}\href{https://doi.org/10.3847/0004-637X/831/1/18}{\detokenize{10.3847/0004-637X/831/1/18}}}.

\bibitem[{Aharonian} \em{et~al.}(2008){Aharonian}, {Akhperjanian}, {Barres de
  Almeida}, {Bazer-Bachi}, {Becherini}, {Behera}, {Benbow}, {Bernl{\"o}hr},
  {Boisson}, {Bochow}, and et~al]{2008PhRvL.101z1104A}
{Aharonian}, F.; {Akhperjanian}, A.G.; {Barres de Almeida}, U.; {Bazer-Bachi},
  A.R.; {Becherini}, Y.; {Behera}, B.; {Benbow}, W.; {Bernl{\"o}hr}, K.;
  {Boisson}, C.; {Bochow}, A.; et~al.
\newblock {Energy Spectrum of Cosmic-Ray Electrons at TeV Energies}.
\newblock {\em Physical Review Letters} {\bf 2008}, {\em 101},~261104,
  \href{http://xxx.lanl.gov/abs/0811.3894}{{\normalfont [0811.3894]}}.
\newblock
  doi:{\changeurlcolor{black}\href{https://doi.org/10.1103/PhysRevLett.101.261104}{\detokenize{10.1103/PhysRevLett.101.261104}}}.

\bibitem[{Aharonian} \em{et~al.}(2009){Aharonian}, {Akhperjanian}, {Anton},
  {Barres de Almeida}, {Bazer-Bachi}, {Becherini}, {Behera}, {Bernl{\"o}hr},
  {Bochow}, {Boisson}, and et~al]{2009A&A...508..561A}
{Aharonian}, F.; {Akhperjanian}, A.G.; {Anton}, G.; {Barres de Almeida}, U.;
  {Bazer-Bachi}, A.R.; {Becherini}, Y.; {Behera}, B.; {Bernl{\"o}hr}, K.;
  {Bochow}, A.; {Boisson}, C.; et~al.
\newblock {Probing the ATIC peak in the cosmic-ray electron spectrum with
  H.E.S.S.}
\newblock {\em \aap} {\bf 2009}, {\em 508},~561--564,
  \href{http://xxx.lanl.gov/abs/0905.0105}{{\normalfont
  [arXiv:astro-ph.HE/0905.0105]}}.
\newblock
  doi:{\changeurlcolor{black}\href{https://doi.org/10.1051/0004-6361/200913323}{\detokenize{10.1051/0004-6361/200913323}}}.

\bibitem[{Fowler} \em{et~al.}(1957){Fowler}, {Waddington}, {Freier}, {Naugle},
  and {Ney}]{1957PMag....2..157F}
{Fowler}, P.H.; {Waddington}, C.J.; {Freier}, P.S.; {Naugle}, J.; {Ney}, E.P.
\newblock {The low energy end of the cosmic ray spectrum of alpha-particles}.
\newblock {\em Philosophical Magazine} {\bf 1957}, {\em 2},~157--175.
\newblock
  doi:{\changeurlcolor{black}\href{https://doi.org/10.1080/14786435708243805}{\detokenize{10.1080/14786435708243805}}}.

\bibitem[{Freier} \em{et~al.}(1958){Freier}, {Ney}, and
  {Fowler}]{1958Natur.181.1319F}
{Freier}, P.S.; {Ney}, E.P.; {Fowler}, P.H.
\newblock {Cosmic Rays and the Sunspot Cycle: Primary {$\alpha$}-Particle
  Intensity at Sunspot Maximum}.
\newblock {\em \nat} {\bf 1958}, {\em 181},~1319--1321.
\newblock
  doi:{\changeurlcolor{black}\href{https://doi.org/10.1038/1811319a0}{\detokenize{10.1038/1811319a0}}}.

\bibitem[{McDonald}(1956)]{1956PhRv..104.1723M}
{McDonald}, F.B.
\newblock {Direct Determination of Primary Cosmic-Ray Alpha-Particle Energy
  Spectrum by New Method}.
\newblock {\em Physical Review} {\bf 1956}, {\em 104},~1723--1729.
\newblock
  doi:{\changeurlcolor{black}\href{https://doi.org/10.1103/PhysRev.104.1723}{\detokenize{10.1103/PhysRev.104.1723}}}.

\bibitem[{McDonald} and {Webber}(1959)]{1959PhRv..115..194M}
{McDonald}, F.B.; {Webber}, W.R.
\newblock {Proton Component of the Primary Cosmic Radiation}.
\newblock {\em Physical Review} {\bf 1959}, {\em 115},~194--205.
\newblock
  doi:{\changeurlcolor{black}\href{https://doi.org/10.1103/PhysRev.115.194}{\detokenize{10.1103/PhysRev.115.194}}}.

\bibitem[{McDonald}(1957)]{1957PhRv..107.1386M}
{McDonald}, F.B.
\newblock {Study of Geomagnetic Cutoff Energies and Temporal Variation of the
  Primary Cosmic Radiation}.
\newblock {\em Physical Review} {\bf 1957}, {\em 107},~1386--1395.
\newblock
  doi:{\changeurlcolor{black}\href{https://doi.org/10.1103/PhysRev.107.1386}{\detokenize{10.1103/PhysRev.107.1386}}}.

\bibitem[{Webber} and {McDonald}(1964)]{1964JGR....69.3097W}
{Webber}, W.R.; {McDonald}, F.B.
\newblock {Cerenkov Scintillation Counter Measurements of the Intensity and
  Modulation of Low Rigidity Cosmic Rays and Features of the Geomagnetic Cutoff
  Rigidity}.
\newblock {\em \jgr} {\bf 1964}, {\em 69},~3097--3114.
\newblock
  doi:{\changeurlcolor{black}\href{https://doi.org/10.1029/JZ069i015p03097}{\detokenize{10.1029/JZ069i015p03097}}}.

\bibitem[{McDonald}(1959)]{1959PhRv..116..462M}
{McDonald}, F.B.
\newblock {Primary Cosmic-Ray Intensity near Solar Maximum}.
\newblock {\em Physical Review} {\bf 1959}, {\em 116},~462--463.
\newblock
  doi:{\changeurlcolor{black}\href{https://doi.org/10.1103/PhysRev.116.462}{\detokenize{10.1103/PhysRev.116.462}}}.

\bibitem[{McDonald} and {Webber}(1960)]{1960JGR....65..767M}
{McDonald}, F.B.; {Webber}, W.R.
\newblock {Changes in the Low-Rigidity Primary Cosmic Radiation during the
  Large Forbush Decrease of May 12, 1959}.
\newblock {\em \jgr} {\bf 1960}, {\em 65},~767.
\newblock
  doi:{\changeurlcolor{black}\href{https://doi.org/10.1029/JZ065i002p00767}{\detokenize{10.1029/JZ065i002p00767}}}.

\bibitem[{Engler} \em{et~al.}(1961{\natexlab{a}}){Engler}, {Kaplon}, {Kernan},
  {Klarmann}, {Fichtel}, and {Friedlander}]{1961NCim...19.1090E}
{Engler}, A.; {Kaplon}, M.F.; {Kernan}, A.; {Klarmann}, J.; {Fichtel}, C.E.;
  {Friedlander}, M.W.
\newblock {Primary cosmic-ray {$\alpha$}-particles I}.
\newblock {\em Il Nuovo Cimento} {\bf 1961}, {\em 19},~1090--1099.
\newblock
  doi:{\changeurlcolor{black}\href{https://doi.org/10.1007/BF02731385}{\detokenize{10.1007/BF02731385}}}.

\bibitem[{Engler} \em{et~al.}(1961{\natexlab{b}}){Engler}, {Foster}, {Green},
  and {Mulvey}]{1961NCim...20.1157E}
{Engler}, A.; {Foster}, F.; {Green}, T.L.; {Mulvey}, J.
\newblock {Primary cosmic ray {$\alpha$}-particles II}.
\newblock {\em Il Nuovo Cimento} {\bf 1961}, {\em 20},~1157--1165.
\newblock
  doi:{\changeurlcolor{black}\href{https://doi.org/10.1007/BF02732525}{\detokenize{10.1007/BF02732525}}}.

\bibitem[{Meyer} and {Vogt}(1963)]{1963PhRv..129.2275M}
{Meyer}, P.; {Vogt}, R.
\newblock {Primary Cosmic Ray and Solar Protons. II}.
\newblock {\em Physical Review} {\bf 1963}, {\em 129},~2275--2279.
\newblock
  doi:{\changeurlcolor{black}\href{https://doi.org/10.1103/PhysRev.129.2275}{\detokenize{10.1103/PhysRev.129.2275}}}.

\bibitem[{Freier} and {Waddington}(1965)]{1965JGR....70.2111F}
{Freier}, P.S.; {Waddington}, C.J.
\newblock {Intensity of 80- to 200-Mev Protons over Fort Churchill on August
  26, 1960}.
\newblock {\em \jgr} {\bf 1965}, {\em 70},~2111--2117.
\newblock
  doi:{\changeurlcolor{black}\href{https://doi.org/10.1029/JZ070i009p02111}{\detokenize{10.1029/JZ070i009p02111}}}.

\bibitem[{Freier} and {Waddington}(1968)]{1968JGR....73.4261F}
{Freier}, P.S.; {Waddington}, C.J.
\newblock {Singly and doubly charged particles in the primary cosmic
  radiation}.
\newblock {\em \jgr} {\bf 1968}, {\em 73},~4261--4271.
\newblock
  doi:{\changeurlcolor{black}\href{https://doi.org/10.1029/JA073i013p04261}{\detokenize{10.1029/JA073i013p04261}}}.

\bibitem[{Fichtel} \em{et~al.}(1964{\natexlab{a}}){Fichtel}, {Guss},
  {Stevenson}, and {Waddington}]{1964PhRv..133..818F}
{Fichtel}, C.E.; {Guss}, D.E.; {Stevenson}, G.R.; {Waddington}, C.J.
\newblock {Cosmic-Ray Hydrogen and Helium Nuclei during a Solar Quiet Time in
  July 1961}.
\newblock {\em Physical Review} {\bf 1964}, {\em 133},~818--827.
\newblock
  doi:{\changeurlcolor{black}\href{https://doi.org/10.1103/PhysRev.133.B818}{\detokenize{10.1103/PhysRev.133.B818}}}.

\bibitem[{Fichtel} \em{et~al.}(1964{\natexlab{b}}){Fichtel}, {Guss}, {Kniffen},
  and {Neelakantan}]{1964JGR....69.3293F}
{Fichtel}, C.E.; {Guss}, D.E.; {Kniffen}, D.A.; {Neelakantan}, K.A.
\newblock {Modulation of Low Energy Galactic Cosmic Ray Hydrogen and Helium}.
\newblock {\em \jgr} {\bf 1964}, {\em 69},~3293--3295.
\newblock
  doi:{\changeurlcolor{black}\href{https://doi.org/10.1029/JZ069i015p03293}{\detokenize{10.1029/JZ069i015p03293}}}.

\bibitem[{O'dell} \em{et~al.}(1965){O'dell}, {Shapiro}, {Silberberg}, and
  {Stiller}]{1965ICRC....1..412O}
{O'dell}, F.W.; {Shapiro}, M.M.; {Silberberg}, R.; {Stiller}, B.
\newblock {$^{3}$He-$^{4}$He in the primary cosmic radiation at Fort Churchill
  1963.}
\newblock  1965, Vol.~1, p. 412.

\bibitem[{Durgaprasad} \em{et~al.}(1967){Durgaprasad}, {Fichtel}, and
  {Guss}]{1967JGR....72.2765D}
{Durgaprasad}, N.; {Fichtel}, C.E.; {Guss}, D.E.
\newblock {Solar Modulation of Cosmic Rays and Its Relationship to Proton and
  Helium Fluxes, Interstellar Travel, and Interstellar Secondary Production}.
\newblock {\em \jgr} {\bf 1967}, {\em 72},~2765.
\newblock
  doi:{\changeurlcolor{black}\href{https://doi.org/10.1029/JZ072i011p02765}{\detokenize{10.1029/JZ072i011p02765}}}.

\bibitem[{Foster} and {Schrautemeier}(1967)]{1967NCimA..47..189F}
{Foster}, F.; {Schrautemeier}, B.E.
\newblock {Energy spectrum and geomagnetic cut-off of primary cosmic-ray
  {$\alpha$}-particles near 41deg N mag}.
\newblock {\em Nuovo Cimento A Serie} {\bf 1967}, {\em 47},~189--194.
\newblock
  doi:{\changeurlcolor{black}\href{https://doi.org/10.1007/BF02818342}{\detokenize{10.1007/BF02818342}}}.

\bibitem[{Anand} \em{et~al.}(1968){Anand}, {Daniel}, {Stephens}, {Bhowmik},
  {Krishna}, {Aditya}, and {Puri}]{1968CaJPS..46..652A}
{Anand}, K.C.; {Daniel}, R.R.; {Stephens}, S.A.; {Bhowmik}, B.; {Krishna},
  C.S.; {Aditya}, P.K.; {Puri}, R.K.
\newblock {Rigidity spectrum of cosmic-ray helium nuclei >= 12 GV}.
\newblock {\em Canadian Journal of Physics Supplement} {\bf 1968}, {\em
  46},~652.

\bibitem[{Bhatia} \em{et~al.}(1977){Bhatia}, {Paruthi}, and
  {Kainth}]{1977JGR....82.2419B}
{Bhatia}, V.S.; {Paruthi}, S.; {Kainth}, G.S.
\newblock {Simultaneous measurements of helium and heavy nuclei fluxes in
  cosmic rays over Fort Churchill}.
\newblock {\em \jgr} {\bf 1977}, {\em 82},~2419--2422.
\newblock
  doi:{\changeurlcolor{black}\href{https://doi.org/10.1029/JA082i016p02419}{\detokenize{10.1029/JA082i016p02419}}}.

\bibitem[{Balasubrahmanyan} and {McDonald}(1964)]{1964JGR....69.3289B}
{Balasubrahmanyan}, V.K.; {McDonald}, F.B.
\newblock {Solar Modulation Effects on the Primary Cosmic Radiation near Solar
  Minimum}.
\newblock {\em \jgr} {\bf 1964}, {\em 69},~3289--3292.
\newblock
  doi:{\changeurlcolor{black}\href{https://doi.org/10.1029/JZ069i015p03289}{\detokenize{10.1029/JZ069i015p03289}}}.

\bibitem[{Balasubrahmanyan} \em{et~al.}(1966){Balasubrahmanyan}, {Hagge},
  {Ludwig}, and {McDonald}]{1966JGR....71.1771B}
{Balasubrahmanyan}, V.K.; {Hagge}, D.E.; {Ludwig}, G.H.; {McDonald}, F.B.
\newblock {The Multiply Charged Primary Cosmic Radiation at Solar Minimum
  1965}.
\newblock {\em \jgr} {\bf 1966}, {\em 71},~1771.

\bibitem[{Ormes} and {Webber}(1964)]{1964PhRvL..13..106O}
{Ormes}, J.; {Webber}, W.R.
\newblock {Measurements of Low-Energy Protons and Alpha Particles in the Cosmic
  Radiation}.
\newblock {\em Physical Review Letters} {\bf 1964}, {\em 13},~106--108.
\newblock
  doi:{\changeurlcolor{black}\href{https://doi.org/10.1103/PhysRevLett.13.106}{\detokenize{10.1103/PhysRevLett.13.106}}}.

\bibitem[{Ormes} and {Webber}(1968)]{1968JGR....73.4231O}
{Ormes}, J.F.; {Webber}, W.R.
\newblock {Proton and helium nuclei cosmic-ray spectra and modulations between
  100 and 2000 Mev/nucleon}.
\newblock {\em \jgr} {\bf 1968}, {\em 73},~4231--4245.
\newblock
  doi:{\changeurlcolor{black}\href{https://doi.org/10.1029/JA073i013p04231}{\detokenize{10.1029/JA073i013p04231}}}.

\bibitem[{Webber} and {Ormes}(1967)]{1967JGR....72.5957W}
{Webber}, W.R.; {Ormes}, J.F.
\newblock {Cerenkov-scintillation counter measurements of nuclei heavier than
  helium in the primary cosmic radiation: 1. Charge composition and energy
  spectra between 200 Mev/nucleon and 5 bev/nucleon}.
\newblock {\em \jgr} {\bf 1967}, {\em 72},~5957--5976.
\newblock
  doi:{\changeurlcolor{black}\href{https://doi.org/10.1029/JZ072i023p05957}{\detokenize{10.1029/JZ072i023p05957}}}.

\bibitem[{von Rosenvinge} \em{et~al.}(1969){von Rosenvinge}, {Ormes}, and
  {Webber}]{1969Ap&SS...3...80V}
{von Rosenvinge}, T.T.; {Ormes}, J.F.; {Webber}, W.R.
\newblock {Measurements of Cosmic-Ray Li, Be and B Nuclei in the Energy Range
  100 MeV/NUC to > 22 BeV/NUC}.
\newblock {\em \apss} {\bf 1969}, {\em 3},~80--101.
\newblock
  doi:{\changeurlcolor{black}\href{https://doi.org/10.1007/BF00649595}{\detokenize{10.1007/BF00649595}}}.

\bibitem[{Courtier} and {Lenney}(1966)]{1966P&SS...14..503C}
{Courtier}, G.M.; {Lenney}, A.D.
\newblock {The flux of the cosmic ray hydrogen and helium nuclei at Kiruna,
  Sweden}.
\newblock {\em \planss} {\bf 1966}, {\em 14},~503--518.
\newblock
  doi:{\changeurlcolor{black}\href{https://doi.org/10.1016/0032-0633(66)90006-7}{\detokenize{10.1016/0032-0633(66)90006-7}}}.

\bibitem[{Hofmann} and {Winckler}(1966)]{1966PhRvL..16..109H}
{Hofmann}, D.J.; {Winckler}, J.R.
\newblock {Isotopic Composition of Low-Energy Helium Nuclei in the Primary
  Cosmic Radiation at Solar Minimum}.
\newblock {\em Physical Review Letters} {\bf 1966}, {\em 16},~109--111.
\newblock
  doi:{\changeurlcolor{black}\href{https://doi.org/10.1103/PhysRevLett.16.109}{\detokenize{10.1103/PhysRevLett.16.109}}}.

\bibitem[{Hofmann} and {Winckler}(1967)]{1967P&SS...15..715H}
{Hofmann}, D.J.; {Winckler}, J.R.
\newblock {The measurement at balloon heights of the low energy hydrogen and
  helium isotopes in the cosmic radiation at solar minimum, 1965}.
\newblock {\em \planss} {\bf 1967}, {\em 15},~715--725.
\newblock
  doi:{\changeurlcolor{black}\href{https://doi.org/10.1016/0032-0633(67)90044-X}{\detokenize{10.1016/0032-0633(67)90044-X}}}.

\bibitem[{Hartman}(1967)]{1967ApJ...150..371H}
{Hartman}, R.C.
\newblock {The Energy Dependence of the Positron-Electron Ratio in the Primary
  Cosmic Radiation in 1965}.
\newblock {\em \apj} {\bf 1967}, {\em 150},~371.
\newblock
  doi:{\changeurlcolor{black}\href{https://doi.org/10.1086/149341}{\detokenize{10.1086/149341}}}.

\bibitem[{Rygg} and {Earl}(1971)]{1971JGR....76.7445R}
{Rygg}, T.A.; {Earl}, J.A.
\newblock {Balloon measurements of cosmic ray protons and helium over half a
  solar cycle 1965-1969}.
\newblock {\em \jgr} {\bf 1971}, {\em 76},~7445.
\newblock
  doi:{\changeurlcolor{black}\href{https://doi.org/10.1029/JA076i031p07445}{\detokenize{10.1029/JA076i031p07445}}}.

\bibitem[{Rygg} \em{et~al.}(1974){Rygg}, {Ogallagher}, and
  {Earl}]{1974JGR....79.4127R}
{Rygg}, T.A.; {Ogallagher}, J.J.; {Earl}, J.A.
\newblock {Modulation of cosmic ray protons and helium nuclei near solar
  maximum}.
\newblock {\em \jgr} {\bf 1974}, {\em 79},~4127--4137.
\newblock
  doi:{\changeurlcolor{black}\href{https://doi.org/10.1029/JA079i028p04127}{\detokenize{10.1029/JA079i028p04127}}}.

\bibitem[{Garrard} \em{et~al.}(1973){Garrard}, {Stone}, and
  {Vogt}]{1973ICRC....2..732G}
{Garrard}, T.L.; {Stone}, E.C.; {Vogt}, R.E.
\newblock {Solar Modulation of Cosmic-Ray Protons and He Nuclei}.
\newblock  1973, Vol.~2, p. 732.

\bibitem[{Webber} and {Chotkowski}(1967)]{1967JGR....72.2783W}
{Webber}, W.R.; {Chotkowski}, C.
\newblock {A determination of the energy spectrum of extraterrestrial electrons
  in the energy range 70-2000 Mev}.
\newblock {\em \jgr} {\bf 1967}, {\em 72},~2783--2802.
\newblock
  doi:{\changeurlcolor{black}\href{https://doi.org/10.1029/JZ072i011p02783}{\detokenize{10.1029/JZ072i011p02783}}}.

\bibitem[{Badhwar} \em{et~al.}(1967){Badhwar}, {Deney}, {Dennis}, and
  {Kaplon}]{1967PhRv..163.1327B}
{Badhwar}, G.D.; {Deney}, C.L.; {Dennis}, B.R.; {Kaplon}, M.F.
\newblock {Measurements of the Low-Energy Cosmic Radiation during the Summer of
  1966}.
\newblock {\em Physical Review} {\bf 1967}, {\em 163},~1327--1342.
\newblock
  doi:{\changeurlcolor{black}\href{https://doi.org/10.1103/PhysRev.163.1327}{\detokenize{10.1103/PhysRev.163.1327}}}.

\bibitem[{Behrnetz} \em{et~al.}(1976){Behrnetz}, {Kristiansson}, {Lindstam},
  and {Soderstrom}]{1976A&A....52..327B}
{Behrnetz}, S.; {Kristiansson}, K.; {Lindstam}, S.; {Soderstrom}, K.
\newblock {Composition of medium energy cosmic rays from silicon to nickel
  measured with nuclear emulsion}.
\newblock {\em \aap} {\bf 1976}, {\em 52},~327--335.

\bibitem[{Singh} and {Bhatia}(1979)]{1979Ap&SS..62..465S}
{Singh}, G.; {Bhatia}, V.S.
\newblock {Charge composition of medium energy cosmic-ray nuclei from neon to
  iron}.
\newblock {\em \apss} {\bf 1979}, {\em 62},~465--475.
\newblock
  doi:{\changeurlcolor{black}\href{https://doi.org/10.1007/BF00645481}{\detokenize{10.1007/BF00645481}}}.

\bibitem[{Bjarle} \em{et~al.}(1979){Bjarle}, {Herrstr{\"o}m}, {Jonsson}, and
  {Kristiansson}]{1979ZPhyA.291..383B}
{Bjarle}, C.; {Herrstr{\"o}m}, N.Y.; {Jonsson}, G.; {Kristiansson}, K.
\newblock {The cosmic ray boron and carbon isotopic composition measured in
  nuclear emulsions.}
\newblock {\em Zeitschrift fur Physik A Hadrons and Nuclei} {\bf 1979}, {\em
  291},~383--390.
\newblock
  doi:{\changeurlcolor{black}\href{https://doi.org/10.1007/BF01408389}{\detokenize{10.1007/BF01408389}}}.

\bibitem[{Webber} \em{et~al.}(1972){Webber}, {Damle}, and
  {Kish}]{1972Ap&SS..15..245W}
{Webber}, W.R.; {Damle}, S.V.; {Kish}, J.
\newblock {Studies of the chemical composition of cosmic rays withZ=3 30 at
  high and low energies}.
\newblock {\em \apss} {\bf 1972}, {\em 15},~245--271.
\newblock
  doi:{\changeurlcolor{black}\href{https://doi.org/10.1007/BF00649920}{\detokenize{10.1007/BF00649920}}}.

\bibitem[{Ryan} \em{et~al.}(1972){Ryan}, {Ormes}, and
  {Balasubrahmanyan}]{1972PhRvL..28..985R}
{Ryan}, M.J.; {Ormes}, J.F.; {Balasubrahmanyan}, V.K.
\newblock {Cosmic-Ray Proton and Helium Spectra above 50 GeV}.
\newblock {\em Physical Review Letters} {\bf 1972}, {\em 28},~985--988.
\newblock
  doi:{\changeurlcolor{black}\href{https://doi.org/10.1103/PhysRevLett.28.985}{\detokenize{10.1103/PhysRevLett.28.985}}}.

\bibitem[{Smith} \em{et~al.}(1973){Smith}, {Buffington}, {Smoot}, {Alvarez},
  and {Wahlig}]{1973ApJ...180..987S}
{Smith}, L.H.; {Buffington}, A.; {Smoot}, D.F.; {Alvarez}, L.W.; {Wahlig}, M.A.
\newblock {A Measurement of Cosmic-Ray Rigidity Spectra above 5 GV/c of
  Elements from Hydrogen to Iron}.
\newblock {\em \apj} {\bf 1973}, {\em 180},~987--1010.
\newblock
  doi:{\changeurlcolor{black}\href{https://doi.org/10.1086/152021}{\detokenize{10.1086/152021}}}.

\bibitem[{Apparao}(1973)]{1973ICRC....1..126A}
{Apparao}, K.M.V.
\newblock {Flux of Cosmic Ray Deuterons with Rigidity Above 16. 8 GV}.
\newblock  International Cosmic Ray Conference,  1973, Vol.~1, {\em
  International Cosmic Ray Conference}, p. 126.

\bibitem[{Juliusson}(1974)]{1974ApJ...191..331J}
{Juliusson}, E.
\newblock {Charge Composition and Energy Spectra of Cosmic-Ray Nuclei at
  Energies above 20 GeV Per Nucleon}.
\newblock {\em \apj} {\bf 1974}, {\em 191},~331--348.
\newblock
  doi:{\changeurlcolor{black}\href{https://doi.org/10.1086/152972}{\detokenize{10.1086/152972}}}.

\bibitem[{Fisher} \em{et~al.}(1976){Fisher}, {Hagen}, {Maehl}, {Ormes}, and
  {Arens}]{1976ApJ...205..938F}
{Fisher}, A.J.; {Hagen}, F.A.; {Maehl}, R.C.; {Ormes}, J.F.; {Arens}, J.F.
\newblock {The isotopic composition of cosmic rays with Z between 5 and 26}.
\newblock {\em \apj} {\bf 1976}, {\em 205},~938--946.
\newblock
  doi:{\changeurlcolor{black}\href{https://doi.org/10.1086/154349}{\detokenize{10.1086/154349}}}.

\bibitem[{Hagen} \em{et~al.}(1977){Hagen}, {Fisher}, and
  {Ormes}]{1977ApJ...212..262H}
{Hagen}, F.A.; {Fisher}, A.J.; {Ormes}, J.F.
\newblock {Be-10 abundance and the age of cosmic rays - A balloon measurement}.
\newblock {\em \apj} {\bf 1977}, {\em 212},~262--277.
\newblock
  doi:{\changeurlcolor{black}\href{https://doi.org/10.1086/155045}{\detokenize{10.1086/155045}}}.

\bibitem[{Maehl} \em{et~al.}(1977){Maehl}, {Ormes}, {Fisher}, and
  {Hagen}]{1977Ap&SS..47..163M}
{Maehl}, R.C.; {Ormes}, J.F.; {Fisher}, A.J.; {Hagen}, F.A.
\newblock {Energy spectra of cosmic ray nuclei - Z of 4 to 26 and E of 0.3 to 2
  GeV/amu}.
\newblock {\em \apss} {\bf 1977}, {\em 47},~163--184.
\newblock
  doi:{\changeurlcolor{black}\href{https://doi.org/10.1007/BF00651365}{\detokenize{10.1007/BF00651365}}}.

\bibitem[{Leech} and {Ogallagher}(1978)]{1978ApJ...221.1110L}
{Leech}, H.W.; {Ogallagher}, J.J.
\newblock {The isotopic composition of cosmic-ray helium from 123 to 279 MeV
  per nucleon - A new measurement and analysis}.
\newblock {\em \apj} {\bf 1978}, {\em 221},~1110--1123.
\newblock
  doi:{\changeurlcolor{black}\href{https://doi.org/10.1086/156114}{\detokenize{10.1086/156114}}}.

\bibitem[{Dwyer}(1978)]{1978ApJ...224..691D}
{Dwyer}, R.
\newblock {The mean mass of the abundant cosmic-ray nuclei from boron to
  silicon at 1.2 GeV per atomic mass unit}.
\newblock {\em \apj} {\bf 1978}, {\em 224},~691--707.
\newblock
  doi:{\changeurlcolor{black}\href{https://doi.org/10.1086/156417}{\detokenize{10.1086/156417}}}.

\bibitem[{Dwyer} and {Meyer}(1987)]{1987ApJ...322..981D}
{Dwyer}, R.; {Meyer}, P.
\newblock {Cosmic-ray elemental abundances from 1 to 10 GeV per AMU for boron
  through nickel}.
\newblock {\em \apj} {\bf 1987}, {\em 322},~981--991.
\newblock
  doi:{\changeurlcolor{black}\href{https://doi.org/10.1086/165793}{\detokenize{10.1086/165793}}}.

\bibitem[{Minagawa}(1981)]{1981ApJ...248..847M}
{Minagawa}, G.
\newblock {The abundances and energy spectra of cosmic ray iron and nickel at
  energies from 1 to 10 GeV per AMU}.
\newblock {\em \apj} {\bf 1981}, {\em 248},~847--855.
\newblock
  doi:{\changeurlcolor{black}\href{https://doi.org/10.1086/159210}{\detokenize{10.1086/159210}}}.

\bibitem[{Scarlett} \em{et~al.}(1978){Scarlett}, {Freier}, and
  {Waddington}]{1978Ap&SS..59..301S}
{Scarlett}, W.R.; {Freier}, P.S.; {Waddington}, C.J.
\newblock {The charge and energy spectra of heavy cosmic-ray nuclei}.
\newblock {\em \apss} {\bf 1978}, {\em 59},~301--311.
\newblock
  doi:{\changeurlcolor{black}\href{https://doi.org/10.1007/BF01023921}{\detokenize{10.1007/BF01023921}}}.

\bibitem[{Lezniak} and {Webber}(1978)]{1978ApJ...223..676L}
{Lezniak}, J.A.; {Webber}, W.R.
\newblock {The charge composition and energy spectra of cosmic-ray nuclei from
  3000 MeV per nucleon to 50 GeV per nucleon}.
\newblock {\em \apj} {\bf 1978}, {\em 223},~676--696.
\newblock
  doi:{\changeurlcolor{black}\href{https://doi.org/10.1086/156301}{\detokenize{10.1086/156301}}}.

\bibitem[{Derrickson} \em{et~al.}(1992){Derrickson}, {Parnell}, {Austin},
  {Selig}, and {Gregory}]{1992IJRAI..20..415D}
{Derrickson}, J.H.; {Parnell}, T.A.; {Austin}, R.W.; {Selig}, W.J.; {Gregory},
  J.C.
\newblock {A measurement of the absolute energy spectra of galactic cosmic rays
  during the 1976-77 solar minimum}.
\newblock {\em International Journal of Radiation Applications and
  Instrumentation D Nuclear Tracks and Radiation Measurements} {\bf 1992}, {\em
  20},~415--421.

\bibitem[{Simon} \em{et~al.}(1980){Simon}, {Spiegelhauer}, {Schmidt}, {Siohan},
  {Ormes}, {Balasubrahmanyan}, and {Arens}]{1980ApJ...239..712S}
{Simon}, M.; {Spiegelhauer}, H.; {Schmidt}, W.K.H.; {Siohan}, F.; {Ormes},
  J.F.; {Balasubrahmanyan}, V.K.; {Arens}, J.F.
\newblock {Energy spectra of cosmic-ray nuclei to above 100 GeV per nucleon}.
\newblock {\em \apj} {\bf 1980}, {\em 239},~712--724.
\newblock
  doi:{\changeurlcolor{black}\href{https://doi.org/10.1086/158157}{\detokenize{10.1086/158157}}}.

\bibitem[{Webber} \em{et~al.}(1987){Webber}, {Golden}, and
  {Stephens}]{1987ICRC....1..325W}
{Webber}, W.R.; {Golden}, R.L.; {Stephens}, S.A.
\newblock {Cosmic ray proton and helium spectra from 5 - 200 GV measured with a
  magnetic spectrometer.}
\newblock  1987, Vol.~1, pp. 325--328.

\bibitem[{Buffington} \em{et~al.}(1978){Buffington}, {Orth}, and
  {Mast}]{1978ApJ...226..355B}
{Buffington}, A.; {Orth}, C.D.; {Mast}, T.S.
\newblock {A measurement of cosmic-ray beryllium isotopes from 200 to 1500 MeV
  per nucleon}.
\newblock {\em \apj} {\bf 1978}, {\em 226},~355--371.
\newblock
  doi:{\changeurlcolor{black}\href{https://doi.org/10.1086/156616}{\detokenize{10.1086/156616}}}.

\bibitem[{Webber} and {Yushak}(1983)]{1983ApJ...275..391W}
{Webber}, W.R.; {Yushak}, S.M.
\newblock {A measurement of the energy spectra and relative abundance of the
  cosmic-ray H and He isotopes over a broad energy range}.
\newblock {\em \apj} {\bf 1983}, {\em 275},~391--404.
\newblock
  doi:{\changeurlcolor{black}\href{https://doi.org/10.1086/161541}{\detokenize{10.1086/161541}}}.

\bibitem[{Webber} and {Kish}(1979)]{1979ICRC....1..389W}
{Webber}, W.R.; {Kish}, J.
\newblock {Further Studies of the Isotopic Composition of Cosmic Ray Li, Be,
  and B Nuclei. Implications for the Cosmic Ray Age}.
\newblock  International Cosmic Ray Conference,  1979, Vol.~1, {\em
  International Cosmic Ray Conference}, p. 389.

\bibitem[{Webber}(1982)]{1982ApJ...252..386W}
{Webber}, W.R.
\newblock {The charge and isotopic composition of Z = 6-14 cosmic ray nuclei at
  their source}.
\newblock {\em \apj} {\bf 1982}, {\em 252},~386--392.
\newblock
  doi:{\changeurlcolor{black}\href{https://doi.org/10.1086/159565}{\detokenize{10.1086/159565}}}.

\bibitem[{Webber}(1981)]{1981ICRC....2...80W}
{Webber}, W.R.
\newblock {The Isotopic Composition of fe and Fragmentation Nuclei with Z =
  20-25}.
\newblock  International Cosmic Ray Conference,  1981, Vol.~2, {\em
  International Cosmic Ray Conference}, p.~80.

\bibitem[{Webber} and {Yushak}(1979)]{1979ICRC...12...51W}
{Webber}, W.R.; {Yushak}, S.M.
\newblock {Comparative Energy Spectra of Z=3-8 Nuclei in the Energy Range 200
  MeV/nuc to 3 GeV/nuc}.
\newblock  International Cosmic Ray Conference,  1979, Vol.~12, {\em
  International Cosmic Ray Conference}, p.~51.

\bibitem[{Jordan}(1985)]{1985ApJ...291..207J}
{Jordan}, S.P.
\newblock {The isotopic composition of helium in the cosmic radiation above 11
  gigavolts}.
\newblock {\em \apj} {\bf 1985}, {\em 291},~207--218.
\newblock
  doi:{\changeurlcolor{black}\href{https://doi.org/10.1086/163058}{\detokenize{10.1086/163058}}}.

\bibitem[{Webber} \em{et~al.}(1985{\natexlab{a}}){Webber}, {Kish}, and
  {Schrier}]{1985ICRC....2...16W}
{Webber}, W.R.; {Kish}, J.C.; {Schrier}, D.A.
\newblock {Cosmic ray charge and energy spectrum measurements using a new large
  area Cerenkov X dE/dx telescope}.
\newblock  International Cosmic Ray Conference; {Jones}, F.C., Ed.,  1985,
  Vol.~2, {\em International Cosmic Ray Conference}, pp. 16--19.

\bibitem[{Webber} \em{et~al.}(1985{\natexlab{b}}){Webber}, {Kish}, and
  {Schrier}]{1985ICRC....2...88W}
{Webber}, W.R.; {Kish}, J.C.; {Schrier}, D.A.
\newblock {Cosmic ray isotope measurements with a new Cerenkov X total energy
  telescope}.
\newblock  International Cosmic Ray Conference; {Jones}, F.C., Ed.,  1985,
  Vol.~2, {\em International Cosmic Ray Conference}, pp. 88--91.

\bibitem[{Ichimura} \em{et~al.}(1993){Ichimura}, {Kogawa}, {Kuramata}, {Mito},
  {Murabayashi}, {Nanjo}, {Nakamura}, {Ohba}, {Ohuchi}, {Ozawa}, and
  et~al]{1993PhRvD..48.1949I}
{Ichimura}, M.; {Kogawa}, M.; {Kuramata}, S.; {Mito}, H.; {Murabayashi}, T.;
  {Nanjo}, H.; {Nakamura}, T.; {Ohba}, K.; {Ohuchi}, T.; {Ozawa}, T.; et~al.
\newblock {Observation of heavy cosmic-ray primaries over the wide energy range
  from \~{}100 GeV/particle to \~{}100 TeV/particle: Is the celebrated ``knee''
  actually so prominent?}
\newblock {\em \prd} {\bf 1993}, {\em 48},~1949--1975.
\newblock
  doi:{\changeurlcolor{black}\href{https://doi.org/10.1103/PhysRevD.48.1949}{\detokenize{10.1103/PhysRevD.48.1949}}}.

\bibitem[{Hatano} \em{et~al.}(1995){Hatano}, {Fukada}, {Saito}, {Oda}, and
  {Yanagita}]{1995PhRvD..52.6219H}
{Hatano}, Y.; {Fukada}, Y.; {Saito}, T.; {Oda}, H.; {Yanagita}, T.
\newblock {Relative abundance of $^{3}$He and $^{4}$He in cosmic rays near 10
  GV}.
\newblock {\em \prd} {\bf 1995}, {\em 52},~6219--6223.
\newblock
  doi:{\changeurlcolor{black}\href{https://doi.org/10.1103/PhysRevD.52.6219}{\detokenize{10.1103/PhysRevD.52.6219}}}.

\bibitem[{Kamioka} \em{et~al.}(1997){Kamioka}, {Hareyama}, {Ichimura},
  {Ishihara}, {Kobayashi}, {Komatsu}, {Kuramata}, {Maruguchi}, {Matsutani},
  {Mihashi}, and et~al]{1997APh.....6..155K}
{Kamioka}, E.; {Hareyama}, M.; {Ichimura}, M.; {Ishihara}, Y.; {Kobayashi}, T.;
  {Komatsu}, H.; {Kuramata}, S.; {Maruguchi}, K.; {Matsutani}, H.; {Mihashi},
  A.; et~al.
\newblock {Azimuthally controlled observation of heavy cosmic-ray primaries by
  means of the balloon-borne emulsion chamber}.
\newblock {\em Astroparticle Physics} {\bf 1997}, {\em 6},~155--167.
\newblock
  doi:{\changeurlcolor{black}\href{https://doi.org/10.1016/S0927-6505(96)00051-5}{\detokenize{10.1016/S0927-6505(96)00051-5}}}.

\bibitem[{Bogomolov} \em{et~al.}(1995){Bogomolov}, {Vasilyev}, {Yu},
  {Krut'kov}, {Stepanov}, and {Shulakova}]{1995ICRC....2..598B}
{Bogomolov}, E.A.; {Vasilyev}, G.I.; {Yu}, S.; {Krut'kov}, S.; {Stepanov},
  S.V.; {Shulakova}, M.S.
\newblock {The Deuterium Cosmic Ray Intensity from Balloon Measurement in
  Energy Range 0.8-1.8 GeV/nucl.}
\newblock  International Cosmic Ray Conference,  1995, Vol.~2, {\em
  International Cosmic Ray Conference}, p. 598.

\bibitem[{Buckley} \em{et~al.}(1994){Buckley}, {Dwyer}, {Mueller}, {Swordy},
  and {Tang}]{1994ApJ...429..736B}
{Buckley}, J.; {Dwyer}, J.; {Mueller}, D.; {Swordy}, S.; {Tang}, K.K.
\newblock {A new measurement of the flux of the light cosmic-ray nuclei at high
  energies}.
\newblock {\em \apj} {\bf 1994}, {\em 429},~736--747.
\newblock
  doi:{\changeurlcolor{black}\href{https://doi.org/10.1086/174357}{\detokenize{10.1086/174357}}}.

\bibitem[{Hesse} \em{et~al.}(1996){Hesse}, {Acharya}, {Heinbach}, {Heinrich},
  {Henkel}, {Luzietti}, {Simon}, {Christian}, {Esposito}, {Balasubrahmanyan},
  and et~al]{1996A&A...314..785H}
{Hesse}, A.; {Acharya}, B.S.; {Heinbach}, U.; {Heinrich}, W.; {Henkel}, M.;
  {Luzietti}, B.; {Simon}, M.; {Christian}, E.R.; {Esposito}, J.A.;
  {Balasubrahmanyan}, V.K.; et~al.
\newblock {Isotopic composition of silicon and iron in the galactic cosmic
  radiation.}
\newblock {\em \aap} {\bf 1996}, {\em 314},~785--794.

\bibitem[{Esposito} \em{et~al.}(1992){Esposito}, {Christian},
  {Balasubrahmanyan}, {Barbier}, {Ormes}, {Streitmatter}, {Acharya},
  {Luzietti}, {Hesse}, and {Heinbach}]{1992APh.....1...33E}
{Esposito}, J.A.; {Christian}, E.R.; {Balasubrahmanyan}, V.K.; {Barbier}, L.M.;
  {Ormes}, J.F.; {Streitmatter}, R.E.; {Acharya}, B.; {Luzietti}, B.; {Hesse},
  A.; {Heinbach}, U.
\newblock {The ALICE instrument and the measured cosmic ray elemental
  abundances}.
\newblock {\em Astroparticle Physics} {\bf 1992}, {\em 1},~33--45.
\newblock
  doi:{\changeurlcolor{black}\href{https://doi.org/10.1016/0927-6505(92)90007-M}{\detokenize{10.1016/0927-6505(92)90007-M}}}.

\bibitem[{Panov} \em{et~al.}(2008){Panov}, {Sokolskaya}, {Adams}, and {et
  al}]{2008ICRC....2....3P}
{Panov}, A.D.; {Sokolskaya}, N.V.; {Adams}, Jr., J.H.; {et al}.
\newblock {Relative abundances of cosmic ray nuclei B-C-N-O in the energy
  region from 10 GeV/n to 300 GeV/n. Results from ATIC-2 (the science flight of
  ATIC)}.
\newblock  International Cosmic Ray Conference,  2008, Vol.~2, {\em
  International Cosmic Ray Conference}, pp. 3--6,
  \href{http://xxx.lanl.gov/abs/0707.4415}{{\normalfont [0707.4415]}}.

\bibitem[{Wang} \em{et~al.}(2002){Wang}, {Seo}, {Anraku}, {Fujikawa}, {Imori},
  {Maeno}, {Matsui}, {Matsunaga}, {Motoki}, {Orito}, and
  et~al]{2002ApJ...564..244W}
{Wang}, J.Z.; {Seo}, E.S.; {Anraku}, K.; {Fujikawa}, M.; {Imori}, M.; {Maeno},
  T.; {Matsui}, N.; {Matsunaga}, H.; {Motoki}, M.; {Orito}, S.; et~al.
\newblock {Measurement of Cosmic-Ray Hydrogen and Helium and Their Isotopic
  Composition with the BESS Experiment}.
\newblock {\em \apj} {\bf 2002}, {\em 564},~244--259.
\newblock
  doi:{\changeurlcolor{black}\href{https://doi.org/10.1086/324140}{\detokenize{10.1086/324140}}}.

\bibitem[{Seo} \em{et~al.}(2001){Seo}, {Wang}, {Matsunaga}, {Anraku}, {Imori},
  {Makida}, {Matsumoto}, {McDonald}, {Mitchell}, and
  {Moiseev}]{2001AdSpR..26.1831S}
{Seo}, E.S.; {Wang}, J.Z.; {Matsunaga}, H.; {Anraku}, K.; {Imori}, M.;
  {Makida}, Y.; {Matsumoto}, H.; {McDonald}, F.B.; {Mitchell}, J.; {Moiseev},
  A.A.
\newblock {Spectra of H and He measured in a series of annual flights}.
\newblock {\em Advances in Space Research} {\bf 2001}, {\em 26},~1831--1834.
\newblock
  doi:{\changeurlcolor{black}\href{https://doi.org/10.1016/S0273-1177(99)01232-6}{\detokenize{10.1016/S0273-1177(99)01232-6}}}.

\bibitem[{Myers} \em{et~al.}(2003){Myers}, {Seo}, {Abe}, {Anraku}, {Imori},
  {Maeno}, {Makida}, {Matsumoto}, {Mitchell}, {Moiseev}, and
  et~al]{2003ICRC....4.1805M}
{Myers}, Z.D.; {Seo}, E.S.; {Abe}, K.; {Anraku}, K.; {Imori}, M.; {Maeno}, T.;
  {Makida}, Y.; {Matsumoto}, H.; {Mitchell}, J.; {Moiseev}, A.; et~al.
\newblock {Cosmic Ray $^{3}$He and $^{4}$He Spectra from BESS 98}.
\newblock  International Cosmic Ray Conference,  2003, Vol.~4, {\em
  International Cosmic Ray Conference}, p. 1805.

\bibitem[{Myers} \em{et~al.}(2005){Myers}, {Seo}, {Wang}, {Alford}, {Abe},
  {Anraku}, {Asaoka}, {Fujikawa}, {Imori}, and {Maeno}]{2005AdSpR..35..151M}
{Myers}, Z.D.; {Seo}, E.S.; {Wang}, J.Z.; {Alford}, R.W.; {Abe}, K.; {Anraku},
  K.; {Asaoka}, Y.; {Fujikawa}, M.; {Imori}, M.; {Maeno}, T.A.
\newblock {Cosmic ray $^{1}$H and $^{2}$H spectra from BESS 98}.
\newblock {\em Advances in Space Research} {\bf 2005}, {\em 35},~151--155.
\newblock
  doi:{\changeurlcolor{black}\href{https://doi.org/10.1016/j.asr.2003.10.050}{\detokenize{10.1016/j.asr.2003.10.050}}}.

\bibitem[{Kim} \em{et~al.}(2013){Kim}, {Abe}, {Fuke}, {Hams}, {Lee}, {Makida},
  {Matsuda}, {Mitchell}, {Nishimura}, and {Ormes}]{2013AdSpR..51..234K}
{Kim}, K.C.; {Abe}, K.; {Fuke}, H.; {Hams}, T.; {Lee}, M.H.; {Makida}, Y.;
  {Matsuda}, S.; {Mitchell}, J.W.; {Nishimura}, J.; {Ormes}, J.F.A.
\newblock {Cosmic ray $^{2}$H/$^{1}$H ratio measured from BESS in 2000 during
  solar maximum}.
\newblock {\em Advances in Space Research} {\bf 2013}, {\em 51},~234--237.
\newblock
  doi:{\changeurlcolor{black}\href{https://doi.org/10.1016/j.asr.2012.01.015}{\detokenize{10.1016/j.asr.2012.01.015}}}.

\bibitem[{Shikaze} \em{et~al.}(2007){Shikaze}, {Haino}, {Abe}, {Fuke}, {Hams},
  {Kim}, {Makida}, {Matsuda}, {Mitchell}, {Moiseev}, and
  et~al]{2007APh....28..154S}
{Shikaze}, Y.; {Haino}, S.; {Abe}, K.; {Fuke}, H.; {Hams}, T.; {Kim}, K.C.;
  {Makida}, Y.; {Matsuda}, S.; {Mitchell}, J.W.; {Moiseev}, A.A.; et~al.
\newblock {Measurements of 0.2 20 GeV/n cosmic-ray proton and helium spectra
  from 1997 through 2002 with the BESS spectrometer}.
\newblock {\em Astroparticle Physics} {\bf 2007}, {\em 28},~154--167,
  \href{http://xxx.lanl.gov/abs/arXiv:astro-ph/0611388}{{\normalfont
  [arXiv:astro-ph/0611388]}}.
\newblock
  doi:{\changeurlcolor{black}\href{https://doi.org/10.1016/j.astropartphys.2007.05.001}{\detokenize{10.1016/j.astropartphys.2007.05.001}}}.

\bibitem[{Sanuki} \em{et~al.}(2000){Sanuki}, {Motoki}, {Matsumoto}, {Seo},
  {Wang}, {Abe}, {Anraku}, {Asaoka}, {Fujikawa}, {Imori}, and
  et~al]{2000ApJ...545.1135S}
{Sanuki}, T.; {Motoki}, M.; {Matsumoto}, H.; {Seo}, E.S.; {Wang}, J.Z.; {Abe},
  K.; {Anraku}, K.; {Asaoka}, Y.; {Fujikawa}, M.; {Imori}, M.; et~al.
\newblock {Precise Measurement of Cosmic-Ray Proton and Helium Spectra with the
  BESS Spectrometer}.
\newblock {\em \apj} {\bf 2000}, {\em 545},~1135--1142,
  \href{http://xxx.lanl.gov/abs/arXiv:astro-ph/0002481}{{\normalfont
  [arXiv:astro-ph/0002481]}}.
\newblock
  doi:{\changeurlcolor{black}\href{https://doi.org/10.1086/317873}{\detokenize{10.1086/317873}}}.

\bibitem[{Haino} \em{et~al.}(2004){Haino}, {Sanuki}, {Abe}, {Anraku}, {Asaoka},
  {Fuke}, {Imori}, {Itasaki}, {Maeno}, {Makida}, and
  et~al]{2004PhLB..594...35H}
{Haino}, S.; {Sanuki}, T.; {Abe}, K.; {Anraku}, K.; {Asaoka}, Y.; {Fuke}, H.;
  {Imori}, M.; {Itasaki}, A.; {Maeno}, T.; {Makida}, Y.; et~al.
\newblock {Measurements of primary and atmospheric cosmic-ray spectra with the
  BESS-TeV spectrometer}.
\newblock {\em Physics Letters B} {\bf 2004}, {\em 594},~35--46,
  \href{http://xxx.lanl.gov/abs/arXiv:astro-ph/0403704}{{\normalfont
  [arXiv:astro-ph/0403704]}}.
\newblock
  doi:{\changeurlcolor{black}\href{https://doi.org/10.1016/j.physletb.2004.05.019}{\detokenize{10.1016/j.physletb.2004.05.019}}}.

\bibitem[{Abe} \em{et~al.}(2015){Abe}, {Fuke}, {Haino}, {Hams}, {Hasegawa},
  {Horikoshi}, {Itazaki}, {Kim}, {Kumazawa}, and
  {Kusumoto}]{2015arXiv150601267A}
{Abe}, K.; {Fuke}, H.; {Haino}, S.; {Hams}, T.; {Hasegawa}, M.; {Horikoshi},
  A.; {Itazaki}, A.; {Kim}, K.C.; {Kumazawa}, T.; {Kusumoto}, A.A.
\newblock {Measurements of cosmic-ray proton and helium spectra from the
  BESS-Polar long-duration balloon flights over Antarctica}.
\newblock {\em ArXiv e-prints} {\bf 2015},
  \href{http://xxx.lanl.gov/abs/1506.01267}{{\normalfont
  [arXiv:astro-ph.HE/1506.01267]}}.

\bibitem[{Boezio} \em{et~al.}(1999){Boezio}, {Carlson}, {Francke}, {Weber},
  {Suffert}, {Hof}, {Menn}, {Simon}, {Stephens}, {Bellotti}, and
  et~al]{1999ApJ...518..457B}
{Boezio}, M.; {Carlson}, P.; {Francke}, T.; {Weber}, N.; {Suffert}, M.; {Hof},
  M.; {Menn}, W.; {Simon}, M.; {Stephens}, S.A.; {Bellotti}, R.; et~al.
\newblock {The Cosmic-Ray Proton and Helium Spectra between 0.4 and 200 GV}.
\newblock {\em \apj} {\bf 1999}, {\em 518},~457--472.
\newblock
  doi:{\changeurlcolor{black}\href{https://doi.org/10.1086/307251}{\detokenize{10.1086/307251}}}.

\bibitem[{Boezio} \em{et~al.}(2003){Boezio}, {Bonvicini}, {Schiavon}, {Vacchi},
  {Zampa}, {Bergstr{\"o}m}, {Carlson}, {Francke}, {Hansen}, {Mocchiutti}, and
  et~al]{2003APh....19..583B}
{Boezio}, M.; {Bonvicini}, V.; {Schiavon}, P.; {Vacchi}, A.; {Zampa}, N.;
  {Bergstr{\"o}m}, D.; {Carlson}, P.; {Francke}, T.; {Hansen}, P.;
  {Mocchiutti}, E.; et~al.
\newblock {The cosmic-ray proton and helium spectra measured with the CAPRICE98
  balloon experiment}.
\newblock {\em Astroparticle Physics} {\bf 2003}, {\em 19},~583--604,
  \href{http://xxx.lanl.gov/abs/arXiv:astro-ph/0212253}{{\normalfont
  [arXiv:astro-ph/0212253]}}.
\newblock
  doi:{\changeurlcolor{black}\href{https://doi.org/10.1016/S0927-6505(02)00267-0}{\detokenize{10.1016/S0927-6505(02)00267-0}}}.

\bibitem[{Mocchiutti} \em{et~al.}(2003){Mocchiutti}
  et~al.]{2003ICRC....4.1809M}
{Mocchiutti}, E.; others.
\newblock {Measurement of High Energy $^{3}$He in Cosmic Rays by the CAPRICE98
  Balloon Experiment}.
\newblock  International Cosmic Ray Conference,  2003, Vol.~4, {\em
  International Cosmic Ray Conference}, p. 1809.

\bibitem[{Papini} \em{et~al.}(2004){Papini}, {Piccardi}, {Spillantini},
  {Vannuccini}, {Ambriola}, {Bellotti}, {Cafagna}, {Ciacio}, {Circella}, {De
  Marzo}, and et~al]{2004ApJ...615..259P}
{Papini}, P.; {Piccardi}, S.; {Spillantini}, P.; {Vannuccini}, E.; {Ambriola},
  M.; {Bellotti}, R.; {Cafagna}, F.; {Ciacio}, F.; {Circella}, M.; {De Marzo},
  C.N.; et~al.
\newblock {High-Energy Deuteron Measurement with the CAPRICE98 Experiment}.
\newblock {\em \apj} {\bf 2004}, {\em 615},~259--274.
\newblock
  doi:{\changeurlcolor{black}\href{https://doi.org/10.1086/424027}{\detokenize{10.1086/424027}}}.

\bibitem[{Ahn} \em{et~al.}(2008){Ahn}, {Allison}, {Bagliesi}, {Beatty},
  {Bigongiari}, {Boyle}, {Brandt}, {Childers}, {Conklin}, {Coutu}, and
  et~al]{2008APh....30..133A}
{Ahn}, H.S.; {Allison}, P.S.; {Bagliesi}, M.G.; {Beatty}, J.J.; {Bigongiari},
  G.; {Boyle}, P.J.; {Brandt}, T.J.; {Childers}, J.T.; {Conklin}, N.B.;
  {Coutu}, S.; et~al.
\newblock {Measurements of cosmic-ray secondary nuclei at high energies with
  the first flight of the CREAM balloon-borne experiment}.
\newblock {\em Astroparticle Physics} {\bf 2008}, {\em 30},~133--141,
  \href{http://xxx.lanl.gov/abs/0808.1718}{{\normalfont [0808.1718]}}.
\newblock
  doi:{\changeurlcolor{black}\href{https://doi.org/10.1016/j.astropartphys.2008.07.010}{\detokenize{10.1016/j.astropartphys.2008.07.010}}}.

\bibitem[{Yoon} \em{et~al.}(2017){Yoon}, {Anderson}, {Barrau}, {Conklin},
  {Coutu}, {Derome}, {Han}, {Jeon}, {Kim}, {Kim}, and
  et~al]{2017ApJ...839....5Y}
{Yoon}, Y.S.; {Anderson}, T.; {Barrau}, A.; {Conklin}, N.B.; {Coutu}, S.;
  {Derome}, L.; {Han}, J.H.; {Jeon}, J.A.; {Kim}, K.C.; {Kim}, M.H.; et~al.
\newblock {Proton and Helium Spectra from the CREAM-III Flight}.
\newblock {\em \apj} {\bf 2017}, {\em 839},~5,
  \href{http://xxx.lanl.gov/abs/1704.02512}{{\normalfont
  [arXiv:astro-ph.HE/1704.02512]}}.
\newblock
  doi:{\changeurlcolor{black}\href{https://doi.org/10.3847/1538-4357/aa68e4}{\detokenize{10.3847/1538-4357/aa68e4}}}.

\bibitem[{Ahn} \em{et~al.}(2009){Ahn}, {Allison}, {Bagliesi}, {Barbier},
  {Beatty}, {Bigongiari}, {Brandt}, {Childers}, {Conklin}, {Coutu}, , and
  et~al]{2009ApJ...707..593A}
{Ahn}, H.S.; {Allison}, P.; {Bagliesi}, M.G.; {Barbier}, L.; {Beatty}, J.J.;
  {Bigongiari}, G.; {Brandt}, T.J.; {Childers}, J.T.; {Conklin}, N.B.; {Coutu},
  S.; .; et~al.
\newblock {Energy Spectra of Cosmic-ray Nuclei at High Energies}.
\newblock {\em \apj} {\bf 2009}, {\em 707},~593--603,
  \href{http://xxx.lanl.gov/abs/0911.1889}{{\normalfont
  [arXiv:astro-ph.HE/0911.1889]}}.
\newblock
  doi:{\changeurlcolor{black}\href{https://doi.org/10.1088/0004-637X/707/1/593}{\detokenize{10.1088/0004-637X/707/1/593}}}.

\bibitem[{Ahn} \em{et~al.}(2010){Ahn}, {Allison}, {Bagliesi}, {Barbier},
  {Beatty}, {Bigongiari}, {Brandt}, {Childers}, {Conklin}, {Coutu}, and
  et~al]{2010ApJ...715.1400A}
{Ahn}, H.S.; {Allison}, P.S.; {Bagliesi}, M.G.; {Barbier}, L.; {Beatty}, J.J.;
  {Bigongiari}, G.; {Brandt}, T.J.; {Childers}, J.T.; {Conklin}, N.B.; {Coutu},
  S.; et~al.
\newblock {Measurements of the Relative Abundances of High-energy Cosmic-ray
  Nuclei in the TeV/Nucleon Region}.
\newblock {\em \apj} {\bf 2010}, {\em 715},~1400--1407.
\newblock
  doi:{\changeurlcolor{black}\href{https://doi.org/10.1088/0004-637X/715/2/1400}{\detokenize{10.1088/0004-637X/715/2/1400}}}.

\bibitem[{Freier} \em{et~al.}(1980){Freier}, {Young}, and
  {Waddington}]{1980ApJ...240L..53F}
{Freier}, P.S.; {Young}, J.S.; {Waddington}, C.J.
\newblock {The neutron-rich isotopes of cosmic-ray neon and magnesium}.
\newblock {\em \apjl} {\bf 1980}, {\em 240},~L53--L58.
\newblock
  doi:{\changeurlcolor{black}\href{https://doi.org/10.1086/183322}{\detokenize{10.1086/183322}}}.

\bibitem[{Young} \em{et~al.}(1981){Young}, {Freier}, {Waddington}, {Brewster},
  and {Fickle}]{1981ApJ...246.1014Y}
{Young}, J.S.; {Freier}, P.S.; {Waddington}, C.J.; {Brewster}, N.R.; {Fickle},
  R.K.
\newblock {The elemental and isotopic composition of cosmic rays - Silicon to
  nickel}.
\newblock {\em \apj} {\bf 1981}, {\em 246},~1014--1030.
\newblock
  doi:{\changeurlcolor{black}\href{https://doi.org/10.1086/158997}{\detokenize{10.1086/158997}}}.

\bibitem[{Gibner} \em{et~al.}(1992){Gibner}, {Mewaldt}, {Schindler}, {Stone},
  and {Webber}]{1992ApJ...391L..89G}
{Gibner}, P.S.; {Mewaldt}, R.A.; {Schindler}, S.M.; {Stone}, E.C.; {Webber},
  W.R.
\newblock {The isotopic composition of cosmic-ray B, C, N, and O - Evidence for
  an overabundance of O-18}.
\newblock {\em \apjl} {\bf 1992}, {\em 391},~L89--L92.
\newblock
  doi:{\changeurlcolor{black}\href{https://doi.org/10.1086/186405}{\detokenize{10.1086/186405}}}.

\bibitem[{Reimer} \em{et~al.}(1998){Reimer}, {Menn}, {Hof}, {Simon}, {Davis},
  {Labrador}, {Mewaldt}, {Schindler}, {Barbier}, {Christian}, {Krombel},
  {Mitchell}, {Ormes}, {Streitmatter}, {Golden}, {Stochaj}, {Webber}, and
  {Rasmussen}]{1998ApJ...496..490R}
{Reimer}, O.; {Menn}, W.; {Hof}, M.; {Simon}, M.; {Davis}, A.J.; {Labrador},
  A.W.; {Mewaldt}, R.A.; {Schindler}, S.M.; {Barbier}, L.M.; {Christian}, E.R.;
  {Krombel}, K.E.; {Mitchell}, J.W.; {Ormes}, J.F.; {Streitmatter}, R.E.;
  {Golden}, R.L.; {Stochaj}, S.J.; {Webber}, W.R.; {Rasmussen}, I.L.
\newblock {The Cosmic-Ray 3He/ 4He Ratio from 200 MeV per Nucleon to 3.7 GeV
  per Nucleon}.
\newblock {\em \apj} {\bf 1998}, {\em 496},~490.
\newblock
  doi:{\changeurlcolor{black}\href{https://doi.org/10.1086/305358}{\detokenize{10.1086/305358}}}.

\bibitem[{Menn} \em{et~al.}(2000){Menn}, {Hof}, {Reimer}, {Simon}, {Davis},
  {Labrador}, {Mewaldt}, {Schindler}, {Barbier}, {Christian}, and
  et~al]{2000ApJ...533..281M}
{Menn}, W.; {Hof}, M.; {Reimer}, O.; {Simon}, M.; {Davis}, A.J.; {Labrador},
  A.W.; {Mewaldt}, R.A.; {Schindler}, S.M.; {Barbier}, L.M.; {Christian}, E.R.;
  et~al.
\newblock {The Absolute Flux of Protons and Helium at the Top of the Atmosphere
  Using IMAX}.
\newblock {\em \apj} {\bf 2000}, {\em 533},~281--297.
\newblock
  doi:{\changeurlcolor{black}\href{https://doi.org/10.1086/308645}{\detokenize{10.1086/308645}}}.

\bibitem[{de Nolfo} \em{et~al.}(2000){de Nolfo}, {Barbier}, {Christian},
  {Davis}, {Golden}, {Hof}, {Krombel}, {Labrador}, {Menn}, {Mewaldt},
  {Mitchell}, {Ormes}, {Rasmussen}, {Reimer}, {Schindler}, {Simon}, {Stochaj},
  {Streitmatter}, and {Webber}]{2000AIPC..528..425D}
{de Nolfo}, G.A.; {Barbier}, L.M.; {Christian}, E.R.; {Davis}, A.J.; {Golden},
  R.L.; {Hof}, M.; {Krombel}, K.E.; {Labrador}, A.W.; {Menn}, W.; {Mewaldt},
  R.A.; {Mitchell}, J.W.; {Ormes}, J.F.; {Rasmussen}, I.L.; {Reimer}, O.;
  {Schindler}, S.M.; {Simon}, M.; {Stochaj}, S.J.; {Streitmatter}, R.E.;
  {Webber}, W.R.
\newblock {A measurement of cosmic ray deuterium from 0.5-2.9 GeV/nucleon}.
\newblock  Acceleration and Transport of Energetic Particles Observed in the
  Heliosphere; {Mewaldt}, R.A.; {Jokipii}, J.R.; {Lee}, M.A.; {M{\"o}bius}, E.;
  {Zurbuchen}, T.H., Eds.,  2000, Vol. 528, {\em American Institute of Physics
  Conference Series}, pp. 425--428.
\newblock
  doi:{\changeurlcolor{black}\href{https://doi.org/10.1063/1.1324352}{\detokenize{10.1063/1.1324352}}}.

\bibitem[{Tarle} \em{et~al.}(1979{\natexlab{a}}){Tarle}, {Ahlen}, and
  {Cartwright}]{1979ApJ...230..607T}
{Tarle}, G.; {Ahlen}, S.P.; {Cartwright}, B.G.
\newblock {Cosmic ray isotope abundances from chromium to nickel}.
\newblock {\em \apj} {\bf 1979}, {\em 230},~607--620.
\newblock
  doi:{\changeurlcolor{black}\href{https://doi.org/10.1086/157119}{\detokenize{10.1086/157119}}}.

\bibitem[{Tarle} \em{et~al.}(1979{\natexlab{b}}){Tarle}, {Ahlen}, {Cartwright},
  and {Solarz}]{1979ApJ...232L.161T}
{Tarle}, G.; {Ahlen}, S.P.; {Cartwright}, B.G.; {Solarz}, M.
\newblock {A measurement of the cosmic-ray source abundance of Ca-40}.
\newblock {\em \apjl} {\bf 1979}, {\em 232},~L161--L164.
\newblock
  doi:{\changeurlcolor{black}\href{https://doi.org/10.1086/183056}{\detokenize{10.1086/183056}}}.

\bibitem[{Hams} \em{et~al.}(2004){Hams}, {Barbier}, {Bremerich}, {Christian},
  {de Nolfo}, {Geier}, {G{\"o}bel}, {Gupta}, {Hof}, {Menn}, and
  et~al]{2004ApJ...611..892H}
{Hams}, T.; {Barbier}, L.M.; {Bremerich}, M.; {Christian}, E.R.; {de Nolfo},
  G.A.; {Geier}, S.; {G{\"o}bel}, H.; {Gupta}, S.K.; {Hof}, M.; {Menn}, W.;
  et~al.
\newblock {Measurement of the Abundance of Radioactive $^{10}$Be and Other
  Light Isotopes in Cosmic Radiation up to 2 GeV Nucleon$^{-1}$ with the
  Balloon-borne Instrument ISOMAX}.
\newblock {\em \apj} {\bf 2004}, {\em 611},~892--905.
\newblock
  doi:{\changeurlcolor{black}\href{https://doi.org/10.1086/422384}{\detokenize{10.1086/422384}}}.

\bibitem[{Asakimori} \em{et~al.}(1998){Asakimori}, {Burnett}, {Cherry},
  {Chevli}, {Christ}, {Dake}, {Derrickson}, {Fountain}, {Fuki}, {Gregory}, and
  et~al]{1998ApJ...502..278A}
{Asakimori}, K.; {Burnett}, T.H.; {Cherry}, M.L.; {Chevli}, K.; {Christ}, M.J.;
  {Dake}, S.; {Derrickson}, J.H.; {Fountain}, W.F.; {Fuki}, M.; {Gregory},
  J.C.; et~al.
\newblock {Cosmic-Ray Proton and Helium Spectra: Results from the JACEE
  Experiment}.
\newblock {\em \apj} {\bf 1998}, {\em 502},~278.
\newblock
  doi:{\changeurlcolor{black}\href{https://doi.org/10.1086/305882}{\detokenize{10.1086/305882}}}.

\bibitem[{Seo} \em{et~al.}(1991){Seo}, {Ormes}, {Streitmatter}, {Stochaj},
  {Jones}, {Stephens}, and {Bowen}]{1991ApJ...378..763S}
{Seo}, E.S.; {Ormes}, J.F.; {Streitmatter}, R.E.; {Stochaj}, S.J.; {Jones},
  W.V.; {Stephens}, S.A.; {Bowen}, T.
\newblock {Measurement of cosmic-ray proton and helium spectra during the 1987
  solar minimum}.
\newblock {\em \apj} {\bf 1991}, {\em 378},~763--772.
\newblock
  doi:{\changeurlcolor{black}\href{https://doi.org/10.1086/170477}{\detokenize{10.1086/170477}}}.

\bibitem[{Webber} \em{et~al.}(1991){Webber}, {Golden}, {Stochaj}, {Ormes}, and
  {Strittmatter}]{1991ApJ...380..230W}
{Webber}, W.R.; {Golden}, R.L.; {Stochaj}, S.J.; {Ormes}, J.F.; {Strittmatter},
  R.E.
\newblock {A measurement of the cosmic-ray H-2 and He-3 spectra and H-2/He-4
  and He-3/He-4 ratios in 1989}.
\newblock {\em \apj} {\bf 1991}, {\em 380},~230--234.
\newblock
  doi:{\changeurlcolor{black}\href{https://doi.org/10.1086/170578}{\detokenize{10.1086/170578}}}.

\bibitem[{Bellotti} \em{et~al.}(1999){Bellotti}, {Cafagna}, {Circella}, {de
  Marzo}, {Golden}, {Stochaj}, {de Pascale}, {Morselli}, {Picozza}, {Stephens},
  {Hof}, and et~al]{1999PhRvD..60e2002B}
{Bellotti}, R.; {Cafagna}, F.; {Circella}, M.; {de Marzo}, C.N.; {Golden},
  R.L.; {Stochaj}, S.J.; {de Pascale}, M.P.; {Morselli}, A.; {Picozza}, P.;
  {Stephens}, S.A.; {Hof}, M.; et~al.
\newblock {Balloon measurements of cosmic ray muon spectra in the atmosphere
  along with those of primary protons and helium nuclei over midlatitude}.
\newblock {\em \prd} {\bf 1999}, {\em 60},~052002,
  \href{http://xxx.lanl.gov/abs/arXiv:hep-ex/9905012}{{\normalfont
  [arXiv:hep-ex/9905012]}}.
\newblock
  doi:{\changeurlcolor{black}\href{https://doi.org/10.1103/PhysRevD.60.052002}{\detokenize{10.1103/PhysRevD.60.052002}}}.

\bibitem[{Zatsepin} \em{et~al.}(1993){Zatsepin}, {Zamchalova}, {Varkovitskaya},
  {Sokolskaya}, {Sazhina}, and {Lazareva}]{1993ICRC....2...13Z}
{Zatsepin}, V.I.; {Zamchalova}, E.A.; {Varkovitskaya}, A.Y.; {Sokolskaya},
  N.V.; {Sazhina}, G.P.; {Lazareva}, T.V.
\newblock {Energy Spectra of Primary Protons and Other Nuclei in Energy Region
  10-100 TeV/nucleus}.
\newblock  International Cosmic Ray Conference,  1993, Vol.~2, {\em
  International Cosmic Ray Conference}, p.~13.

\bibitem[{Diehl} \em{et~al.}(2003){Diehl}, {Ellithorpe}, {M{\"u}ller}, and
  {Swordy}]{2003APh....18..487D}
{Diehl}, E.; {Ellithorpe}, D.; {M{\"u}ller}, D.; {Swordy}, S.P.
\newblock {The energy spectrum of cosmic-ray protons and helium near 100 GeV}.
\newblock {\em Astroparticle Physics} {\bf 2003}, {\em 18},~487--500.
\newblock
  doi:{\changeurlcolor{black}\href{https://doi.org/10.1016/S0927-6505(02)00157-3}{\detokenize{10.1016/S0927-6505(02)00157-3}}}.

\bibitem[{Ahlen} \em{et~al.}(2000){Ahlen}, {Greene}, {Loomba}, {Mitchell},
  {Bower}, {Heinz}, {Mufson}, {Musser}, {Pitts}, {Spiczak}, and
  et~al]{2000ApJ...534..757A}
{Ahlen}, S.P.; {Greene}, N.R.; {Loomba}, D.; {Mitchell}, J.W.; {Bower}, C.R.;
  {Heinz}, R.M.; {Mufson}, S.L.; {Musser}, J.; {Pitts}, J.J.; {Spiczak}, G.M.;
  et~al.
\newblock {Measurement of the Isotopic Composition of Cosmic-Ray Helium,
  Lithium, Beryllium, and Boron up to 1700 MEV per Atomic Mass Unit}.
\newblock {\em \apj} {\bf 2000}, {\em 534},~757--769.
\newblock
  doi:{\changeurlcolor{black}\href{https://doi.org/10.1086/308762}{\detokenize{10.1086/308762}}}.

\bibitem[{Wefel} \em{et~al.}(1995){Wefel}, {Ahlen}, {Beatty}, {Bower}, {Clem},
  {Ficenec}, {Greene}, {Guzik}, {Heinz}, {Lijowski}, {Loomba}, {McKee},
  {Mitchell}, {Mufson}, {Musser}, {Nutter}, {Spiczak}, {Tarle}, and
  {Tomasch}]{1995ICRC....2..630W}
{Wefel}, J.P.; {Ahlen}, S.P.; {Beatty}, J.J.; {Bower}, C.R.; {Clem}, J.;
  {Ficenec}, D.J.; {Greene}, N.; {Guzik}, T.G.; {Heinz}, R.M.; {Lijowski}, M.;
  {Loomba}, D.; {McKee}, S.; {Mitchell}, J.W.; {Mufson}, S.L.; {Musser}, J.;
  {Nutter}, S.; {Spiczak}, G.M.; {Tarle}, G.; {Tomasch}, A.
\newblock {Measurements of Cosmic Ray Helium During the 1991 Solar Maximum}.
\newblock  1995, Vol.~2, p. 630.

\bibitem[{Gahbauer} \em{et~al.}(2004){Gahbauer}, {Hermann}, {H{\"o}randel},
  {M{\"u}ller}, and {Radu}]{2004ApJ...607..333G}
{Gahbauer}, F.; {Hermann}, G.; {H{\"o}randel}, J.R.; {M{\"u}ller}, D.; {Radu},
  A.A.
\newblock {A New Measurement of the Intensities of the Heavy Primary Cosmic-Ray
  Nuclei around 1 TeV amu$^{-1}$}.
\newblock {\em \apj} {\bf 2004}, {\em 607},~333--341.
\newblock
  doi:{\changeurlcolor{black}\href{https://doi.org/10.1086/383304}{\detokenize{10.1086/383304}}}.

\bibitem[{Ave} \em{et~al.}(2008){Ave}, {Boyle}, {Gahbauer}, {H{\"o}ppner},
  {H{\"o}randel}, {Ichimura}, {M{\"u}ller}, and
  {Romero-Wolf}]{2008ApJ...678..262A}
{Ave}, M.; {Boyle}, P.J.; {Gahbauer}, F.; {H{\"o}ppner}, C.; {H{\"o}randel},
  J.R.; {Ichimura}, M.; {M{\"u}ller}, D.; {Romero-Wolf}, A.
\newblock {Composition of Primary Cosmic-Ray Nuclei at High Energies}.
\newblock {\em \apj} {\bf 2008}, {\em 678},~262--273,
  \href{http://xxx.lanl.gov/abs/0801.0582}{{\normalfont [0801.0582]}}.
\newblock
  doi:{\changeurlcolor{black}\href{https://doi.org/10.1086/529424}{\detokenize{10.1086/529424}}}.

\bibitem[{Obermeier} \em{et~al.}(2011){Obermeier}, {Ave}, {Boyle},
  {H{\"o}ppner}, {H{\"o}randel}, and {M{\"u}ller}]{2011ApJ...742...14O}
{Obermeier}, A.; {Ave}, M.; {Boyle}, P.; {H{\"o}ppner}, C.; {H{\"o}randel}, J.;
  {M{\"u}ller}, D.
\newblock {Energy Spectra of Primary and Secondary Cosmic-Ray Nuclei Measured
  with TRACER}.
\newblock {\em \apj} {\bf 2011}, {\em 742},~14,
  \href{http://xxx.lanl.gov/abs/1108.4838}{{\normalfont
  [arXiv:astro-ph.HE/1108.4838]}}.
\newblock
  doi:{\changeurlcolor{black}\href{https://doi.org/10.1088/0004-637X/742/1/14}{\detokenize{10.1088/0004-637X/742/1/14}}}.

\bibitem[{George} \em{et~al.}(2009){George}, {Lave}, {Wiedenbeck}, {Binns},
  {Cummings}, {Davis}, {de Nolfo}, {Hink}, {Israel}, {Leske}, {Mewaldt},
  {Scott}, {Stone}, {von Rosenvinge}, and {Yanasak}]{2009ApJ...698.1666G}
{George}, J.S.; {Lave}, K.A.; {Wiedenbeck}, M.E.; {Binns}, W.R.; {Cummings},
  A.C.; {Davis}, A.J.; {de Nolfo}, G.A.; {Hink}, P.L.; {Israel}, M.H.; {Leske},
  R.A.; {Mewaldt}, R.A.; {Scott}, L.M.; {Stone}, E.C.; {von Rosenvinge}, T.T.;
  {Yanasak}, N.E.
\newblock {Elemental Composition and Energy Spectra of Galactic Cosmic Rays
  During Solar Cycle 23}.
\newblock {\em \apj} {\bf 2009}, {\em 698},~1666--1681.
\newblock
  doi:{\changeurlcolor{black}\href{https://doi.org/10.1088/0004-637X/698/2/1666}{\detokenize{10.1088/0004-637X/698/2/1666}}}.

\bibitem[{Lave} \em{et~al.}(2013){Lave}, {Wiedenbeck}, {Binns}, {Christian},
  {Cummings}, {Davis}, {de Nolfo}, {Israel}, {Leske}, {Mewaldt}, and
  et~al]{2013ApJ...770..117L}
{Lave}, K.A.; {Wiedenbeck}, M.E.; {Binns}, W.R.; {Christian}, E.R.; {Cummings},
  A.C.; {Davis}, A.J.; {de Nolfo}, G.A.; {Israel}, M.H.; {Leske}, R.A.;
  {Mewaldt}, R.A.; et~al.
\newblock {Galactic Cosmic-Ray Energy Spectra and Composition during the
  2009-2010 Solar Minimum Period}.
\newblock {\em \apj} {\bf 2013}, {\em 770},~117.
\newblock
  doi:{\changeurlcolor{black}\href{https://doi.org/10.1088/0004-637X/770/2/117}{\detokenize{10.1088/0004-637X/770/2/117}}}.

\bibitem[{Wiedenbeck} \em{et~al.}(1999){Wiedenbeck}, {Binns}, {Christian},
  {Cummings}, {Dougherty}, {Hink}, {Klarmann}, {Leske}, {Lijowski}, {Mewaldt},
  {Stone}, {Thayer}, {von Rosenvinge}, and {Yanasak}]{1999ApJ...523L..61W}
{Wiedenbeck}, M.E.; {Binns}, W.R.; {Christian}, E.R.; {Cummings}, A.C.;
  {Dougherty}, B.L.; {Hink}, P.L.; {Klarmann}, J.; {Leske}, R.A.; {Lijowski},
  M.; {Mewaldt}, R.A.; {Stone}, E.C.; {Thayer}, M.R.; {von Rosenvinge}, T.T.;
  {Yanasak}, N.E.
\newblock {Constraints on the Time Delay between Nucleosynthesis and Cosmic-Ray
  Acceleration from Observations of \^{}59NI and \^{}59CO}.
\newblock {\em \apjl} {\bf 1999}, {\em 523},~L61--L64.
\newblock
  doi:{\changeurlcolor{black}\href{https://doi.org/10.1086/312242}{\detokenize{10.1086/312242}}}.

\bibitem[{Yanasak} \em{et~al.}(2001){Yanasak}, {Wiedenbeck}, {Mewaldt},
  {Davis}, {Cummings}, {George}, {Leske}, {Stone}, {Christian}, {von
  Rosenvinge}, and et~al]{2001ApJ...563..768Y}
{Yanasak}, N.E.; {Wiedenbeck}, M.E.; {Mewaldt}, R.A.; {Davis}, A.J.;
  {Cummings}, A.C.; {George}, J.S.; {Leske}, R.A.; {Stone}, E.C.; {Christian},
  E.R.; {von Rosenvinge}, T.T.; et~al.
\newblock {Measurement of the Secondary Radionuclides $^{10}$Be, $^{26}$Al,
  $^{36}$Cl, $^{54}$Mn, and $^{14}$C and Implications for the Galactic
  Cosmic-Ray Age}.
\newblock {\em \apj} {\bf 2001}, {\em 563},~768--792.
\newblock
  doi:{\changeurlcolor{black}\href{https://doi.org/10.1086/323842}{\detokenize{10.1086/323842}}}.

\bibitem[{Ogliore} \em{et~al.}(2009){Ogliore}, {Stone}, {Leske}, {Mewaldt},
  {Wiedenbeck}, {Binns}, {Israel}, {von Rosenvinge}, {de Nolfo}, and
  {Moskalenko}]{2009ApJ...695..666O}
{Ogliore}, R.C.; {Stone}, E.C.; {Leske}, R.A.; {Mewaldt}, R.A.; {Wiedenbeck},
  M.E.; {Binns}, W.R.; {Israel}, M.H.; {von Rosenvinge}, T.T.; {de Nolfo},
  G.A.; {Moskalenko}, I.V.
\newblock {The Phosphorus, Sulfur, Argon, and Calcium Isotopic Composition of
  the Galactic Cosmic Ray Source}.
\newblock {\em \apj} {\bf 2009}, {\em 695},~666--678.
\newblock
  doi:{\changeurlcolor{black}\href{https://doi.org/10.1088/0004-637X/695/1/666}{\detokenize{10.1088/0004-637X/695/1/666}}}.

\bibitem[{Binns} \em{et~al.}(2005){Binns}, {Wiedenbeck}, {Arnould}, {Cummings},
  {George}, {Goriely}, {Israel}, {Leske}, {Mewaldt}, {Meynet}, {Scott},
  {Stone}, and {von Rosenvinge}]{2005ApJ...634..351B}
{Binns}, W.R.; {Wiedenbeck}, M.E.; {Arnould}, M.; {Cummings}, A.C.; {George},
  J.S.; {Goriely}, S.; {Israel}, M.H.; {Leske}, R.A.; {Mewaldt}, R.A.;
  {Meynet}, G.; {Scott}, L.M.; {Stone}, E.C.; {von Rosenvinge}, T.T.
\newblock {Cosmic-Ray Neon, Wolf-Rayet Stars, and the Superbubble Origin of
  Galactic Cosmic Rays}.
\newblock {\em \apj} {\bf 2005}, {\em 634},~351--364,
  \href{http://xxx.lanl.gov/abs/astro-ph/0508398}{{\normalfont
  [astro-ph/0508398]}}.
\newblock
  doi:{\changeurlcolor{black}\href{https://doi.org/10.1086/496959}{\detokenize{10.1086/496959}}}.

\bibitem[{Binns} \em{et~al.}(2001){Binns}, {Wiedenbeck}, {Christian},
  {Cummings}, {George}, {Hink}, {Israel}, {Klarmann}, {Leske}, {Lijowski},
  {Mewaldt}, {Stone}, {von Rosenvinge}, and {Yanasak}]{2001AdSpR..27..767B}
{Binns}, W.R.; {Wiedenbeck}, M.E.; {Christian}, E.R.; {Cummings}, A.C.;
  {George}, J.S.; {Hink}, P.L.; {Israel}, M.H.; {Klarmann}, J.; {Leske}, R.A.;
  {Lijowski}, M.; {Mewaldt}, R.A.; {Stone}, E.C.; {von Rosenvinge}, T.T.;
  {Yanasak}, N.E.
\newblock {Galactic cosmic ray neon isotopic abundances measured by the cosmic
  ray isotope spectrometer (cris) on ace}.
\newblock {\em Advances in Space Research} {\bf 2001}, {\em 27},~767--772.
\newblock
  doi:{\changeurlcolor{black}\href{https://doi.org/10.1016/S0273-1177(01)00119-3}{\detokenize{10.1016/S0273-1177(01)00119-3}}}.

\bibitem[{de Nolfo} \em{et~al.}(2006){de Nolfo}, {Moskalenko}, {Binns},
  {Christian}, {Cummings}, {Davis}, {George}, {Hink}, {Israel}, {Leske},
  {Lijowski}, {Mewaldt}, {Stone}, {Strong}, {von Rosenvinge}, {Wiedenbeck}, and
  {Yanasak}]{2006AdSpR..38.1558D}
{de Nolfo}, G.A.; {Moskalenko}, I.V.; {Binns}, W.R.; {Christian}, E.R.;
  {Cummings}, A.C.; {Davis}, A.J.; {George}, J.S.; {Hink}, P.L.; {Israel},
  M.H.; {Leske}, R.A.; {Lijowski}, M.; {Mewaldt}, R.A.; {Stone}, E.C.;
  {Strong}, A.W.; {von Rosenvinge}, T.T.; {Wiedenbeck}, M.E.; {Yanasak}, N.E.
\newblock {Observations of the Li, Be, and B isotopes and constraints on
  cosmic-ray propagation}.
\newblock {\em Advances in Space Research} {\bf 2006}, {\em 38},~1558--1564,
  \href{http://xxx.lanl.gov/abs/arXiv:astro-ph/0611301}{{\normalfont
  [arXiv:astro-ph/0611301]}}.
\newblock
  doi:{\changeurlcolor{black}\href{https://doi.org/10.1016/j.asr.2006.09.008}{\detokenize{10.1016/j.asr.2006.09.008}}}.

\bibitem[{Alcaraz} \em{et~al.}(2000{\natexlab{a}}){Alcaraz}, {Alpat},
  {Ambrosi}, {Anderhub}, {Ao}, {Arefiev}, {Azzarello}, {Babucci}, {Baldini},
  {Basile}, and et~al]{2000PhLB..490...27A}
{Alcaraz}, J.; {Alpat}, B.; {Ambrosi}, G.; {Anderhub}, H.; {Ao}, L.; {Arefiev},
  A.; {Azzarello}, P.; {Babucci}, E.; {Baldini}, L.; {Basile}, M.; et~al.
\newblock {Cosmic protons}.
\newblock {\em Physics Letters B} {\bf 2000}, {\em 490},~27--35.
\newblock
  doi:{\changeurlcolor{black}\href{https://doi.org/10.1016/S0370-2693(00)00970-9}{\detokenize{10.1016/S0370-2693(00)00970-9}}}.

\bibitem[{Alcaraz} \em{et~al.}(2000{\natexlab{b}}){Alcaraz}, {Alpat},
  {Ambrosi}, {Anderhub}, {Ao}, {Arefiev}, {Azzarello}, {Babucci}, {Baldini},
  {Basile}, and et~al]{2000PhLB..494..193A}
{Alcaraz}, J.; {Alpat}, B.; {Ambrosi}, G.; {Anderhub}, H.; {Ao}, L.; {Arefiev},
  A.; {Azzarello}, P.; {Babucci}, E.; {Baldini}, L.; {Basile}, M.; et~al.
\newblock {Helium in near Earth orbit}.
\newblock {\em Physics Letters B} {\bf 2000}, {\em 494},~193--202.
\newblock
  doi:{\changeurlcolor{black}\href{https://doi.org/10.1016/S0370-2693(00)01193-X}{\detokenize{10.1016/S0370-2693(00)01193-X}}}.

\bibitem[{Xiong} \em{et~al.}(2003){Xiong}, {Chen}, {Yang}, {Yang}, {Chen},
  {Chen}, {L{\"u}}, {Zhuang}, and {Tang}]{2003JHEP...11..048X}
{Xiong}, Z.; {Chen}, H.; {Yang}, C.; {Yang}, M.; {Chen}, G.; {Chen}, G.;
  {L{\"u}}, Y.; {Zhuang}, H.; {Tang}, X.
\newblock {Measurement of $^{3}$He/$^{4}$He ratio in cosmic rays with the AMS
  experiment}.
\newblock {\em Journal of High Energy Physics} {\bf 2003}, {\em 11},~48.
\newblock
  doi:{\changeurlcolor{black}\href{https://doi.org/10.1088/1126-6708/2003/11/048}{\detokenize{10.1088/1126-6708/2003/11/048}}}.

\bibitem[{Aguilar} \em{et~al.}(2010){Aguilar}, {Alcaraz}, {Allaby}, {Alpat},
  {Ambrosi}, {Anderhub}, {Ao}, {Arefiev}, {Arruda}, {Azzarello}, {Basile}, and
  et~al]{2010ApJ...724..329A}
{Aguilar}, M.; {Alcaraz}, J.; {Allaby}, J.; {Alpat}, B.; {Ambrosi}, G.;
  {Anderhub}, H.; {Ao}, L.; {Arefiev}, A.; {Arruda}, L.; {Azzarello}, P.;
  {Basile}, M.; et~al.
\newblock {Relative Composition and Energy Spectra of Light Nuclei in Cosmic
  Rays: Results from AMS-01}.
\newblock {\em \apj} {\bf 2010}, {\em 724},~329--340.
\newblock
  doi:{\changeurlcolor{black}\href{https://doi.org/10.1088/0004-637X/724/1/329}{\detokenize{10.1088/0004-637X/724/1/329}}}.

\bibitem[{Aguilar} \em{et~al.}(2011){Aguilar}, {Alcaraz}, {Allaby}, {Alpat},
  {Ambrosi}, {Anderhub}, {Ao}, {Arefiev}, {Arruda}, {Azzarello}, and
  et~al]{2011ApJ...736..105A}
{Aguilar}, M.; {Alcaraz}, J.; {Allaby}, J.; {Alpat}, B.; {Ambrosi}, G.;
  {Anderhub}, H.; {Ao}, L.; {Arefiev}, A.; {Arruda}, L.; {Azzarello}, P.;
  et~al.
\newblock {Isotopic Composition of Light Nuclei in Cosmic Rays: Results from
  AMS-01}.
\newblock {\em \apj} {\bf 2011}, {\em 736},~105,
  \href{http://xxx.lanl.gov/abs/1106.2269}{{\normalfont
  [arXiv:astro-ph.HE/1106.2269]}}.
\newblock
  doi:{\changeurlcolor{black}\href{https://doi.org/10.1088/0004-637X/736/2/105}{\detokenize{10.1088/0004-637X/736/2/105}}}.

\bibitem[{Battiston}(2014)]{2014PDU.....4....6B}
{Battiston}, R.
\newblock {Precision measurements of e$^{+}$e$^{-}$ in Cosmic Ray with the
  Alpha Magnetic Spectrometer on the ISS}.
\newblock {\em Physics of the Dark Universe} {\bf 2014}, {\em 4},~6--9.
\newblock
  doi:{\changeurlcolor{black}\href{https://doi.org/10.1016/j.dark.2014.05.002}{\detokenize{10.1016/j.dark.2014.05.002}}}.

\bibitem[{Aguilar} \em{et~al.}(2016){Aguilar}, {Ali Cavasonza}, {Ambrosi},
  {Arruda}, {Attig}, {Aupetit}, {Azzarello}, {Bachlechner}, {Barao}, {Barrau},
  and et~al]{2016PhRvL.117w1102A}
{Aguilar}, M.; {Ali Cavasonza}, L.; {Ambrosi}, G.; {Arruda}, L.; {Attig}, N.;
  {Aupetit}, S.; {Azzarello}, P.; {Bachlechner}, A.; {Barao}, F.; {Barrau}, A.;
  et~al.
\newblock {Precision Measurement of the Boron to Carbon Flux Ratio in Cosmic
  Rays from 1.9 GV to 2.6 TV with the Alpha Magnetic Spectrometer on the
  International Space Station}.
\newblock {\em Physical Review Letters} {\bf 2016}, {\em 117},~231102.
\newblock
  doi:{\changeurlcolor{black}\href{https://doi.org/10.1103/PhysRevLett.117.231102}{\detokenize{10.1103/PhysRevLett.117.231102}}}.

\bibitem[{Aguilar} \em{et~al.}(2019){Aguilar}, {Ali Cavasonza}, {Ambrosi},
  {Arruda}, {Attig}, {Azzarello}, {Bachlechner}, {Barao}, {Barrau}, {Barrin},
  and et~al]{2019PhRvL.122d1102A}
{Aguilar}, M.; {Ali Cavasonza}, L.; {Ambrosi}, G.; {Arruda}, L.; {Attig}, N.;
  {Azzarello}, P.; {Bachlechner}, A.; {Barao}, F.; {Barrau}, A.; {Barrin}, L.;
  et~al.
\newblock {Towards Understanding the Origin of Cosmic-Ray Positrons}.
\newblock {\em \prl} {\bf 2019}, {\em 122},~041102.
\newblock
  doi:{\changeurlcolor{black}\href{https://doi.org/10.1103/PhysRevLett.122.041102}{\detokenize{10.1103/PhysRevLett.122.041102}}}.

\bibitem[{Swordy} \em{et~al.}(1990){Swordy}, {Mueller}, {Meyer}, {L'Heureux},
  and {Grunsfeld}]{1990ApJ...349..625S}
{Swordy}, S.P.; {Mueller}, D.; {Meyer}, P.; {L'Heureux}, J.; {Grunsfeld}, J.M.
\newblock {Relative abundances of secondary and primary cosmic rays at high
  energies}.
\newblock {\em \apj} {\bf 1990}, {\em 349},~625--633.
\newblock
  doi:{\changeurlcolor{black}\href{https://doi.org/10.1086/168349}{\detokenize{10.1086/168349}}}.

\bibitem[{Mueller} \em{et~al.}(1991){Mueller}, {Swordy}, {Meyer}, {L'Heureux},
  and {Grunsfeld}]{1991ApJ...374..356M}
{Mueller}, D.; {Swordy}, S.P.; {Meyer}, P.; {L'Heureux}, J.; {Grunsfeld}, J.M.
\newblock {Energy spectra and composition of primary cosmic rays}.
\newblock {\em \apj} {\bf 1991}, {\em 374},~356--365.
\newblock
  doi:{\changeurlcolor{black}\href{https://doi.org/10.1086/170125}{\detokenize{10.1086/170125}}}.

\bibitem[{Duvernois} \em{et~al.}(1996){Duvernois}, {Garcia-Munoz}, {Pyle},
  {Simpson}, and {Thayer}]{1996ApJ...466..457D}
{Duvernois}, M.A.; {Garcia-Munoz}, M.; {Pyle}, K.R.; {Simpson}, J.A.; {Thayer},
  M.R.
\newblock {The Isotopic Composition of Galactic Cosmic-Ray Elements from Carbon
  to Silicon: The Combined Release and Radiation Effects Satellite
  Investigation}.
\newblock {\em \apj} {\bf 1996}, {\em 466},~457.
\newblock
  doi:{\changeurlcolor{black}\href{https://doi.org/10.1086/177524}{\detokenize{10.1086/177524}}}.

\bibitem[{Clayton} \em{et~al.}(2000){Clayton}, {Guzik}, and
  {Wefel}]{2000SoPh..195..175C}
{Clayton}, E.G.; {Guzik}, T.G.; {Wefel}, J.P.
\newblock {CRRES Measurements of Energetic Helium During the 1990-1991 Solar
  Maximum}.
\newblock {\em \solphys} {\bf 2000}, {\em 195},~175--194.
\newblock
  doi:{\changeurlcolor{black}\href{https://doi.org/10.1023/A:1005251630568}{\detokenize{10.1023/A:1005251630568}}}.

\bibitem[{Stone}(1964)]{1964JGR....69.3939S}
{Stone}, E.C.
\newblock {A Measurement of the Primary Proton Flux from 10 to 130 Million
  Electron Volts}.
\newblock {\em \jgr} {\bf 1964}, {\em 69},~3939--3945.
\newblock
  doi:{\changeurlcolor{black}\href{https://doi.org/10.1029/JZ069i019p03939}{\detokenize{10.1029/JZ069i019p03939}}}.

\bibitem[{Marquardt} and {Heber}(2019)]{2019A&A...625A.153M}
{Marquardt}, J.; {Heber}, B.
\newblock {Galactic cosmic ray hydrogen spectra and radial gradients in the
  inner heliosphere measured by the HELIOS Experiment 6}.
\newblock {\em \aap} {\bf 2019}, {\em 625},~A153,
  \href{http://xxx.lanl.gov/abs/1905.01052}{{\normalfont
  [arXiv:physics.space-ph/1905.01052]}}.
\newblock
  doi:{\changeurlcolor{black}\href{https://doi.org/10.1051/0004-6361/201935413}{\detokenize{10.1051/0004-6361/201935413}}}.

\bibitem[{K{\"u}hl} \em{et~al.}(2016){K{\"u}hl}, {G{\'o}mez-Herrero}, and
  {Heber}]{2016SoPh..291..965K}
{K{\"u}hl}, P.; {G{\'o}mez-Herrero}, R.; {Heber}, B.
\newblock {Annual Cosmic Ray Spectra from 250 MeV up to 1.6 GeV from 1995 -
  2014 Measured with the Electron Proton Helium Instrument onboard SOHO}.
\newblock {\em \solphys} {\bf 2016}, {\em 291},~965--974,
  \href{http://xxx.lanl.gov/abs/1603.00676}{{\normalfont
  [arXiv:physics.space-ph/1603.00676]}}.
\newblock
  doi:{\changeurlcolor{black}\href{https://doi.org/10.1007/s11207-016-0879-0}{\detokenize{10.1007/s11207-016-0879-0}}}.

\bibitem[{Bryant} \em{et~al.}(1962){Bryant}, {Cline}, {Desai}, and
  {McDonald}]{1962JGR....67.4983B}
{Bryant}, D.A.; {Cline}, T.L.; {Desai}, U.D.; {McDonald}, F.B.
\newblock {Explorer 12 Observations of Solar Cosmic Rays and Energetic Storm
  Particles after the Solar Flare of September 28, 1961}.
\newblock {\em \jgr} {\bf 1962}, {\em 67},~4983--5000.
\newblock
  doi:{\changeurlcolor{black}\href{https://doi.org/10.1029/JZ067i013p04983}{\detokenize{10.1029/JZ067i013p04983}}}.

\bibitem[{Durgaprasad} \em{et~al.}(1970){Durgaprasad}, {Fichtel}, {Guss},
  {Reames}, {O'dell}, {Shapiro}, {Silberberg}, {Stiller}, and
  {Tsao}]{1970PhRvD...1.1021D}
{Durgaprasad}, N.; {Fichtel}, C.E.; {Guss}, D.E.; {Reames}, D.V.; {O'dell},
  F.W.; {Shapiro}, M.M.; {Silberberg}, R.; {Stiller}, B.; {Tsao}, C.H.
\newblock {Chemical Composition of Relativistic Cosmic Rays Detected above the
  Atmosphere}.
\newblock {\em \prd} {\bf 1970}, {\em 1},~1021--1028.
\newblock
  doi:{\changeurlcolor{black}\href{https://doi.org/10.1103/PhysRevD.1.1021}{\detokenize{10.1103/PhysRevD.1.1021}}}.

\bibitem[{Ferrando} \em{et~al.}(1988){Ferrando}, {Engelmann}, {Goret},
  {Koch-Miramond}, and {Petrou}]{1988A&A...193...69F}
{Ferrando}, P.; {Engelmann}, J.J.; {Goret}, P.; {Koch-Miramond}, L.; {Petrou},
  N.
\newblock {Measurement of the isotopic composition of cosmic-rays at 3 GeV/n
  using a new geomagnetic method}.
\newblock {\em \aap} {\bf 1988}, {\em 193},~69--80.

\bibitem[{Engelmann} \em{et~al.}(1990){Engelmann}, {Ferrando}, {Soutoul},
  {Goret}, and {Juliusson}]{1990A&A...233...96E}
{Engelmann}, J.J.; {Ferrando}, P.; {Soutoul}, A.; {Goret}, P.; {Juliusson}, E.
\newblock {Charge composition and energy spectra of cosmic-ray nuclei for
  elements from Be to NI - Results from HEAO-3-C2}.
\newblock {\em \aap} {\bf 1990}, {\em 233},~96--111.

\bibitem[{Gloeckler}(1965)]{1965JGR....70.5333G}
{Gloeckler}, G.
\newblock {Solar Modulation of the Low-Energy Galactic Helium Spectrum as
  Observed on the Imp 1 Satellite}.
\newblock {\em \jgr} {\bf 1965}, {\em 70},~5333--5343.
\newblock
  doi:{\changeurlcolor{black}\href{https://doi.org/10.1029/JZ070i021p05333}{\detokenize{10.1029/JZ070i021p05333}}}.

\bibitem[{Fan} \em{et~al.}(1965){Fan}, {Gloeckler}, and
  {Simpson}]{1965JGR....70.3515F}
{Fan}, C.Y.; {Gloeckler}, G.; {Simpson}, J.A.
\newblock {Cosmic Radiation Helium Spectrum below 90 Mev per Nucleon Measured
  on Imp 1 Satellite}.
\newblock {\em \jgr} {\bf 1965}, {\em 70},~3515--3527.
\newblock
  doi:{\changeurlcolor{black}\href{https://doi.org/10.1029/JZ070i015p03515}{\detokenize{10.1029/JZ070i015p03515}}}.

\bibitem[{McDonald} and {Ludwig}(1964)]{1964PhRvL..13..783M}
{McDonald}, F.B.; {Ludwig}, G.H.
\newblock {Measurement of Low-Energy Primary Cosmic-Ray Protons on IMP-1
  Satellite}.
\newblock {\em Physical Review Letters} {\bf 1964}, {\em 13},~783--785.
\newblock
  doi:{\changeurlcolor{black}\href{https://doi.org/10.1103/PhysRevLett.13.783}{\detokenize{10.1103/PhysRevLett.13.783}}}.

\bibitem[{Fan} \em{et~al.}(1966{\natexlab{a}}){Fan}, {Gloeckler}, {Hsieh}, and
  {Simpson}]{1966PhRvL..16..813F}
{Fan}, C.Y.; {Gloeckler}, G.; {Hsieh}, K.C.; {Simpson}, J.A.
\newblock {Isotopic Abundances and Energy Spectra of He$^{3}$ and He$^{4}$
  Above 40 MeV per Nucleon from the Galaxy}.
\newblock {\em Physical Review Letters} {\bf 1966}, {\em 16},~813--817.
\newblock
  doi:{\changeurlcolor{black}\href{https://doi.org/10.1103/PhysRevLett.16.813}{\detokenize{10.1103/PhysRevLett.16.813}}}.

\bibitem[{Fan} \em{et~al.}(1966{\natexlab{b}}){Fan}, {Gloeckler}, and
  {Simpson}]{1966PhRvL..17..329F}
{Fan}, C.Y.; {Gloeckler}, G.; {Simpson}, J.A.
\newblock {Galactic Deuterium and its Energy Spectrum above 20 MeV per
  Nucleon}.
\newblock {\em Physical Review Letters} {\bf 1966}, {\em 17},~329--333.
\newblock
  doi:{\changeurlcolor{black}\href{https://doi.org/10.1103/PhysRevLett.17.329}{\detokenize{10.1103/PhysRevLett.17.329}}}.

\bibitem[{Hsieh}(1970)]{1970ApJ...159...61H}
{Hsieh}, K.C.
\newblock {Study of Solar Modulation of Low-Energy Cosmic Rays Using
  Differential Spectra of Protons, \^{}$\{$3$\}$He, and \^{}$\{$4$\}$He at E >
  100 MeV Per Nucleon during the Quiet Time in 1965 and 1967}.
\newblock {\em \apj} {\bf 1970}, {\em 159},~61.
\newblock
  doi:{\changeurlcolor{black}\href{https://doi.org/10.1086/150290}{\detokenize{10.1086/150290}}}.

\bibitem[{Hsieh} \em{et~al.}(1971){Hsieh}, {Mason}, and
  {Simpson}]{1971ApJ...166..221H}
{Hsieh}, K.C.; {Mason}, G.M.; {Simpson}, J.A.
\newblock {Cosmic-Ray \^{}$\{$2$\}$H from Satellite Measurements, 1965-1969}.
\newblock {\em \apj} {\bf 1971}, {\em 166},~221.
\newblock
  doi:{\changeurlcolor{black}\href{https://doi.org/10.1086/150951}{\detokenize{10.1086/150951}}}.

\bibitem[{Mason}(1972)]{1972ApJ...171..139M}
{Mason}, G.M.
\newblock {Interstellar Propagation of Galactic Cosmic-Ray Nuclei 2 <= Z<= 8 IN
  the Energy Range 10-1000 MeV Per Nucleon}.
\newblock {\em \apj} {\bf 1972}, {\em 171},~139.
\newblock
  doi:{\changeurlcolor{black}\href{https://doi.org/10.1086/151266}{\detokenize{10.1086/151266}}}.

\bibitem[{Garcia-Munoz} \em{et~al.}(1975){Garcia-Munoz}, {Mason}, and
  {Simpson}]{1975ApJ...202..265G}
{Garcia-Munoz}, M.; {Mason}, G.M.; {Simpson}, J.A.
\newblock {The anomalous He-4 component in the cosmic-ray spectrum at below
  approximately 50 MeV per nucleon during 1972-1974}.
\newblock {\em \apj} {\bf 1975}, {\em 202},~265--275.
\newblock
  doi:{\changeurlcolor{black}\href{https://doi.org/10.1086/153973}{\detokenize{10.1086/153973}}}.

\bibitem[{Teegarden} \em{et~al.}(1975){Teegarden}, {von Rosenvinge},
  {McDonald}, {Trainor}, and {Webber}]{1975ApJ...202..815T}
{Teegarden}, B.J.; {von Rosenvinge}, T.T.; {McDonald}, F.B.; {Trainor}, J.H.;
  {Webber}, W.R.
\newblock {Measurement of the fluxes of galactic cosmic-ray H-2 and He-3 in
  1972-1973}.
\newblock {\em \apj} {\bf 1975}, {\em 202},~815--822.
\newblock
  doi:{\changeurlcolor{black}\href{https://doi.org/10.1086/154036}{\detokenize{10.1086/154036}}}.

\bibitem[{Garcia-Munoz} \em{et~al.}(1979){Garcia-Munoz}, {Simpson}, and
  {Wefel}]{1979ApJ...232L..95G}
{Garcia-Munoz}, M.; {Simpson}, J.A.; {Wefel}, J.P.
\newblock {The isotopes of neon in the galactic cosmic rays}.
\newblock {\em \apjl} {\bf 1979}, {\em 232},~L95--L99.
\newblock
  doi:{\changeurlcolor{black}\href{https://doi.org/10.1086/183043}{\detokenize{10.1086/183043}}}.

\bibitem[{Mewaldt} \em{et~al.}(1976){Mewaldt}, {Stone}, and
  {Vogt}]{1976ApJ...206..616M}
{Mewaldt}, R.A.; {Stone}, E.C.; {Vogt}, R.E.
\newblock {The isotopic composition of hydrogen and helium in low-energy cosmic
  rays}.
\newblock {\em \apj} {\bf 1976}, {\em 206},~616--621.
\newblock
  doi:{\changeurlcolor{black}\href{https://doi.org/10.1086/154418}{\detokenize{10.1086/154418}}}.

\bibitem[{Garcia-Munoz} \em{et~al.}(1977{\natexlab{a}}){Garcia-Munoz}, {Mason},
  and {Simpson}]{1977ApJ...217..859G}
{Garcia-Munoz}, M.; {Mason}, G.M.; {Simpson}, J.A.
\newblock {The age of the galactic cosmic rays derived from the abundance of
  Be-10}.
\newblock {\em \apj} {\bf 1977}, {\em 217},~859--877.
\newblock
  doi:{\changeurlcolor{black}\href{https://doi.org/10.1086/155632}{\detokenize{10.1086/155632}}}.

\bibitem[{Garcia-Munoz} \em{et~al.}(1977{\natexlab{b}}){Garcia-Munoz}, {Mason},
  and {Simpson}]{1977ICRC....1..301G}
{Garcia-Munoz}, M.; {Mason}, G.M.; {Simpson}, J.A.
\newblock {The isotopic composition of galactic cosmic ray lithium, beryllium
  and boron}.
\newblock  1977, Vol.~1, pp. 301--306.

\bibitem[{Guzik}(1981)]{1981ApJ...244..695G}
{Guzik}, T.G.
\newblock {The low-energy galactic cosmic ray carbon, nitrogen, and oxygen
  isotopic composition}.
\newblock {\em \apj} {\bf 1981}, {\em 244},~695--710.
\newblock
  doi:{\changeurlcolor{black}\href{https://doi.org/10.1086/158747}{\detokenize{10.1086/158747}}}.

\bibitem[{Garcia-Munoz} \em{et~al.}(1981){Garcia-Munoz}, {Simpson}, and
  {Wefel}]{1981ICRC....2...72G}
{Garcia-Munoz}, M.; {Simpson}, J.A.; {Wefel}, J.P.
\newblock {The propagation lifetime of galactic cosmic rays determined from the
  measurement of the beryllium isotopes}.
\newblock  International Cosmic Ray Conference,  1981, Vol.~2, {\em
  International Cosmic Ray Conference}, pp. 72--75.

\bibitem[{Beatty} \em{et~al.}(1985){Beatty}, {Garcia-Munoz}, and
  {Simpson}]{1985ApJ...294..455B}
{Beatty}, J.J.; {Garcia-Munoz}, M.; {Simpson}, J.A.
\newblock {The cosmic-ray spectra of H-1, H-2, and He-4 as a test of the origin
  of the hydrogen superfluxes at solar minimum modulation}.
\newblock {\em \apj} {\bf 1985}, {\em 294},~455--462.
\newblock
  doi:{\changeurlcolor{black}\href{https://doi.org/10.1086/163311}{\detokenize{10.1086/163311}}}.

\bibitem[{Garcia-Munoz} \em{et~al.}(1987){Garcia-Munoz}, {Simpson}, {Guzik},
  {Wefel}, and {Margolis}]{1987ApJS...64..269G}
{Garcia-Munoz}, M.; {Simpson}, J.A.; {Guzik}, T.G.; {Wefel}, J.P.; {Margolis},
  S.H.
\newblock {Cosmic-ray propagation in the Galaxy and in the heliosphere - The
  path-length distribution at low energy}.
\newblock {\em \apjs} {\bf 1987}, {\em 64},~269--304.
\newblock
  doi:{\changeurlcolor{black}\href{https://doi.org/10.1086/191197}{\detokenize{10.1086/191197}}}.

\bibitem[{Evenson} \em{et~al.}(1983){Evenson}, {Garcia-Munoz}, {Meyer}, {Pyle},
  and {Simpson}]{1983ApJ...275L..15E}
{Evenson}, P.; {Garcia-Munoz}, M.; {Meyer}, P.; {Pyle}, K.R.; {Simpson}, J.A.
\newblock {A quantitative test of solar modulation theory - The proton, helium,
  and electron spectra from 1965 through 1979}.
\newblock {\em \apjl} {\bf 1983}, {\em 275},~L15--L18.
\newblock
  doi:{\changeurlcolor{black}\href{https://doi.org/10.1086/184162}{\detokenize{10.1086/184162}}}.

\bibitem[{Mewaldt} \em{et~al.}(1980){Mewaldt}, {Spalding}, {Stone}, and
  {Vogt}]{1980ApJ...236L.121M}
{Mewaldt}, R.A.; {Spalding}, J.D.; {Stone}, E.C.; {Vogt}, R.E.
\newblock {The isotopic composition of galactic cosmic-ray iron nuclei}.
\newblock {\em \apjl} {\bf 1980}, {\em 236},~L121--L125.
\newblock
  doi:{\changeurlcolor{black}\href{https://doi.org/10.1086/183211}{\detokenize{10.1086/183211}}}.

\bibitem[{Mewaldt} \em{et~al.}(1981){Mewaldt}, {Spalding}, {Stone}, and
  {Vogt}]{1981ApJ...251L..27M}
{Mewaldt}, R.A.; {Spalding}, J.D.; {Stone}, E.C.; {Vogt}, R.E.
\newblock {The isotropic composition of cosmic ray B, C, N, and O nuclei}.
\newblock {\em \apjl} {\bf 1981}, {\em 251},~L27--L31.
\newblock
  doi:{\changeurlcolor{black}\href{https://doi.org/10.1086/183686}{\detokenize{10.1086/183686}}}.

\bibitem[{Mewaldt}(1986)]{1986ApJ...311..979M}
{Mewaldt}, R.A.
\newblock {He-3 in galactic cosmic rays}.
\newblock {\em \apj} {\bf 1986}, {\em 311},~979--983.
\newblock
  doi:{\changeurlcolor{black}\href{https://doi.org/10.1086/164835}{\detokenize{10.1086/164835}}}.

\bibitem[{Mewaldt} \em{et~al.}(1980){Mewaldt}, {Spalding}, {Stone}, and
  {Vogt}]{1980ApJ...235L..95M}
{Mewaldt}, R.A.; {Spalding}, J.D.; {Stone}, E.C.; {Vogt}, R.E.
\newblock {High resolution measurements of galactic cosmic-ray neon, magnesium,
  and silicon isotopes}.
\newblock {\em \apjl} {\bf 1980}, {\em 235},~L95--L99.
\newblock
  doi:{\changeurlcolor{black}\href{https://doi.org/10.1086/183166}{\detokenize{10.1086/183166}}}.

\bibitem[{Kroeger}(1986)]{1986ApJ...303..816K}
{Kroeger}, R.
\newblock {Measurements of hydrogen and helium isotopes in Galactic cosmic rays
  from 1978 through 1984}.
\newblock {\em \apj} {\bf 1986}, {\em 303},~816--828.
\newblock
  doi:{\changeurlcolor{black}\href{https://doi.org/10.1086/164130}{\detokenize{10.1086/164130}}}.

\bibitem[{Wiedenbeck} and {Greiner}(1980)]{1980ApJ...239L.139W}
{Wiedenbeck}, M.E.; {Greiner}, D.E.
\newblock {A cosmic-ray age based on the abundance of Be-10}.
\newblock {\em \apjl} {\bf 1980}, {\em 239},~L139--L142.
\newblock
  doi:{\changeurlcolor{black}\href{https://doi.org/10.1086/183310}{\detokenize{10.1086/183310}}}.

\bibitem[{Wiedenbeck} and {Greiner}(1981{\natexlab{a}})]{1981ApJ...247L.119W}
{Wiedenbeck}, M.E.; {Greiner}, D.E.
\newblock {High-resolution observations of the isotopic composition of carbon
  and silicon in the galactic cosmic rays}.
\newblock {\em \apjl} {\bf 1981}, {\em 247},~L119--L122.
\newblock
  doi:{\changeurlcolor{black}\href{https://doi.org/10.1086/183603}{\detokenize{10.1086/183603}}}.

\bibitem[{Wiedenbeck} and {Greiner}(1981{\natexlab{b}})]{1981PhRvL..46..682W}
{Wiedenbeck}, M.E.; {Greiner}, D.E.
\newblock {Isotopic anomalies in the galactic cosmic-ray source}.
\newblock {\em Physical Review Letters} {\bf 1981}, {\em 46},~682--685.
\newblock
  doi:{\changeurlcolor{black}\href{https://doi.org/10.1103/PhysRevLett.46.682}{\detokenize{10.1103/PhysRevLett.46.682}}}.

\bibitem[{Wiedenbeck}(1983)]{1983ICRC....9..147W}
{Wiedenbeck}, M.E.
\newblock {The abundance of the radioactive isotope Al-26 in galactic cosmic
  rays}.
\newblock  International Cosmic Ray Conference,  1983, Vol.~9, {\em
  International Cosmic Ray Conference}, pp. 147--150.

\bibitem[{Krombel} and {Wiedenbeck}(1988)]{1988ApJ...328..940K}
{Krombel}, K.E.; {Wiedenbeck}, M.E.
\newblock {Isotopic composition of cosmic-ray boron and nitrogen}.
\newblock {\em \apj} {\bf 1988}, {\em 328},~940--953.
\newblock
  doi:{\changeurlcolor{black}\href{https://doi.org/10.1086/166348}{\detokenize{10.1086/166348}}}.

\bibitem[{Leske}(1993)]{1993ApJ...405..567L}
{Leske}, R.A.
\newblock {The elemental and isotopic composition of Galactic cosmic-ray nuclei
  from scandium through nickel}.
\newblock {\em \apj} {\bf 1993}, {\em 405},~567--583.
\newblock
  doi:{\changeurlcolor{black}\href{https://doi.org/10.1086/172388}{\detokenize{10.1086/172388}}}.

\bibitem[{Grebenyuk} \em{et~al.}(2019){Grebenyuk}, {Karmanov}, {Kovalev},
  {Kudryashov}, {Kurganov}, {Panov}, {Podorozhny}, {Tkachenko}, {Tkachev},
  {Turundaevskiy}, {Vasiliev}, and {Voronin}]{2019AdSpR..64.2559G}
{Grebenyuk}, V.; {Karmanov}, D.; {Kovalev}, I.; {Kudryashov}, I.; {Kurganov},
  A.; {Panov}, A.; {Podorozhny}, D.; {Tkachenko}, A.; {Tkachev}, L.;
  {Turundaevskiy}, A.; {Vasiliev}, O.; {Voronin}, A.
\newblock {Secondary cosmic rays in the NUCLEON space experiment}.
\newblock {\em Advances in Space Research} {\bf 2019}, {\em 64},~2559--2563,
  \href{http://xxx.lanl.gov/abs/1809.09665}{{\normalfont
  [arXiv:astro-ph.HE/1809.09665]}}.
\newblock
  doi:{\changeurlcolor{black}\href{https://doi.org/10.1016/j.asr.2019.06.030}{\detokenize{10.1016/j.asr.2019.06.030}}}.

\bibitem[{Comstock} \em{et~al.}(1969){Comstock}, {Fan}, and
  {Simpson}]{1969ApJ...155..609C}
{Comstock}, G.M.; {Fan}, C.Y.; {Simpson}, J.A.
\newblock {Energy Spectra and Abundances of the Cosmic-Ray Nuclei Helium to
  Iron from the Ogo-I Satellite Experiment}.
\newblock {\em \apj} {\bf 1969}, {\em 155},~609.
\newblock
  doi:{\changeurlcolor{black}\href{https://doi.org/10.1086/149895}{\detokenize{10.1086/149895}}}.

\bibitem[{Teegarden} \em{et~al.}(1970){Teegarden}, {McDonald}, and
  {Balasubrahmanyan}]{1970ICRC....1..345T}
{Teegarden}, B.J.; {McDonald}, F.B.; {Balasubrahmanyan}, V.K.
\newblock {Spectra and charge composition of the low energy galactic cosmic
  radiation from Z=2 to 14}.
\newblock  1970, Vol.~1, p. 345.

\bibitem[{Adriani} \em{et~al.}(2014){Adriani}, {Barbarino}, {Bazilevskaya},
  {Bellotti}, {Boezio}, {Bogomolov}, {Bongi}, {Bonvicini}, {Bottai}, {Bruno},
  and et~al]{2014ApJ...791...93A}
{Adriani}, O.; {Barbarino}, G.C.; {Bazilevskaya}, G.A.; {Bellotti}, R.;
  {Boezio}, M.; {Bogomolov}, E.A.; {Bongi}, M.; {Bonvicini}, V.; {Bottai}, S.;
  {Bruno}, A.; et~al.
\newblock {Measurement of Boron and Carbon Fluxes in Cosmic Rays with the
  PAMELA Experiment}.
\newblock {\em \apj} {\bf 2014}, {\em 791},~93,
  \href{http://xxx.lanl.gov/abs/1407.1657}{{\normalfont
  [arXiv:astro-ph.HE/1407.1657]}}.
\newblock
  doi:{\changeurlcolor{black}\href{https://doi.org/10.1088/0004-637X/791/2/93}{\detokenize{10.1088/0004-637X/791/2/93}}}.

\bibitem[{Adriani} \em{et~al.}(2011){Adriani}, {Barbarino}, {Bazilevskaya},
  {Bellotti}, {Boezio}, {Bogomolov}, {Bonechi}, {Bongi}, {Bonvicini},
  {Borisov}, and et~al]{2011Sci...332...69A}
{Adriani}, O.; {Barbarino}, G.C.; {Bazilevskaya}, G.A.; {Bellotti}, R.;
  {Boezio}, M.; {Bogomolov}, E.A.; {Bonechi}, L.; {Bongi}, M.; {Bonvicini}, V.;
  {Borisov}, S.; et~al.
\newblock {PAMELA Measurements of Cosmic-Ray Proton and Helium Spectra}.
\newblock {\em Science} {\bf 2011}, {\em 332},~69--,
  \href{http://xxx.lanl.gov/abs/1103.4055}{{\normalfont
  [arXiv:astro-ph.HE/1103.4055]}}.
\newblock
  doi:{\changeurlcolor{black}\href{https://doi.org/10.1126/science.1199172}{\detokenize{10.1126/science.1199172}}}.

\bibitem[{Adriani} \em{et~al.}(2016){Adriani}, {Barbarino}, {Bazilevskaya},
  {Bellotti}, {Boezio}, {Bogomolov}, {Bongi}, {Bonvicini}, {Bottai}, {Bruno},
  and et~al]{2016ApJ...818...68A}
{Adriani}, O.; {Barbarino}, G.C.; {Bazilevskaya}, G.A.; {Bellotti}, R.;
  {Boezio}, M.; {Bogomolov}, E.A.; {Bongi}, M.; {Bonvicini}, V.; {Bottai}, S.;
  {Bruno}, A.; et~al.
\newblock {Measurements of Cosmic-Ray Hydrogen and Helium Isotopes with the
  PAMELA Experiment}.
\newblock {\em \apj} {\bf 2016}, {\em 818},~68,
  \href{http://xxx.lanl.gov/abs/1512.06535}{{\normalfont
  [arXiv:astro-ph.HE/1512.06535]}}.
\newblock
  doi:{\changeurlcolor{black}\href{https://doi.org/10.3847/0004-637X/818/1/68}{\detokenize{10.3847/0004-637X/818/1/68}}}.

\bibitem[{Menn} \em{et~al.}(2018){Menn}, {Bogomolov}, {Simon}, {Vasilyev},
  {Adriani}, {Barbarino}, {Bazilevskaya}, {Bellotti}, {Boezio}, {Bongi}, and
  et~al]{2018ApJ...862..141M}
{Menn}, W.; {Bogomolov}, E.A.; {Simon}, M.; {Vasilyev}, G.; {Adriani}, O.;
  {Barbarino}, G.C.; {Bazilevskaya}, G.A.; {Bellotti}, R.; {Boezio}, M.;
  {Bongi}, M.; et~al.
\newblock {\em \apj} {\bf 2018}, {\em 862},~141,
  \href{http://xxx.lanl.gov/abs/1806.10948}{{\normalfont
  [arXiv:astro-ph.HE/1806.10948]}}.
\newblock
  doi:{\changeurlcolor{black}\href{https://doi.org/10.3847/1538-4357/aacf89}{\detokenize{10.3847/1538-4357/aacf89}}}.

\bibitem[{Adriani} \em{et~al.}(2013){Adriani}, {Barbarino}, {Bazilevskaya},
  {Bellotti}, {Boezio}, {Bogomolov}, {Bongi}, {Bonvicini}, {Borisov}, {Bottai},
  and et~al]{2013ApJ...770....2A}
{Adriani}, O.; {Barbarino}, G.C.; {Bazilevskaya}, G.A.; {Bellotti}, R.;
  {Boezio}, M.; {Bogomolov}, E.A.; {Bongi}, M.; {Bonvicini}, V.; {Borisov}, S.;
  {Bottai}, S.; et~al.
\newblock {Measurement of the Isotopic Composition of Hydrogen and Helium
  Nuclei in Cosmic Rays with the PAMELA Experiment}.
\newblock {\em \apj} {\bf 2013}, {\em 770},~2,
  \href{http://xxx.lanl.gov/abs/1304.5420}{{\normalfont
  [arXiv:astro-ph.HE/1304.5420]}}.
\newblock
  doi:{\changeurlcolor{black}\href{https://doi.org/10.1088/0004-637X/770/1/2}{\detokenize{10.1088/0004-637X/770/1/2}}}.

\bibitem[{Lezniak} and {Webber}(1971)]{1971JGR....76.1605L}
{Lezniak}, J.A.; {Webber}, W.R.
\newblock {Solar modulation of cosmic ray protons, helium nuclei, and
  electrons: A comparison of experiment with theory}.
\newblock {\em \jgr} {\bf 1971}, {\em 76},~1605.
\newblock
  doi:{\changeurlcolor{black}\href{https://doi.org/10.1029/JA076i007p01605}{\detokenize{10.1029/JA076i007p01605}}}.

\bibitem[{Webber} and {McDonald}(1994)]{1994ApJ...435..464W}
{Webber}, W.R.; {McDonald}, F.B.
\newblock {The cosmic-ray oxygen and helium spectra measured at Pioneer 10 over
  the time of the 1987 modulation minimum, and implications for the He/O source
  ratio}.
\newblock {\em \apj} {\bf 1994}, {\em 435},~464--468.
\newblock
  doi:{\changeurlcolor{black}\href{https://doi.org/10.1086/174828}{\detokenize{10.1086/174828}}}.

\bibitem[{Ivanenko} \em{et~al.}(1993){Ivanenko}, {Shestoperov}, {Chikova},
  {Fateeva}, {Khein}, {Podoroznyi}, {Rapoport}, {Samsonov}, {Sobinyakov},
  {Turundaevskyi}, and {Yashin}]{1993ICRC....2...17I}
{Ivanenko}, I.P.; {Shestoperov}, V.Y.; {Chikova}, L.O.; {Fateeva}, I.M.;
  {Khein}, L.A.; {Podoroznyi}, D.M.; {Rapoport}, I.D.; {Samsonov}, G.A.;
  {Sobinyakov}, V.A.; {Turundaevskyi}, A.N.; {Yashin}, I.V.
\newblock {Energy Spectra of Cosmic Rays above 2 TeV as Measured by the 'SOKOL'
  Apparatus}.
\newblock  International Cosmic Ray Conference,  1993, Vol.~2, {\em
  International Cosmic Ray Conference}, p.~17.

\bibitem[{Turundaevskiy} and {Podorozhnyi}(2017)]{2017AdSpR..59..496T}
{Turundaevskiy}, A.; {Podorozhnyi}, D.
\newblock {High energy deuterons in cosmic rays registered by the SOKOL
  satellite experiment}.
\newblock {\em Advances in Space Research} {\bf 2017}, {\em 59},~496--501.
\newblock
  doi:{\changeurlcolor{black}\href{https://doi.org/10.1016/j.asr.2016.08.028}{\detokenize{10.1016/j.asr.2016.08.028}}}.

\bibitem[{Westphal} \em{et~al.}(1996){Westphal}, {Afanasyev}, {Price},
  {Solarz}, {Akimov}, {Rodin}, and {Shvets}]{1996ApJ...468..679W}
{Westphal}, A.J.; {Afanasyev}, V.G.; {Price}, P.B.; {Solarz}, M.; {Akimov},
  V.V.; {Rodin}, V.G.; {Shvets}, N.I.
\newblock {Measurement of the Isotopic Composition of Manganese, Iron, and
  Nickel in the Galactic Cosmic Rays}.
\newblock {\em \apj} {\bf 1996}, {\em 468},~679.
\newblock
  doi:{\changeurlcolor{black}\href{https://doi.org/10.1086/177725}{\detokenize{10.1086/177725}}}.

\bibitem[{Duvernois} and {Thayer}(1996)]{1996ApJ...465..982D}
{Duvernois}, M.A.; {Thayer}, M.R.
\newblock {The Elemental Composition of the Galactic Cosmic-Ray Source: ULYSSES
  High-Energy Telescope Results}.
\newblock {\em \apj} {\bf 1996}, {\em 465},~982.
\newblock
  doi:{\changeurlcolor{black}\href{https://doi.org/10.1086/177483}{\detokenize{10.1086/177483}}}.

\bibitem[{Duvernois} \em{et~al.}(1996){Duvernois}, {Simpson}, and
  {Thayer}]{1996A&A...316..555D}
{Duvernois}, M.A.; {Simpson}, J.A.; {Thayer}, M.R.
\newblock {Interstellar propagation of cosmic rays: analysis of the ULYSSES
  primary and secondary elemental abundances.}
\newblock {\em \aap} {\bf 1996}, {\em 316},~555--563.

\bibitem[{Connell} and {Simpson}(1997)]{1997ApJ...475L..61C}
{Connell}, J.J.; {Simpson}, J.A.
\newblock {Isotopic Abundances of Fe and Ni in Galactic Cosmic-Ray Sources}.
\newblock {\em \apjl} {\bf 1997}, {\em 475},~L61.
\newblock
  doi:{\changeurlcolor{black}\href{https://doi.org/10.1086/310452}{\detokenize{10.1086/310452}}}.

\bibitem[{Duvernois}(1997)]{1997ApJ...481..241D}
{Duvernois}, M.A.
\newblock {Galactic Cosmic-Ray Manganese: ULYSSES High Energy Telescope
  Results}.
\newblock {\em \apj} {\bf 1997}, {\em 481},~241.
\newblock
  doi:{\changeurlcolor{black}\href{https://doi.org/10.1086/304032}{\detokenize{10.1086/304032}}}.

\bibitem[{Thayer}(1997)]{1997ApJ...482..792T}
{Thayer}, M.R.
\newblock {An Investigation into Sulfur Isotopes in the Galactic Cosmic Rays}.
\newblock {\em \apj} {\bf 1997}, {\em 482},~792.
\newblock
  doi:{\changeurlcolor{black}\href{https://doi.org/10.1086/304173}{\detokenize{10.1086/304173}}}.

\bibitem[{Simpson} and {Connell}(1998)]{1998ApJ...497L..85S}
{Simpson}, J.A.; {Connell}, J.J.
\newblock {Cosmic-Ray 26 A(l and Its Decay in the Galaxy}.
\newblock {\em \apjl} {\bf 1998}, {\em 497},~L85.
\newblock
  doi:{\changeurlcolor{black}\href{https://doi.org/10.1086/311290}{\detokenize{10.1086/311290}}}.

\bibitem[{Connell}(1997)]{1997ICRC....3..381C}
{Connell}, J.
\newblock {High Resolution Measurements of the Isotopic Composition of Galactic
  Cosmic Ray C, N, O, Ne, Mg and Si from the Ulysses HET}.
\newblock  1997, Vol.~3, p. 381.

\bibitem[{Connell}(1998)]{1998ApJ...501L..59C}
{Connell}, J.J.
\newblock {Galactic Cosmic-Ray Confinement Time: ULYSSES High Energy Telescope
  Measurements of the Secondary Radionuclide 10Be}.
\newblock {\em \apjl} {\bf 1998}, {\em 501},~L59.
\newblock
  doi:{\changeurlcolor{black}\href{https://doi.org/10.1086/311437}{\detokenize{10.1086/311437}}}.

\bibitem[{Connell} \em{et~al.}(1998){Connell}, {Duvernois}, and
  {Simpson}]{1998ApJ...509L..97C}
{Connell}, J.J.; {Duvernois}, M.A.; {Simpson}, J.A.
\newblock {The Cosmic-Ray Radioactive Nuclide \^{}36CL and Its Propagation in
  the Galaxy}.
\newblock {\em \apjl} {\bf 1998}, {\em 509},~L97--L100.
\newblock
  doi:{\changeurlcolor{black}\href{https://doi.org/10.1086/311773}{\detokenize{10.1086/311773}}}.

\bibitem[{Connell}(1999)]{1999ICRC....3...33C}
{Connell}, J.
\newblock {Ulysses HET Measurements of Electron-capture Secondary Isotopes:
  Testing the Role of Cosmic Ray Reacceleration}.
\newblock  International Cosmic Ray Conference,  1999, Vol.~3, {\em
  International Cosmic Ray Conference}, p.~33.

\bibitem[{Webber} and {Higbie}(2009)]{2009JGRA..11402103W}
{Webber}, W.R.; {Higbie}, P.R.
\newblock {Galactic propagation of cosmic ray nuclei in a model with an
  increasing diffusion coefficient at low rigidities: A comparison of the new
  interstellar spectra with Voyager data in the outer heliosphere}.
\newblock {\em Journal of Geophysical Research (Space Physics)} {\bf 2009},
  {\em 114},~2103.
\newblock
  doi:{\changeurlcolor{black}\href{https://doi.org/10.1029/2008JA013689}{\detokenize{10.1029/2008JA013689}}}.

\bibitem[{Lukasiak} \em{et~al.}(1994{\natexlab{a}}){Lukasiak}, {Ferrando},
  {McDonald}, and {Webber}]{1994ApJ...423..426L}
{Lukasiak}, A.; {Ferrando}, P.; {McDonald}, F.B.; {Webber}, W.R.
\newblock {The isotopic composition of cosmic-ray beryllium and its implication
  for the cosmic ray's age}.
\newblock {\em \apj} {\bf 1994}, {\em 423},~426--431.
\newblock
  doi:{\changeurlcolor{black}\href{https://doi.org/10.1086/173818}{\detokenize{10.1086/173818}}}.

\bibitem[{Lukasiak} \em{et~al.}(1994{\natexlab{b}}){Lukasiak}, {McDonald}, and
  {Webber}]{1994ApJ...430L..69L}
{Lukasiak}, A.; {McDonald}, F.B.; {Webber}, W.R.
\newblock {Voyager measurements of the isotopic composition of cosmic-ray
  aluminum and implications for the propagation of cosmic rays}.
\newblock {\em \apjl} {\bf 1994}, {\em 430},~L69--L72.
\newblock
  doi:{\changeurlcolor{black}\href{https://doi.org/10.1086/187440}{\detokenize{10.1086/187440}}}.

\bibitem[{Webber} \em{et~al.}(1996){Webber}, {Lukasiak}, {McDonald}, and
  {Ferrando}]{1996ApJ...457..435W}
{Webber}, W.R.; {Lukasiak}, A.; {McDonald}, F.B.; {Ferrando}, P.
\newblock {New High-Statistical--High-Resolution Measurements of the Cosmic-Ray
  CNO Isotopes from a 17 Year Study Using the Voyager 1 and 2 Spacecraft}.
\newblock {\em \apj} {\bf 1996}, {\em 457},~435.
\newblock
  doi:{\changeurlcolor{black}\href{https://doi.org/10.1086/176743}{\detokenize{10.1086/176743}}}.

\bibitem[{Lukasiak} \em{et~al.}(1997{\natexlab{a}}){Lukasiak}, {McDonald}, and
  {Webber}]{1997ApJ...488..454L}
{Lukasiak}, A.; {McDonald}, F.B.; {Webber}, W.R.
\newblock {Voyager Measurements of the Mass Composition of Cosmic-Ray Ca
  through Fe Nuclei}.
\newblock {\em \apj} {\bf 1997}, {\em 488},~454.
\newblock
  doi:{\changeurlcolor{black}\href{https://doi.org/10.1086/304677}{\detokenize{10.1086/304677}}}.

\bibitem[{Lukasiak} \em{et~al.}(1997{\natexlab{b}}){Lukasiak}, {McDonald}, and
  {Webber}]{1997ICRC....3..389L}
{Lukasiak}, A.; {McDonald}, F.B.; {Webber}, W.R.
\newblock {Voyager Measurements of the Isotopic Composition of Li, Be and B
  Nuclei}.
\newblock  International Cosmic Ray Conference,  1997, Vol.~3, {\em
  International Cosmic Ray Conference}, p. 389.

\bibitem[{Webber} \em{et~al.}(1997){Webber}, {Lukasiak}, and
  {McDonald}]{1997ApJ...476..766W}
{Webber}, W.R.; {Lukasiak}, A.; {McDonald}, F.B.
\newblock {Voyager Measurements of the Mass Composition of Cosmic-Ray Ne, Mg,
  Si, and S Nuclei}.
\newblock {\em \apj} {\bf 1997}, {\em 476},~766.
\newblock
  doi:{\changeurlcolor{black}\href{https://doi.org/10.1086/303662}{\detokenize{10.1086/303662}}}.

\bibitem[{Lukasiak}(1999)]{1999ICRC....3...41L}
{Lukasiak}, A.
\newblock {Voyager Measurements of the Charge and Isotopic Composition of
  Cosmic Ray Li, Be and B Nuclei and Implications for Their Production in the
  Galaxy}.
\newblock  International Cosmic Ray Conference,  1999, Vol.~3, {\em
  International Cosmic Ray Conference}, p.~41.

\bibitem[{Lukasiak} \em{et~al.}(1994){Lukasiak}, {Ferrando}, {McDonald}, and
  {Webber}]{1994ApJ...426..366L}
{Lukasiak}, A.; {Ferrando}, P.; {McDonald}, F.B.; {Webber}, W.R.
\newblock {Cosmic-ray isotopic composition of C, N, O, Ne, Mg, SI nuclei in the
  energy range 50-200 MeV per nucleon measured by the Voyager spacecraft during
  the solar minimum period}.
\newblock {\em \apj} {\bf 1994}, {\em 426},~366--372.
\newblock
  doi:{\changeurlcolor{black}\href{https://doi.org/10.1086/174072}{\detokenize{10.1086/174072}}}.

\bibitem[{Seo} and {McDonald}(1995)]{1995ApJ...451L..33S}
{Seo}, E.S.; {McDonald}, F.B.
\newblock {Cosmic-Ray H and He Isotopes in the Outer Heliosphere in 1994}.
\newblock {\em \apjl} {\bf 1995}, {\em 451},~L33.
\newblock
  doi:{\changeurlcolor{black}\href{https://doi.org/10.1086/309681}{\detokenize{10.1086/309681}}}.

\bibitem[{Ferrando} \em{et~al.}(1991){Ferrando}, {Lal}, {McDonald}, and
  {Webber}]{1991A&A...247..163F}
{Ferrando}, P.; {Lal}, N.; {McDonald}, F.B.; {Webber}, W.R.
\newblock {Studies of low-energy Galactic cosmic-ray composition at 22 AU. I -
  Secondary/primary ratios}.
\newblock {\em \aap} {\bf 1991}, {\em 247},~163--172.

\bibitem[{Seo} \em{et~al.}(1994){Seo}, {McDonald}, {Lal}, and
  {Webber}]{1994ApJ...432..656S}
{Seo}, E.S.; {McDonald}, F.B.; {Lal}, N.; {Webber}, W.R.
\newblock {Study of cosmic-ray H and He isotopes at 23 AU}.
\newblock {\em \apj} {\bf 1994}, {\em 432},~656--664.
\newblock
  doi:{\changeurlcolor{black}\href{https://doi.org/10.1086/174604}{\detokenize{10.1086/174604}}}.

\bibitem[{Wissel}(2010)]{2010PhDT........37W}
{Wissel}, S.A.
\newblock {Observations of direct Cerenkov light in ground-based telescopes and
  the flux of iron nuclei at TeV energies}.
\newblock PhD thesis, The University of Chicago,  2010.

\bibitem[{Archer} \em{et~al.}(2018){Archer}, {Benbow}, {Bird}, {Brose},
  {Buchovecky}, {Bugaev}, {Connolly}, {Cui}, {Daniel}, {Falcone}, and
  et~al]{2018PhRvD..98b2009A}
{Archer}, A.; {Benbow}, W.; {Bird}, R.; {Brose}, R.; {Buchovecky}, M.;
  {Bugaev}, V.; {Connolly}, M.P.; {Cui}, W.; {Daniel}, M.K.; {Falcone}, A.;
  et~al.
\newblock {Measurement of the iron spectrum in cosmic rays by VERITAS}.
\newblock {\em \prd} {\bf 2018}, {\em 98},~022009,
  \href{http://xxx.lanl.gov/abs/1807.08010}{{\normalfont
  [arXiv:astro-ph.HE/1807.08010]}}.
\newblock
  doi:{\changeurlcolor{black}\href{https://doi.org/10.1103/PhysRevD.98.022009}{\detokenize{10.1103/PhysRevD.98.022009}}}.

\bibitem[{Rauch} \em{et~al.}(2009){Rauch}, {Link}, {Lodders}, {Israel},
  {Barbier}, {Binns}, {Christian}, {Cummings}, {de Nolfo}, {Geier}, and
  et~al]{2009ApJ...697.2083R}
{Rauch}, B.F.; {Link}, J.T.; {Lodders}, K.; {Israel}, M.H.; {Barbier}, L.M.;
  {Binns}, W.R.; {Christian}, E.R.; {Cummings}, J.R.; {de Nolfo}, G.A.;
  {Geier}, S.; et~al.
\newblock {Cosmic Ray origin in OB Associations and Preferential Acceleration
  of Refractory Elements: Evidence from Abundances of Elements $_{26}$Fe
  through $_{34}$Se}.
\newblock {\em \apj} {\bf 2009}, {\em 697},~2083--2088,
  \href{http://xxx.lanl.gov/abs/0906.2021}{{\normalfont
  [arXiv:astro-ph.HE/0906.2021]}}.
\newblock
  doi:{\changeurlcolor{black}\href{https://doi.org/10.1088/0004-637X/697/2/2083}{\detokenize{10.1088/0004-637X/697/2/2083}}}.

\bibitem[{Binns} \em{et~al.}(1982){Binns}, {Israel}, {Klarmann}, {Fickle},
  {Waddington}, {Garrard}, and {Stone}]{1982ApJ...261L.117B}
{Binns}, W.R.; {Israel}, M.H.; {Klarmann}, J.; {Fickle}, R.K.; {Waddington},
  C.J.; {Garrard}, T.L.; {Stone}, E.C.
\newblock {The abundance of the actinides in the cosmic radiation as measured
  on HEAO 3}.
\newblock {\em \\apjl} {\bf 1982}, {\em 261},~L117--L120.
\newblock
  doi:{\changeurlcolor{black}\href{https://doi.org/10.1086/183899}{\detokenize{10.1086/183899}}}.

\bibitem[{Binns} \em{et~al.}(1983){Binns}, {Israel}, {Klarmann}, {Fickle},
  {Waddington}, {Garrard}, {Krombel}, and {Stone}]{1983ApJ...267L..93B}
{Binns}, W.R.; {Israel}, M.H.; {Klarmann}, J.; {Fickle}, R.K.; {Waddington},
  C.J.; {Garrard}, T.L.; {Krombel}, K.E.; {Stone}, E.C.
\newblock {Cosmic-ray abundances of Sn, Te, Xe, and BA nuclei measured on HEAO
  3}.
\newblock {\em \\apjl} {\bf 1983}, {\em 267},~L93--L96.
\newblock
  doi:{\changeurlcolor{black}\href{https://doi.org/10.1086/184010}{\detokenize{10.1086/184010}}}.

\bibitem[{Binns} \em{et~al.}(1985){Binns}, {Israel}, {Brewster}, {Fixsen}, and
  {Garrard}]{1985ApJ...297..111B}
{Binns}, W.R.; {Israel}, M.H.; {Brewster}, N.R.; {Fixsen}, D.J.; {Garrard},
  T.L.
\newblock {Lead, platinum, and other heavy elements in the primary cosmic
  radiation - HEAO 3 results}.
\newblock {\em \\apj} {\bf 1985}, {\em 297},~111--118.
\newblock
  doi:{\changeurlcolor{black}\href{https://doi.org/10.1086/163508}{\detokenize{10.1086/163508}}}.

\end{thebibliography}
\end{document}